\documentclass[preprint]{aastex61}
\pdfoutput=1

\usepackage{amsmath}
\usepackage{amsfonts}
\usepackage{dsfont}
\usepackage{amsxtra}
\usepackage{hyperref}
\usepackage{amssymb}
\usepackage{upgreek}

\usepackage{multirow}
\usepackage{url}

\usepackage{graphicx,epsfig}

\usepackage{xspace} 

\usepackage{color}
\usepackage{xcolor}

\usepackage{lineno}
\renewcommand{\baselinestretch}{1.2}

\newcommand{\be}{\begin{equation}}
\newcommand{\ee}{\end{equation}}
\newcommand{\bea}{\begin{eqnarray}}
\newcommand{\eea}{\end{eqnarray}}
\newcommand{\beaa}{\begin{eqnarray*}}
\newcommand{\eeaa}{\end{eqnarray*}}
\newcommand{\ba}{\begin{array}}
\newcommand{\ea}{\end{array}}
\newcommand{\bi}{\begin{itemize}}
\newcommand{\ei}{\end{itemize}}
\newcommand{\ben}{\begin{enumerate}}
\newcommand{\een}{\end{enumerate}}

\newcommand{\bra}{\langle}
\newcommand{\ket}{\rangle}

\newcommand{\lb}{\label}
\newcommand{\g}{\ensuremath{\gamma}\xspace}

\newcommand{\al}{\alpha}

\newcommand{\dl}{\delta}

\newcommand{\vp}{\varphi}

\newcommand{\sm}{\sigma}

\newcommand{\La}{{\mathcal{L}}}

\newcommand{\HI}{H~\textsc{i}\xspace}
\newcommand{\Htwo}{$\mathrm{H}_2$\xspace}
\newcommand{\hi}{$\mathrm{H\,\scriptstyle{I}}$\xspace}

\newcommand{\hd}{$\mathrm{H}_2$\xspace}
\newcommand{\xco}{$X_\mathrm{CO}$\xspace}

\newcommand{\Fermi}{\textit{Fermi}\xspace}

\newcommand{\Planck}{\textsl{Planck}\xspace}

\newcommand{\SM}{Sample Model\xspace}

\newcommand{\sigmav}{\ensuremath{\langle \sigma v \rangle}\xspace}
\newcommand{\bbbar}{\ensuremath{b \bar b}\xspace}
\newcommand{\tautau}{\ensuremath{\tau^{+}\tau^{-}}\xspace}

\newcommand{\beff}{\ensuremath{b_{\rm eff}}\xspace}
\newcommand{\DM}{\ensuremath{\mathrm{DM}}}
\newcommand{\mDM}{\ensuremath{m_\DM}\xspace}

\newcommand{\onepic}{0.45}
\newcommand{\twopic}{0.38}

\newcommand{\threepic}{0.25}

\definecolor{darkgreen}{rgb}{0.0, 0.7, 0.0}

\shorttitle{\Fermi Galactic center GeV excess}

\begin{document}

\title{The \Fermi Galactic Center GeV Excess and Implications for Dark Matter}
 
\author{M.~Ackermann}
\affiliation{Deutsches Elektronen Synchrotron DESY, D-15738 Zeuthen, Germany}
\author{M.~Ajello}
\affiliation{Department of Physics and Astronomy, Clemson University, Kinard Lab of Physics, Clemson, SC 29634-0978, USA}
\author{A.~Albert}
\affiliation{Los Alamos National Laboratory, Los Alamos, NM 87545, USA}
\author{W.~B.~Atwood}
\affiliation{Santa Cruz Institute for Particle Physics, Department of Physics and Department of Astronomy and Astrophysics, University of California at Santa Cruz, Santa Cruz, CA 95064, USA}
\author{L.~Baldini}
\affiliation{Universit\`a di Pisa and Istituto Nazionale di Fisica Nucleare, Sezione di Pisa I-56127 Pisa, Italy}
\author{J.~Ballet}
\affiliation{Laboratoire AIM, CEA-IRFU/CNRS/Universit\'e Paris Diderot, Service d'Astrophysique, CEA Saclay, F-91191 Gif sur Yvette, France}
\author{G.~Barbiellini}
\affiliation{Istituto Nazionale di Fisica Nucleare, Sezione di Trieste, I-34127 Trieste, Italy}
\affiliation{Dipartimento di Fisica, Universit\`a di Trieste, I-34127 Trieste, Italy}
\author{D.~Bastieri}
\affiliation{Istituto Nazionale di Fisica Nucleare, Sezione di Padova, I-35131 Padova, Italy}
\affiliation{Dipartimento di Fisica e Astronomia ``G. Galilei'', Universit\`a di Padova, I-35131 Padova, Italy}
\author{R.~Bellazzini}
\affiliation{Istituto Nazionale di Fisica Nucleare, Sezione di Pisa, I-56127 Pisa, Italy}
\author{E.~Bissaldi}
\affiliation{Istituto Nazionale di Fisica Nucleare, Sezione di Bari, I-70126 Bari, Italy}
% ...

\collaboration{(The \textit{Fermi}-LAT collaboration)}
\noaffiliation

\correspondingauthor{A.~Albert}
\email{aalbert@slac.stanford.edu}
\correspondingauthor{E.~Charles}
\email{echarles@slac.stanford.edu}
\correspondingauthor{A.~Franckowiak}
\email{afrancko@slac.stanford.edu}
\correspondingauthor{D.~Malyshev}
\email{dvmalyshev@gmail.com}
\correspondingauthor{L.~Tibaldo}
\email{ltibaldo@slac.stanford.edu}

\begin{abstract}
The region around the Galactic center (GC) is now well established to be brighter at energies of a few GeV than expected from conventional models of diffuse gamma-ray emission
and catalogs of known gamma-ray sources.
We study the GeV excess using 6.5 years of data from the \Fermi Large Area Telescope. 
We characterize the uncertainty of the GC excess spectrum and morphology due to
uncertainties in cosmic-ray source distributions and propagation, uncertainties in the distribution of interstellar gas in the Milky Way, and uncertainties due to a potential contribution from the \Fermi bubbles.  We also evaluate uncertainties in the excess properties due to resolved point sources of gamma rays. The Galactic center is of particular interest as it would be expected to have the brightest signal from annihilation of weakly interacting massive dark matter particles.
However, control regions along the Galactic plane, where a dark-matter signal is not expected, show excesses of similar amplitude relative to the local background. 
Based on the magnitude of the systematic uncertainties,
we conservatively report upper limits for the annihilation cross section as function of particle mass and annihilation channel.

\end{abstract}

\keywords{
astroparticle physics --- 
dark matter ---
Galaxy: center ---
gamma rays: diffuse background
}

\tableofcontents

\section{Introduction}

The region around the Galactic center  (GC) is one of the richest in the gamma-ray sky. Gamma-ray emission in this direction includes the products of interactions between cosmic rays (CRs) with interstellar gas (from nucleon-nucleon inelastic collisions and electron/positron bremsstrahlung) and radiation fields (from inverse Compton scattering of electrons and positrons), as well as many individual sources such as pulsars, binary systems, and supernova remnants.
Some of the most compelling theories advocate dark matter (DM) to consist of  weakly interacting massive particles (WIMPs) that can self-annihilate to produce gamma rays in the final states \citep[for a review see, e.g.,][]{2005PhR...405..279B, 2012AnP...524..479B}.
A hypothetical signal in gamma~rays from WIMP annihilation  is expected to be brightest towards the Galactic center 
\citep[e.g.,][]{2008Natur.456...73S,2009Sci...325..970K}.

Based on data from the Large Area Telescope (LAT) onboard the \textit{Fermi Gamma-ray Space Telescope} \citep{2009ApJ...697.1071A}, several groups reported the detection of excess emission at energies of a few GeV near the GC on top of a variety of models for interstellar gamma-ray emission and point sources \citep[e.g.,][]{Goodenough:2009gk,2009arXiv0912.3828V,Hooper:2010mq,2012PhRvD..86h3511A,2013PDU.....2..118H,Gordon:2013vta,Daylan:2014rsa,Calore:2014xka,2015PhRvD..91l3010Z,2016ApJ...819...44A},
while other groups refuted the existence of an excess after considering the uncertainties in the modeling of sources and interstellar emission 
\citep[e.g.,][]{2011PhLB..705..165B}.
Some studies claim that the excess appears to have a spherical morphology centered at the GC and spectral characteristics consistent with DM annihilation \citep[e.g.,][]{Daylan:2014rsa, 2016JCAP...04..030H}, while
\cite{2016arXiv161008926D} find that the GC excess is correlated with the distribution of molecular clouds. 
\cite{2016A&A...589A.117Y} and \cite{2016arXiv161106644M} have argued that 
the morphology of the excess has a bi-lobed structure, which is expected for a continuation of the \Fermi bubbles.
\citet{Calore:2014xka} investigated uncertainties of foreground/background models using both a model-driven and a data-driven approach, and concluded that an excess is present, but uncertainties due to interstellar emission modeling are too large to conclusively prove a DM origin. 
\citet{2016ApJ...819...44A} studied in detail the different components of gamma-ray emission toward the GC, and confirmed that a residual component is consistently found above interstellar emission and sources, and that its spectrum is highly dependent on the choice of interstellar emission model.  More recently, some groups advocated that data favor an origin of the excess from a population of yet undetected gamma-ray sources such as millisecond pulsars \citep[MSPs,][]{2016PhRvL.116e1102B, 2016PhRvL.116e1103L, 2015ApJ...812...15B}, although MSPs may be insufficient to explain the excess 
if aging is taken into account \citep{2015JCAP...02..023P, 2016JCAP...08..018H}.
Additional sources of cosmic rays near the GC may also significantly affect the properties of the excess
\citep{2015JCAP...12..005C, 2015JCAP...12..056G, 2015arXiv151004698C}.

This paper revisits the \Fermi GC GeV excess using data from 6.5 years of observations. The new Pass~8 event-level analysis \citep{2013arXiv1303.3514A} provides improved direction and energy reconstruction, better background rejection, a wider energy range, and significantly increased effective area especially at the high- and low-energy ends. In addition, information provided by the Pass~8 dataset enables the user to subselect events based on the quality of their energy or direction reconstruction (i.e., based on energy dispersion and point-spread function, PSF).

The two main goals of this work are to study the range of aspects in the modeling of interstellar emission and discrete sources  in the vicinity of the GC that can possibly affect the characterization of the GC GeV excess, and to explore the implications for DM accounting at the best for these sources of uncertainty.
The paper is organized as follows.

In Section \ref{sec:data_baseline}, we describe the dataset used and the generalities of the analysis procedure.  A sample interstellar emission model is constructed and fit to the data along with a list of sources to test the presence of an excess at the GC and derive its spectrum.

Sections \ref{sec:analysis_setup}--\ref{sec:PS} are dedicated to exploring the impact on the determination of the properties of the GC excess of several aspects of the foreground/background emission models and analysis features, with emphasis on the spectrum of the excess. Section~\ref{sec:analysis_setup} considers the choice of dataset and region of the sky analyzed. Section \ref{sec:CRmodel} focuses on modeling choices related to interstellar emission, including distribution of CR sources and targets, and assumptions about CR transport in the Milky Way. In Section \ref{sec:SCA} we use a spectral component analysis (SCA) technique \citep{2012arXiv1202.1034M, 2014ApJ...793...64A}
to derive spatial templates for some gamma-ray emission components in the vicinity of the GC, i.e., the \textit{Fermi} bubbles \citep{2010ApJ...724.1044S} and the GC excess itself. In Section \ref{sec:PS}, we study the impact on the GC excess of different choices concerning the modeling of individual gamma-ray sources.

Section~\ref{sec:spectrum} summarizes our results concerning the spectrum of the GC excess. Section~\ref{sec:morphology} focuses on the morphology of the GC excess in the light of previously discussed sources of uncertainty. We consider some key properties of the excess morphology, contrasting the hypotheses of spherical and bipolar morphology, and studying the radial steepness and the excess centroid location.

The conclusion from this first part of the analysis is that the excess emission remains significant around a few GeV
in all the model variations that we have tested. 
However, it is practically impossible to consider an exhaustive set of models that encompass all the uncertainties in foreground/background emission. 
Therefore, in Section \ref{sec:DMlimits}, we consider an empirical approach to test the robustness of a DM interpretation of the excess. Control regions along the Galactic plane, where no DM signal is expected, are used for this purpose. Combining the results with those previously obtained from varying modeling and analysis features, we set constraints on DM annihilation in the GC from a variety of candidates.
Our conclusions are summarized in Section \ref{sec:concl}.

In Appendix \ref{app:SLext} we provide details of the derivation of alternative distribution of gas along the line of sight using
starlight extinction by dust.
We calculate the ratio of the GC excess signal in the Sample Model to statistical and systematic uncertainty maps in
Appendix \ref{app:unc_maps},
while in Appendix \ref{app:beff} we give details of the derivation of the DM limits.

\section{Data Selection, Analysis Methodology, and Sample Model Fit to the Data}
\lb{sec:data_baseline}

\subsection{Data Selection}
\lb{sec:data}

The analysis is based on 6.5 years of {\Fermi}-LAT data recorded between 2008 August 4 and 2015 January 31 ({\Fermi} Mission Elapsed Time 239557418\,s--444441067\,s). 
We select the standard good-time intervals, e.g., excluding calibration runs.
In order not to be biased by residual backgrounds in all-sky or large-scale analysis, we select events belonging to the Pass~8 UltraCleanVeto class, that provides the highest purity against contamination from charged particles misclassified as gamma rays.
Additionally, to minimize the contamination from emission from the Earth atmosphere, we select events with an angle $\theta < 90^{\circ}$ with respect to the local zenith.

We use events with measured energies between 100 MeV and 1 TeV in 27 logarithmic energy bins, which is about seven bins per decade.
For each energy bin events are binned spatially using HEALPix\footnote{\url{http://sourceforge.net/projects/healpix/}} \citep{2005ApJ...622..759G} with a pixelization of order 6 ($\approx 0\degr\!\!.92$ pixel size) or order 7 ($\approx 0\degr\!\!.46$ pixel size). 
For all-sky fitting we use maps with adaptive resolution that have order 7 pixels in areas with large statistics 
(near the Galactic plane or close to bright sources) and order 6 in areas with fewer counts.
The pixelization is determined based on the count map between 1.1 and 1.6 GeV at order 6, for which we  further subdivide pixels with more than 100 photons to order 7.
In the resulting maps we also mask 200 point sources with largest flux above 1 GeV from the  
Fermi LAT third source catalog~\citep[3FGL,][]{2015ApJS..218...23A}
within a radius of $1\degr$
(we mask pixels with centers within $1^\circ + a \sin(\pi/4)$ from the position of point sources (PS), where $a \approx 0\degr\!\!.46$ is the size of pixels at order 7).
Masking the bright PS effectively excludes pixels within about $2^\circ$ from the GC 
(the fraction of masked pixels within $4^\circ$ is about 50\%, within $10^\circ$ it is about 20\%).
We test the effect of the PS mask on the GC excess in Section \ref{sec:PS_refitting}, where we refit PS within $10^\circ$ from the GC
and only mask 200 brightest PS outside of $10^\circ$.
The order 6 pixels at high latitudes are masked if any of the underlying order 7 pixels are masked.

We calculate the exposure and PSF using the standard {\Fermi} LAT Science Tools package version 
10-01-01 available from the {\Fermi} Science Support Center\footnote{\url{http://fermi.gsfc.nasa.gov/ssc/data/analysis/}} 
using the P8R2\_ULTRACLEANVETO\_V6 instrument response functions.

\subsection{Sample Model}
\label{sec:baseline}

The emission measured by the LAT in any direction on the sky can be separated into individually detected sources, most of which are point-like sources, and diffuse emission.
The majority of diffuse gamma-ray emission at GeV energies arises from inelastic hadronic collisions, mostly through the decay of neutral pions ($\pi^0$). This component is produced in interactions of CR nuclei with interstellar gas, therefore it is spatially correlated with the distribution of gas in the Milky Way.
Another interstellar emission component, that becomes dominant at the highest and lowest gamma-ray energies,  is due to inverse Compton (IC) scattering of leptonic CRs (electrons and positrons) interacting with the low-energy interstellar radiation field (ISRF). The ISRF can be considered to consist of three components: starlight, infrared light emitted by dust, and the cosmic microwave background (CMB) radiation. The IC contribution is expected to be less structured compared to the hadronic component. At energies $\lesssim 10$ GeV, 
bremsstrahlung emission from electrons and positrons interacting with interstellar gas can become important. 
All of these three components of Galactic interstellar emission are brighter in the direction of the Galactic disk.
Additionally, there is a diffuse emission component with approximately isotropic intensity over the sky. It is made of residual contamination from interactions of charged particles in the LAT misclassified as gamma rays, unresolved (i.e., not detected individually) extragalactic sources, and, possibly, truly diffuse extragalactic gamma-ray emission.  

Throughout the paper, we will model Galactic interstellar emission from the large-scale CR populations in the Galaxy starting from predictions  obtained through the GALPROP code\footnote{\url{http://galprop.stanford.edu}} 
\citep{Moskalenko:1997gh, Strong:1998fr, Strong:2004de, Ptuskin:2005ax, 2007ARNPS..57..285S, Porter:2008ve,Vladimirov:2010aq}. We use the GALPROP package v54.1 unless mentioned otherwise. 
GALPROP calculates the propagation and interactions of CRs in the Galaxy by numerically solving the transport equations given a model for the CR source distribution, a
CR injection spectrum, and a model of targets for CR interactions. Parameters of the model are constrained to reproduce various CR observables, including CR secondary abundances and spectra obtained from direct measurements in the solar system, and diffuse gamma-ray and synchrotron emission. GALPROP is used to generate spatial templates for the gamma-ray emission produced in CR interactions with interstellar gas and radiation fields, that are then fitted to the data as described below. 

The GALPROP models we employ in this paper assume CR diffusion with a Kolmogorov spectrum of interstellar turbulence plus reacceleration, and no convection. The diffusion coefficient is assumed to be constant and isotropic in the Galaxy. Additionally, unless otherwise mentioned, calculations assume azimuthal symmetry of the CR density with respect to the GC. For the Sample Model described in this section we chose one of the models from \cite{FermiLAT:2012aa}. It assumes a CR source distribution traced by the measured 
distribution of pulsars \citep[][from now on referred to as Lorimer]{Lorimer:2006qs},
the CR confinement volume has a height of 10 kpc and a radius of 20 kpc. 
It should be stressed that the parameters selected for the Sample Model represent only one of the possible choices.
The goal of our analysis is not to find the best model but, rather, to estimate the uncertainty in the GC excess due to the choice 
of parameters and analysis procedure.
A study of the dependence of the results on propagation parameters and on the distribution of CR sources is presented in Section \ref{sec:CRmodel}. 
Also, note that the Sample Model, as most of the models considered in the paper, is derived from solving the transport equation in two dimensions 
(Galactocentric radius, height over the Galactic plane). This speeds up the derivation of the model, but it is worth noting that cylindrical coordinates have a coordinate singularity at the GC. In our case, the modeling of interstellar emission is meant mainly for estimating the foreground/background emission (which dominates over emission from the region near the GC) and this is not a source of concern. In Section~\ref{sec:GC_CR_sources} we will employ some three-dimensional GALPROP models to address the case of CR sources near the GC.

The distribution of target gas is based on multiwavelength surveys. For the Sample Model described in this section, the calculation of gamma-ray fluxes from CR interactions employs the LAB survey~\citep{2005A&A...440..775K} of the 21-cm line of H~\textsc{i} and the survey of 2.6-mm line of CO (a tracer of H$_2$) by~\citet{2001ApJ...547..792D} to evaluate the distribution of atomic and molecular gas, respectively, in Galactocentric annuli.
The partitioning of the interstellar gas into Galactocentric annuli based on the Doppler shifts of the lines is particularly uncertain at longitudes within about 10$\degr$ of the Galactic center, for latitudes within a few degrees of the Galactic equator.  This is because the velocity from circular motion is almost perpendicular to the line of sight, therefore Doppler shifts of the H~\textsc{i} and CO lines are small relative to random and streaming motions of the interstellar medium in this range.
The Sample Model taken from  \cite{FermiLAT:2012aa} assumes H~\textsc{i} column densities derived from the 21-cm line intensities for a spin temperature of 150\,K. 
The dust reddening map of \citet{Schlegel:1997yv} is used to correct the H~\textsc{i} maps to account for the presence of dark neutral gas not traced by the combination of H~\textsc{i} and CO surveys \citep{2005Sci...307.1292G,FermiLAT:2012aa}. We neglect the contribution from ionized gas.  
The impact of the choice of input data for modeling the interstellar gas distribution is addressed in Section~\ref{sec:gasunc}.

Since the distribution of CR densities in the Galaxy is not well constrained a priori, we fit the templates for the emission from interstellar gas split in Galactocentric annuli to the gamma-ray data, independently for each energy bin, using the procedure described later.
In the fit we use five independent annuli: three inner annuli spanning Galactocentric radii 
0--1.5\,kpc, 1.5--3.5\,kpc, and 3.5--8\,kpc; a local ring 8--10\,kpc, and an outer ring 10--50\,kpc. Also, H~\textsc{i} and CO maps are fitted independently to the data so that assuming an a-priori CO-to-H$_2$ ratio is not required.
In each ring, we add the bremsstrahlung and the hadronic components together in one template.

Furthermore, we separately fit to the data templates for the three IC components from CMB, dust infrared emission, and starlight as models for the last two have significant uncertainties.
Note that since we fit the model maps to the data in several independent energy bins, and the morphology of gamma-ray emission from gas is determined mainly by the distribution of interstellar gas, the CR transport modeling in GALPROP in our case affects the analysis mainly through the morphology predicted for IC emission.

The LAT data revealed the presence of diffuse emission components extending over large fractions of the sky, that are not represented in the GALPROP templates that we use to model gamma-ray emission resulting from CR interactions in the Galaxy.
Loop~I is a giant radio loop spanning $100^{\circ}$ on the 
sky~\citep{1962MNRAS.124..405L}, which was also detected in LAT data~\citep{Casandjian:2009wq}. The origin of Loop~I is an open question. 
It may be a local object, produced either by a nearby supernova explosion or by the wind activity of the Scorpio-Centaurus OB association at a distance of 170 pc~\citep{Wolleben:2007pq, 2015ApJ...811...40S}. Alternatively, it may be interpreted as the result of a large-scale outflow from the Galactic center~\citep{2013ApJ...779...57K}. In the Sample Model we account for Loop~I using a geometric model \citep[e.g., Figure 2 of][]{2014ApJ...793...64A}
based on a polarization survey at 1.4 GHz~\citep{Wolleben:2007pq}. The geometric Loop~I model assumes synchrotron emission from two shells. 
Each shell is described by 5 parameters: the center coordinates $\ell$, $b$; the distance to the center $d$;  
and the inner ($r_{in}$) and outer ($r_{out}$) radius of the shell. 
The parameters are set to: $\ell_{1} = 341^{\circ}$, $b_{1} = 3^{\circ}$, $d_{1} = 78$\;pc, $r_{in,1} = 62$\;pc, $r_{out,1} = 81$\;pc, $\ell_{2} = 332^{\circ}$, $b_{2} = 37^{\circ}$, $d_{2}=95$\;pc, $r_{in,2} = 58$\;pc, $r_{out,2} = 82$\;pc. 

An additional large-scale extended emission component is represented by the so-called \Fermi bubbles, two large gamma-ray lobes seen above and below the Galactic center~\citep{2010ApJ...724.1044S,2014ApJ...793...64A}. While the \Fermi bubbles are well studied at high latitudes, a careful characterization of their properties close to the Galactic plane is complicated by large systematic uncertainties introduced by the modeling of other bright components of Galactic interstellar emission in this region. In the Sample Model we include a flat intensity spatial model of the \Fermi bubbles 
at $|b| > 10^\circ$~\citep[Figure 5 of][]{2014ApJ...793...64A}.
In Section~\ref{sec:SCA} we will derive an alternative template of the Fermi bubbles that includes emission at low latitudes. 

We model the emission from the Sun \citep{2007Ap&SS.309..359O, 2006ApJ...652L..65M, 2008A&A...480..847O}
and Moon, which is trailed along the ecliptic in our long data set, using
templates derived with the \Fermi Science Tools\footnote{\url{http://fermi.gsfc.nasa.gov/ssc/data/analysis/scitools/solar_template.html}} following the description in~\citet{2013arXiv1307.0197J}.

For the Sample Model, we use the 3FGL catalog~\citep{2015ApJS..218...23A}.
We add all point sources in a single template in each energy bin.
To construct the template, we use the parameterized spectra from the catalog and convolve with the PSF in each energy bin.
Extended sources, except for the Large Magellanic cloud (LMC) and Cygnus region, are assembled in a separate template.
Since the LMC and Cygnus are the brightest extended sources, we have independent templates.
We describe extended sources based on the templates used in 3FGL. 
Unresolved Galactic sources, which may amount to up to $\sim$10\% of the Galactic diffuse component \citep{2015ApJS..218...23A} are not explicitly accounted for, but their spatial distribution is assumed to be similar to other Galactic components, and so the corresponding emission will be taken up by the other components in the fit (such as $\pi^0$, bremsstrahlung, and IC).

We tentatively 
include in the Sample Model an additional component with the spatial distribution expected from annihilation of DM that follows in the inner Galaxy a generalized Navarro-Frenk-White (gNFW) profile with index $\gamma = 1.25$ and scaling radius $r_{\rm s} = 20\;{\rm kpc}$.
The NFW profile is an approximation to the equilibrium configuration of DM produced in simulations of collisionless particles \citep{1997ApJ...490..493N}. Modified NFW profiles with $\gamma > 1$ are expected from numerical simulations of DM halos including baryons \citep[e.g.,][]{2011ApJ...742...76G}. However, in our Sample Model the modified NFW profile is mainly motivated by earlier analyses of the GeV excess \citep{Goodenough:2009gk,Calore:2014xka,Abazajian:2014fta}, that found it to be a good representation for the residual emission, with values of $\gamma$ varying around $\sim$1.25. We will consider different $\gamma$ values in Section~\ref{sec:NFWindex}.

The components of the \SM\ are summarized in Table~\ref{tab:SM}.
\begin{table}
\footnotesize
\centering
\begin{tabular}[t]{ll}
\hline
Component & Definition\\
\hline
\hline
Hadronic interactions and bremsstrahlung & GALPROP, 5 rings\\
\hline
Inverse Compton scattering & GALPROP, 3 components (CMB, starlight, infrared) \\
\hline
Loop I & Geometric template based on radio data \citep{Wolleben:2007pq} \\
\hline
\Fermi bubbles & Flat template from \cite{2014ApJ...793...64A} \\
\hline
Point sources & Template derived from 3FGL catalog \\
\hline
Extended sources, Cygnus, LMC & Templates derived from 3FGL catalog \\
\hline
Isotropic emission & Proportional to \Fermi-LAT exposure \\
\hline
Sun and Moon templates & Derived with \Fermi LAT Science Tools \\
\hline
GC excess & gNFW annihilation template with $\gamma = 1.25$ \\
\hline
\end{tabular}
\caption{\small
\label{tab:SM}
Components of the \SM. 
Input to GALPROP: pulsars as a tracer of CR production \citep{Lorimer:2006qs}; $z = 10$ kpc, $R = 20$ kpc propagation halo;
H~\textsc{i}  spin temperature 150 K (see text for details on the choice of the input parameters).
}
\end{table}

\subsection{Fitting Procedure}
\lb{sec:data_fitting}

We simultaneously fit the different components of diffuse emission (including an isotropic component) and a combined map of point sources to the {\Fermi}-LAT maps independently in each energy bin by maximizing the likelihood function based on Poisson statistics
\be
\lb{eq:chi2}
\log \La =  \sum_i   \left(d_{i} \log \mu_{i} - \mu_{i}  - \log (d_i!) \right)
\ee
where $d_{i}$ represents the photon counts in the spatial pixel with index $i$ , 
$\mu_{i}$ the model counts in the same bin (since the fit is performed in each energy bin independently, we omit the energy bin index in this and the following equations).
The model is constructed as a linear combination of templates
\be
\mu_{i} = \sum_m f_{m} P^{(m)}_{i},
\ee
where $m$ labels the components of emission, $P^{(m)}_{i}$ is the spatial template of component $m$ in the appropriate energy bin corrected for exposure and convolved with the LAT PSF.
The coefficients $f_{m}$ are adjusted to maximize the likelihood.

Occasionally the best solution has negative normalization coefficients associated with some of the templates. Since this is unphysical, in such case the corresponding template is removed and the fitting procedure is repeated. From comparing residual maps to the templates 
it seems most likely that this behavior is due to incompleteness or imperfections of the models. The normalization of some of the templates are either over- or underestimated to compensate for such defects, and other templates' normalization in turn may react to this. 

Our fitting strategy differs from previous examples in the literature \citep[e.g.,][]{Calore:2014xka,2016ApJ...819...44A}. We summarize below the main distinctive features:
\begin{itemize}
\item we fit the various emission templates independently in fine energy bins; this enables us to mitigate the impact on the results of several assumptions related to modeling background/foreground emission; we reiterate that the spectra of the various components, e.g., Figure~\ref{fig:baseline_spectra}, show that our procedure results in stable and physically plausible spectra;
\item we perform an all-sky fit of all the component templates simultaneously; this provides us with a fast and simple procedure, that can also be consistently applied to other regions in the sky as a control sample to assess systematic uncertainties related to the GC results (Section~\ref{sec:DMlimits}). We characterize the impact of the analysis region choice on the excess spectrum in Section~\ref{sec:ROI}.
\end{itemize}

\subsection{Results from the Analysis with the Sample Model}
\lb{sec:sample_model}

\begin{figure}[htbp]
\begin{center}
\includegraphics[scale=\twopic]{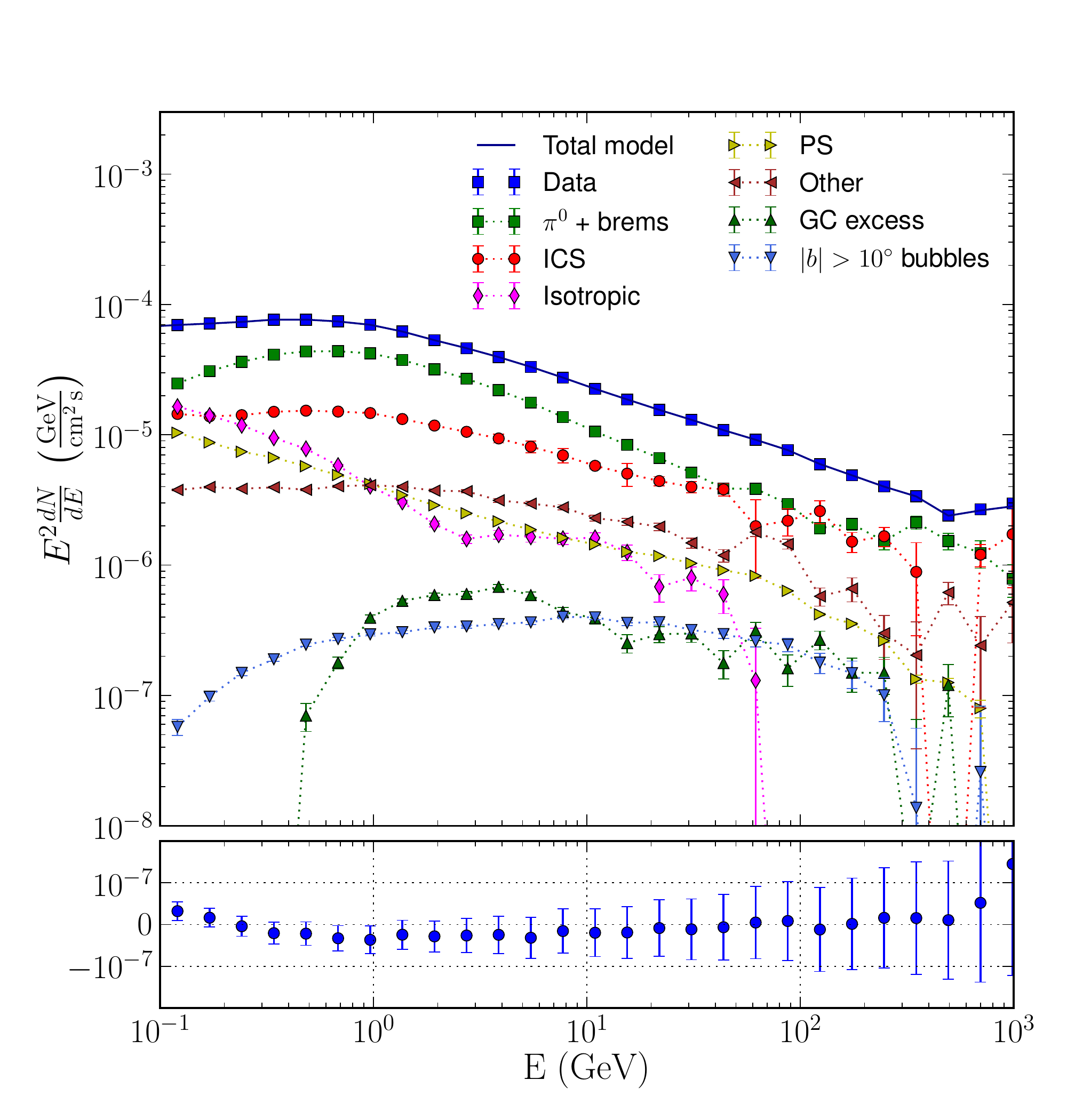}
\includegraphics[scale=\twopic]{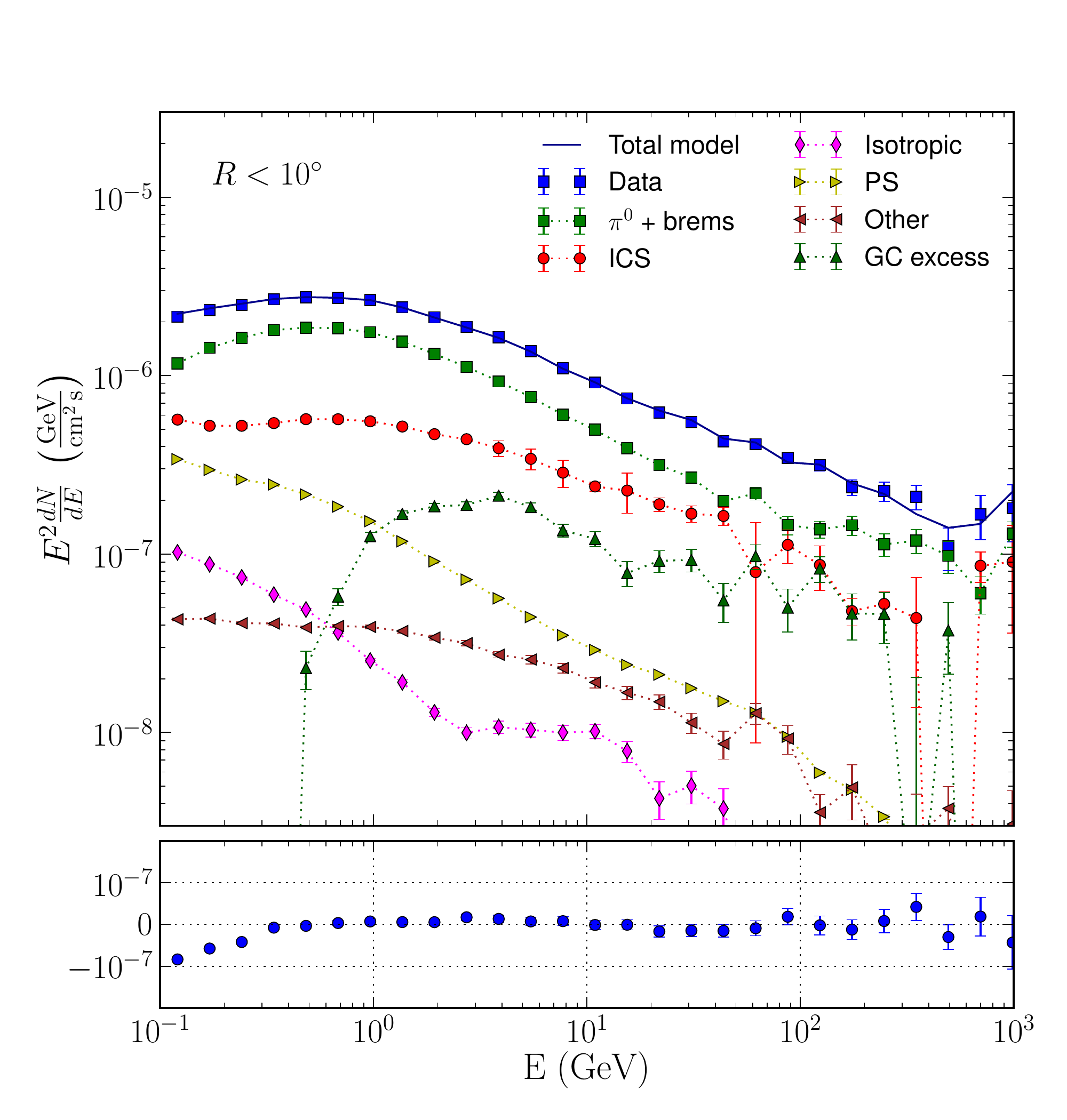}
\noindent
\caption{\small 
Flux of the components of the Sample Model (\ref{sec:baseline}) fitted to the all-sky data. 
Some templates are summed together in several groups for presentation.
``$\pi^0$ + brems" includes the hadronic and bremsstrahlung components.
``ICS" includes the three IC templates corresponding to the three radiation fields.
``Other" includes Loop I, Sun, Moon, and extended sources.
GC excess is modeled by gNFW template with index $\gamma = 1.25$.
Left: the fluxes of the components integrated over the whole sky except for the PS mask.
Right: the flux of the components integrated inside 10 deg radius from the GC, the model is the same as in the left plot,
the only difference is the area of integration for the flux.
The bubbles are not present in the right plot, since the Sample Model includes the bubbles template defined at latitudes $|b| > 10^\circ$.
}
\label{fig:baseline_spectra}
\end{center}
\end{figure}

\begin{figure}[htbp]
\begin{center}
\includegraphics[scale=\threepic]{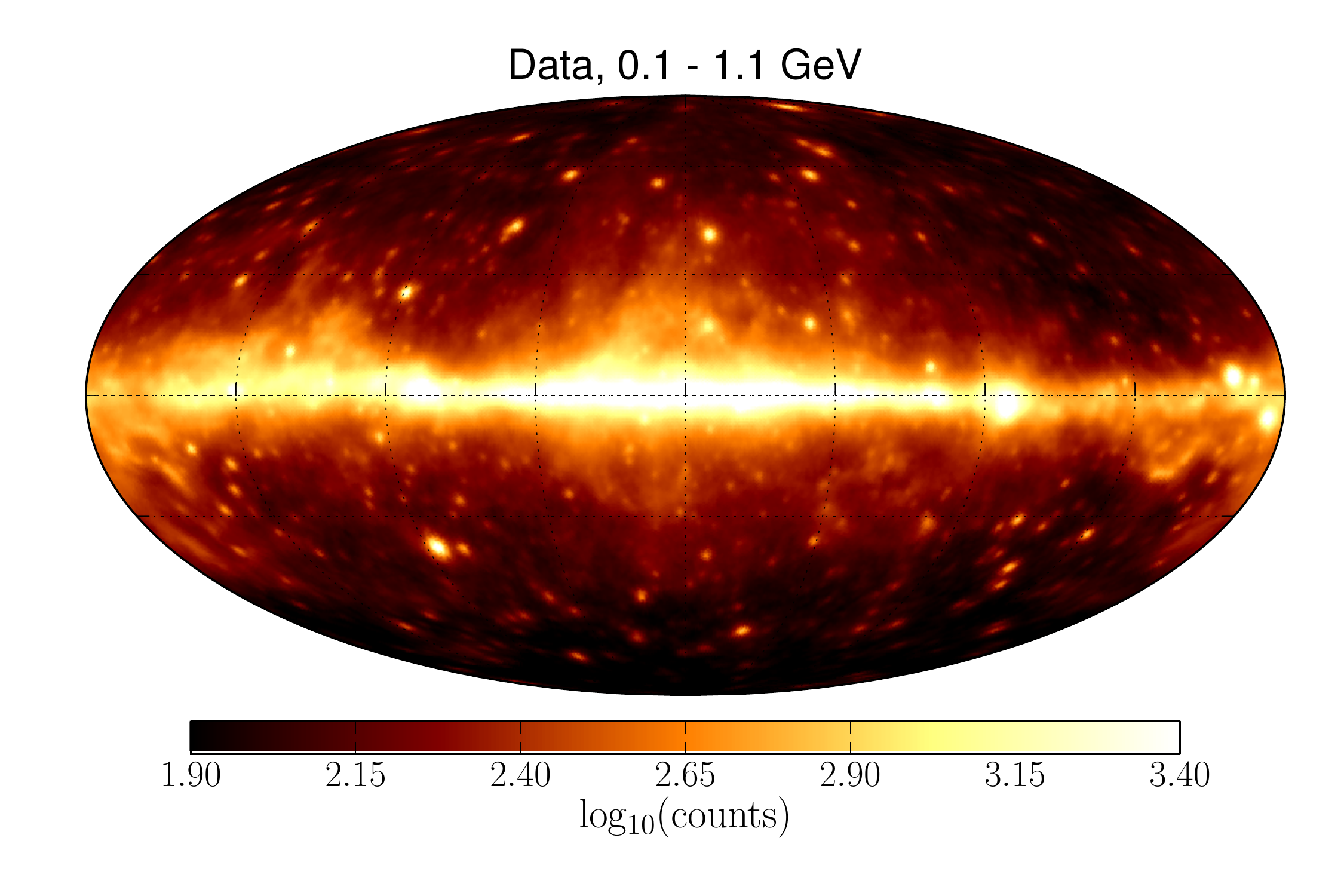}
\includegraphics[scale=\threepic]{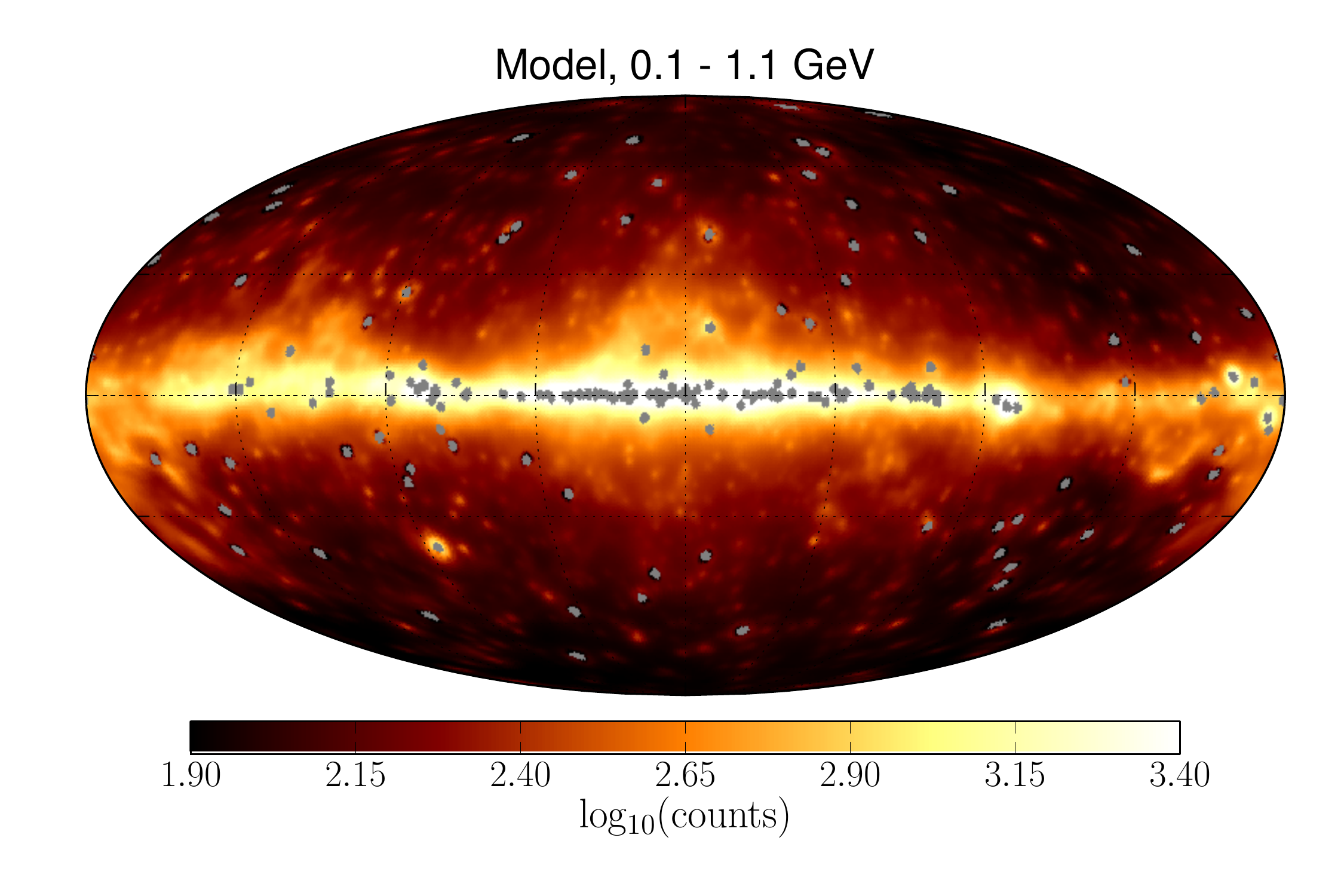}
\includegraphics[scale=\threepic]{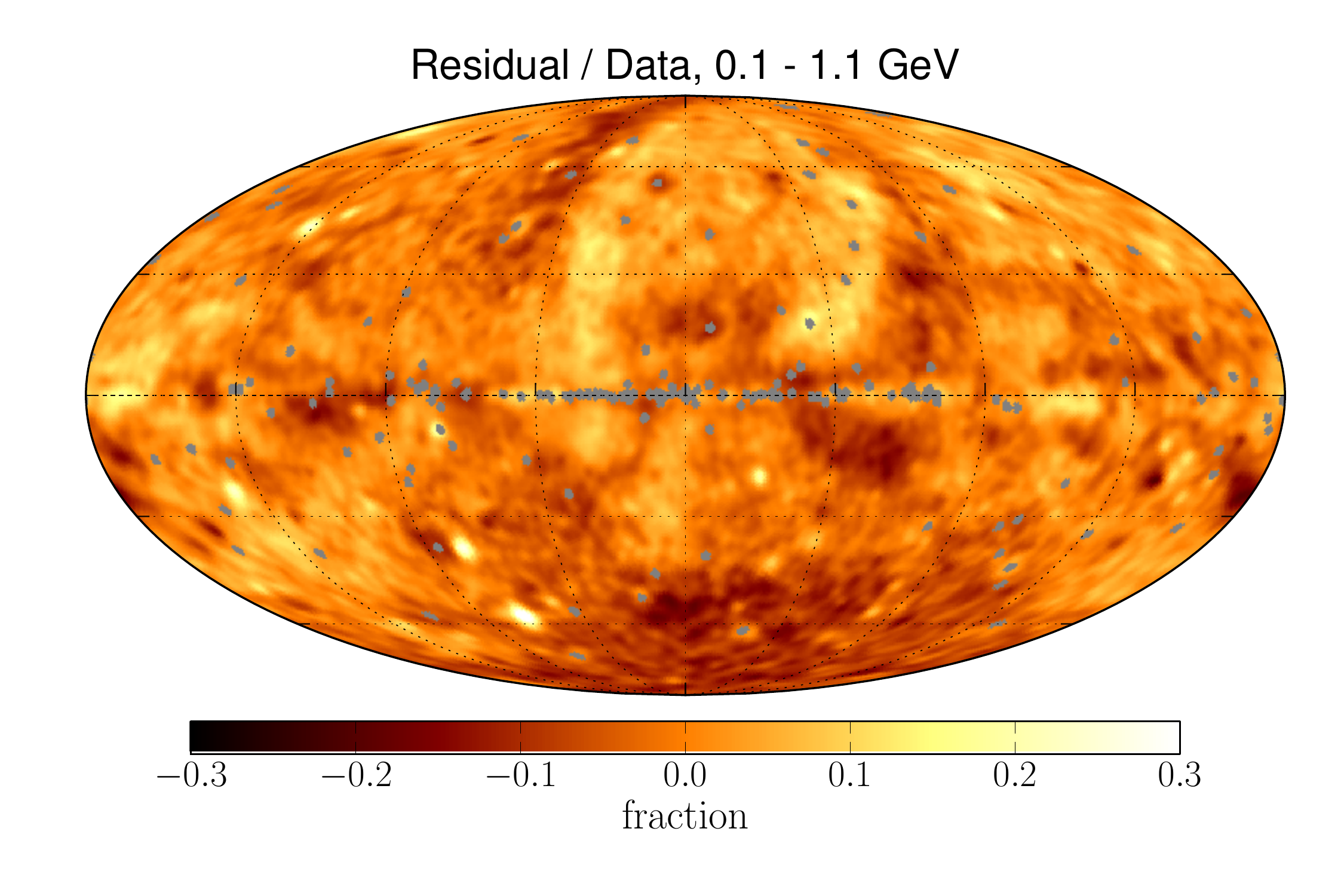} \\
\includegraphics[scale=\threepic]{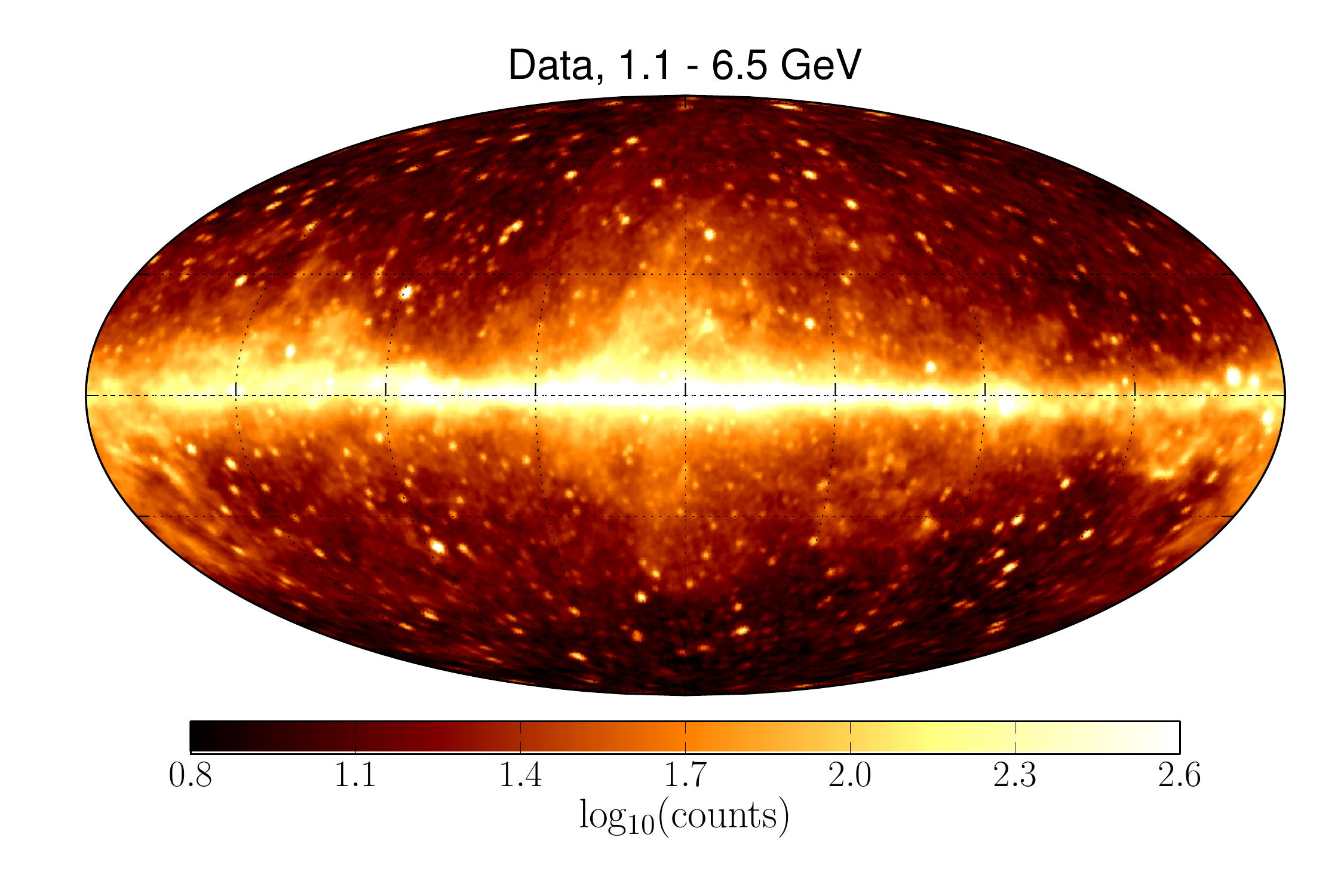}
\includegraphics[scale=\threepic]{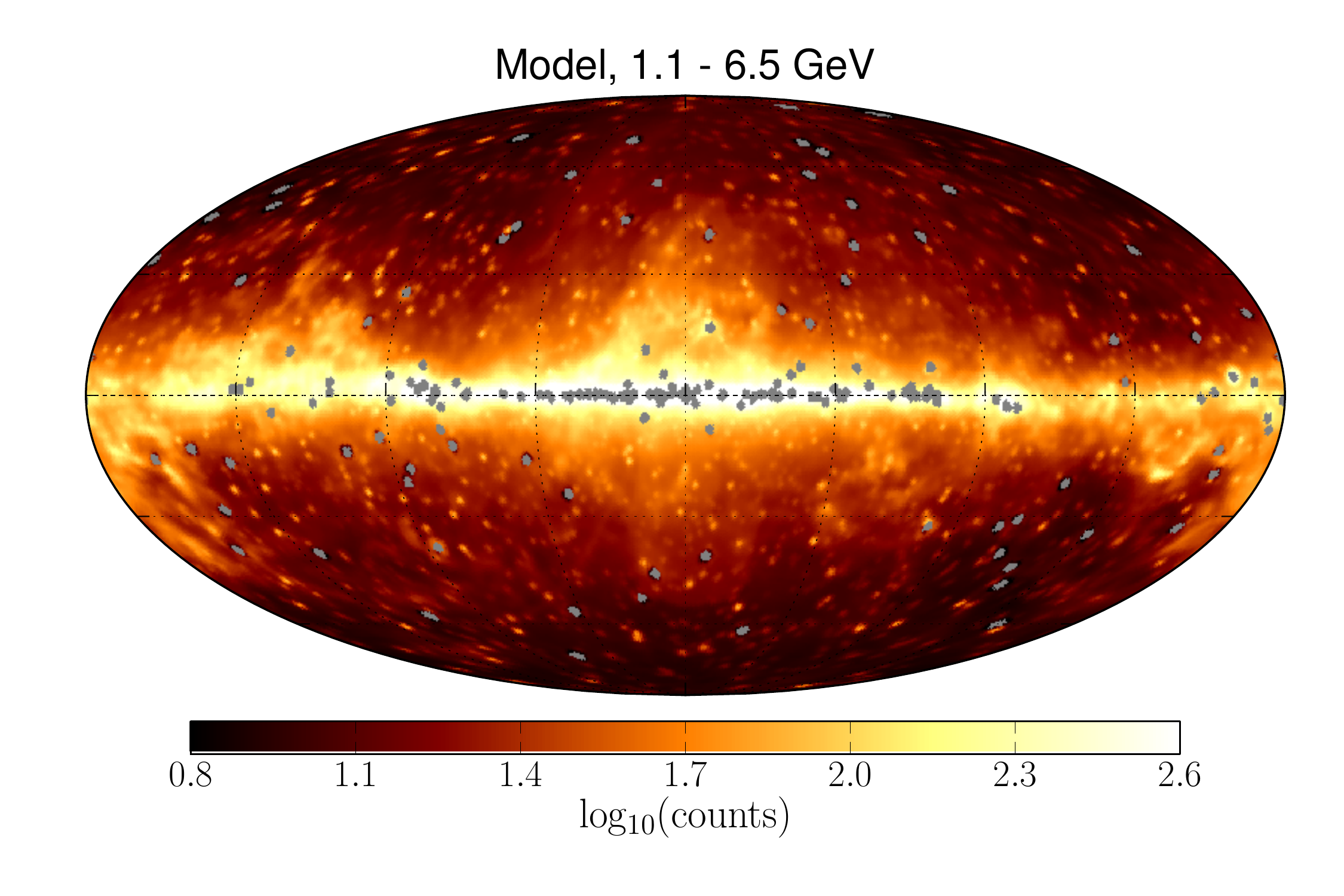}
\includegraphics[scale=\threepic]{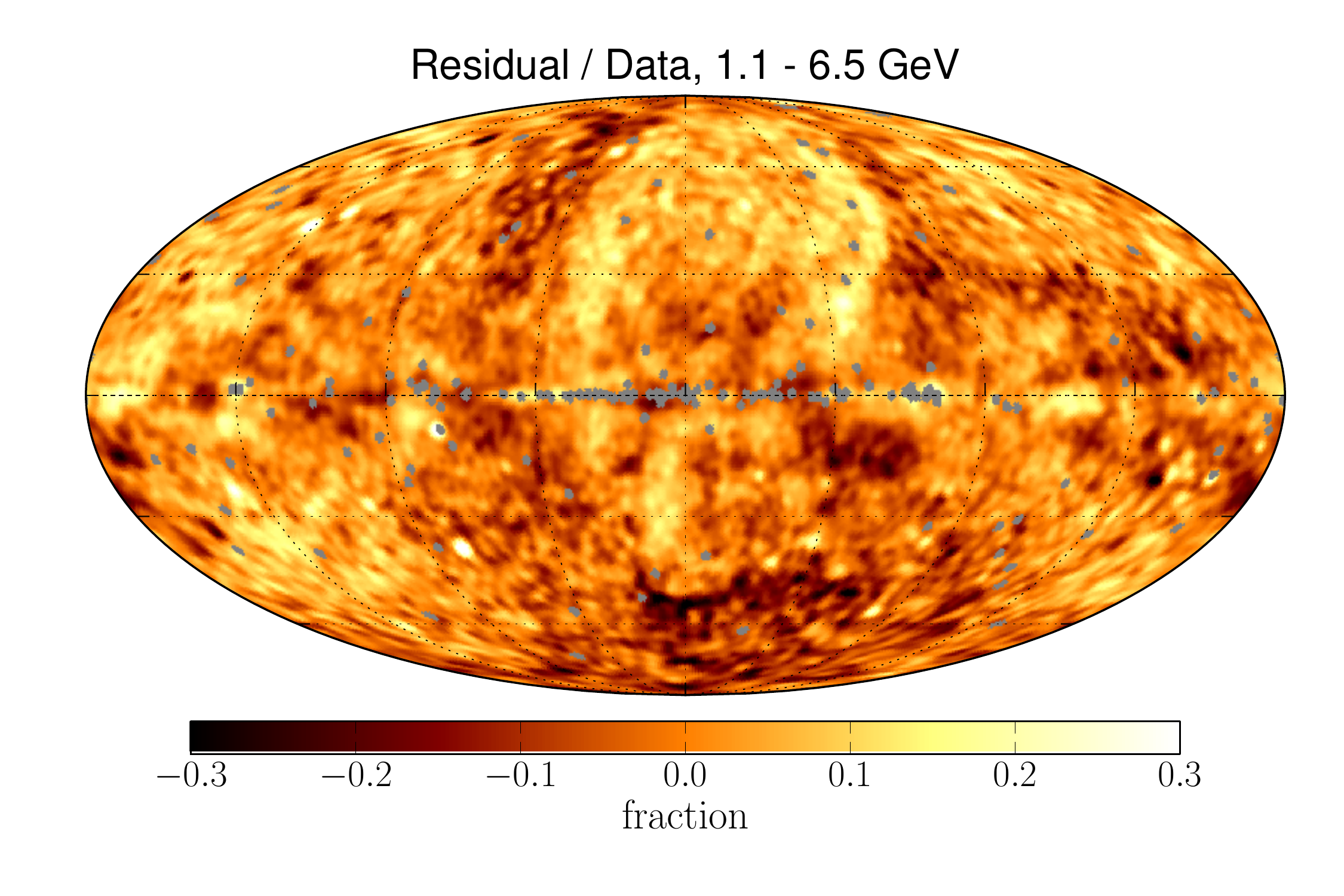} \\
\includegraphics[scale=\threepic]{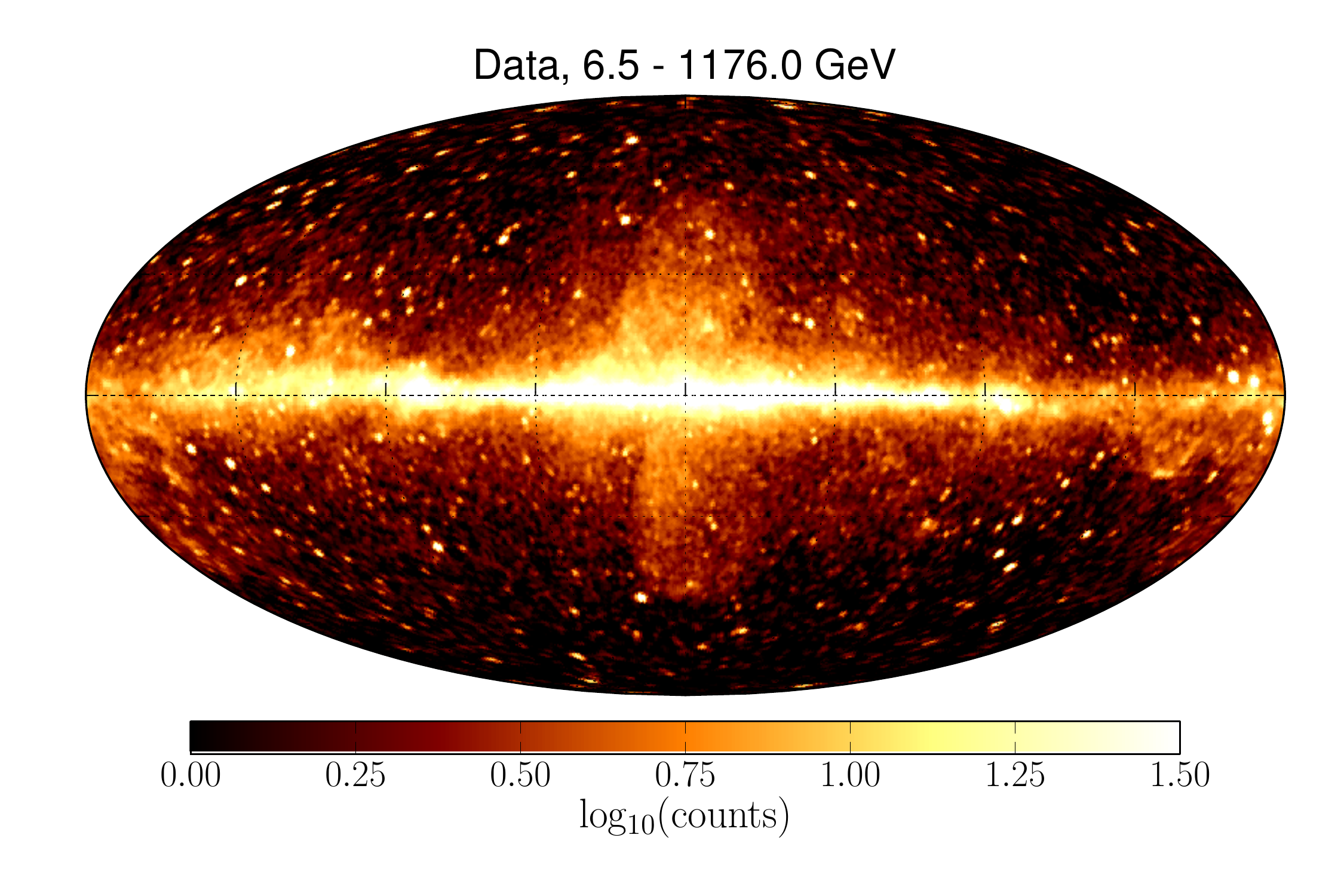}
\includegraphics[scale=\threepic]{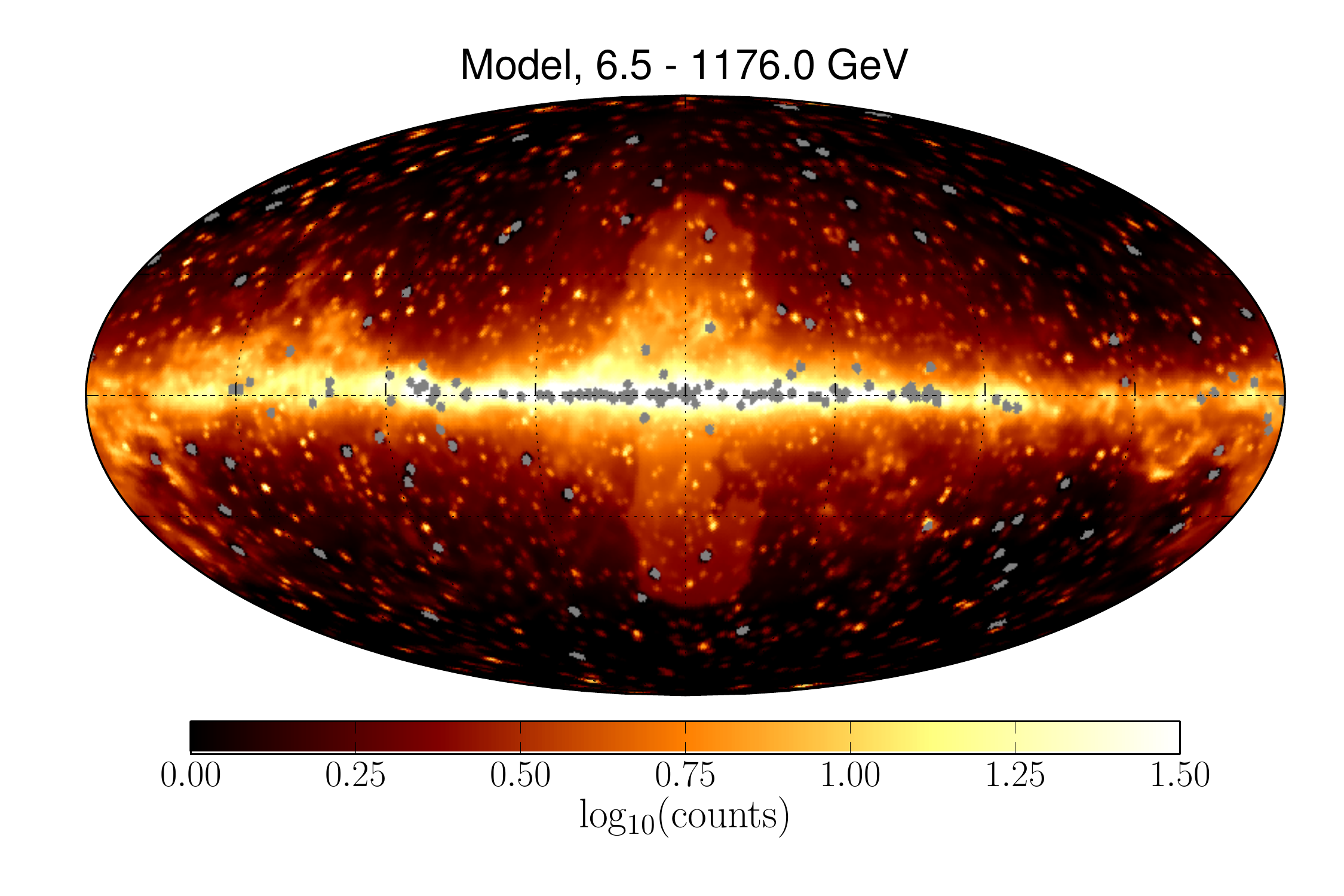}
\includegraphics[scale=\threepic]{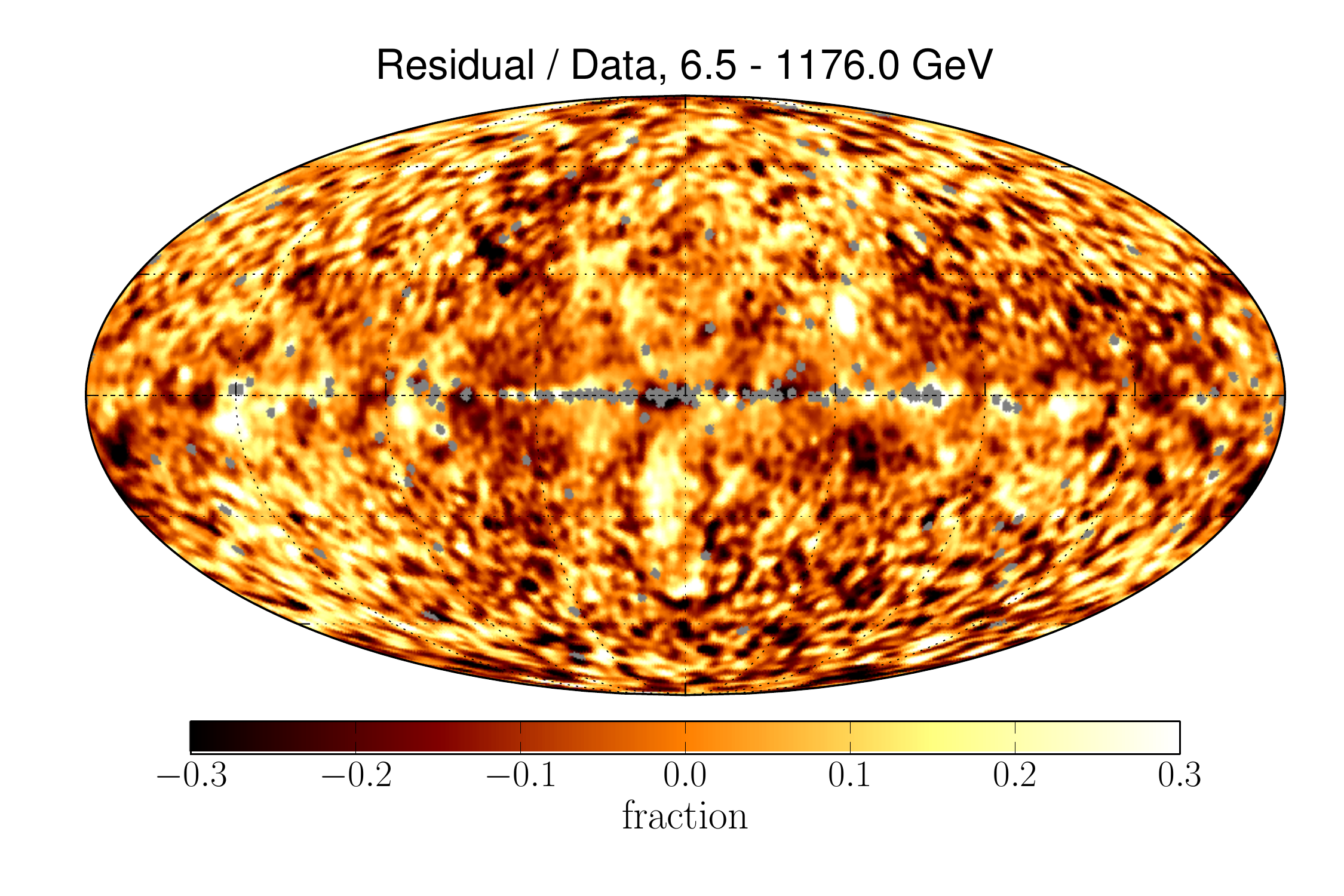} 
\noindent
\caption{\small 
Sample Model fit to the data.
Gamma-ray data (left), total model (middle) and fractional residual (right) 
maps summed over several energy bins: 7 energy bins between 100 MeV and 1.1 GeV (top row),
5 energy bins between 1.1 GeV and 6.5 GeV (middle row), 15 energy bins between 6.5 GeV and 1.2 TeV (bottom row).
The grey circles for the model and residual maps correspond to the 
mask constructed for the 200 highest-flux ($>1$ GeV) 3FGL sources (see Section \ref{sec:data}).
The pixel size is about $0\degr\!\!.46$ corresponding to HEALPix nside = 128 
(we will use the same pixel size for all all-sky plots in this paper).
}
\label{fig:data_model_resid}
\end{center}
\end{figure}

\begin{figure}[htbp]
\begin{center}
\includegraphics[scale=\twopic]{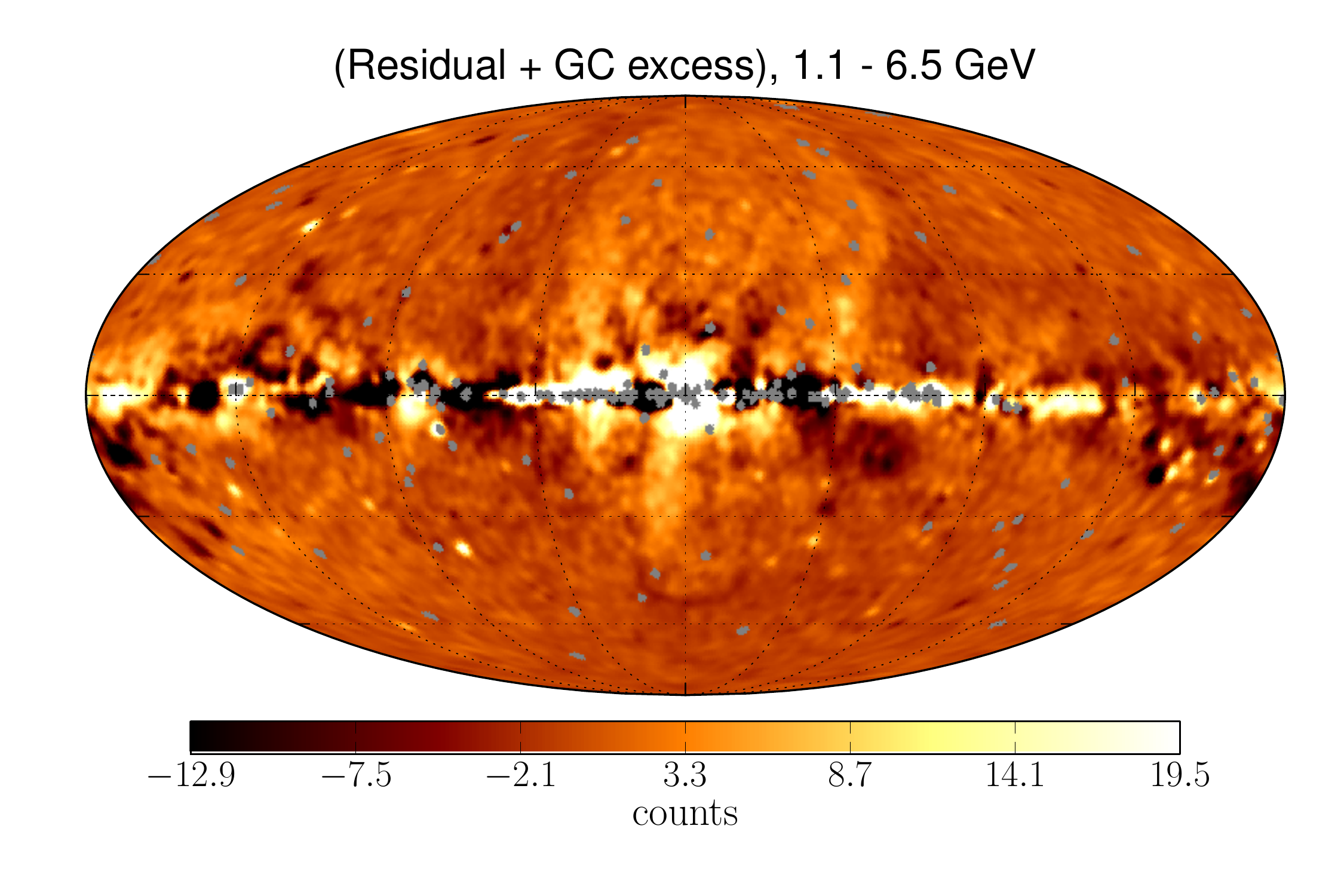}
\includegraphics[scale=\twopic]{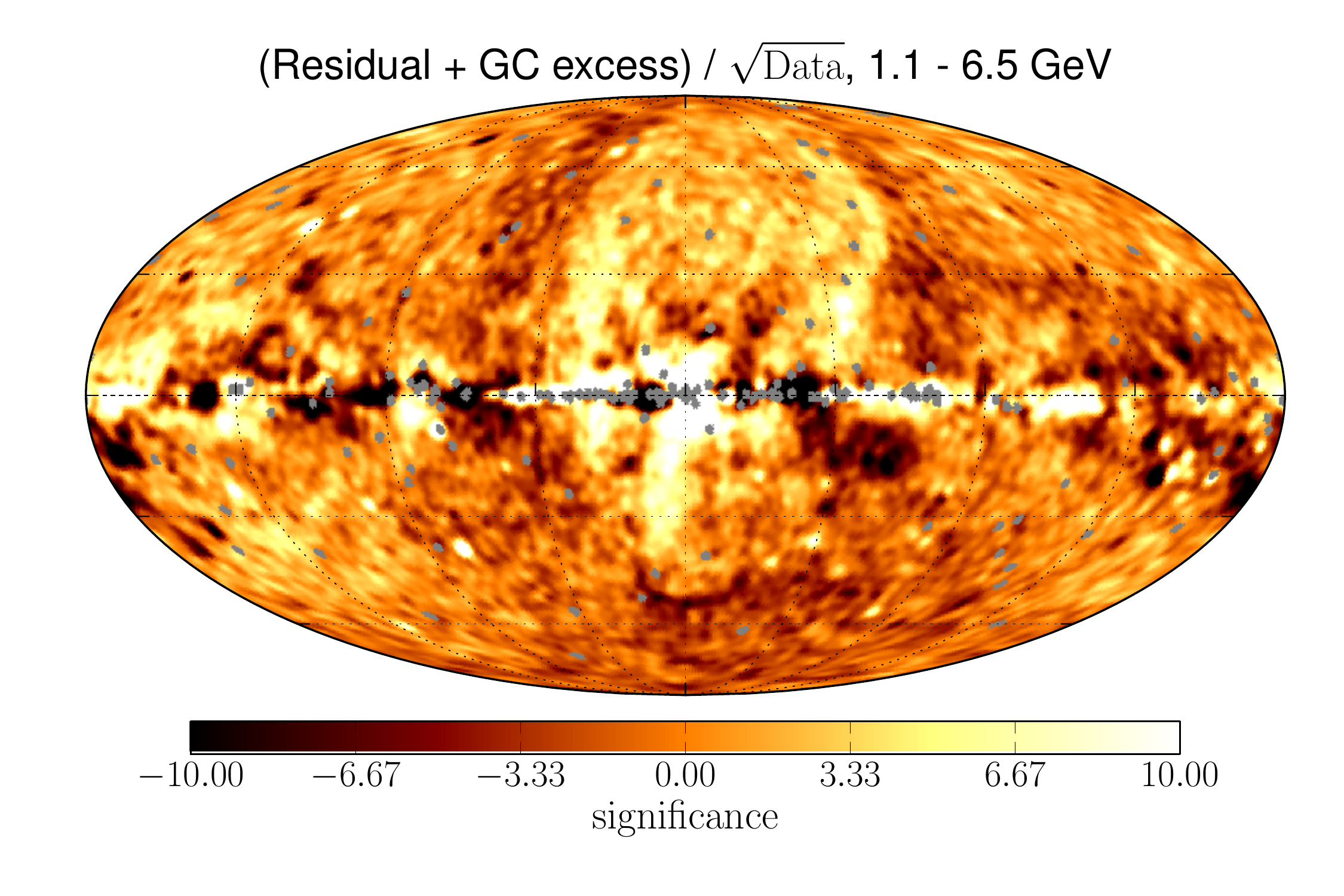} \\
\includegraphics[scale=\twopic]{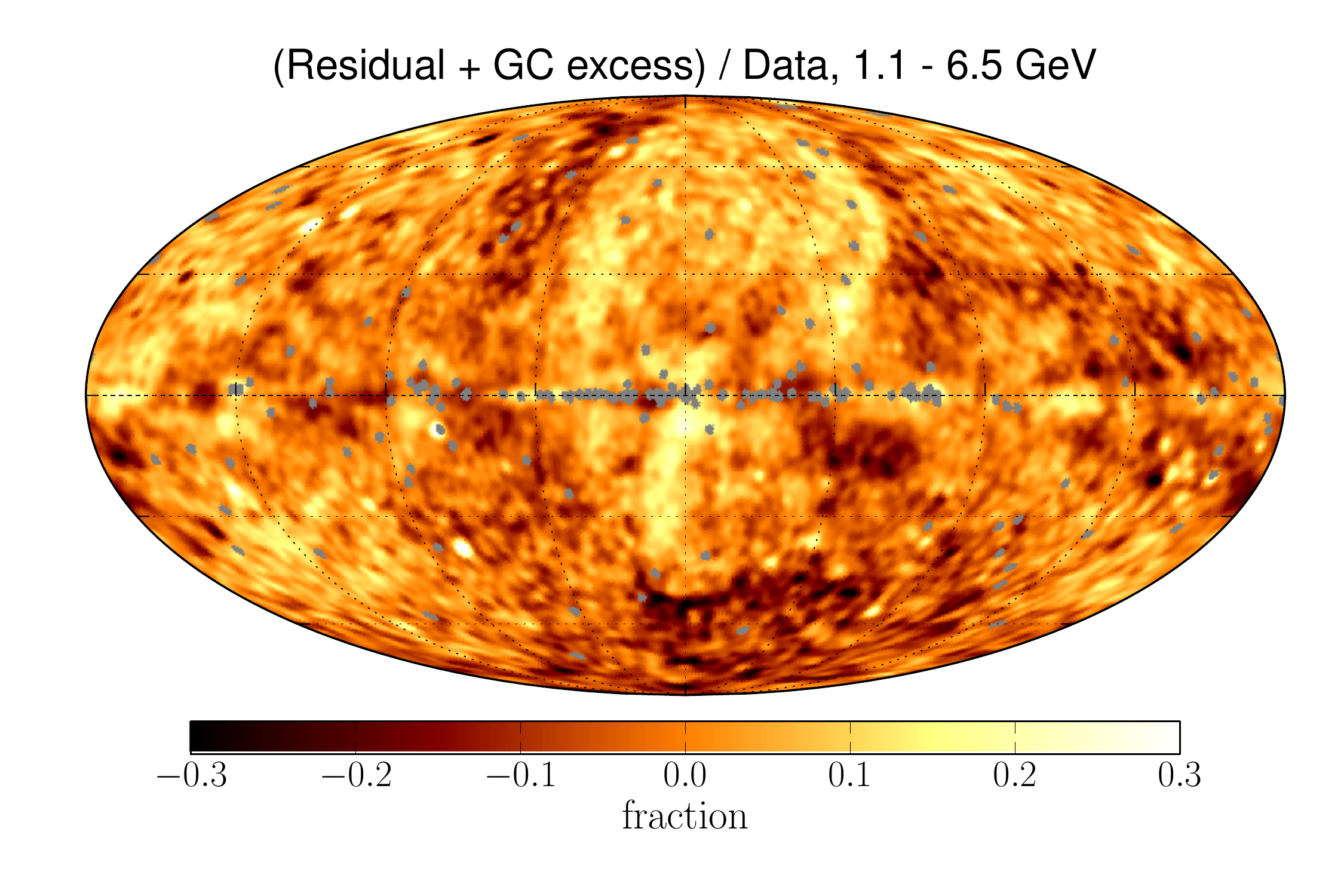}
\includegraphics[scale=0.33]{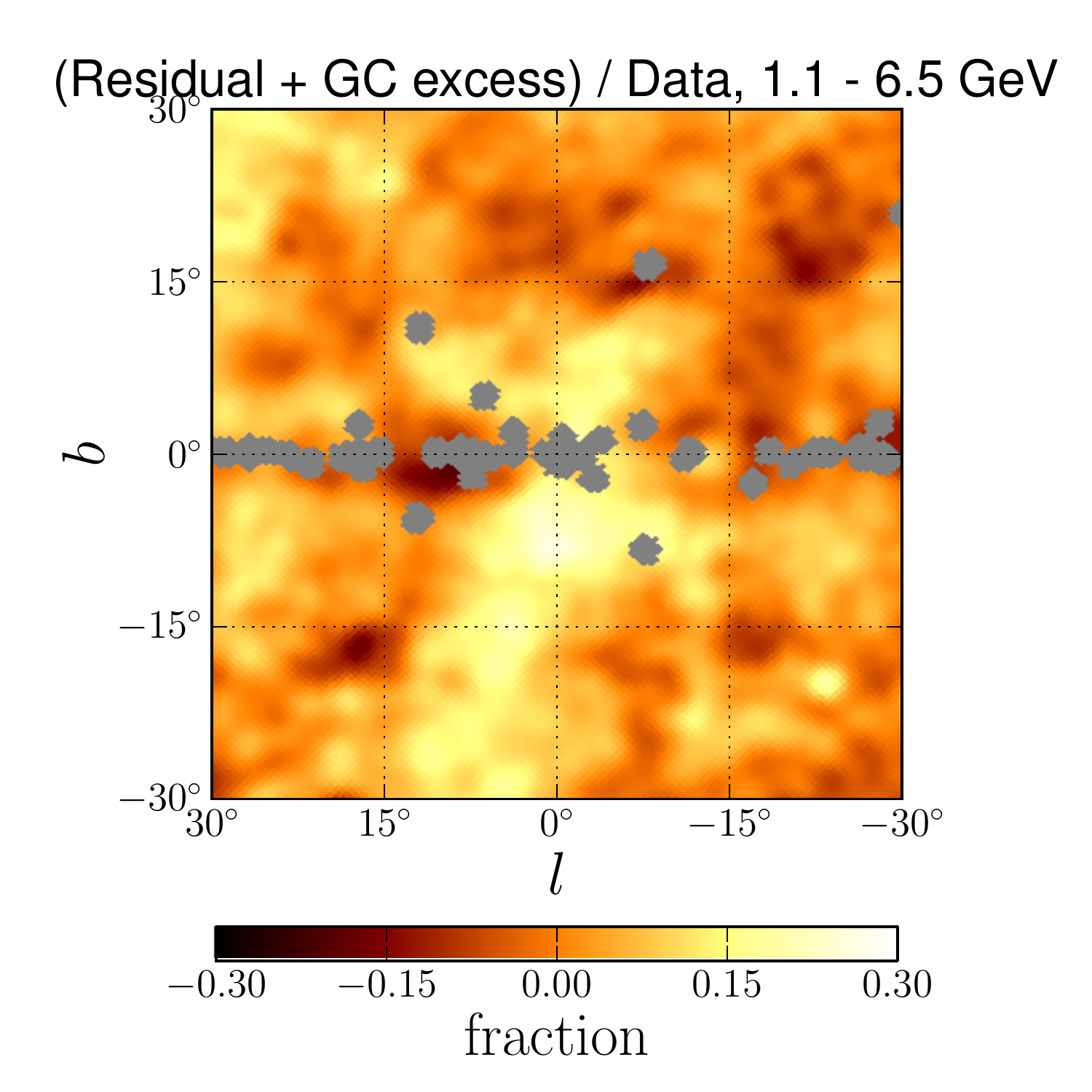}
\noindent
\caption{\small 
Residuals after fitting the \SM (see Figure \ref{fig:baseline_spectra} and text for details)
where we add back the GC excess modeled by the gNFW annihilation profile with $\gamma = 1.25$.
Top left: GC excess plus residual counts. 
Top right: GC excess plus residual counts divided by the square root of the total data counts.
Bottom left: GC excess plus residual counts divided by the total data counts.
Bottom right: 
enlarged scale residual map for the region around the GC.
The data in the denominator of the fractional residual and the residual significance is the smoothed data
that we used to determine the statistical fluctuations (see discussion after Equation \ref{eq:chi2}).
The counts in the maps are summed between 1.1 GeV and 6.5 GeV.
}
\label{fig:baseline_resid}
\end{center}
\end{figure}

The spectra of the components of the Sample Model fitted to the all-sky data are shown in Figure \ref{fig:baseline_spectra}. 
The Galactic center excess spectrum peaks around 3 GeV and extends up to about 100 GeV.
The corresponding data maps, total model maps and fractional residuals summed over several energy bins are shown in Figure 
\ref{fig:data_model_resid}.
Although the Sample Model approximately reproduces the data, many excesses are evident.
There is a clear residual associated with Loop I at energies below a few GeV in spite of including the geometrical template in the Sample Model.
There are also residuals associated with substructures inside the \Fermi bubbles.
Furthermore, many excesses are seen along the Galactic plane.
In Figure \ref{fig:baseline_resid} we show the GC excess modeled by the gNFW annihilation template added back to the residual
summed over energy bins between 1.1 GeV and 6.5 GeV.

The analysis with the \SM also serves to confirm through inspection of the likelihood hessian matrix that the large number of degrees of freedom does not create degeneracy between the model components, i.e., there is enough information in the gamma-ray data to separate them. However, some of the components that would be assigned negative fluxes are set to zero. As discussed before, this is most likely due to imperfections or incompleteness of the model. Although the procedure results in stable  and physically plausible spectra for most of the various components, in a few instances this is not the case (e.g., in the \SM, for the gas rings between 1.5~kpc and 3.5 kpc, the reason of which is discussed later in \ref{sec:gasunc}) . This has limited impact on the determination of the GC excess properties as the overall fore/background model is physically sound (Figure~\ref{fig:baseline_spectra}).

\section{Uncertainties from the Analysis Setup}
\lb{sec:analysis_setup}

This section is dedicated to assess the impact on the results of some key aspects of the analysis procedure, namely the selection of the data sample and of the region of interest (ROI).

\subsection{Dataset Selection}
\lb{sec:data_selection}

We start by testing the systematic uncertainty related to selection of the data sample. 
As an alternative to the sample analysis we use the Clean event class (P8R2\_CLEAN\_V6 instrument response functions) with a selection on zenith $< 100^\circ$.
By considering the Clean class instead of the UltraCleanVeto,
we estimate the magnitude of the residual CR contamination, which is larger for Clean class events compared to the UltraCleanVeto events.
By using a larger zenith angle cut, we estimate a possible effect of emission from the Earth limb at $\sim$112$^\circ$.
In general, residual Earth limb emission, from gamma rays in the tails of the point-spread function, becomes more important at lower energies, where the tails are broadest.
Using the Clean class events with the zenith angle cut $< 100^\circ$ has relatively small influence on the GC excess spectrum (Figure \ref{fig:GC_escess_vars} top left): the GC excess spectra are consistent within the statistical uncertainties.

We also sub-select gamma-ray events with the best angular resolution: PSF classes 2 and 3 (Section \ref{sec:data}).  
We then convolve the Sample Model components with the respective instrument response functions (notably, PSF) independently
and perform a joint fit of the two datasets.
The comparison of the GC excess flux with the Sample Model is again shown in Figure \ref{fig:GC_escess_vars} top left. There is a moderate effect on the spectrum at low energies only, where the LAT PSF gets worse.  

\subsection{Region of Interest Selection}
\lb{sec:ROI}

One of the limitations of the template fitting approach we use is that  to model gamma-ray emission from gas we assume that the CR densities depend only on Galactocentric radius and distance from the Galactic plane, and we rely on GALPROP to accurately predict the morphology of IC emission at each energy.
Therefore, variations of the CR spectrum or mismodeling in one part of the Galaxy can lead to oversubtraction or unmodeled excesses in other regions.

One way to moderate this type of effect is to restrict the fitting procedure to a smaller region of interest around the GC, so that there is more freedom to reproduce the features in the data for this specific part of the sky.
To gauge the effect on the spectrum of the GC excess, we repeat the analysis in \S\ref{sec:baseline} restricting the ROI to some square regions:  $|b|, |\ell| < 10\arcdeg, 20\arcdeg, 30\arcdeg$.
In this subsection we use maps with order 7 resolution (for all-sky fits we use adaptive resolution as discussed in 
\S\ref{sec:data}), which gives more than one thousand pixels even for $|b|, |\ell| < 10\degr$ case.
This is generally sufficient to resolve the gas correlated templates.
However, the IC templates are rather smooth and may be degenerate in a small ROI.
For this reason we combine the three IC templates in the Sample Model into a single template for fits in small ROIs.
We also do not have the bubbles template in the $|b|, |\ell| < 10\degr$ case, because it is defined only at
$|b| > 10\degr$.

\begin{figure}[htbp]
\begin{center}
\includegraphics[scale=\twopic]{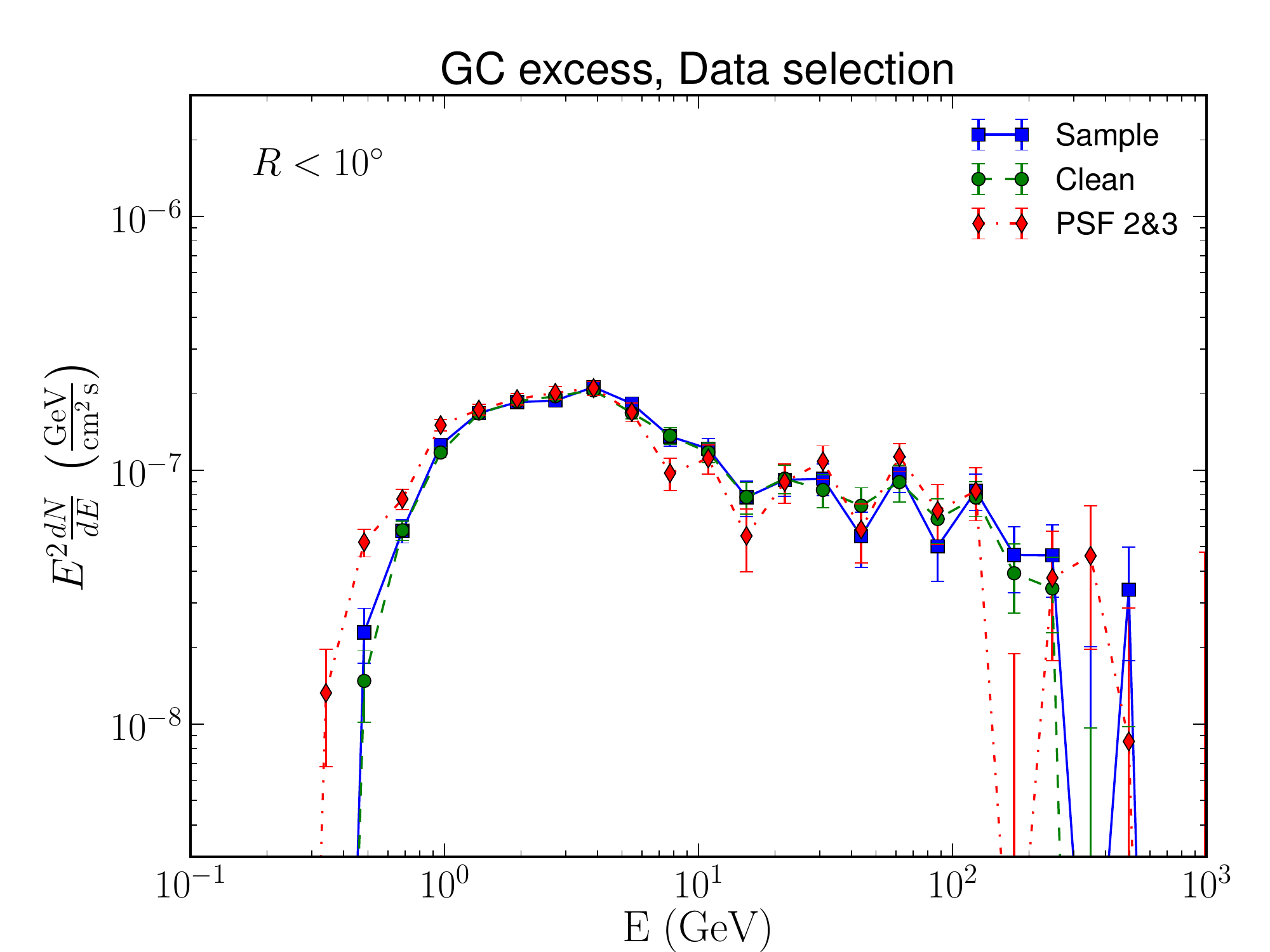}
\includegraphics[scale=\twopic]{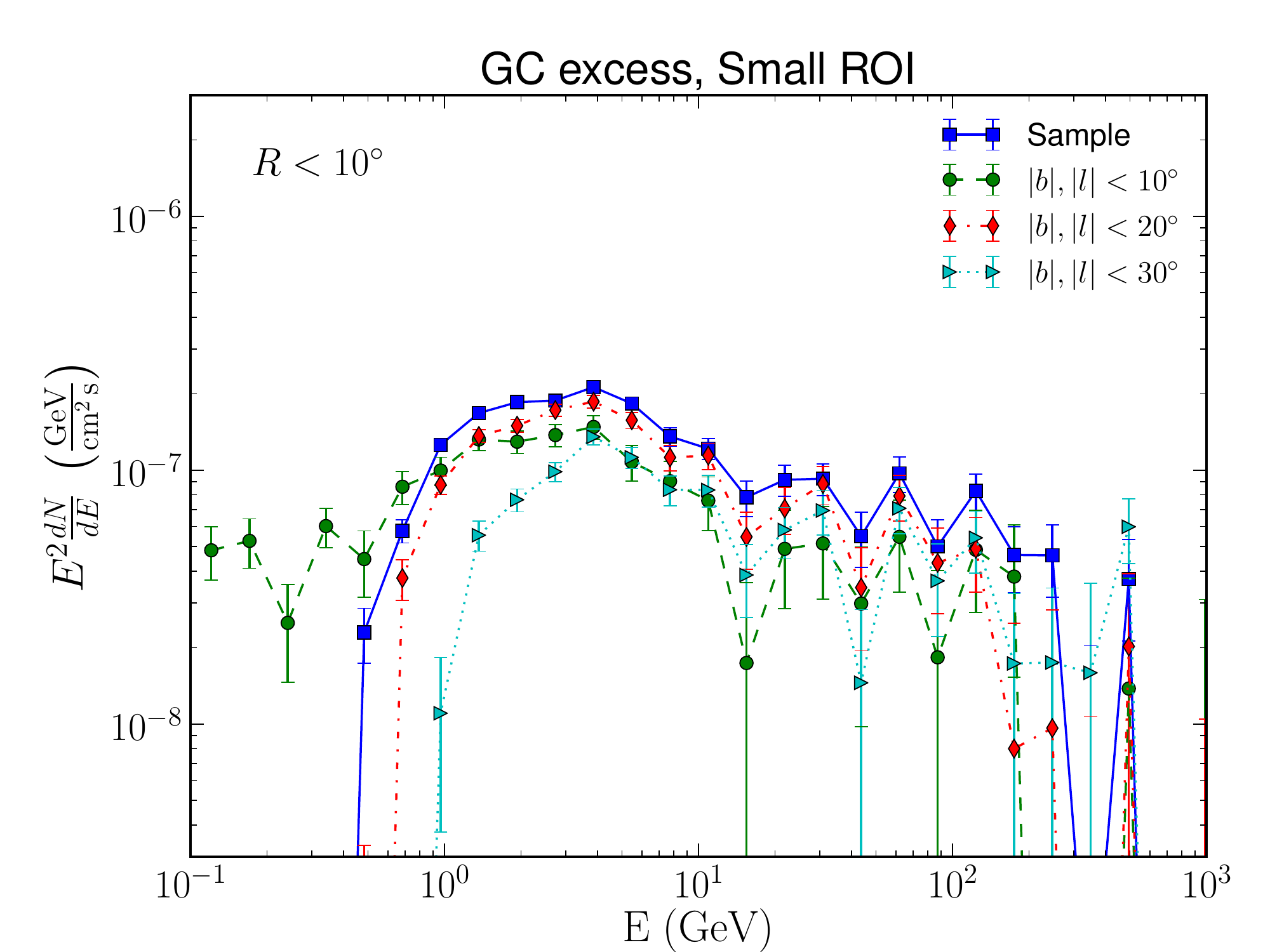}
\includegraphics[scale=\twopic]{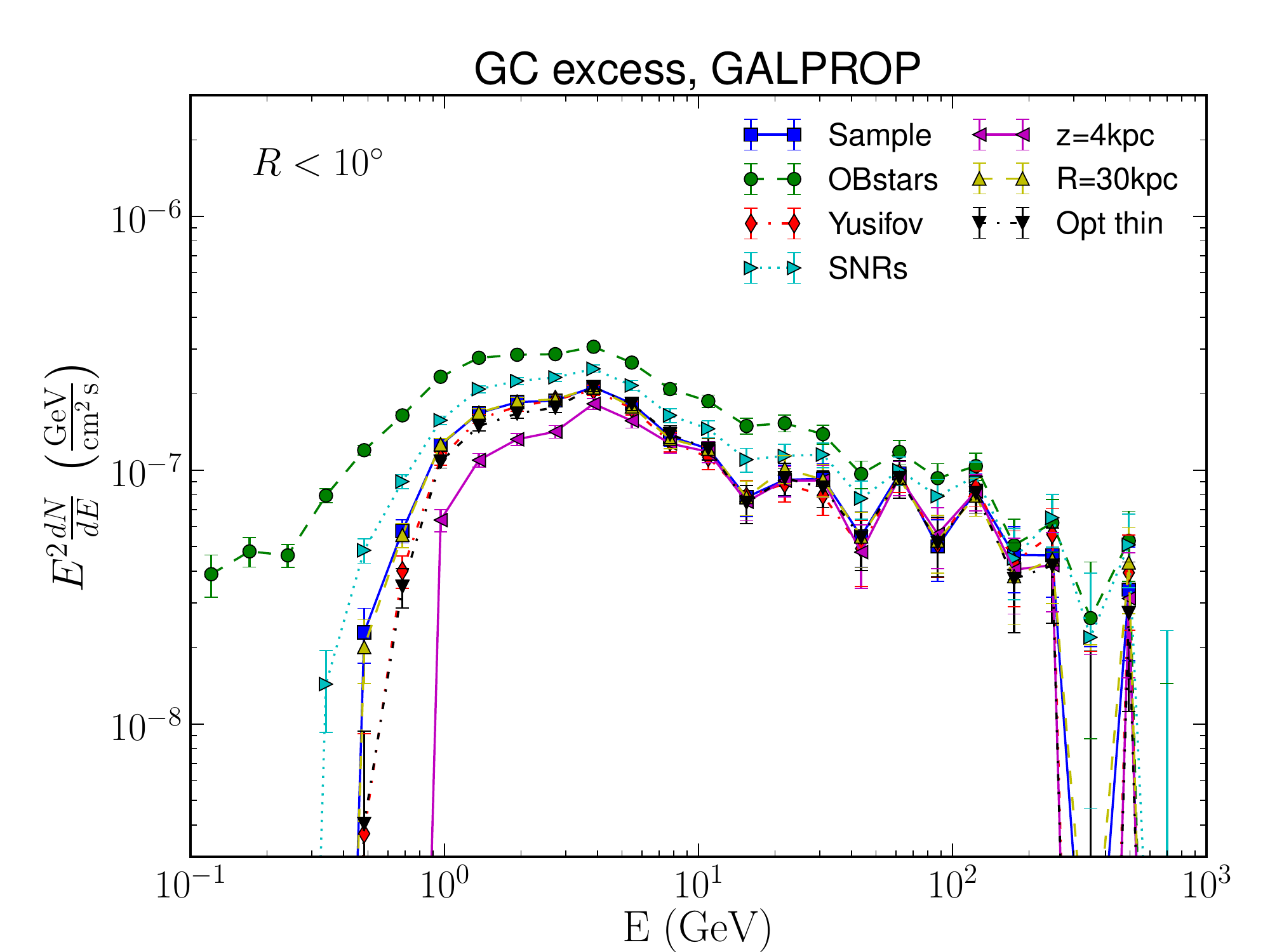}
\includegraphics[scale=\twopic]{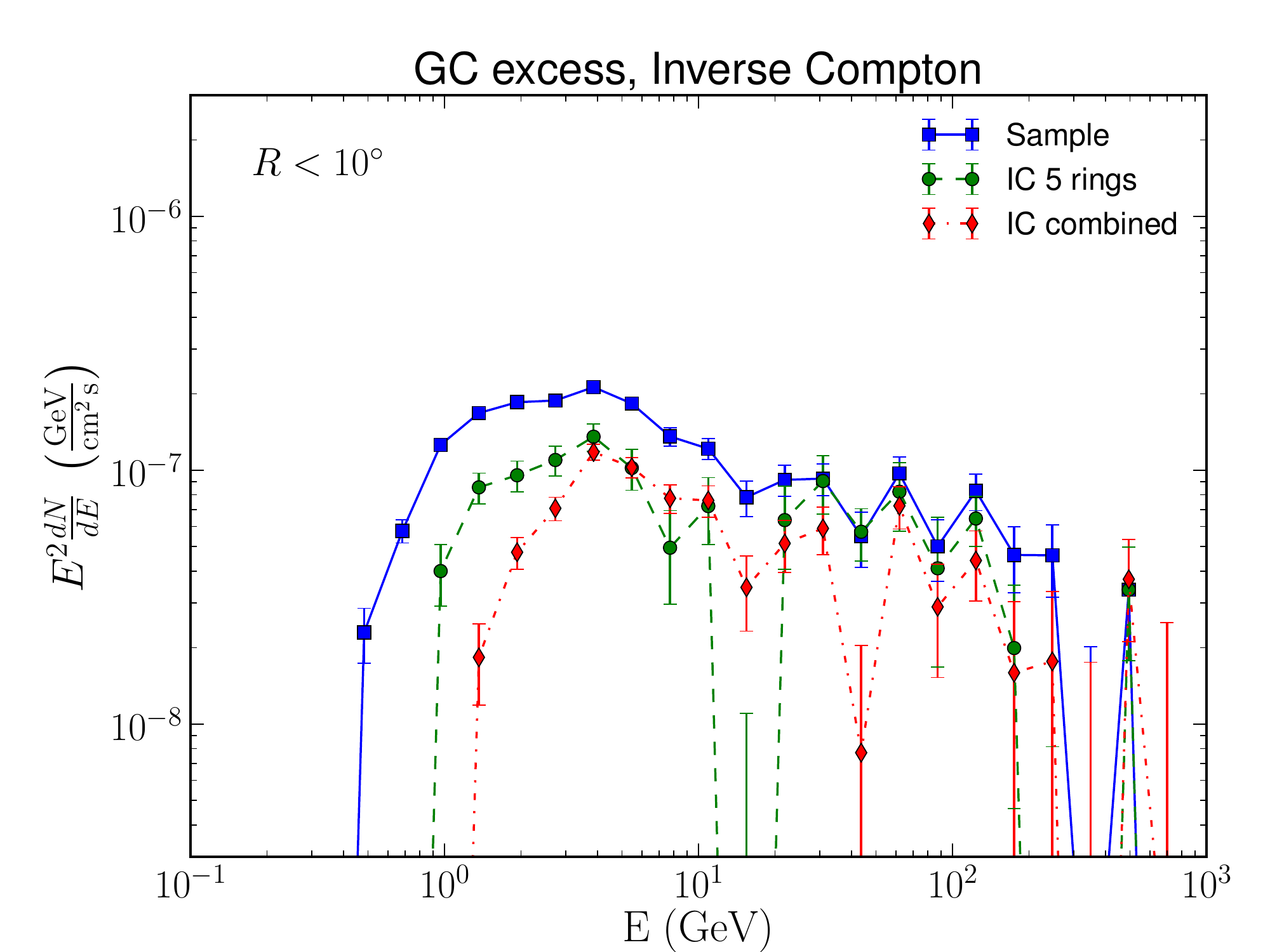}
\includegraphics[scale=\twopic]{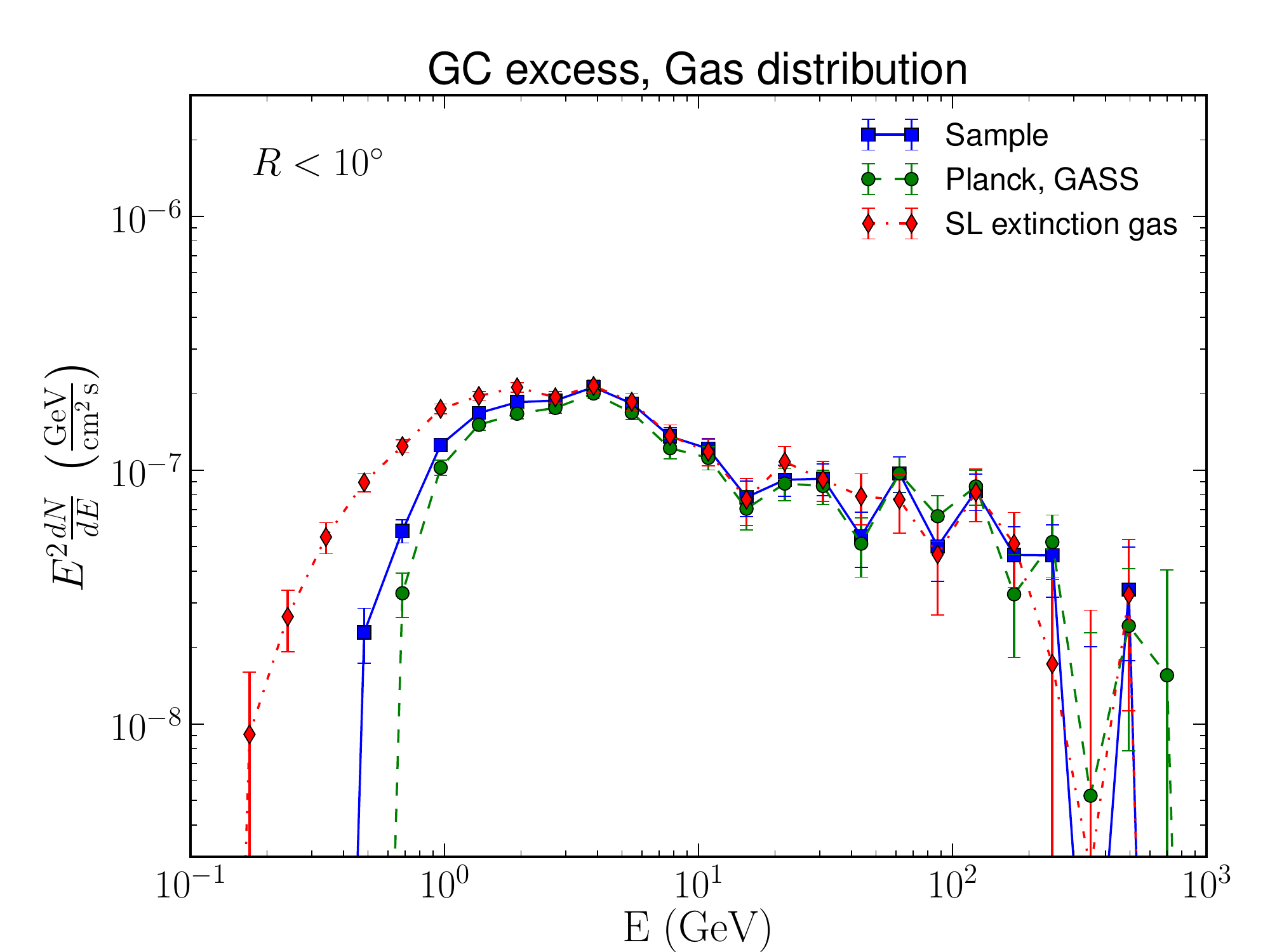}
\includegraphics[scale=\twopic]{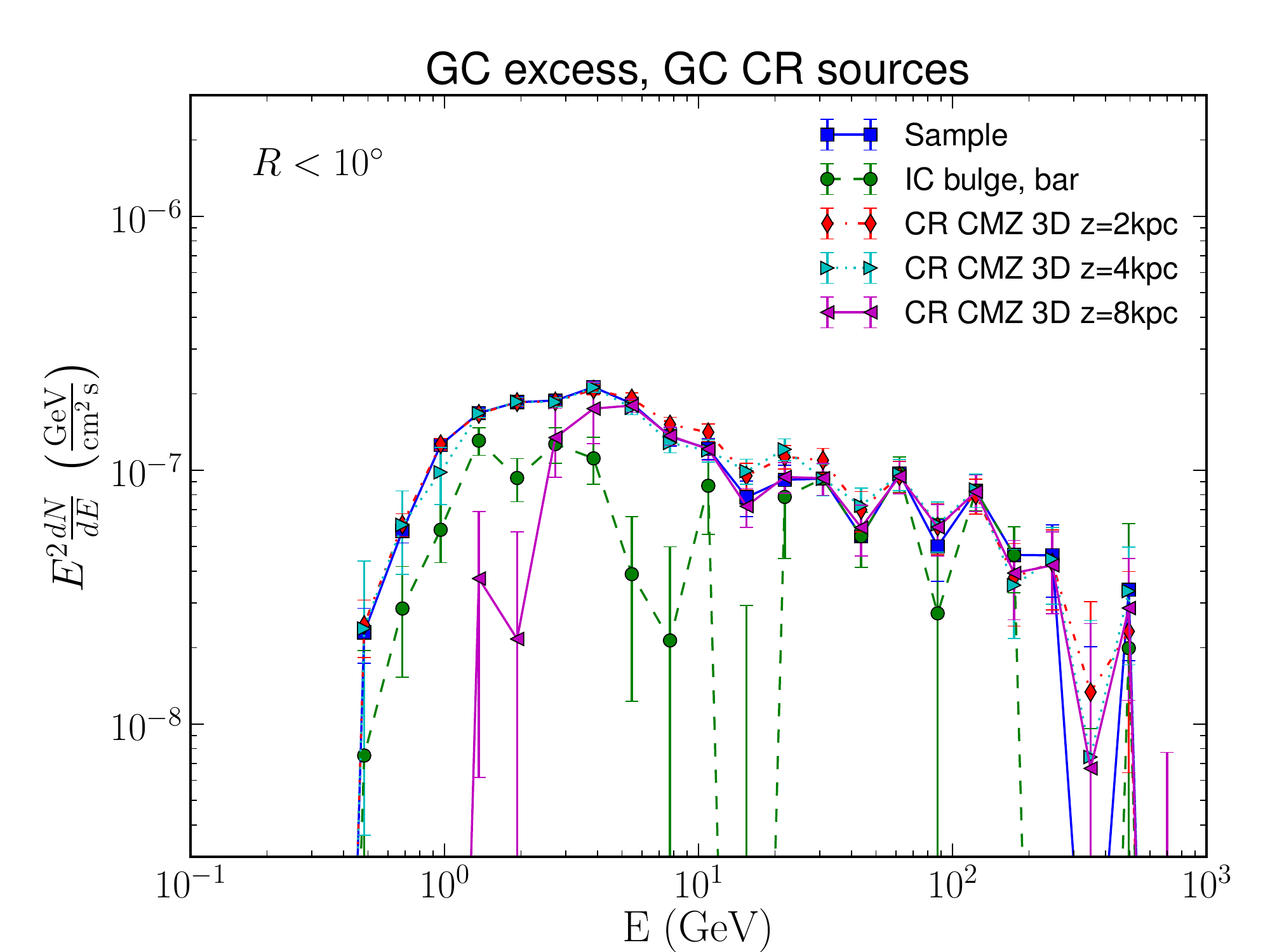}
\noindent
\caption{\small 
Comparison of the GC excess spectrum in the Sample Model (\S\ref{sec:baseline})
and different choices for data selection, ROI, and the Galactic interstellar emission model.
Top left: choice of the data sample (\S\ref{sec:data_selection}).
Top right: size of the ROI used for fitting the model components to the data (\S\ref{sec:ROI}).
Middle left: CR source tracers and confinement halo height (\S\ref{sec:galprop_vars}).
Middle right: fitting of the IC template (\S\ref{sec:IC}).
Bottom left: tracers of interstellar gas and partition of gas column densities along the line of sight (\S\ref{sec:gasunc}).
Bottom right: additional sources of CRs near the GC and variation of the propagation halo height in models with 
an additional source of CR correlated with the CMZ (\S\ref{sec:GC_CR_sources}).
The flux is obtained by integrating over the circle $R < 10^\circ$ from the GC excluding the PS mask.
}
\label{fig:GC_escess_vars}
\end{center}
\end{figure}

The results are shown in Figure \ref{fig:GC_escess_vars} top right.
We note that gNFW cusp profile remains non-degenerate with the other components of emission even in the small ROI,
because the degeneracies would result in large error bars, while the error bars on the GC excess flux remain reasonably small below 10 GeV.
The intensity of the GC excess is generally reduced for the fits in smaller ROIs.
For $10\arcdeg$ ROI the GC excess continues to be significant at energies below 400 MeV.
While for $30\arcdeg$ ROI the excess cuts off below 1 GeV.
The change in the GC excess flux for different ROI sizes is likely due to mismodeling of Galactic diffuse components.

\section{Uncertainties from the Modeling of Galactic Interstellar Emission}
\label{sec:CRmodel}

This section is devoted to exploration of the uncertainties in the spectrum of the GC excess due to the modeling
of Galactic interstellar emission. We consider the following aspects:
\ben
\item
Definition of the distribution of CR sources, size of the CR confinement halo, and spin temperature of atomic hydrogen (for the derivation of gas column densities from the 21-cm line data) used in GALPROP;
\item
Handling of the IC component in the fit to the gamma-ray data;
\item
Selection of the tracers of interstellar gas, and distribution of gas column densities along the line of sight;
\item
Possible additional sources of CRs near the GC.
\een

\subsection{GALPROP Parameters}
\label{sec:galprop_vars}

\citet{FermiLAT:2012aa} explored the effects of varying several parameters of the GALPROP models that we use to create templates for interstellar gamma-ray emission. They concluded that the parameters with the largest impact on the predictions for gamma rays are 1) distribution of CR sources in the Galaxy, 2) height of the CR confinement halo, and 3) spin temperature used in deriving the atomic gas column densities from the 21-cm H~\textsc{i} line intensities.

\begin{figure}[htbp]
\begin{center}
\includegraphics[scale=\twopic]{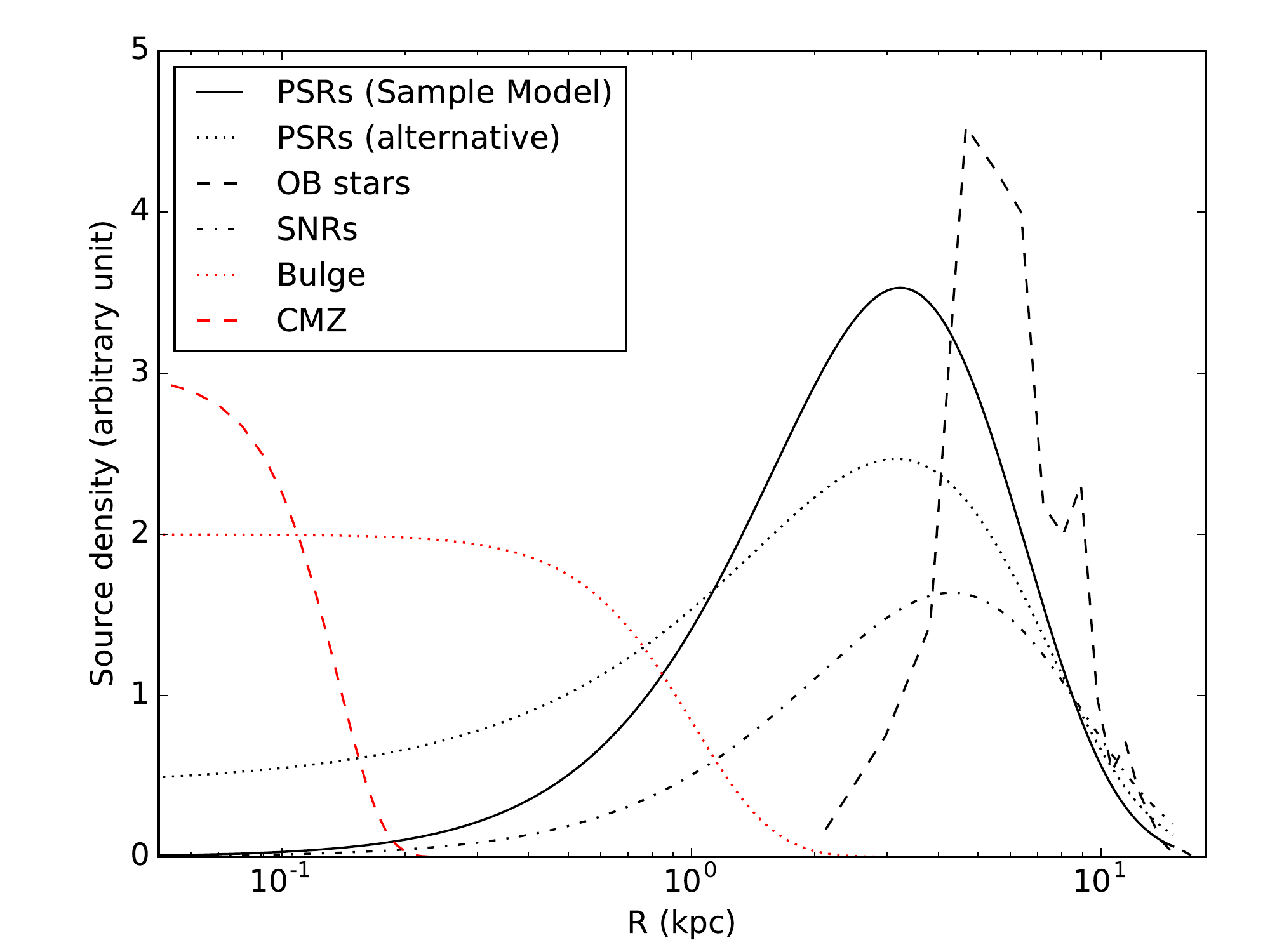}
\noindent
\caption{\small Radial distribution in the Galactic plane of the CR source models employed in this work. In black, distributions from \citet{FermiLAT:2012aa}: the pulsars (PSRs) distribution by \citet{Lorimer:2006qs} used in the Sample Model, the alternative PSRs distribution by \citet{2004A&A...422..545Y}, the OB stars distribution from \citet{2000A&A...358..521B}, and the SNRs distribution from \citet{Case:1998qg}. In red the source models in the inner Galaxy introduced in our work: sources in the bulge (azimuthal average) following the distribution of the old stellar population as in model B from \citet{2012A&A...538A.106R}, and sources in the central molecular zone (CMZ) following the distribution of molecular gas from \citet{2007A&A...467..611F}. The source distributions are independently normalized for display.
}
\label{fig:CRsources}
\end{center}
\end{figure}

Our Sample Model in \S\ref{sec:baseline} uses the Lorimer pulsar distribution as a tracer of CR sources (supposedly SNRs, whose distribution is more difficult to determine from observations), a CR confinement height of 10~kpc, and radius of 20~kpc and an H~\textsc{i} spin temperature of 150~K. In order to quantify the impact of these choices on the spectrum of the GC excess we use a subset of models in~\citet{FermiLAT:2012aa}. We have used different CR source distributions: an alternative pulsar distribution \citep[][hereafter referred to as Yusifov]{2004A&A...422..545Y}, the distribution of SNRs\footnote{We note that 
the derivation of the Galactic supernova remnant distribution in 
\citet{Case:1998qg} is subject to uncertainties and the results are discordant with some later 
works \citep[e.g.,][]{2015MNRAS.454.1517G}. In our study, though, we use it only as a way 
to probe the uncertainties due to the modeling of CR propagation relying on the 
previous work by \citet{FermiLAT:2012aa} who made extensive comparisons to gamma-ray data.}~\citep{Case:1998qg}, and the distribution of OB stars~\citep{2000A&A...358..521B}. Radial distributions of these CR source models are shown in Figure~\ref{fig:CRsources}. We changed the CR confinement height from 10~kpc to 4~kpc and its radius from 20~kpc to 30~kpc. In addition we derived the H~\textsc{i} column densities from the 21-cm line intensities assuming an optically thin medium, which we formally modeled by setting the spin temperature to $10^5$~K.

The resulting spectra for the GC excess are presented in Figure \ref{fig:GC_escess_vars} middle row, left. The largest effect is observed from the OB star source distribution model, which leads to an overall increase in the GC excess flux, while a decrease of the CR confinement height to 4~kpc leads to reduction of the flux at energies below a few GeV.

\subsection{Inverse Compton Emission}
\lb{sec:IC}

IC emission is subdominant at GeV energies with respect to the gas-correlated components, especially near the Galactic plane.
Its spatial distribution is expected to be smooth, but depends on indirect knowledge of the ISRF and the calculated distribution of CR electrons.
As a result, the IC emission is very difficult to model, especially near the GC, which can lead to significant uncertainties in the GC excess.
In the Sample Model we use three IC components corresponding to the three seed ISRF components (CMB, starlight and infrared), which are fitted to the gamma-ray data in independent energy bins. This procedure reduces the impact of our imperfect knowledge of the ISRF and CR electron spatial/spectral distribution.
As an alternative approach,
here we use a combined (i.e., summed over the three ISRF spectral bands) IC emission template, and we split the total IC emission in five Galactocentric rings with the same boundaries as the gas templates\footnote{To produce the IC templates split in Galactocentric rings we have used the development branch of the GALPROP code available from \url{http://galprop.stanford.edu/}, as described in~\citet{2016ApJ...819...44A}.}.
The results are shown in Figure \ref{fig:GC_escess_vars} middle row, right. 
We find that there is a significant reduction of the GC excess flux between the models with combined ISRF components compared to models with ISRF components separated in different templates.
We confirm that the IC emission can have a strong effect on the GC excess, which was previously discussed in, e.g., \citet{2016ApJ...819...44A}.

\subsection{Gas Maps Derived with Starlight Extinction Data and Planck and GASS Maps}
\label{sec:gasunc}

Uncertainties in the 3D models of the gas distribution in the Galaxy are important contributors to the uncertainties in the models of diffuse gamma-ray emission. In addition to the different values of the \hi spin temperature considered in \S\ref{sec:galprop_vars}, 
in this section we explore uncertainties due to 1) method used to partition the gas along the line of sight, as well as
2) uncertainties related to the input interstellar tracer data and their angular resolution.

In the construction of the maps of interstellar gas used in Sample Model, Doppler shifts of the \hi and CO lines are used to partition the gas column densities in annuli along the line of sight under the assumption of circular velocity around the Galactic center. This method is not applicable toward the Galactic center (and anticenter). For the Sample Model, the gas contents of the ring maps in this range are interpolated and renormalized as described in Appendix B of \citet{FermiLAT:2012aa}. Also, in the case of CO, the line widths toward the Galactic center are stretched by large noncircular motions \citep[e.g.,][]{2001ApJ...547..792D}.
Therefore, all gas traced by CO at high velocities is assigned to the innermost ring based on the assumption that it is in the central molecular zone (CMZ).  Furthermore, the \hi maps in \citet{FermiLAT:2012aa} are augmented to incorporate dark neutral gas, i.e., neutral gas that is not traced by the combination of the \hi and CO lines \citep{2005Sci...307.1292G}. At low latitudes, where more than one ring map can have substantial column densities of interstellar gas, the inferred column densities of dark gas are distributed proportionally to the relative H~\textsc{i} column densities in the rings.  As noted in \cite{FermiLAT:2012aa} the correction for the dark gas component was limited to directions with $E(B - V) < 5$ mag, and special procedures were applied to handle large negative corrections. These considerations make the inferred column densities in the inner Galaxy  relatively uncertain in the Sample Model.

Although \citet{FermiLAT:2012aa} assessed that this did not change the results of their large-scale analysis, we investigate here the impact on the properties of low-level residual emission seen toward the Galactic center. Hence, we developed an alternative procedure to partition the gas column densities in the region at $|\ell|<10\arcdeg$ (Appendix \ref{app:SLext})
that employs complementary information from starlight (SL) extinction due to interstellar dust. Dust grains are thought to be well mixed with gas in the cold and warm phases of the interstellar medium, hence to be tracing the total (atomic and molecular) column densities \citep[e.g.,][]{bohlin1978}. \citet{marshall2006} proposed a method that combined infrared surveys with stellar population synthesis models to derive the distribution of dust in 3D. For our test we use the maps for the region of the Galactic bulge derived using this method by \citet{schultheis2014} based on the VISTA Variables in the Via Lactea survey\footnote{\url{https://vvvsurvey.org/}} \citep{2010NewA...15..433M}.

To further investigate the uncertainties related to the datasets used to build the gas maps, we considered alternative datasets that became recently available, and, among other advantages, provide superior angular resolution. The \hi maps in \citet{FermiLAT:2012aa} are based on the LAB survey \citep{2005A&A...440..775K} that has an effective angular resolution of $\sim$0$\fdg6$ which is larger than the LAT angular resolution at energies $\gtrsim 2$~GeV.
We produced alternative high-resolution maps by using the GASS survey \citep{gass2010}. 
In the region of the sky where GASS data are available, including around the Galactic center, they provide an angular resolution of $\sim 0\fdg25$.

Furthermore, the dark-gas correction in \citet{FermiLAT:2012aa} was based on the dust reddening map by \citet{Schlegel:1997yv}.
The map by \citet{Schlegel:1997yv} traces dust reddening based on dust thermal emission measured using \textit{IRAS} and corrected for temperature variations using data from \textit{COBE}/DIRBE. The latter has an angular resolution of $\sim$1$\arcdeg$. For the alternative high-resolution maps 
we applied the same correction as in \citet{FermiLAT:2012aa}, but based instead on the dust extinction map from \citet{2014A&A...571A..11P} that is built using {\it IRAS} and \Planck data (\Planck public data release \texttt{R1.20}). The limit in reddening of 2 or 5 magnitudes, above which no correction is applied in \citet{FermiLAT:2012aa}, in our case was replaced by a comparable limit of $3 \times 10^{-4}$ in dust optical depth at 353~GHz.

The effect of the alternative gas maps on the GC excess spectrum is shown in
Figure \ref{fig:GC_escess_vars} bottom row, left. While the high-resolution gas maps only modestly change the excess spectrum, the alternative procedure to divide the gas into Galactocentric rings yields an increase in the excess flux below a few GeV. The latter also results in more plausible spectra for the gas rings between 1.5~kpc and 3.5~kpc that are set to zero in several energy bins in the \SM (\ref{sec:sample_model}). The derivation of the gas distribution from dust extinction, however, has its own set of systematic and modeling uncertainties, which requires further investigation beyond the scope of the current analysis. Therefore, here we only use it to estimate the possible effect on the GC excess.

\subsection{Additional Sources of Cosmic Rays near the Galactic Center}
\lb{sec:GC_CR_sources}

The large-scale distributions of CR sources that we consider
peak at a few kpc from the GC, in correspondence with the so-called molecular ring and the main segment of the Scutum-Centaurus spiral arm. Some of them, notably the distribution used for the Sample Model,
go to zero at the GC. However, there is evidence that a source of CR up to PeV energies exists at the GC \citep{2016Natur.531..476H, 2017arXiv170201124G}.
\citet{2015JCAP...12..056G} and \citet{2015arXiv151004698C} argued that taking into account additional sources near the GC may substantially change the significance and spectrum of the GC excess. We test two additional steady sources of CR: one associated with the bulge/bar in the central kpc of the Milky Way and one associated with the central molecular zone (CMZ) in the innermost few hundred pc.

The stellar population of the Galactic bulge is older than 5 Gyr (e.g., \citealt{2012A&A...538A.106R}, and references therein).
CRs that were accelerated by SNRs associated with the star-formation activity in the bulge have either escaped the Galaxy
or lost their energy (in the case of electrons).
However, a possible source of CRs at the present time in the bulge is a population of MSPs that could potentially accelerate electrons and positrons to hundreds of GeV \citep[e.g.,][]{2015JCAP...02..023P}.
To model a possible population of MSPs in the bulge we assume that their distribution is traced by the old stellar population in the bulge
which we take from
\citet{2012A&A...538A.106R}. 
We parametrize the bulge as an ellipsoid with  an orientation of $7.1^\circ$ with respect to the Sun-Galactic center direction 
\citep[model B in][]{2012A&A...538A.106R}. 

As an alternative, we consider a second possible population of CR sources distributed like the dense interstellar gas in the CMZ.
The CMZ contains very dense molecular clouds which can host intensive star formation
\citep[e.g.,][]{2013MNRAS.429..987L},
and, as a result, a significant rate of supernovae explosions.
The star formation rate (SFR) is rather uncertain in the CMZ and can vary from a few percent 
of the total SFR in the Galaxy if traced by the free-free emission
\citep{2013MNRAS.429..987L} up to 10--13\% if traced by young stellar objects \citep{2009ApJ...702..178Y, 2012A&A...537A.121I}
or Wolf-Rayet stars \citep{2015MNRAS.447.2322R}.

As a tracer of the CR production in the CMZ we use the distribution of molecular gas,
which we model by 
a simplified axisymmetric version of Equation (18) in~\citet{2007A&A...467..611F}. The radial distribution is described as 
\be
f(R) \propto \exp(-(R/L_c)^4) \times \exp(-|z|/H_c)
\ee
where $R$ is the radial distance from the GC and $z$ is the height above the Galactic plane. The two scaling factors were chosen to be $L_c = 137$\,pc and $H_c = 18$\,pc.

The additional source distributions, illustrated in Figure~\ref{fig:CRsources}, are implemented in the GALPROP code to calculate the resulting gamma-ray emission. Owing to large uncertainty in their contributions, we treat these components
independently from the rest of the templates.
In the case of the bulge source, we add the IC emission from the additional electrons and positrons as an extra component together with the components of the Sample Model in the fit to the data.
In the case of the CMZ source, we add the IC emission template and the gas-correlated components associated with 
\HI and \Htwo in the first four rings in the Sample Model (Section \ref{sec:baseline}) omitting the outer ring, i.e., we use 4 out of 5 rings in the Sample Model
(nine additional parameters in each bin relative to the Sample model).

Throughout our paper we are using GALPROP to model the particle propagation in two dimensions (for cylindrical symmetry in the Galaxy) with a 1 kpc resolution in radius and 100 pc resolution perpendicular to the disk. 
To have a more accurate description of the CR distribution in the CMZ case, 
we perform some 3D GALPROP runs for the CMZ source distribution with a resolution of 100~pc in all coordinates
(the difference of the GC excess spectra for 2D and 3D GALPROP runs is less than about 2--3$\sigma$ statistical uncertainties around a few GeV).
For the CMZ source, we test different sizes of the propagation halo $z = $ 2, 4, 8 kpc and $R = 10$ kpc.
The rest of the components are derived with a 2D GALPROP calculation the same halo height and $R = 20$ kpc.
The results are shown in Figure \ref{fig:GC_escess_vars} bottom, right. 
The CMZ source of CRs with $z = $ 2, 4 kpc propagation halo height has little effect on the GC excess flux.
The CMZ source of CRs with $z = 8$ kpc has a significant effect on the GC excess spectrum at energies below $\sim 4$ GeV,
while the bulge source of CR takes up a significant part of the GC excess around $\sim 10$ GeV.

\section{Spectral Components Analysis of the \Fermi Bubbles and the Galactic Center Excess}
\label{sec:SCA}

An important source of uncertainty in the derivation of the GC excess is the contribution to the emission near the GC from the \Fermi bubbles. 
The bubbles do not have a clear counterpart in other frequencies that can be used as a  template.
As a result, neither the spectrum nor the shape of the bubbles is known near the GC.
Above $|b| = 10\degr$ the spectrum of the bubbles is approximately uniform as a function of the latitude
\citep{2010ApJ...724.1044S, 2014ApJ...793...64A}.
Therefore, in this section we will assume that the spectrum of the bubbles at low latitudes is the same as at high latitudes in a limited energy range,  between 1 GeV and 10 GeV. 
Based on this assumption, we will derive an all-sky template for the \Fermi bubbles in Section \ref{sec:FBtmpl}.

Then,  in Section \ref{sec:3comp_sca}, we will derive a template for the GC excess itself, using the same technique, and based on different assumptions on its spectrum. We will consider an MSP-like spectrum, since a population of MSPs is expected to contribute to gamma-ray emission near the GC  \citep{2011JCAP...03..010A, 2013A&A...554A..62G, Gordon:2013vta, 2013MNRAS.436.2461M, 2014JHEAp...3....1Y,  2015JCAP...02..023P, 2015ApJ...812...15B, 2016JCAP...08..018H}, as well as estimates of the GC excess spectrum from earlier works.

\subsection{\Fermi Bubbles Template}
\lb{sec:FBtmpl}

We derive the \Fermi bubbles template using
the spectral components analysis (SCA) procedure used to extract the \Fermi bubbles component at high latitudes in
\citet{2014ApJ...793...64A}.
In this derivation we will use the ROI $|b| < 60\degr$, $|\ell| < 45\degr$.

\subsubsection{Subtraction of Gas-correlated Emission and Point Sources from the Data}
\lb{sec:gas_PS_model}

The first step in modeling the \Fermi bubbles is to subtract the gas-correlated emission and PS from the data. 
We fit the data with a combination of gas-correlated emission components, PS template obtained by adding 3FGL point sources (the overall normalization is free in each energy bin),
and a combination of smooth templates.
The smooth components are introduced as a proxy for the other components of emission, such as IC, \Fermi bubbles, Loop I, extended sources,
GC excess. 
They are required to avoid biasing the determination of the contribution from the gas-correlated templates in the fit.
As a basis of smooth functions we use spherical harmonics (calculated using the HEALPix package, \citealt{2005ApJ...622..759G}).
The general basis of smooth functions makes it possible to model non-gas-related emission
without a pre-defined template.

\begin{figure}[htbp]
\begin{center}
\includegraphics[scale=\onepic]{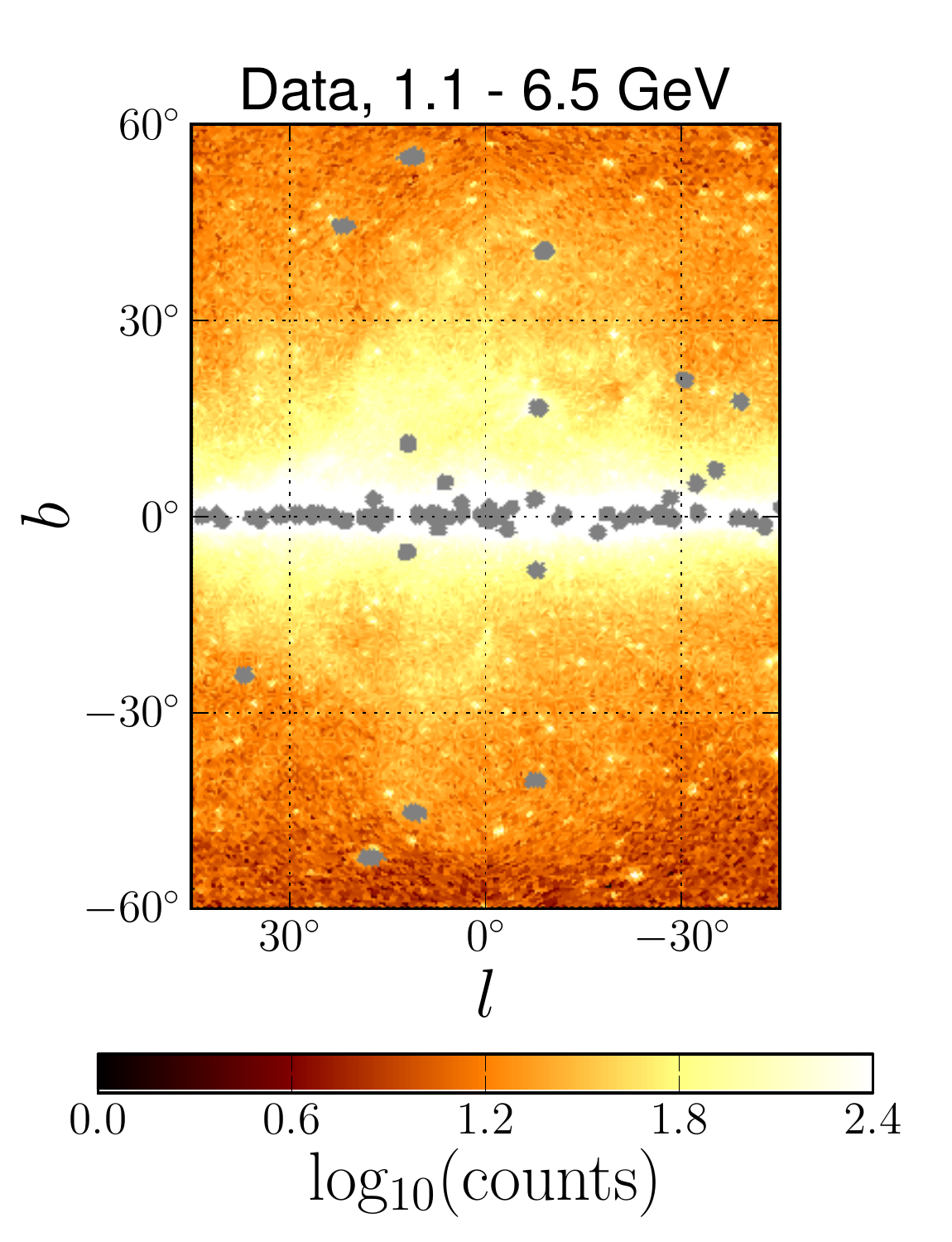}
\includegraphics[scale=\onepic]{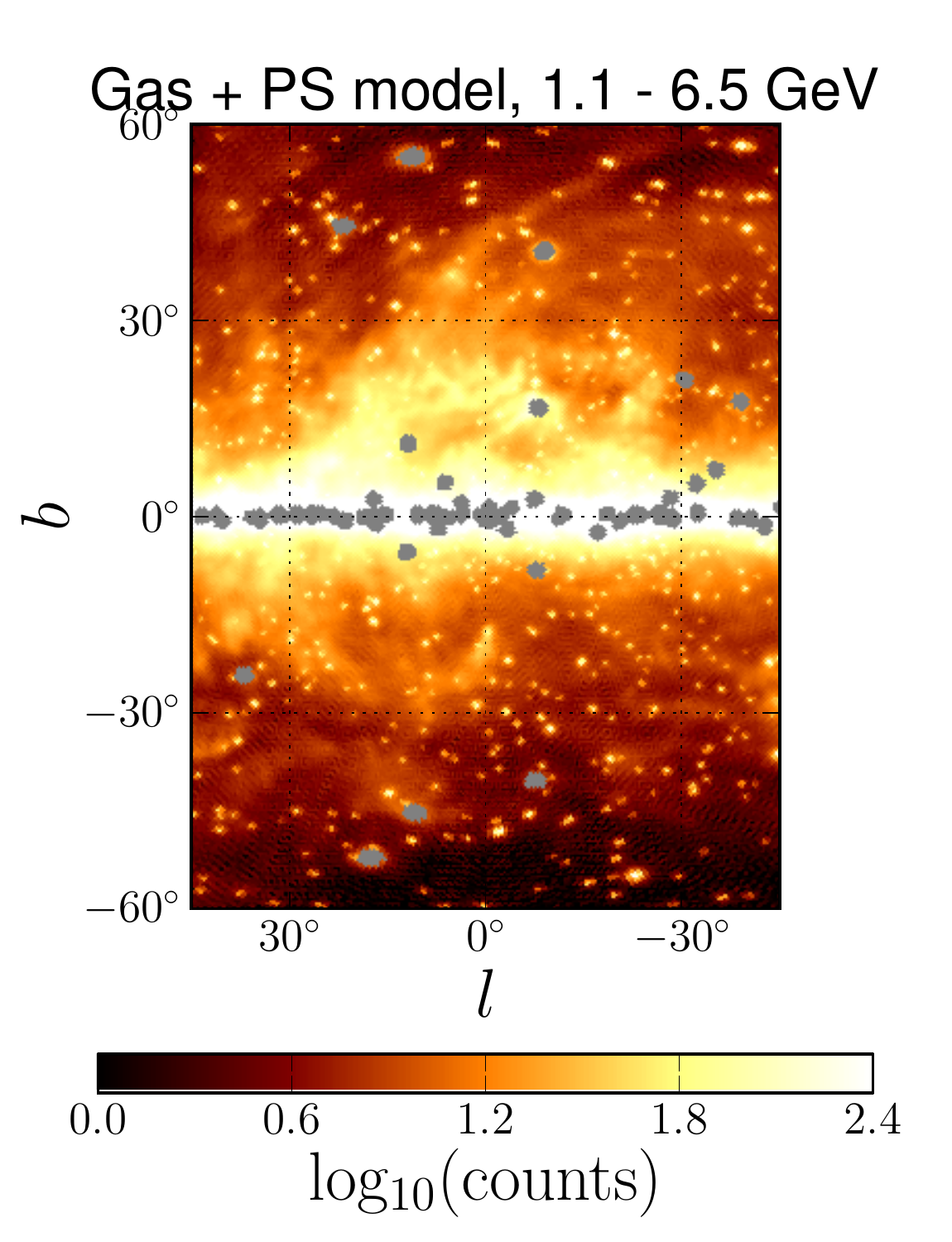} \\
\includegraphics[scale=\onepic]{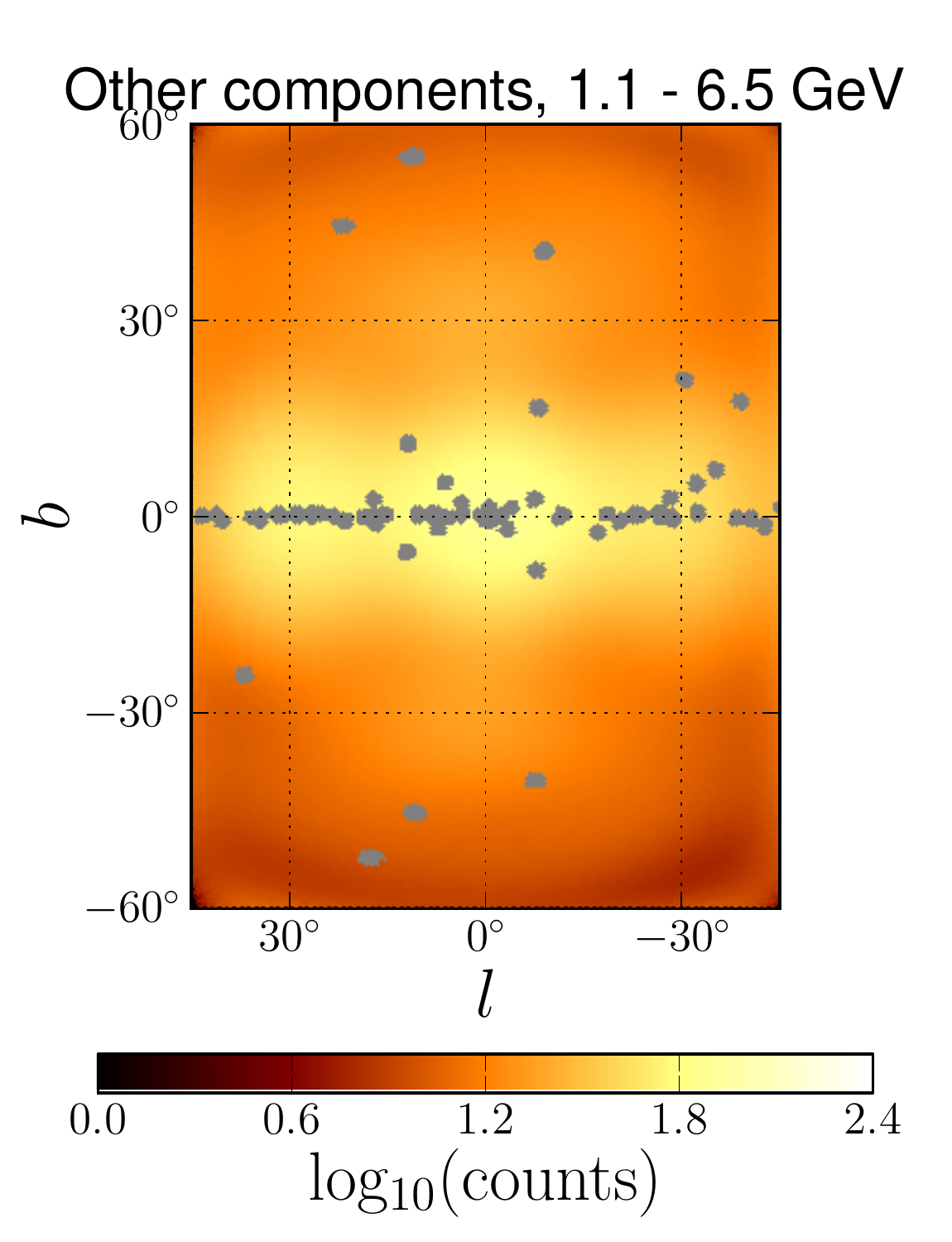}
\includegraphics[scale=\onepic]{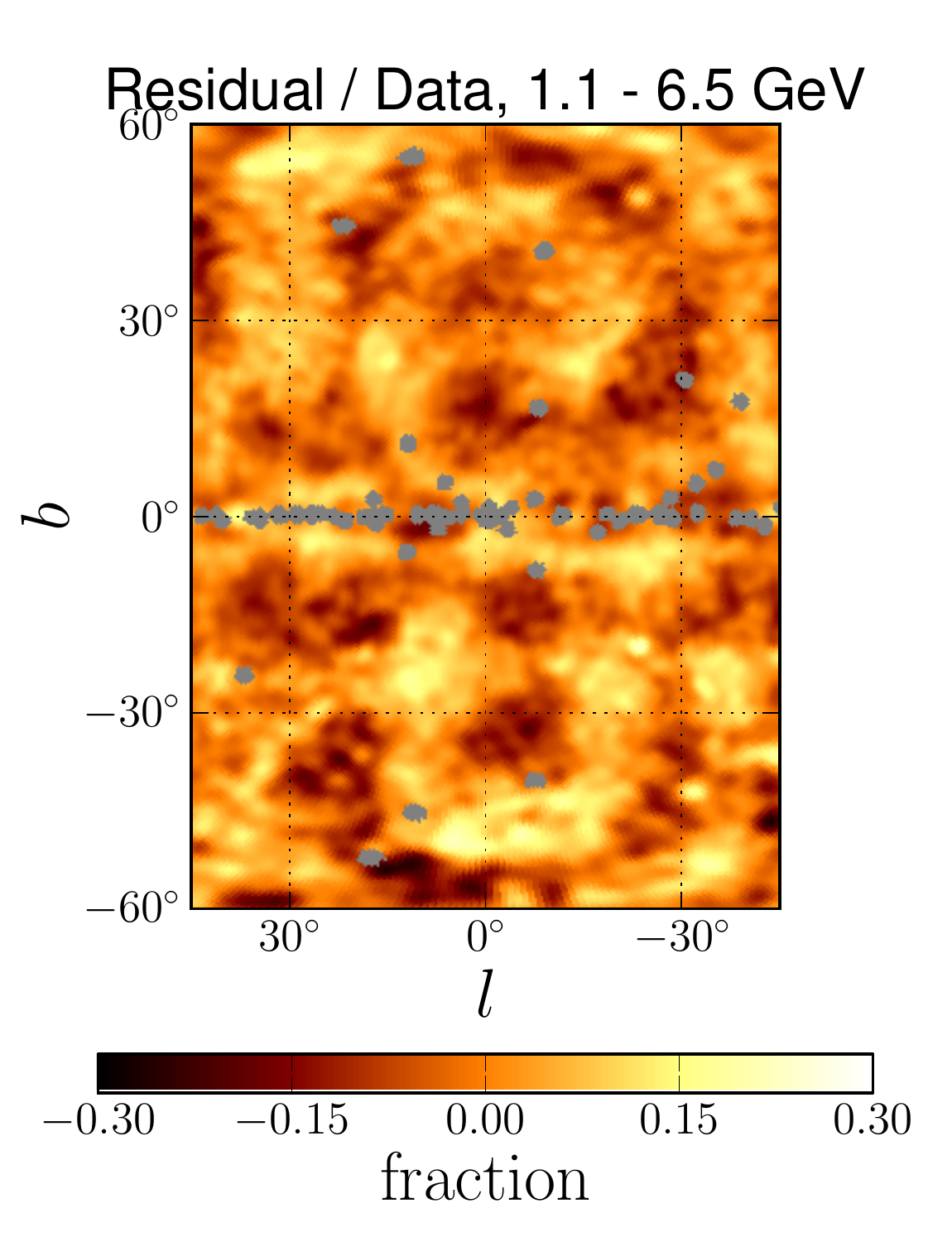}
\noindent
\caption{\small 
Example of modeling the data by a combination of gas-correlated components, 
PS and spherical harmonics.
Top left: data summed in energy bins between 1.1 GeV and 6.5 GeV. Top right: a combination of gas-correlated components and PS.
Bottom left: a combination of the 30 most-significant spherical components out of the first 100 ($Y_{l m}$ with $l \leq 9$) 
that model the remaining components of gamma-ray emission.
Bottom right: residual after subtraction of gas-correlated, PS, and spherical harmonics model from the data
as a fraction of the data counts.
}
\label{fig:harmonics}
\end{center}
\end{figure}

As a basis of smooth templates, 
we select the 30 spherical harmonics that provide the largest improvement in likelihood out of the first 100, 
i.e., $Y_{l m}(\theta, \vp)$ with degree $l \leq 9$ (angular resolution $\approx 20^\circ$).
In Section \ref{sec:Fermibubbletemplate} we will 
test the consistency of the derivation by selecting the 60 most-significant harmonics out of the first 225 
(degree $l \leq 14$, angular resolution $\approx 14^\circ$)
and the 90 most-significant harmonics out of 400 (degree $l \leq 19$, angular resolution $\approx 10^\circ$). 
An example of data fitting by a combination of gas-correlated emission components, PS and spherical harmonics
is shown in Figure \ref{fig:harmonics}.

To speed up the calculations in this sub-section,
we use a quadratic approximation to the log likelihood in Equation (\ref{eq:chi2})
\be
\lb{eq:chi2_quad}
\chi^2 =  \sum_{i} \frac{(d_{i} - \mu_{i})^2}{\sm^2_{i}},
\ee
where $d_{i}$ is the photon counts in pixel $i$ and $\mu_{i}$ is the model counts.
The statistical uncertainty is calculated by smoothing the data count maps in each energy bin, $\sm^2_{i} = \tilde{d}_{i}$, 
to avoid bias in using either data or the model as an estimator of standard deviation \citep[see, e.g., Appendix A in][]{2014ApJ...793...64A}.
The smoothing radius $R$ is chosen in each energy bin independently, such that there are at least 100 photons on average inside a circle of radius $R$. The minimum smoothing radius is $1\degr$, while the maximum smoothing radius is $20\degr$.
For smoothing, photon counts inside the PS mask are approximated by an average of the neighboring pixels outside the mask.

\subsubsection{Decomposition into Spectral Components}
\lb{sec:SCA_comps}

\begin{figure}[htbp]
\begin{center}
\includegraphics[scale=\onepic]{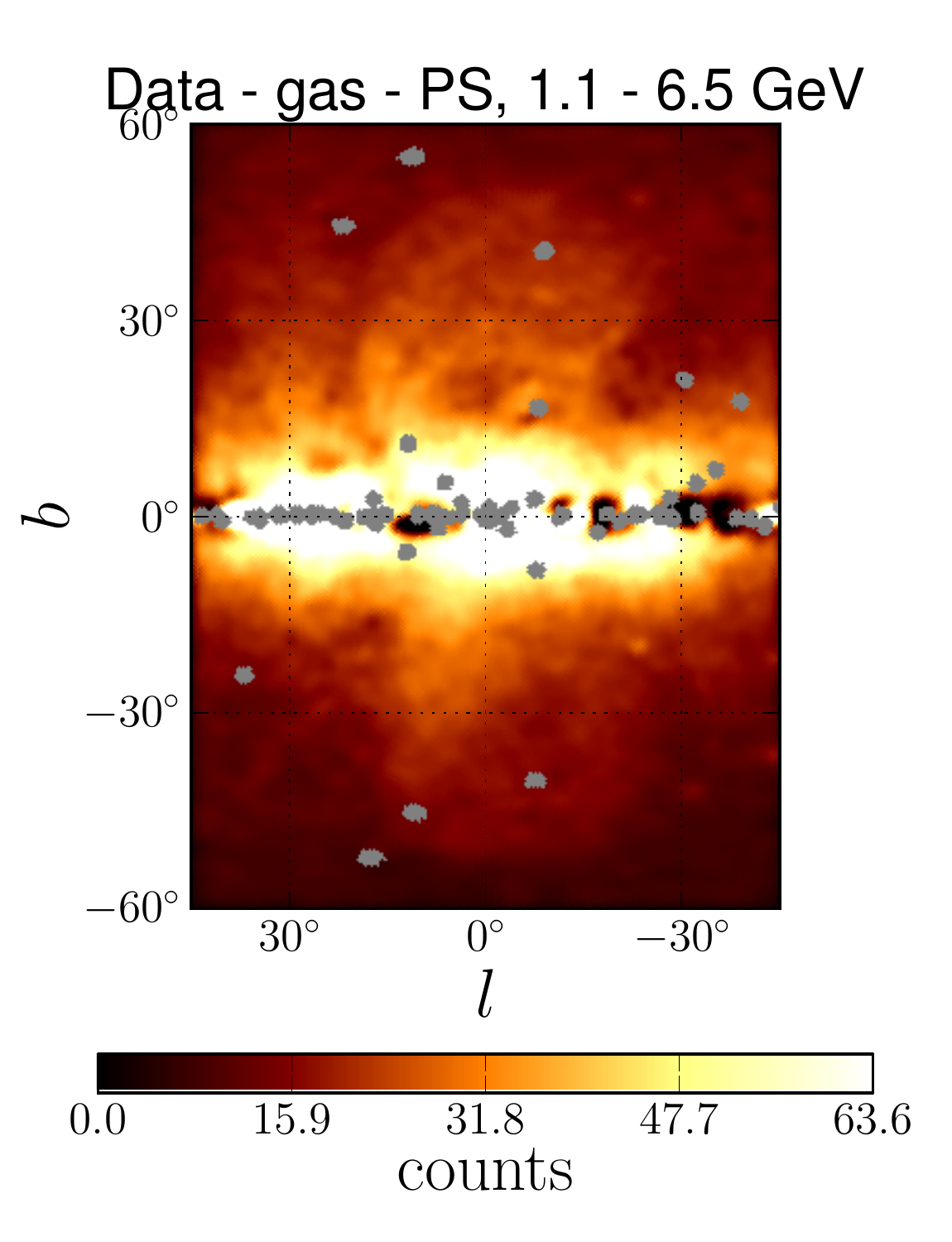}
\includegraphics[scale=\onepic]{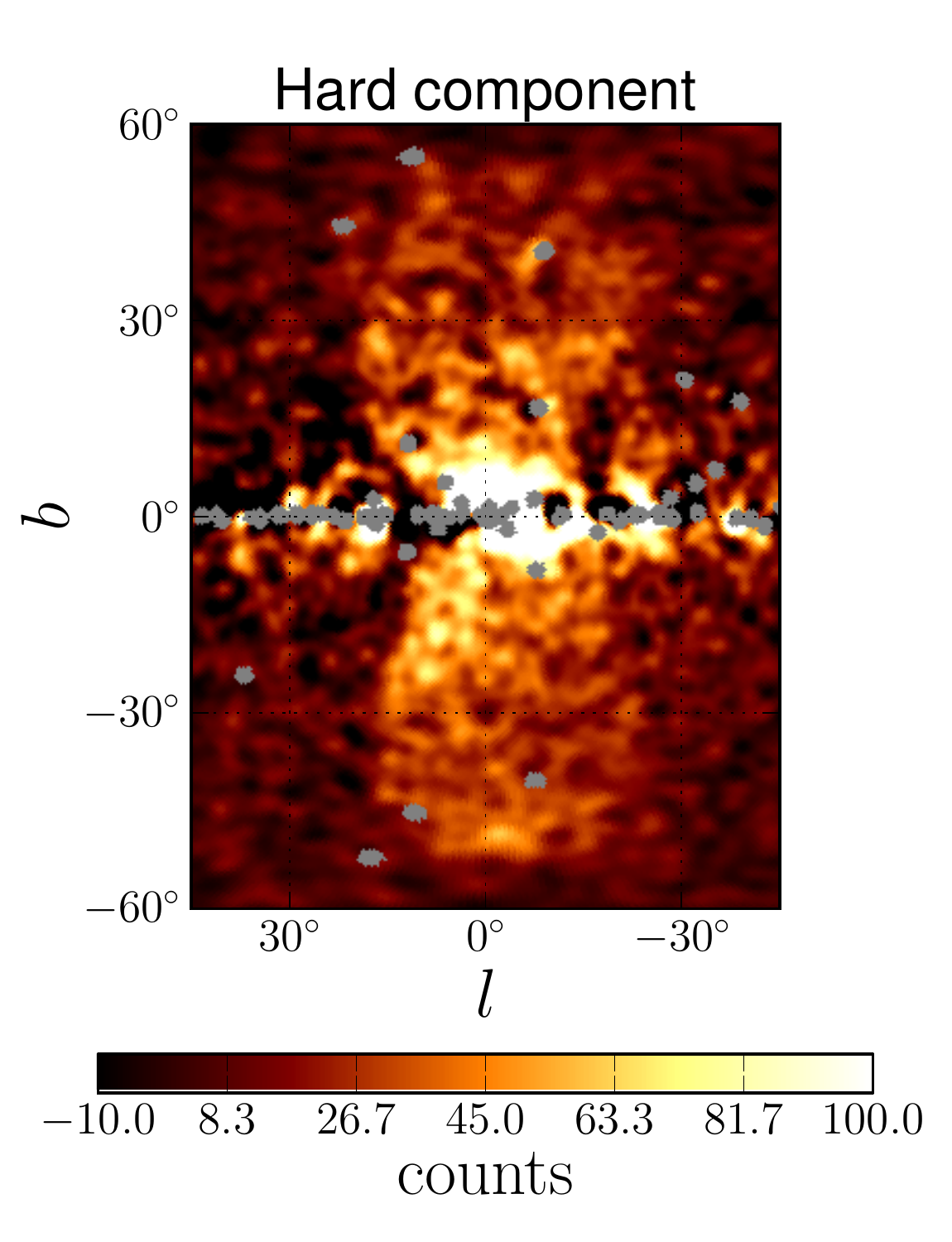}
\includegraphics[scale=\onepic]{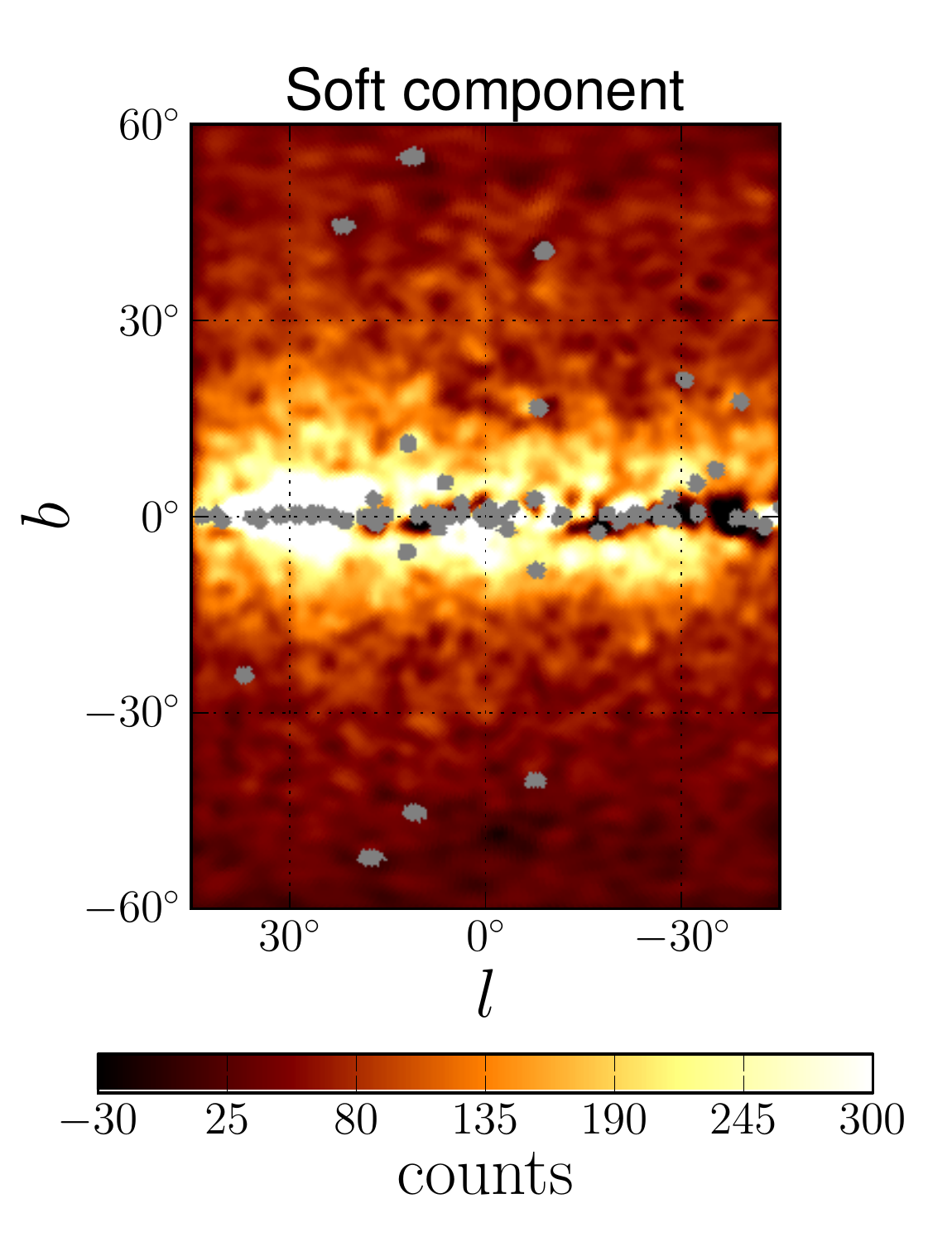}
\noindent
\caption{\small 
Decomposition of residuals after subtracting the gas-correlated emission and PS
from the data between 1 and 10 GeV into spectral components.
Left: residual after subtracting the gas-correlated components and PS from the data (Figure \ref{fig:harmonics}). 
Middle: hard spectral component correlated with spectrum
$\propto E^{-1.9}$ -- determined from the spectrum of the \Fermi bubbles at $|b| > 10^\circ$.
Right: soft spectral component $\propto E^{-2.4}$ --
determined from fitting the sum of IC, isotropic and Loop I components in the Sample Model (Section \ref{sec:baseline}).
The hard and soft components are introduced in Equation (\ref{eq:hs_comps}).
The maps are smoothed with $1^\circ$ Gaussian kernel.
}
\label{fig:SCA_components}
\end{center}
\end{figure}

We use the results of the previous subsection to subtract the gas-correlated emission and PS from the data.
An example of the residual map summed over energy bins between 1.1 GeV and 6.5 GeV is shown in Figure~\ref{fig:SCA_components}, left.
These residuals primarily consist of the \Fermi bubbles, IC emission, isotropic background, and Loop I.

At latitudes $|b| > 10^\circ$ the bubbles have an approximately uniform spectrum 
\citep[e.g.,][]{2013PDU.....2..118H, 2014ApJ...793...64A}.
We will assume that the spectrum of the bubbles at low latitudes is the same as at high latitudes
and use the difference between the spectrum of the bubbles and that of other components to determine a template for the bubbles
at low latitudes.
The assumption about the bubbles spectrum at low latitudes is a limitation of the current method,
but as we will see below, we will only need to use the spectrum in a relatively small energy range between 1 GeV and 10 GeV.
In this energy range the \Fermi bubbles have a spectrum markedly different from the other gamma-ray emission components  (see, e.g., Figure \ref{fig:baseline_spectra}), and the LAT PSF is relatively good compared to energies below 1 GeV.

We further decompose the residuals, obtained by subtracting the gas-correlated emission and PS from the data,
as a combination of components correlated with the spectrum of the \Fermi bubbles at high latitudes
and the spectrum of the sum of IC, isotropic, and Loop I components.
Between 1 GeV and 10 GeV the high-latitude bubbles spectrum is well fit by a power law $\propto E^{-1.9}$,
while the sum of IC, isotropic, and loop I components obtained in the Sample Model (Section \ref{sec:baseline})
is $\propto E^{-2.4}$.
In this section we don't include a model for the GC excess component, we will take it into account as a separate spectral component in the next section.
The model is determined as a combination of two spectral components
\be
\lb{eq:hs_comps}
M_{\al i} = (E_\al / E_0)^{-1.9} H_i + (E_\al / E_0)^{-2.4} S_i
\ee
where $\al$ is the energy bin index for energies between 1 and 10 GeV, $i$ is the pixel index, and
$H$ and $S$ are defined as a hard and a soft template.
The reference energy is taken to be $E_0 = 1$ GeV, in this case the values of $H$ and $S$ maps correspond to contribution at 1 GeV.
The maps $H_i$ and $S_i$ are found by minimizing the $\chi^2$ similar to the $\chi^2$ in
Equation (\ref{eq:chi2_quad}):
\be
\lb{eq:chi2_sca}
\chi^2 = \sum_{ 1 {\rm GeV} < E_\alpha < 10 {\rm GeV}} \sum_i^{\rm pixels} \frac{(R_{\al i} - M_{\al i})^2}{\sm^2_{\al i}},
\ee
where $R_{\al i}$ are the residual maps obtained by subtracting gas-correlated emission and PS from the data 
(Figure~\ref{fig:SCA_components}, left).
The statistical uncertainty $\sm^2_{\al i}$ is estimated from smoothed counts maps (Section~\ref{sec:data_fitting}).
We represent the maps $H_i$ and $S_i$ as linear combinations of residual maps: 
$H_i = \sum_\al f^\al_{\rm H} R_{\al i}$ and $S_i = \sum_\al f^\al_{\rm S} R_{\al i}$.
The coefficients $f^\al_{\rm H}$ and $f^\al_{\rm S}$ are found by minimizing the $\chi^2$ in Equation (\ref{eq:chi2_sca}).
The statistical uncertainties of $H$ and $S$ are calculated by propagating the uncertainties of the maps $R_{\al i}$, e.g.,
$\sigma_{\rm H_i}^2 = \sum_\al {f^\al_{\rm H}}^2 \sigma^2_{\al i}$.
The hard ($H_i$) and the soft ($S_i$) component maps are presented in Figure~\ref{fig:SCA_components}.
The hard component primarily contains the \Fermi bubbles, and is used below to derive an all-sky model of the bubbles; the soft component contains isotropic background, IC emission, and Loop I. 

\subsubsection{Derivation of the \Fermi Bubbles Template}
\label{sec:Fermibubbletemplate}

\begin{figure}[htbp]
\begin{center}
\includegraphics[scale=\onepic]{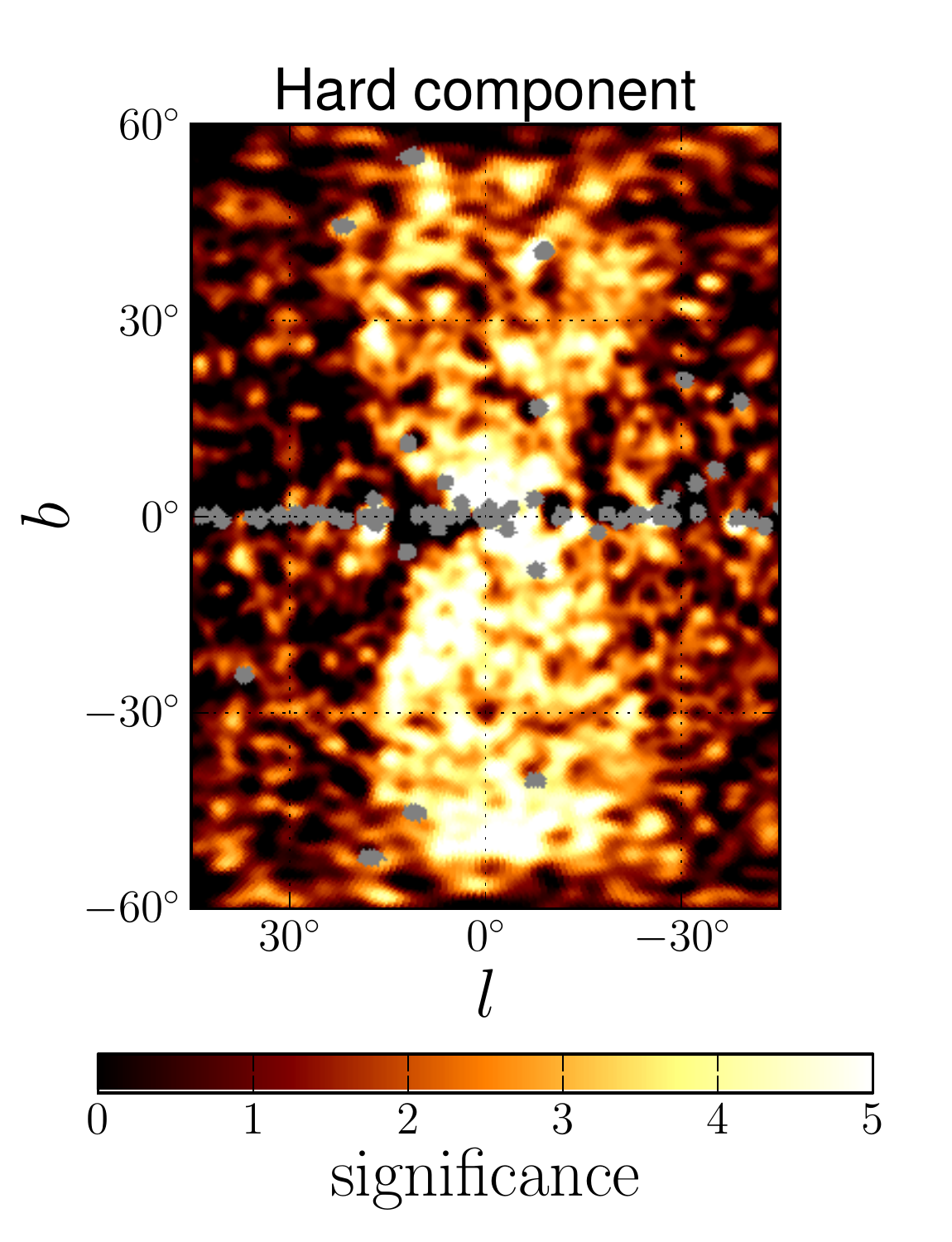}
\includegraphics[scale=\onepic]{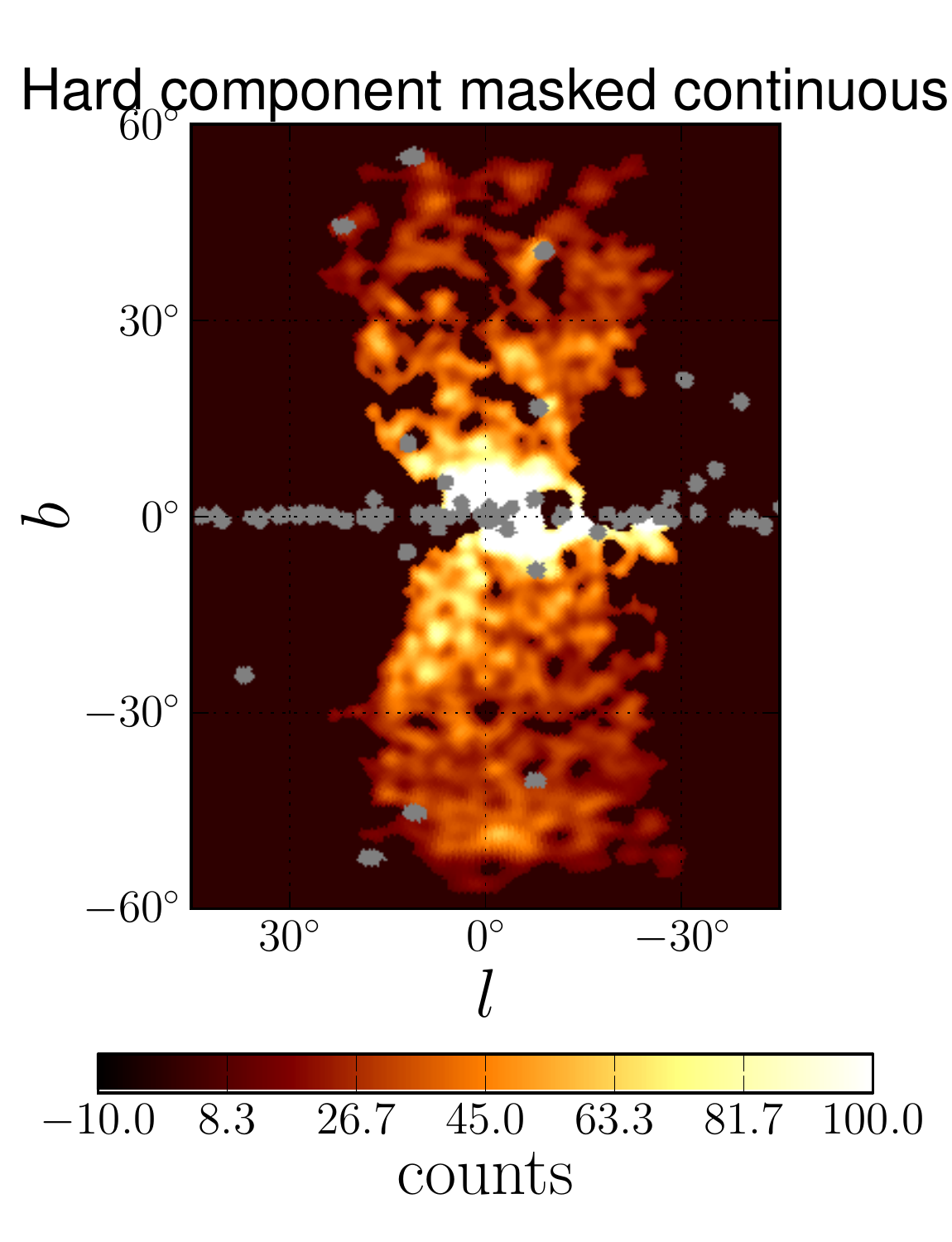} \\
\includegraphics[scale=\onepic]{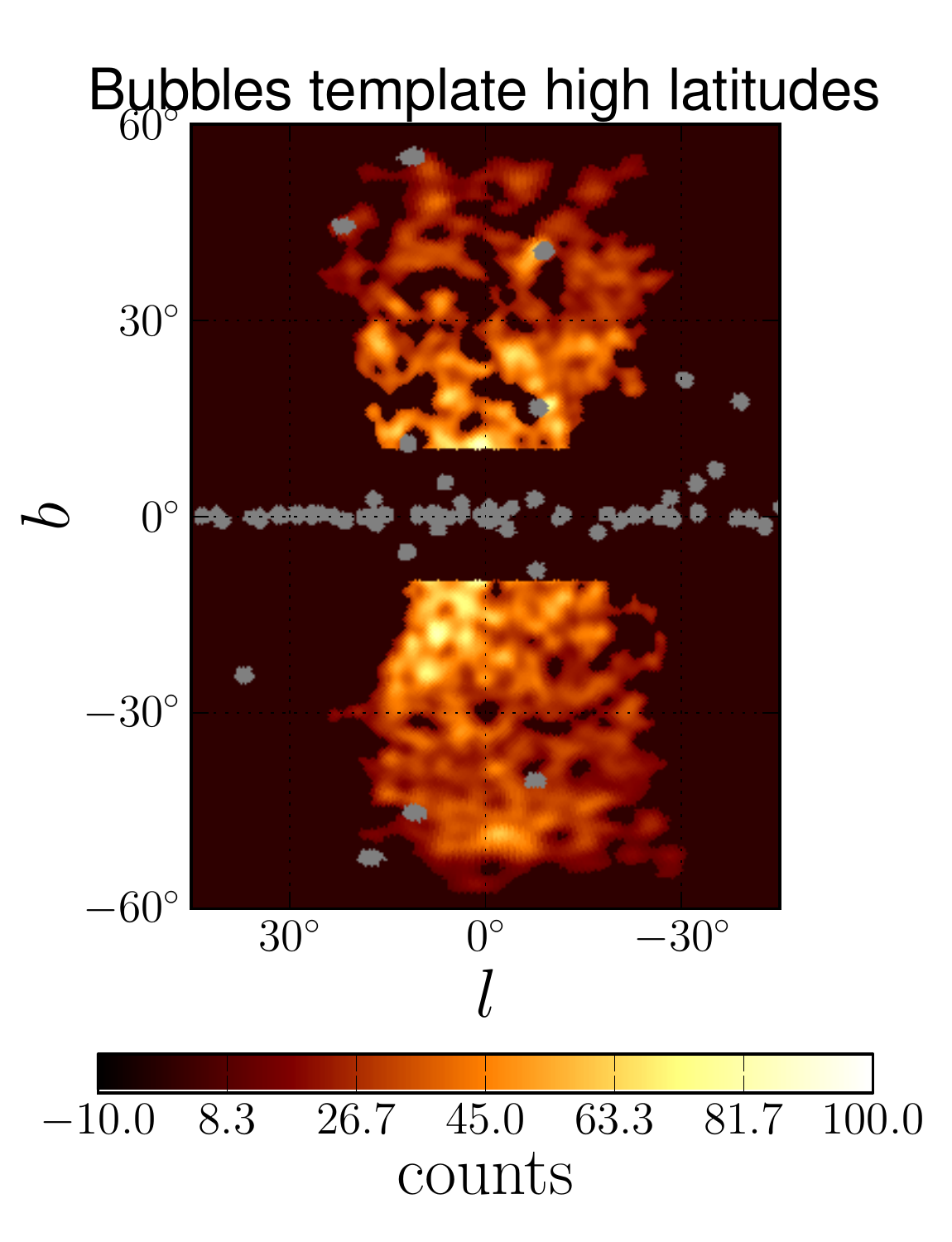}
\includegraphics[scale=\onepic]{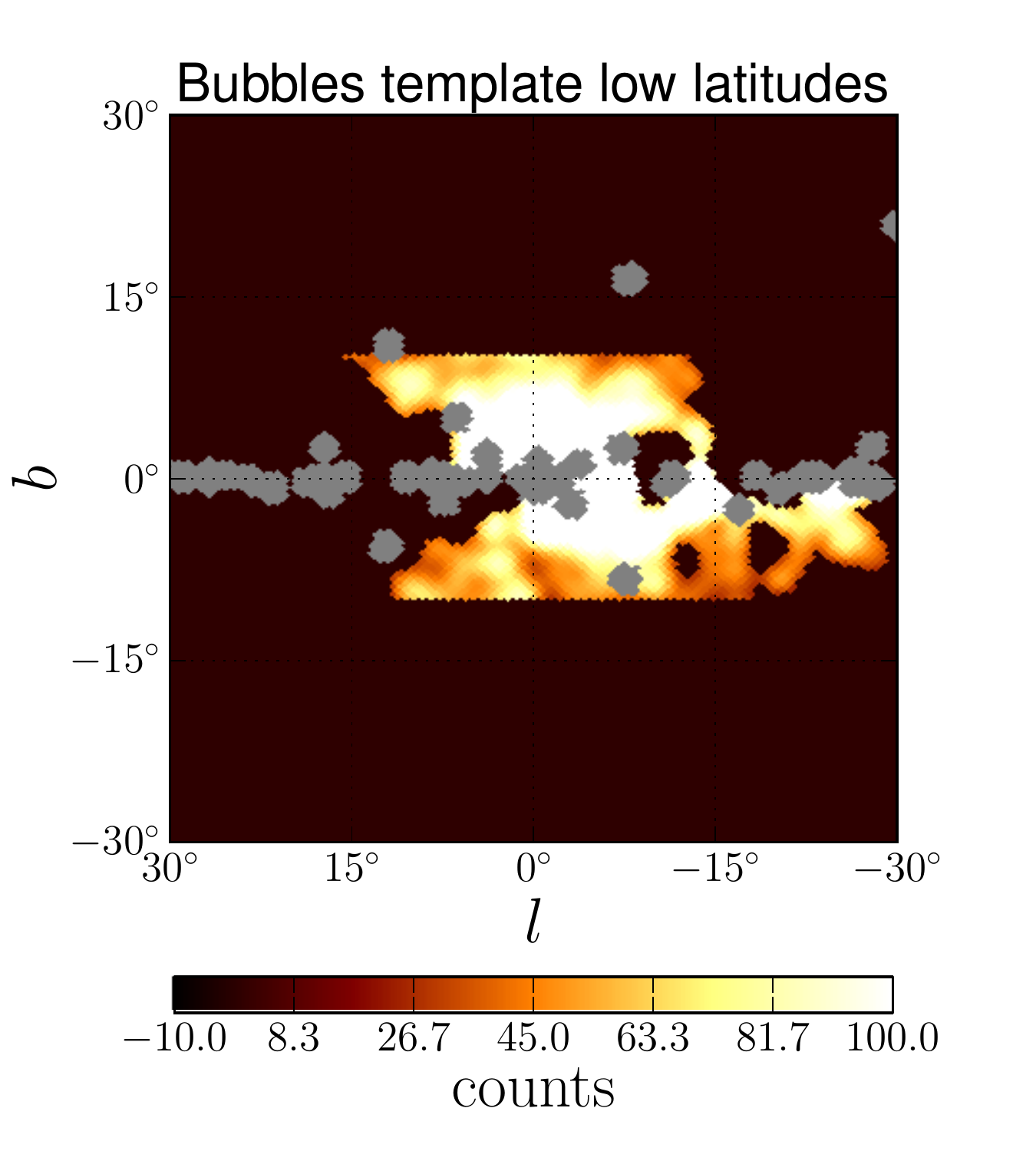} 
\noindent
\caption{\small 
Derivation of the \Fermi bubbles template at low latitudes. Top left: the hard component defined in
Equation (\ref{eq:hs_comps}) in significance units.
Top right: connected part of the hard components after applying a $2 \sigma$ cut in significance. 
Bottom left and right: \Fermi bubbles templates above and below $|b| = 10\degr$ derived 
by splitting the masked hard component in the top right plot.
}
\label{fig:SCA_bubbles}
\end{center}
\end{figure}

To derive the \Fermi bubbles template we take the map of the hard spectral 
component $\propto E^{-1.9}$ in significance units smoothed with $1^\circ$ Gaussian kernel (Figure \ref{fig:SCA_bubbles} top left)
and cut in significance at the level of $2 \sigma$ 
relative to the statistical uncertainty of the $H_i$ map discussed after Equation~\ref{eq:chi2_sca}.
As one can see in Figure \ref{fig:SCA_bubbles}, the emission from the bubbles has a high significance and the cut at $2 \sigma$ level
keeps most of the area of the bubbles.
To eliminate the fluctuations and residuals outside of the \Fermi bubbles, 
we select only the pixels that are above the threshold and are continuously
connected to each other.
The resulting \Fermi bubbles templates are shown in Figure \ref{fig:SCA_bubbles}.

\begin{figure}[htbp]
\begin{center}
\includegraphics[scale=\twopic]{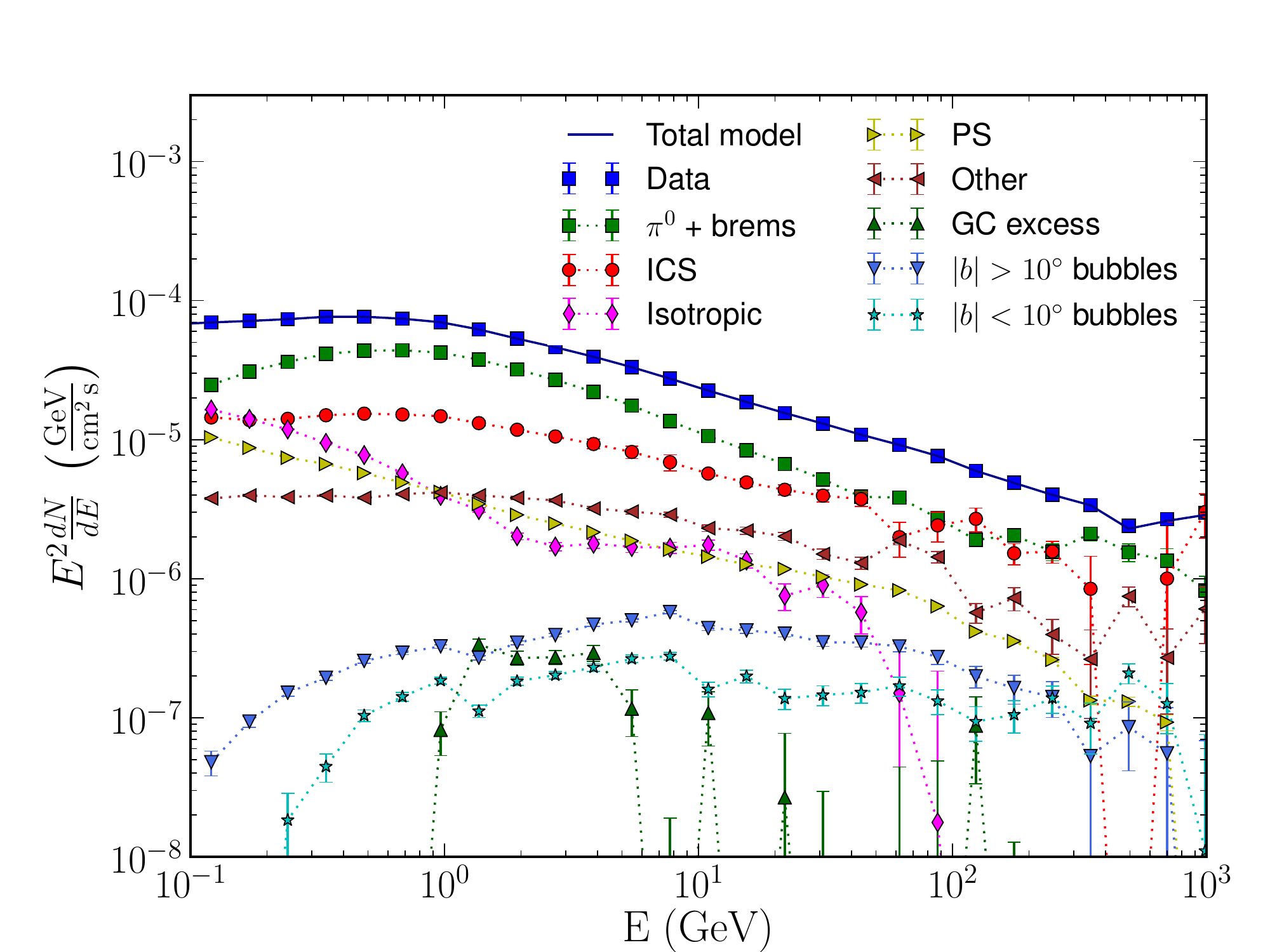}
\includegraphics[scale=\twopic]{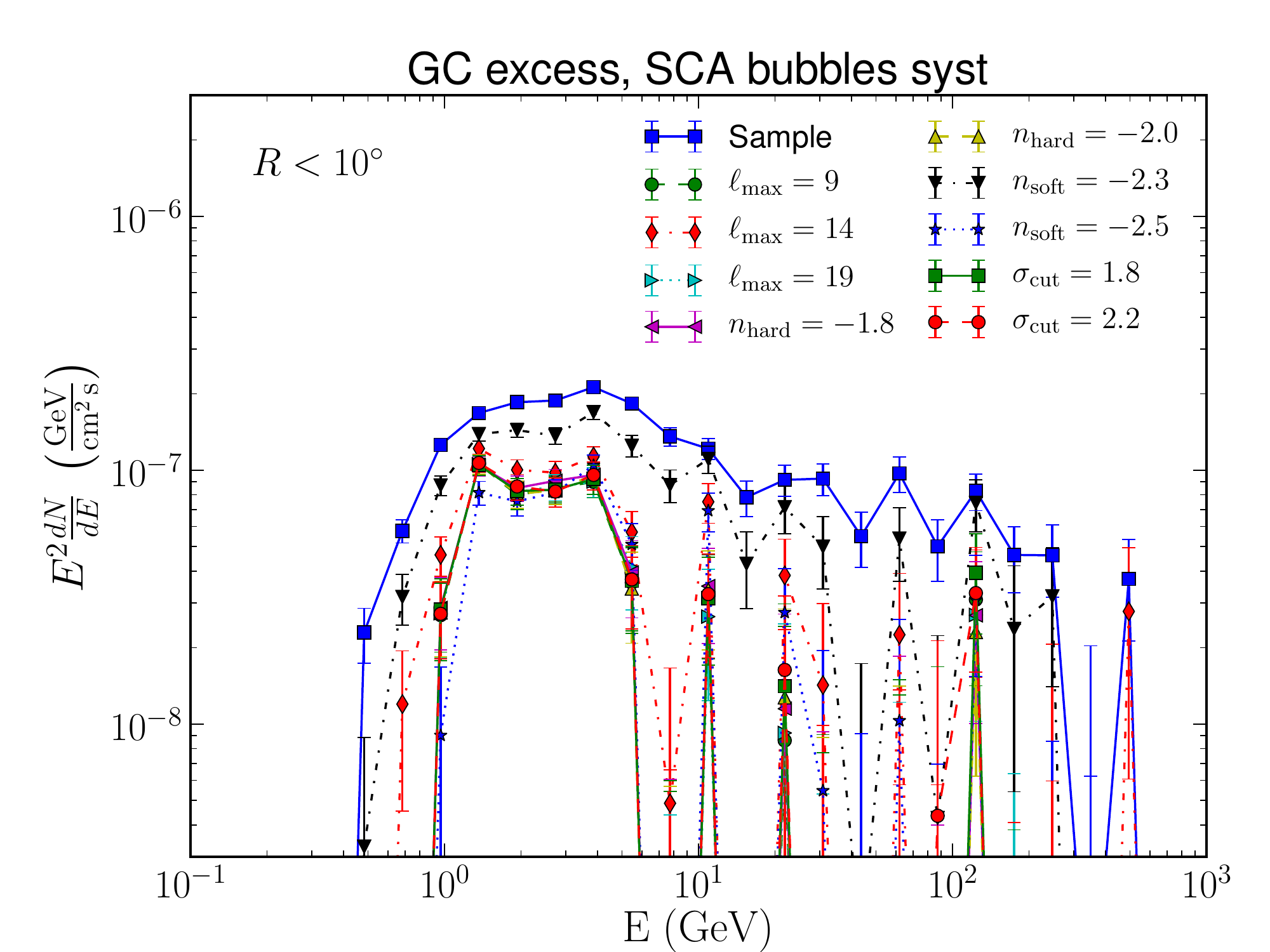}
\noindent
\caption{\small 
Components of gamma-ray emission and the GC excess spectrum in the presence of high and low-latitude
\Fermi bubbles.
Left: spectra of components; the templates are the same as in the Sample Model, except for the \Fermi bubbles
templates, which are shown in Figure \ref{fig:SCA_bubbles}.
Right: comparison of the GC excess spectrum in the presence of the high and low-latitude bubbles templates
with the Sample Model for different parameters in the determination of the bubbles template.
The main effect comes from the variation of the index of the soft component $n_{\rm soft} = -2.3$, 
all of the other alternative cases overlap and are hard to distinguish on the plot 
(see text for the definition of parameters $\ell_{\rm max}$, $n_{\rm hard}$, $n_{\rm soft}$, and $\sm_{\rm cut}$).
}
\label{fig:bubbles_spectra}
\end{center}
\end{figure}

One of the main assumptions in this derivation of the \Fermi bubbles is that their spectrum between 1 GeV and 10 GeV 
below $|b| = 10\degr$ is the same as the spectrum above $|b| = 10\degr$.
To reduce the dependence on this assumption we split the derived \Fermi bubbles template into two templates: high latitude ($|b| > 10\degr$) 
and low latitude ($|b| < 10\degr$).
The corresponding templates are shown in Figure \ref{fig:SCA_bubbles} at the bottom.
The spectra of components derived with the new \Fermi bubbles templates in the Sample Model
are shown in Figure \ref{fig:bubbles_spectra}.
The spectrum of the low-latitude bubbles is similar to the spectrum of the bubbles at high latitudes
between $\sim$ 100 MeV and $\sim$ 100 GeV, which supports the hypothesis of the homogeneous spectrum
of the bubbles as a function of latitude.
However, above 100 GeV the low-latitude bubbles spectrum continues to be hard,
while the high-latitude spectrum of the bubbles softens.

The effect of the introduction of the low-latitude bubbles template on the GC excess spectrum is shown in Figure \ref{fig:bubbles_spectra}
at the right. Note, that the \Fermi bubbles template in the Sample Model is determined only for $|b| > 10^\circ$.
The GC excess above 10 GeV is taken up by the bubbles template,
while between 1 GeV and 10 GeV the GC excess is reduced by a factor of 2 or more. 

To test the robustness of the bubbles template derivation and the effect on the GC excess flux
we also show the results for choosing different basis of smooth functions 
$\ell_{\rm max} = 9, 14, 19$ (Section \ref{sec:gas_PS_model});
different indices for the hard component $n_{\rm hard} = -1.8,\;  -2.0$,
different indices for the soft component $n_{\rm soft} = -2.3,\;  -2.5$ (Section \ref{sec:SCA_comps});
and different significance threshold in the derivation of the bubbles template $\sigma_{\rm cut} = 1.8,\; 2.2$.
The largest effect comes from the change in the soft components index $n_{\rm soft} = -2.3$.
The reason is that with harder spectrum of the soft component a part of the bubbles template is now attributed to the 
soft component. As a result, the bubbles template has a smaller area and it has a less significant influence on the GC excess flux.

\begin{figure}[htbp]
\begin{center}
\includegraphics[scale=\twopic]{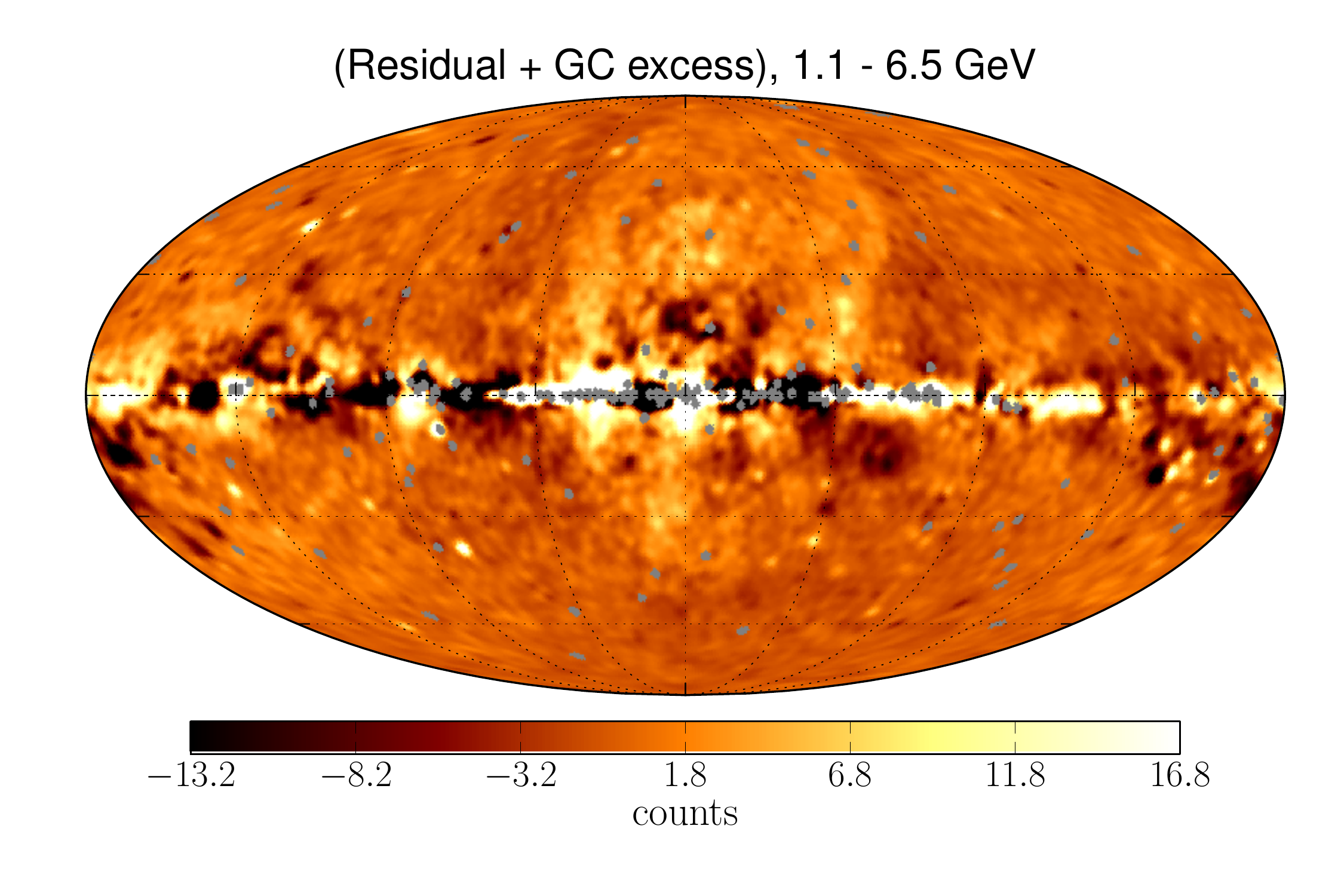}
\includegraphics[scale=\twopic]{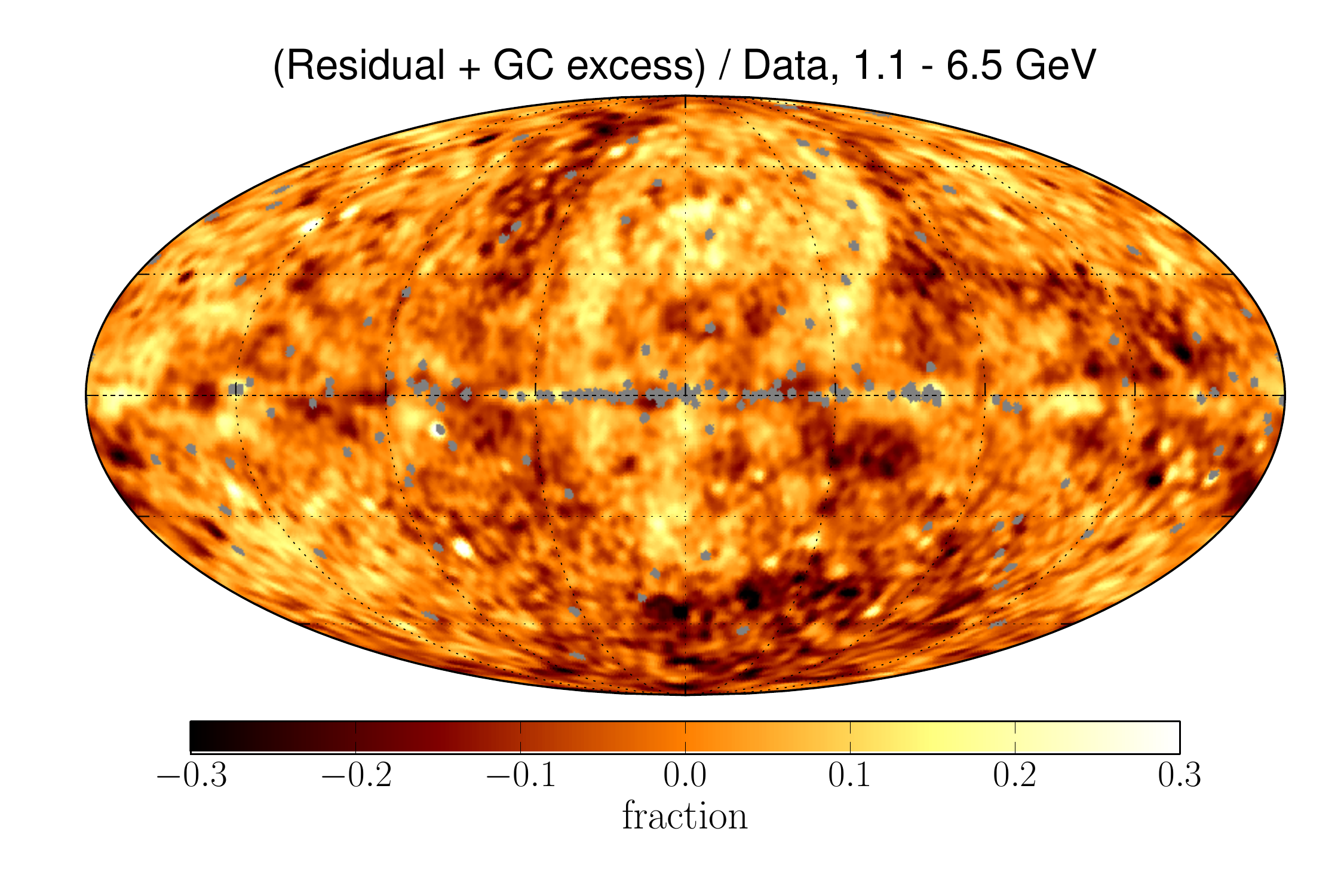} \\
\includegraphics[scale=\twopic]{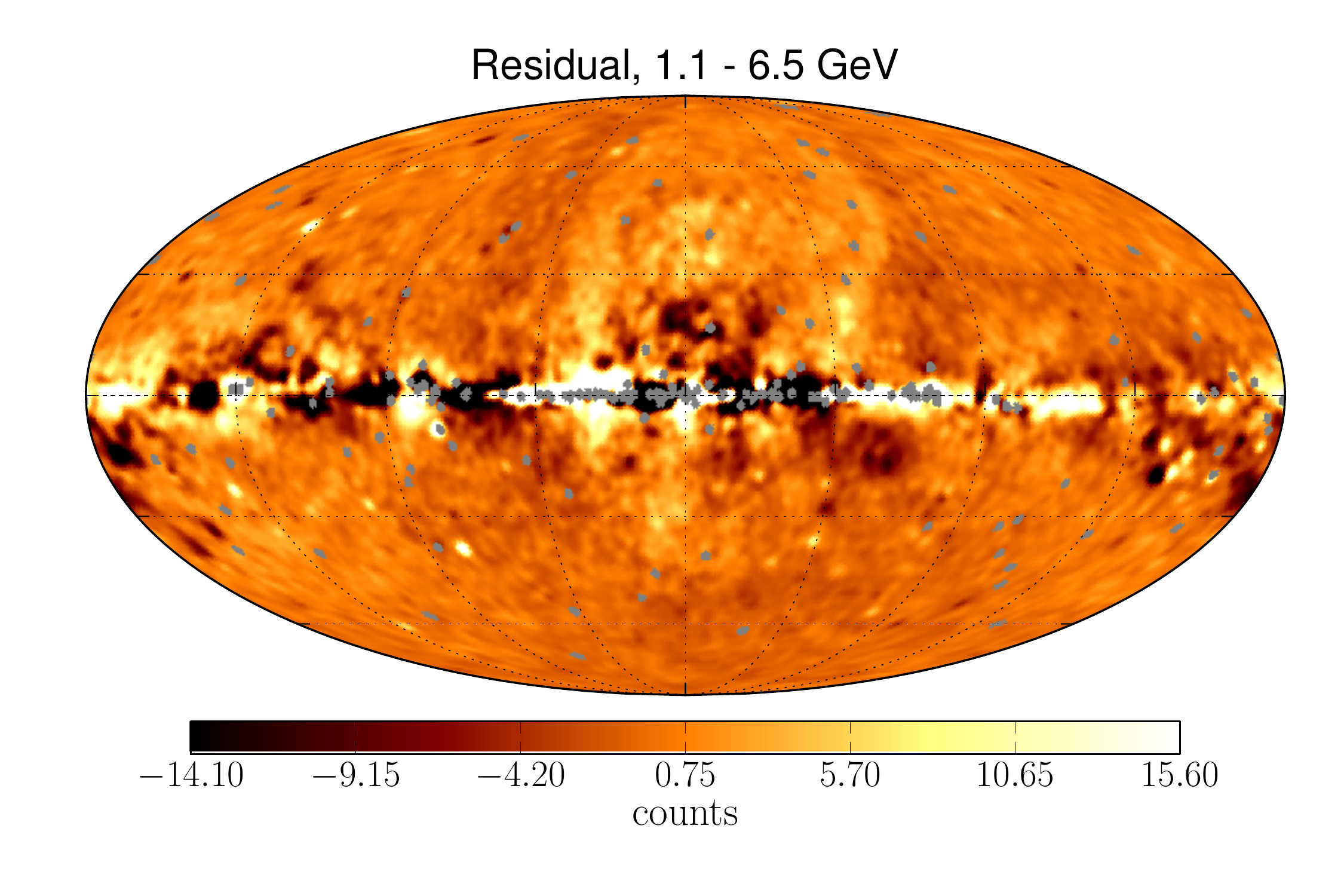}
\includegraphics[scale=\twopic]{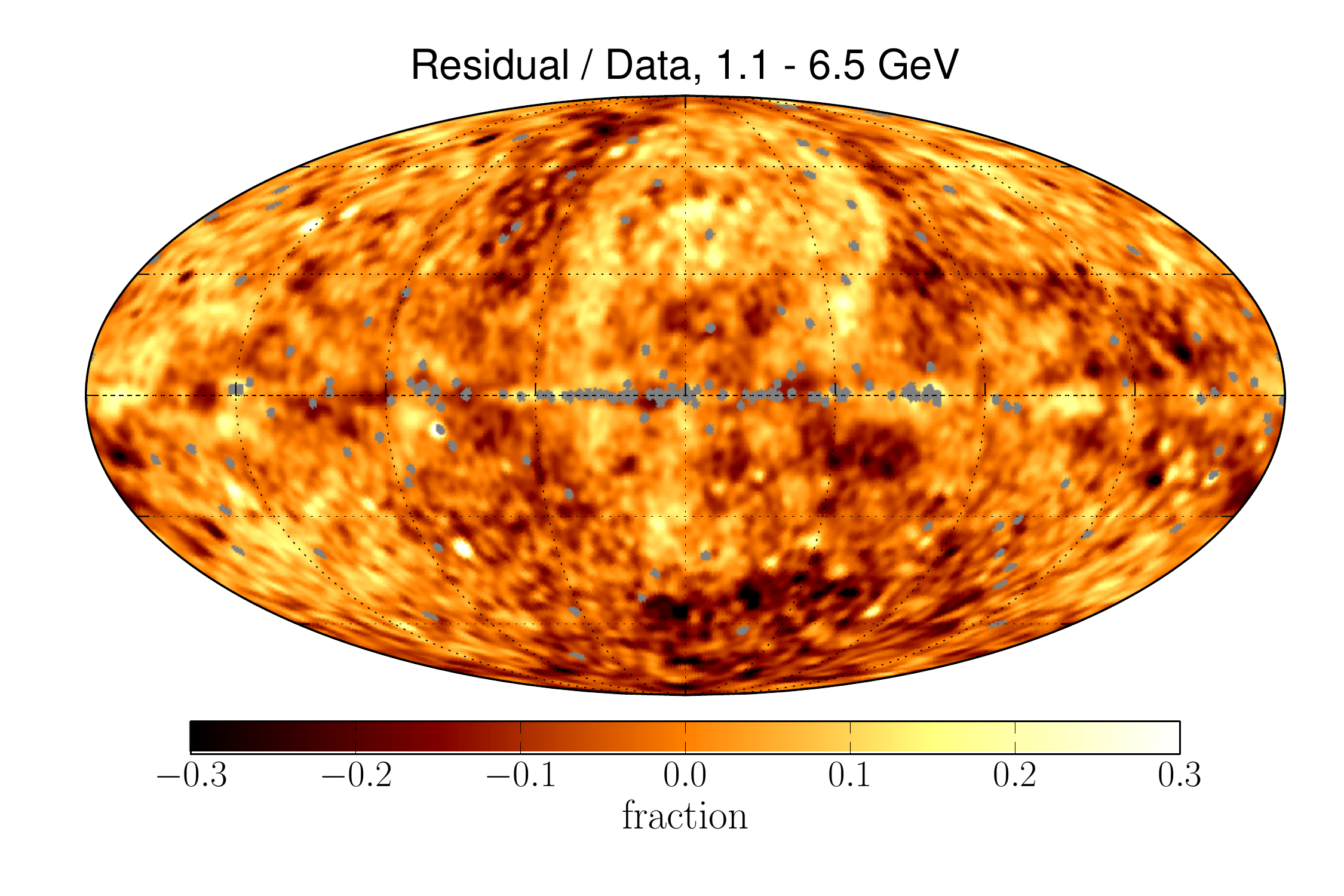}
\noindent
\caption{\small 
Residuals in the model with an all-sky bubbles template.
Top: residuals plus GC excess for the model in Figure \ref{fig:bubbles_spectra} left.
Bottom: residuals in the model with all-sky bubbles but without a gNFW template to model the GC excess.
}
\label{fig:SCAbubbles_resid}
\end{center}
\end{figure}

In Figure \ref{fig:SCAbubbles_resid} we show the residuals plus the GC excess modeled by the gNFW template
with index $\g=1.25$.
We also show residuals in the model with all-sky bubbles without including a template for the GC excess.
The excess remains in the presence of the all-sky bubbles template, but it is reduced compared to the 
residuals in Figure \ref{fig:baseline_resid}.
We note that \citet{2016ApJ...819...44A} modeled the \Fermi bubbles as an isotropic emission component within a $15^\circ \times 15^\circ$ region around the GC, which led to a limited effect on the GC excess.
This differs from our analysis, in which the \Fermi bubbles have non-uniform intensity, and become increasingly brighter near the Galactic plane, as derived from the SCA analysis.
In conclusion, we find that the \Fermi bubbles can significantly reduce the GC excess or even explain it completely above 10 GeV.

\subsection{Galactic Center Excess Template Derivation}
\lb{sec:3comp_sca}

\begin{figure}[htbp]
\begin{center}
\includegraphics[scale=\onepic]{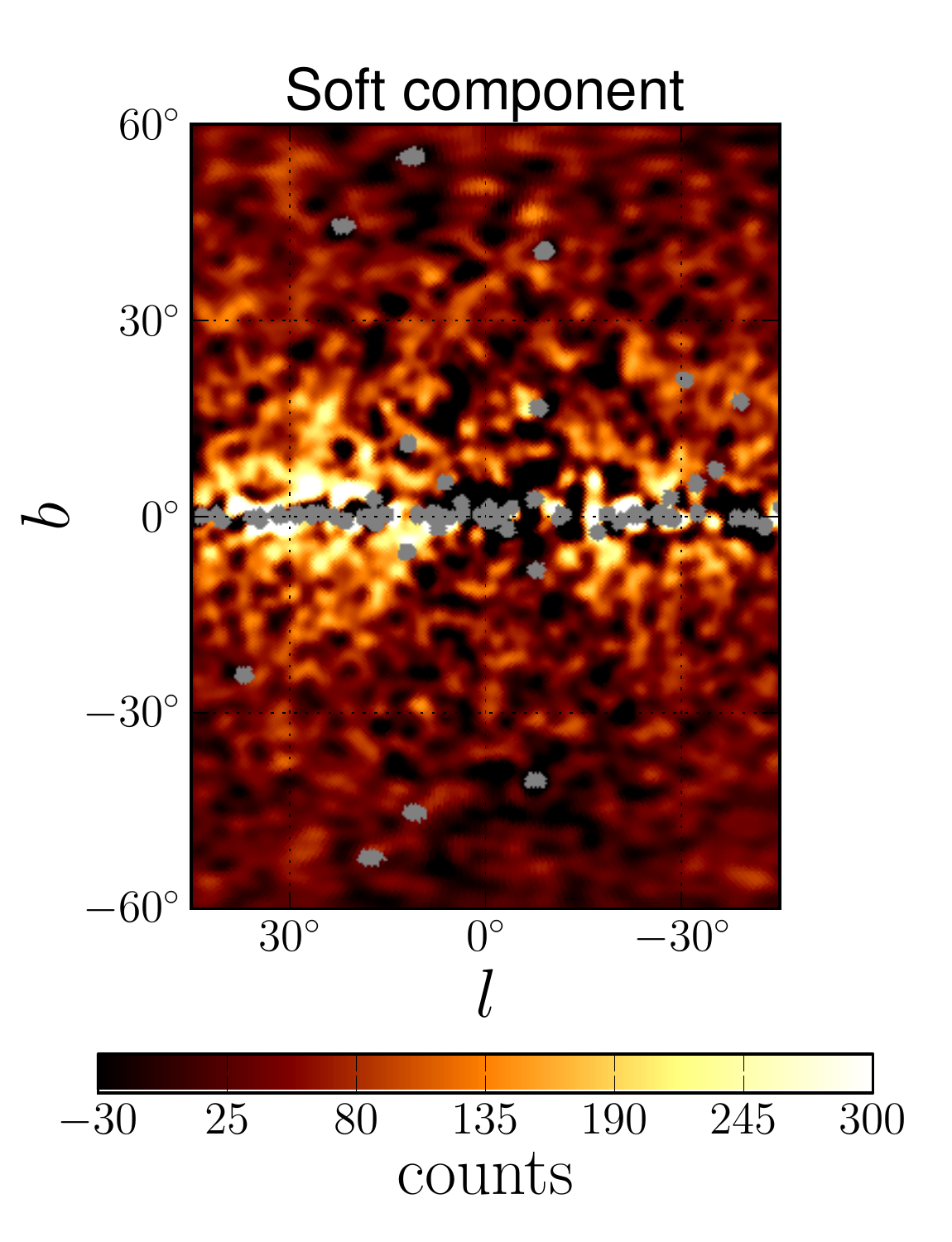}
\includegraphics[scale=\onepic]{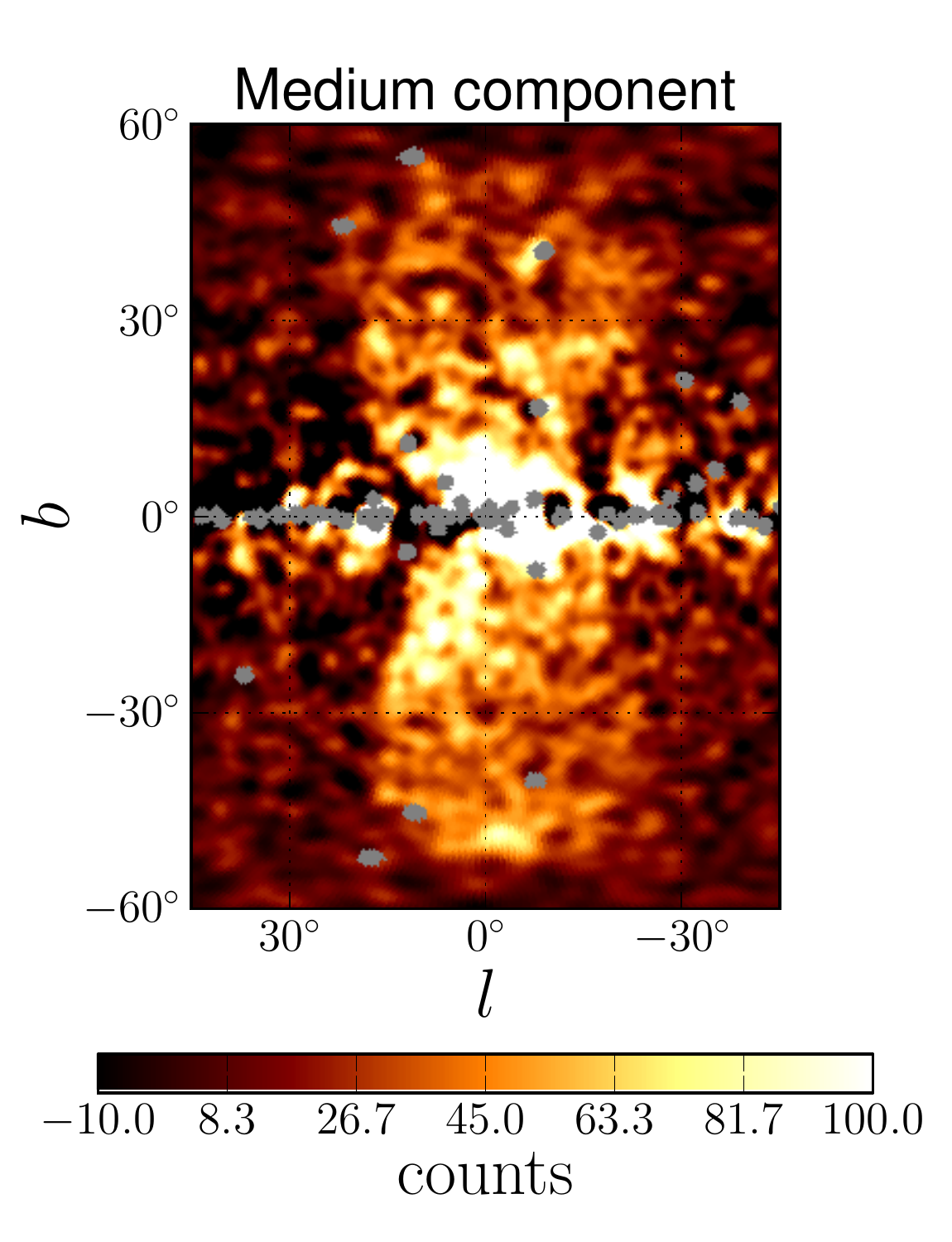}
\includegraphics[scale=\onepic]{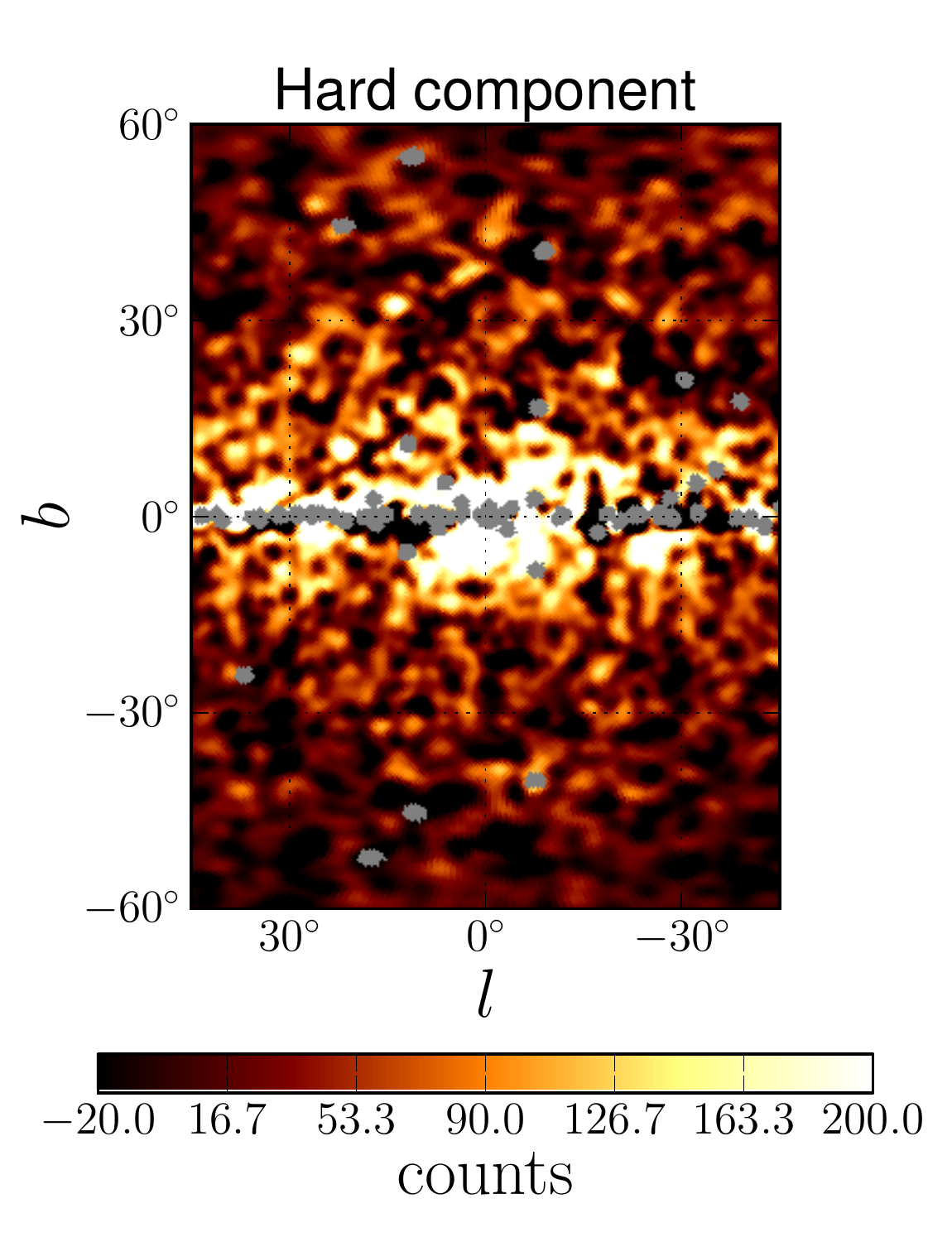}
\noindent
\caption{\small 
Spectral components templates in the three-component SCA model (Section~\ref{sec:3comp_sca}).
The templates are derived from the residuals after subtracting the gas-correlated emission and PS
between 1 and 10 GeV (Section~\ref{sec:gas_PS_model})
assuming the following correlation of spectra:
soft $\propto E^{-2.4}$, medium $\propto E^{-1.9}$, hard $\propto E^{-1.6} e^{-E/4\:{\rm GeV}}$.
}
\label{fig:SCA_cutoff_components}
\end{center}
\end{figure}

In this Section we apply the SCA technique to derive a template for the GC excess itself.
Decomposition into spectral components was used previously by several groups to determine the morphology of the excess,
in particular, \cite{2016arXiv161008926D} found that the excess emission resembles the distribution of molecular clouds near the GC,
while \cite{2016JCAP...04..030H} argued that the excess morphology is spherical.
Motivated by the possibility that the excess comes from a population of MSPs \citep{2015ApJ...812...15B},
we add the third spectral component with an average spectrum of observed MSPs 
$\propto E^{-1.6} e^{-E/4\:{\rm GeV}}$ \citep[e.g.,][]{2014arXiv1407.5583C, 2015ApJ...804...86M}.
Consequently, we fit the residuals obtained after subtracting the gas-correlated emission and PS in Section~\ref{sec:gas_PS_model}
between 1 GeV and 10 GeV with three spectral components, 
hard $\propto E^{-1.6} e^{-E/4\:{\rm GeV}}$ (MSP like), medium $\propto E^{-1.9}$ (bubble like), and soft $\propto E^{-2.4}$.
The derivation of the templates for the components is analogous to Equations (\ref{eq:hs_comps}) and (\ref{eq:chi2_sca})
except that now there are three components instead of two.
The maps of the templates are shown in Figure \ref{fig:SCA_cutoff_components}.

\begin{figure}[htbp]
\begin{center}
\includegraphics[scale=\onepic]{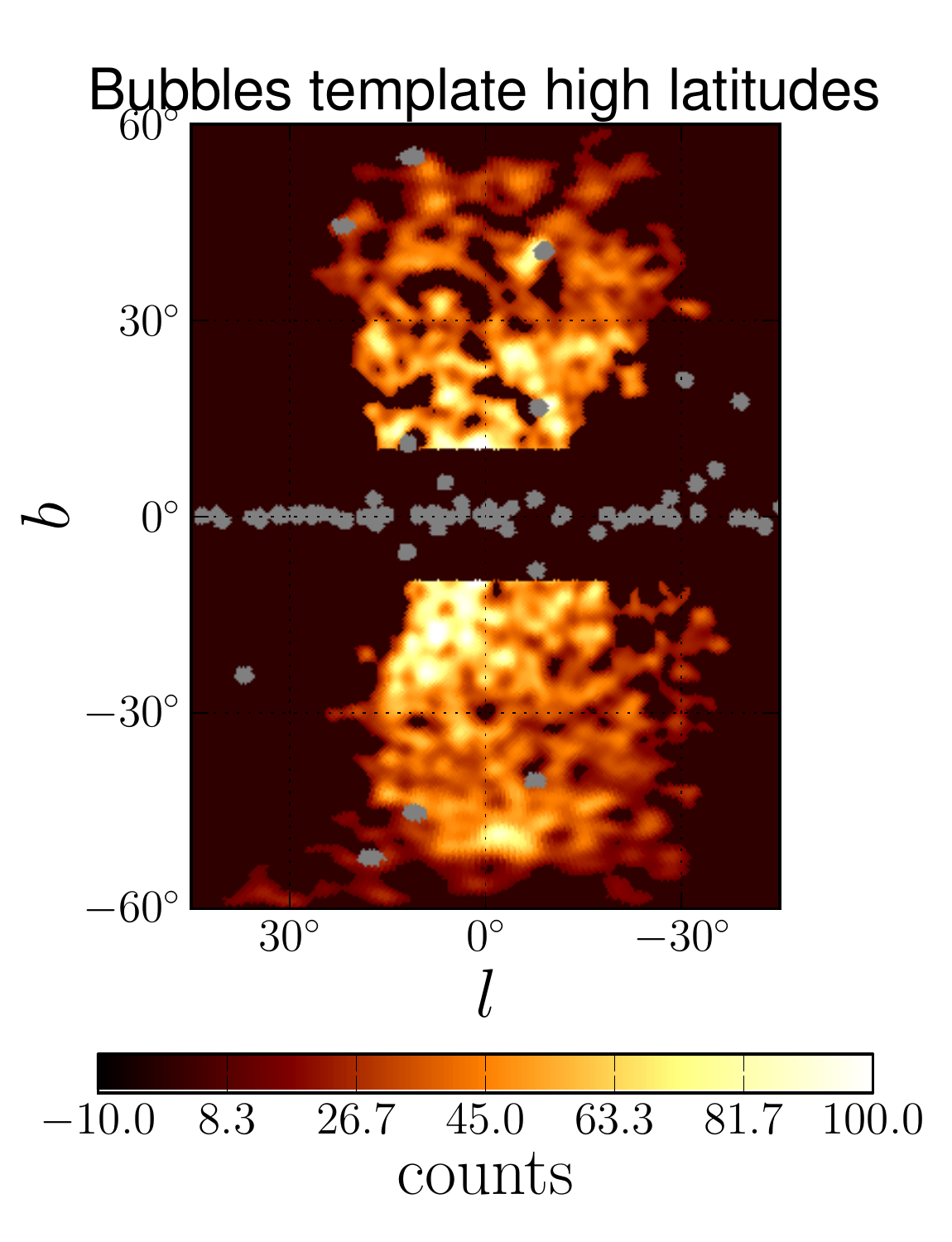}
\includegraphics[scale=\onepic]{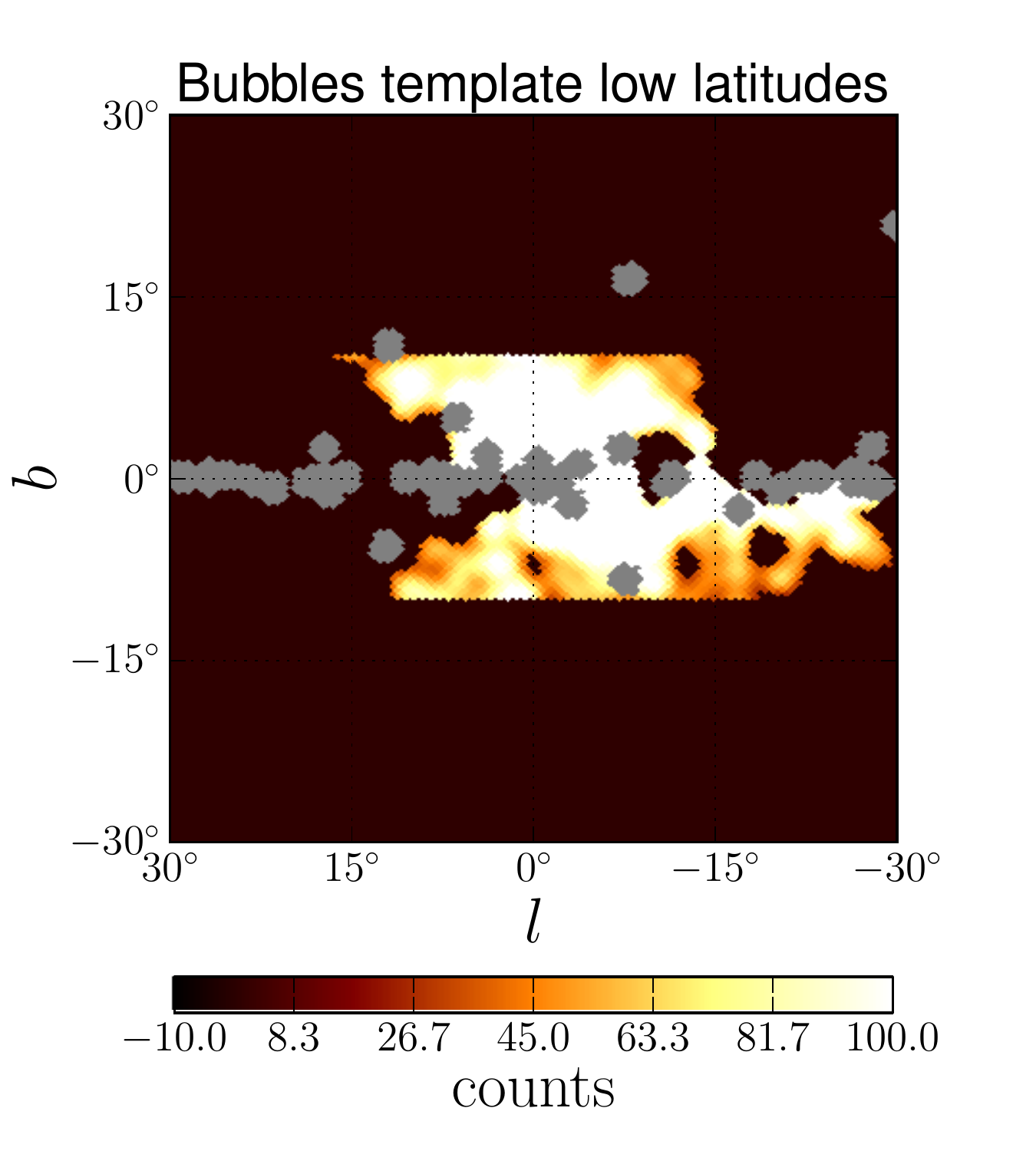}
\includegraphics[scale=\onepic]{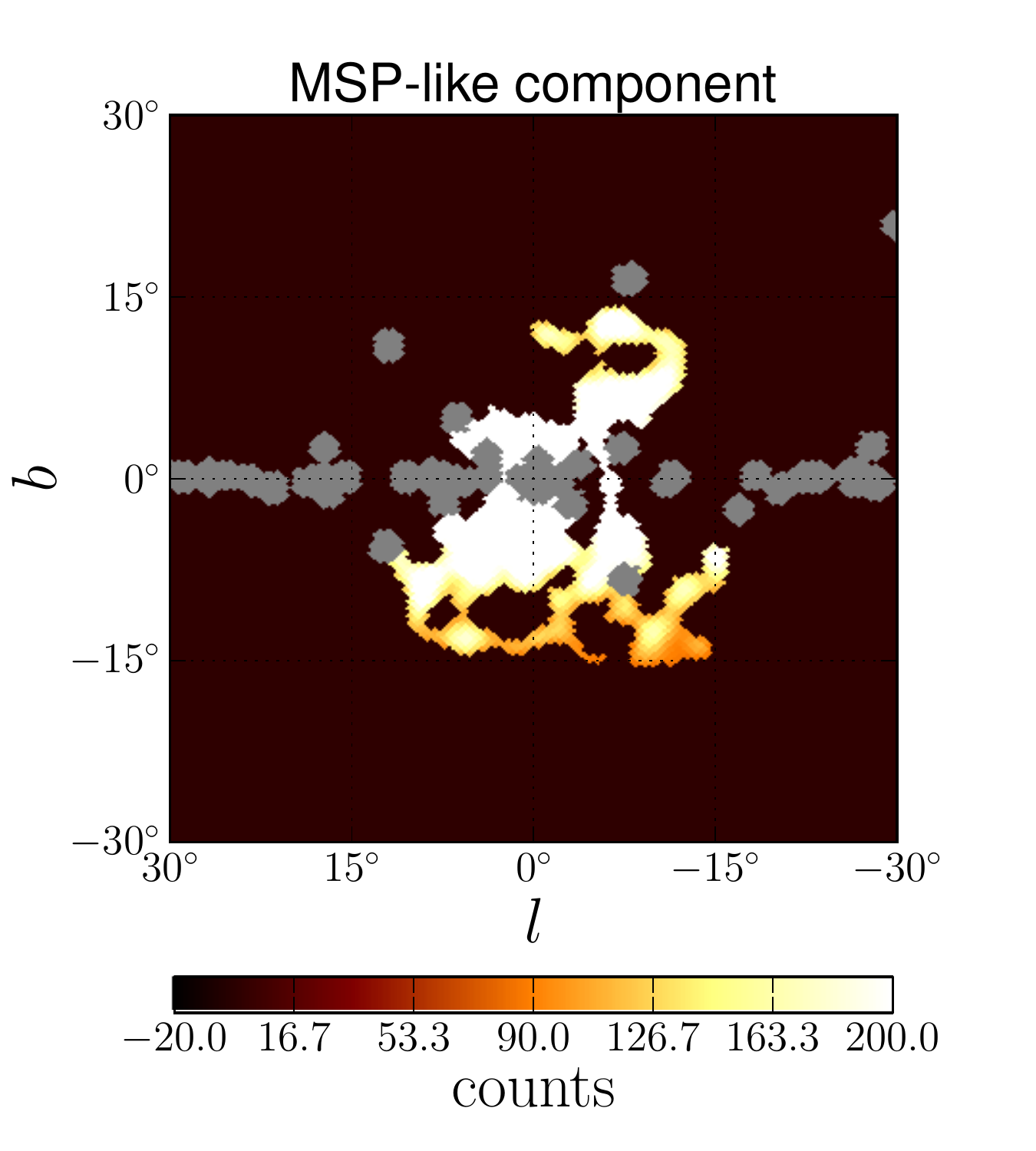}
\noindent
\caption{\small 
\Fermi bubbles and GC excess templates derived from the spectral components in Figure \ref{fig:SCA_cutoff_components}.
The \Fermi bubbles templates are derived similarly to the derivation in Figure \ref{fig:SCA_bubbles},
but with a cut of $1.5 \sigma$ on the significance of the medium spectral component in Figure \ref{fig:SCA_cutoff_components}.
The GC excess template is derived from the hard spectral component in Figure \ref{fig:SCA_cutoff_components}
by applying a cut of $2 \sigma$ in significance. 
}
\label{fig:SCA_cutoff_bubbles_excess}
\end{center}
\end{figure}

The templates for the \Fermi bubbles are derived by applying a cut in significance at $1.5 \sigma$ to the medium spectral component.
Due to the presence of the third spectral component, the bubbles component becomes relatively less significant; thus, 
we choose the $1.5 \sigma$ cut rather than the $2 \sigma$ cut used in the previous subsection.
The template for the GC excess is derived by applying a $2 \sigma$ cut
in the hard component (Figure \ref{fig:SCA_cutoff_bubbles_excess}).
The statistical uncertainties of the spectral components maps are derived by propagating the statistical uncertainties in the data maps (see discussion after Equation~\ref{eq:chi2_sca}).
As before, we also split the \Fermi bubbles template into high and low-latitude bubbles.

\begin{figure}[htbp]
\begin{center}
\includegraphics[scale=\twopic]{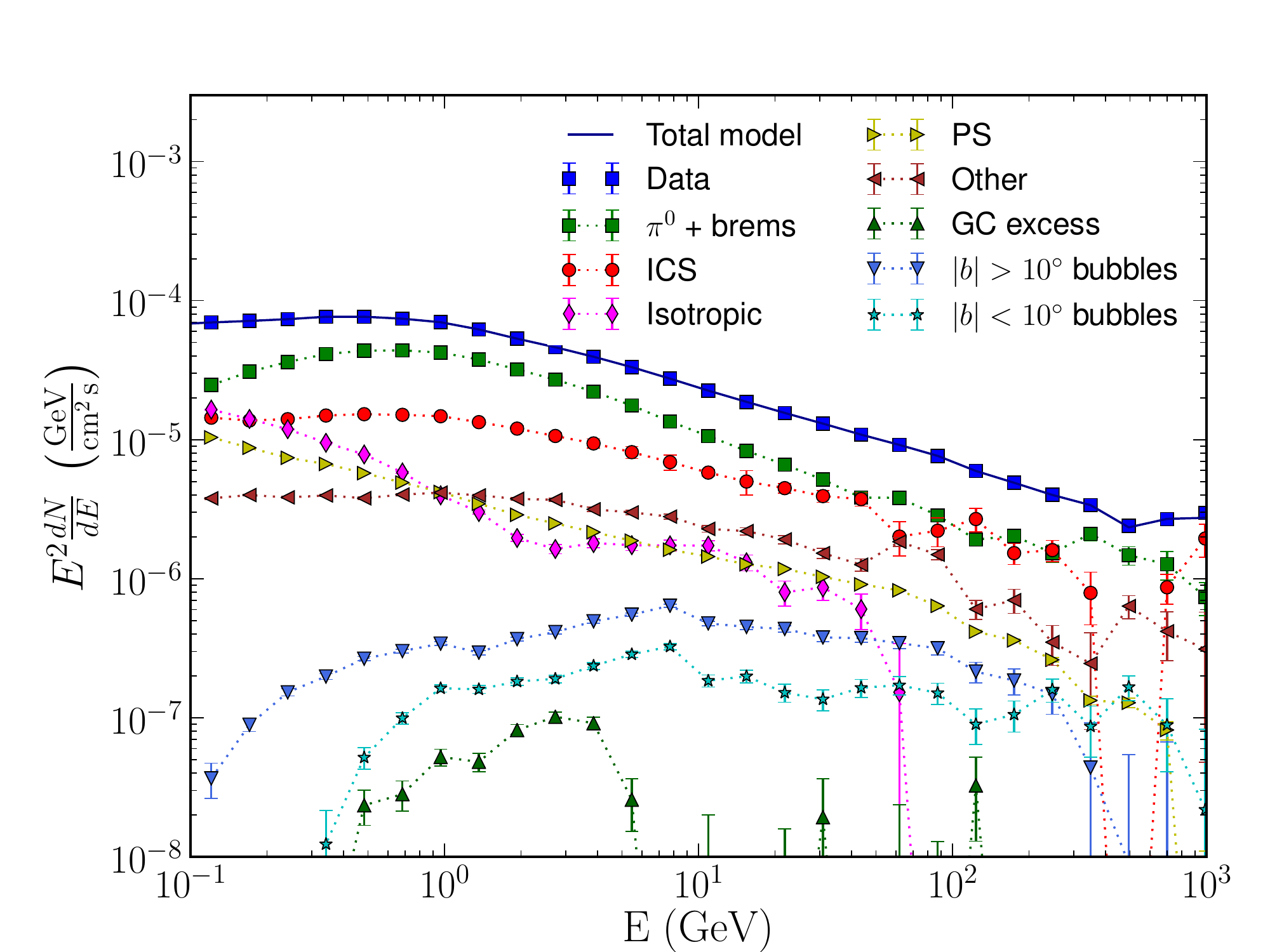}
\includegraphics[scale=\twopic]{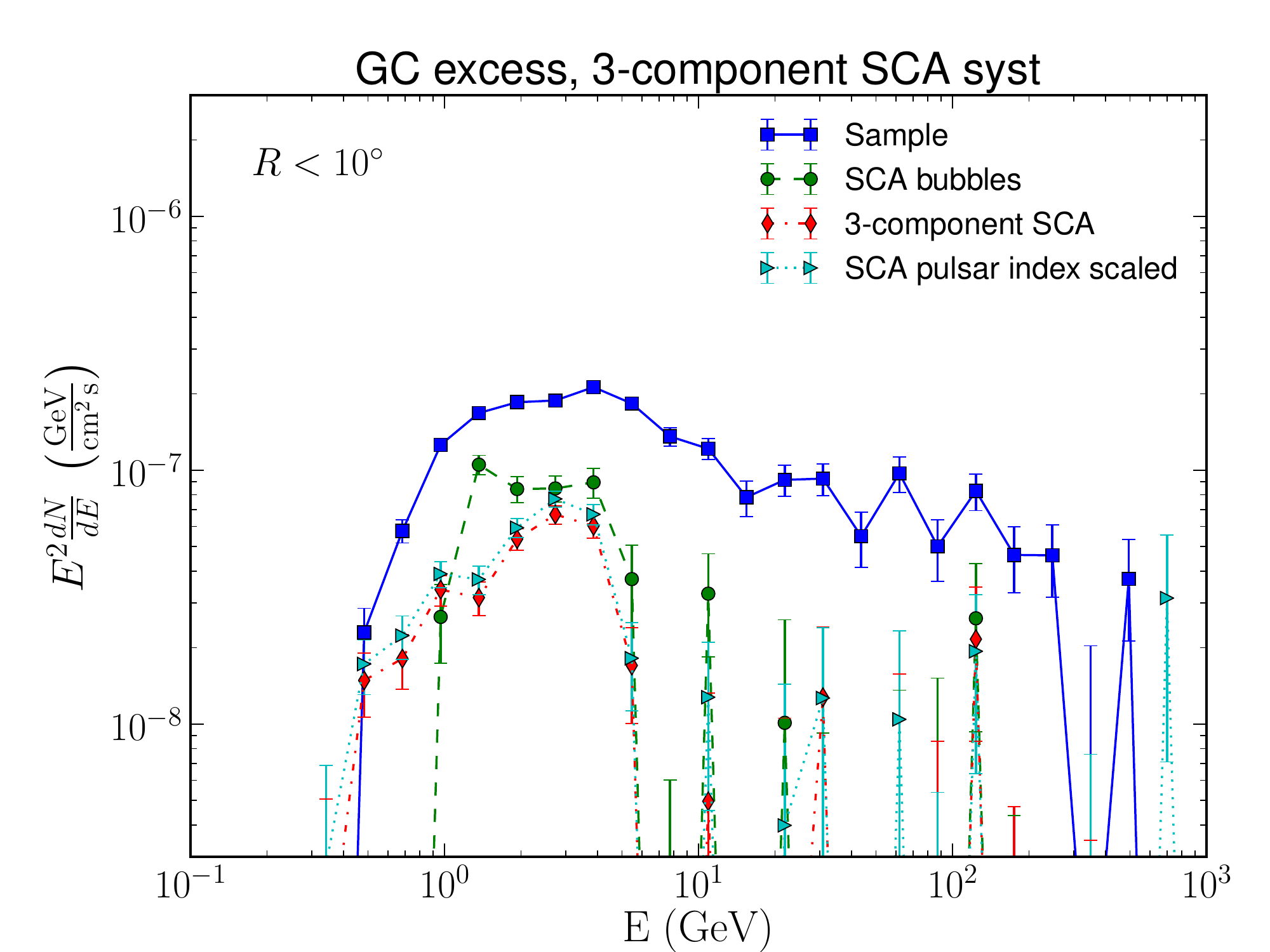}
\noindent
\caption{\small 
\Fermi bubbles and GC excess spectra.
The templates of the components are the same as in the Sample Model but with the 
bubbles and the GC excess templates derived in Figure \ref{fig:SCA_cutoff_bubbles_excess}.
In the right plot, the GC excess spectrum labeled as ``SCA bubbles" is the same as in Figure \ref{fig:bubbles_spectra} right.
The GC excess labeled as ``3-component SCA" is the same as the GC excess spectrum in the left plot; 
it is derived using the template in Figure \ref{fig:SCA_cutoff_bubbles_excess}.
The ``SCA pulsar index scaled" spectrum is determined with the GC excess template assuming one of the spectra for the GC excess 
derived by \cite{2016ApJ...819...44A} (see text for details).
}
\label{fig:SCA_cutoff_spectra}
\end{center}
\end{figure}

The corresponding spectra for the GC excess, high and low-latitude bubbles are shown in Figure \ref{fig:SCA_cutoff_spectra}.
The spectra of the bubbles at high and low latitudes are consistent with each other between $\sim$1 GeV and $\sim$100 GeV.
At energies $<$ 10 GeV, the GC excess spectrum derived with the gNFW profile and the two-component SCA model of the bubbles
is similar to the GC excess spectrum derived in the 3-component SCA model (Figure \ref{fig:SCA_cutoff_spectra} right).

As an alternative derivation of the GC excess template, we use the spectrum $\propto E^{0.5} e^{-E/1.1\:{\rm GeV}}$
derived in \citet{2016ApJ...819...44A} from the LAT data in the case of diffuse models with variable index and CR sources traced by distribution of pulsars. In this case the spectral shape is derived using a phenomenological spectral function to fit the LAT data, and is not based on any specific scenario for the origin of the excess.
The resulting spectrum for the alternative excess template is very similar to the spectrum derived with the template
for the MSP-like spectrum.

\section{Modeling of Point Sources}
\label{sec:PS}

In this section we assess the impact of the modeling of PS on the GC excess, with emphasis on the spectrum.
Difficulties in modeling the PS near the GC include confusion between
PS and features of interstellar emission from CR interactions with gas or radiation fields that are 
not modeled accurately.
Spurious PS included in a model may absorb a part of the GC excess signal while
the flux from non-detected PS may be attributed to the GC excess.

For this purpose we consider the 3FGL catalog \citep{2015ApJS..218...23A} and 
the First \Fermi-LAT Inner Galaxy point source list (1FIG),
which was created in a dedicated study of diffuse gamma-ray emission and PS near the GC \citep{2016ApJ...819...44A}.
The 3FGL catalog is based on 4 years of Pass 7 reprocessed Source class events in the 
energy range between 100 MeV and 300 GeV \citep{2015ApJS..218...23A}.
The 1FIG list is derived using 5 years and 2 months of Pass 7 reprocessed Clean class events
in the energy range between 1 GeV and 100 GeV \citep{2016ApJ...819...44A}. 
We refer the reader to the respective papers for descriptions of the methodology employed 
to derive the source lists and the diffuse emission models.
Additionally, we derive two new lists of PS using the same dataset as for our study of the GC excess
(details are described in the following section).

\subsection{Source Finding Procedures}
\label{sec:PS_finding}

In this section we present two PS search methods that were applied to the same datasets and diffuse models
used in this work.
The data selection is the same as for the Sample Model: 6.5 years of UltraCleanVeto events with zenith angle cut $\theta < 90^\circ$.
The goal is to test how much the selection of a PS detection algorithm can affect the inferred properties of the GC excess. 
Although both algorithms are based on a local likelihood method, there are differences in how the PS are selected and localized.
In both cases this is an iterative procedure from bright sources to faint ones, but the details are different and we describe them in this subsection.

The first source detection algorithm is the same used in the production of the \Fermi LAT source catalogs based on the \textit{pointlike} package \citep{2010PhDT.......147K}, and described in \cite{2015ApJS..218...23A}.
The data are binned in energy, in 14 bands from 100 MeV to 316 GeV (or four per decade), and separated into front and back event types. For each band and event type, the photons are binned using HEALPix, with pixel sizes selected to be small compared with the PSF. 
The log-likelihood is then computed summing over the energy bands and event types. 
As described for 3FGL \citep[\S 3.1.2 of][]{2015ApJS..218...23A}, the contributions to the likelihood function are `unweighted' for the lower energies, 
to account for systematics of the diffuse background spectrum. 
For the diffuse model, we use the Sample Model from Section \ref{sec:baseline} fitted to the data but without adding the gNFW template.
 
The data are fitted using a diffuse model template, an isotropic template, and point sources in small ROIs covering the whole sky.
The centers of the ROIs are determined by the centers of HEALPix pixels with nside = 12 
(1728 tiles in total, average distance between the centers of ROIs is about $5^\circ$)
and the radius of the ROIs is $5^\circ$.
The pixel size depends on the energy. At energies below 10 GeV it is about 1/5th of the 68\% containment radius, e.g., for back-entering events around 100 MeV the pixel size is about $1^\circ$ (nside = 52). 
Note, that nside = 12 and nside = 52 are non-standard nside values which are typically powers of 2. 
Above 10 GeV the analysis is unbinned.
The likelihood for the data within the radius is optimized with respect to the spectral parameters of the sources located within the tile. Correlations with sources outside the tile, but contributing to the likelihood, are accounted for by iterating until changes of the likelihood for each tile are small.
 
In the search for new point sources, we calculate the likelihood ratio, expressed as a Test Statistic 
($TS = 2 \Delta \log \La $) for an additional point source, 
assuming a power-law spectrum with a fixed spectral index of 2.3 but variable flux, at each of the positions defined by nside~=~512 (3.2M total). This is done for all pixels within each ROI. A clustering analysis is applied to the resulting map of the pixels with $TS > 25$. All clusters with more than one pixel are used to define seeds for inclusion in the model. As for all sources, the spectral index is now optimized and the source is localized.
If the power-law spectrum does not fit the data well, 
then the spectrum is described by a log parabola \citep[e.g.,][]{2015ApJS..218...23A}.
A source candidate is accepted for inclusion if its optimized TS is greater than 25 and the localization process converged properly.
 
This source-finding procedure relies on the model being an accurate description of the data, given the set of sources and diffuse components. Thus, the set of sources needs to be fairly complete, so that new sources are weak and do not strongly affect the current model. We have found that it is necessary to rerun the procedure several times after adding new sources.

For the determination of the second new list of PS, we use the \textit{Fermipy} package, a set of Python tools built around the \Fermi~LAT  \textit{Science Tools} that automate and enhance their functionalities\footnote{\url{http://fermipy.readthedocs.io/en/latest/}}. 
In this case we use data between 300 MeV and 550 GeV binned with 5 bins per decade.
The ROI is $|\ell|, |b| < 22^\circ$, and we bin the data in $0^\circ\!.08$ pixels (on a square grid).
As a preliminary step, we start with the 3FGL PS and reoptimize their positions. We then perform a fit to the ROI with those sources and delete all 3FGL sources with TS~$< 49$.
Then we build a TS map (map of TS of a PS candidate at each position in the spatial grid) and select maxima with TS~$> 64$ and separation from other sources greater than $0^\circ\!.5$.
The best positions and location uncertainties of source candidates are derived from the likelihood profile in the nine pixels around the TS map peak by 
fitting it with a paraboloid.
We repeat the selection of TS maxima and PS localizations two more times for $TS > 36$ and separation greater than $ 0^\circ\!.4$,
and with $TS > 25$ and separation larger than $0^\circ\!.3$.
After this third iteration we have a list of sources with $TS > 25$ and we perform a final fit to the ROI to determine the PS spectra.

Discussing the properties of the source candidates in the new lists is beyond the scope of this article. Instead, we focus in the following section on the effects that the choice of a PS list has on the determination of the GC excess spectrum.

\subsection{Refitting Point Sources near the Galactic Center}
\label{sec:PS_refitting}

\begin{figure}[htbp]
\begin{center}
\includegraphics[scale=\twopic]{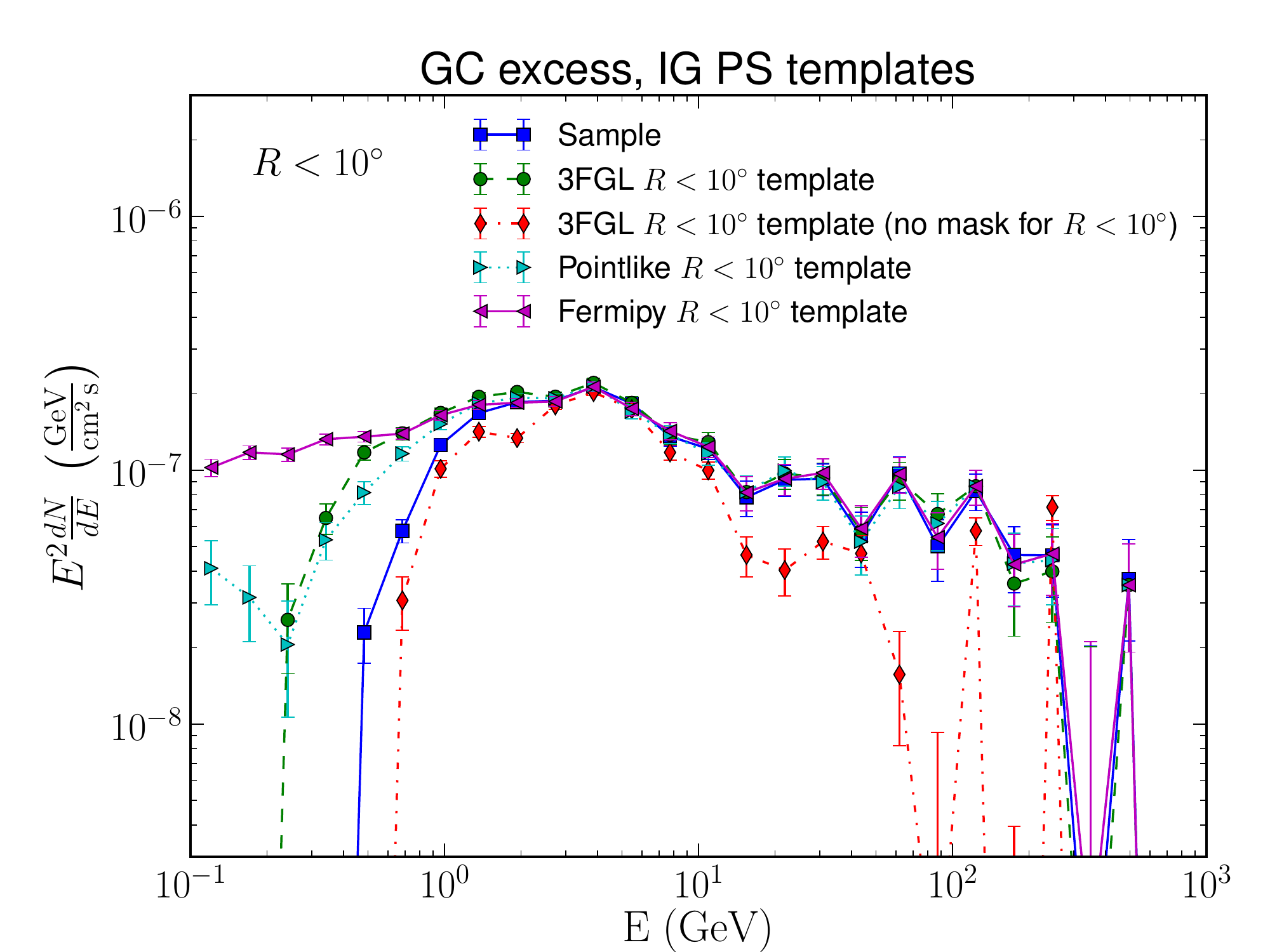}
\includegraphics[scale=\twopic]{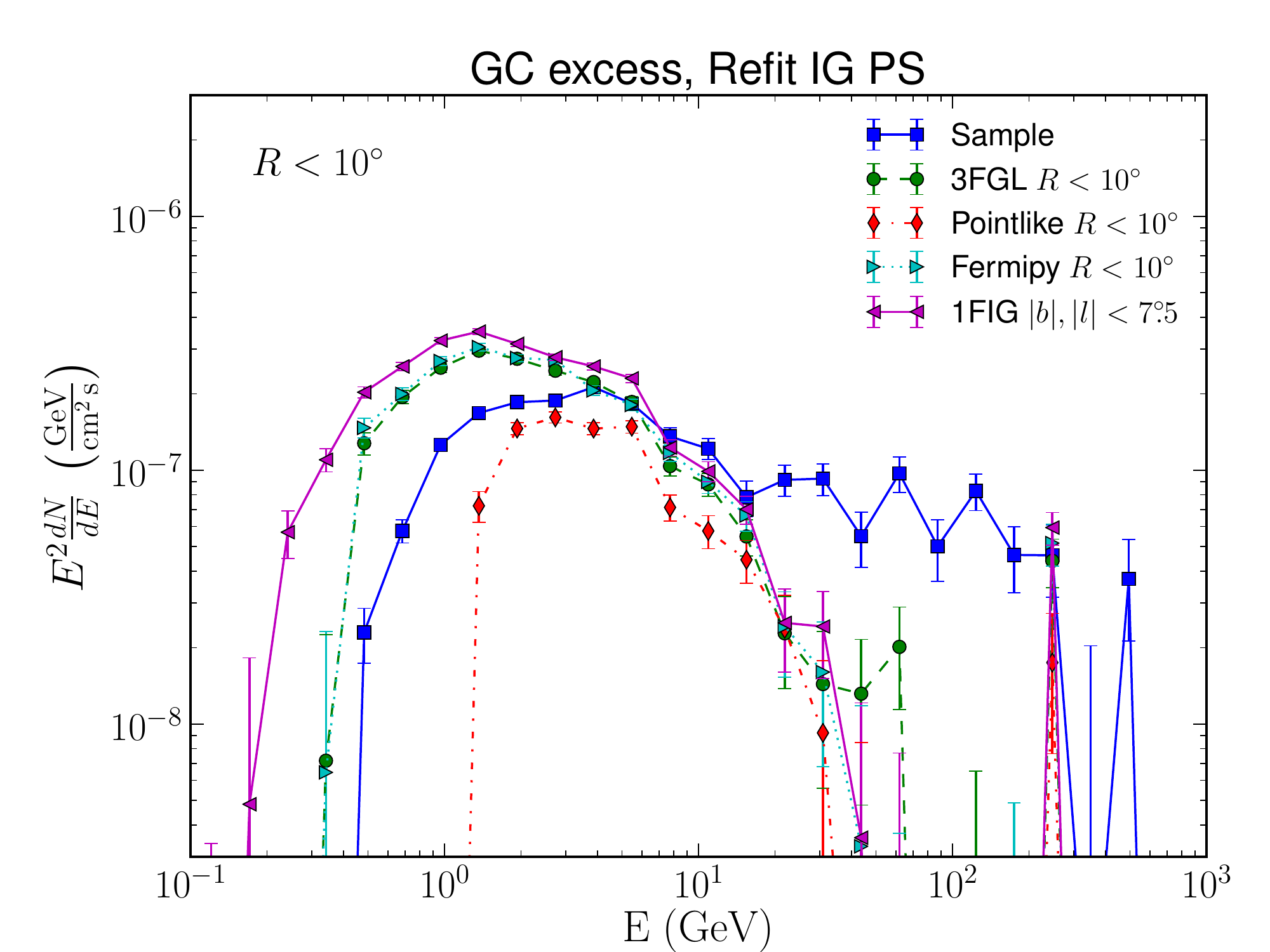}
\noindent
\caption{\small 
GC excess spectra in models with adjusted spectra for PS within $10\degr$ of the GC. Left: all sources within $10\degr$ from the GC are combined in a single template.
Right: PS within $10^\circ$ ($|b|,|\ell | < 7^\circ\!\!.5$ in the case of 1FIG) from the GC are fitted independently in each energy bin.
The curves correspond to 3FGL catalog, 1FIG list, and the PS lists derived with our dataset and diffuse models using \textit{pointlike} and \textit{Fermipy}.
}
\label{fig:PS_syst}
\end{center}
\end{figure}

We start the analysis of the effect of PS characterization near the GC by
combining PS within $10^\circ$ from the GC into independent templates, using the spectra provided by the
3FGL catalog, \textit{pointlike} or \textit{Fermipy}.
We use one of these templates at a time together with the GC excess template, the other diffuse emission components, and the PS template determined with sources outside of $10^\circ$ from the GC.
The effect on the GC excess spectrum is shown in Figure \ref{fig:PS_syst} on the left.
An independent template for sources inside $10^\circ$ generally leads to a softer GC excess flux at energies below 1 GeV,
while above 1 GeV the GC excess flux is not affected significantly by the introduction of the independent PS template for sources
within $10^\circ$.
To test the effect of the PS mask we also include a model with 3FGL templates where we do not mask PS within $10^\circ$ from the GC. In this case we find an overall reduction of the flux attributed to the GC excess.
The 1FIG list is not considered in this step because the spectra are provided only for energies larger than 1 GeV. The spectra for the \textit{Fermipy} sources are extrapolated from the energies above 300 MeV used to determine the source list, which is likely the cause for the deviations of the GC excess spectrum at lower energies with respect to the other determinations.

To allow more freedom in PS modeling,
we also refit the spectra of the PS within $10\degr$ from the GC with a free normalization
for each source in each energy bin independently.
The 3FGL Catalog has 76 sources within this region, and so our model has 76 additional free parameters in each energy
bin compared to the Sample Model.
The 1FIG catalog has 48 sources inside $|b|,|\ell | < 7^\circ\!\!.5$. 
\textit{Pointlike} provides 104 PS candidates
and \textit{Fermipy} has 127 PS candidates inside $R < 10^\circ$.
Due to limited statistics at high energies, we restrict the energy range to 100 MeV--300 GeV (23 energy bins) for these fits.
Similarly to the Sample Model, point sources more than $10\degr$ away from the GC (outside of $|b|,|\ell | < 7^\circ\!\!.5$ in the case of 1FIG)
are combined in a single template based on the 3FGL catalog. 
Diffuse emission components, extended sources, the LMC, and
Cygnus also have independent templates: all these templates have free overall normalization in each energy bin.
We mask regions of $1\degr$ radius for the 200 brightest PS {\it outside} of $10\degr$ from the GC.

The spectra derived for the GC excess are shown in Figure \ref{fig:PS_syst} on the right. The spectrum below a few GeV is clearly dependent on the dataset, diffuse emission model, and method used to derive the PS source list. In several cases the flux attributed to the GC excess is larger than in the Sample Model. In the case of the 3FGL Catalog, this could be attributed to an overestimation of the PS fluxes near the GC stemming from not having accounted for an excess at the GC explicitly. The largest effect is observed for refitting 1FIG sources which were derived using data above 1 GeV, without accounting in any way for a GC excess, and 1FIG has the fewest number of PS near the GC due to rather strict selection criteria \citep{2016ApJ...819...44A}. 
On the other hand, refitting sources individually above 10 GeV
reduces the flux attributed to the excess for all PS lists.
This is consistent with the analysis of \citet{2016PhRvD..94j3013L}, who find hat the GC excess above $\sim$ 10 GeV can be explained by a contribution of point sources.

In the derivation of the new lists of PS, we did not account for large-scale residuals, as done, e.g, in \citet{2016ApJS..223...26A} for the derivation of the 3FGL catalog, or specifically for an excess near the GC.
This is likely to result in several spurious sources that absorb positive residuals due to underprediction of diffuse emission.
We use them as a conservative starting point in the derivation of the GC excess
component, since some part of the emission from this component will be attributed
to spurious PS. By allowing free normalizations for both the gNFW template and the point sources, 
we have also tested whether the gNFW template is preferred relative to a combination of resolved PS from a statistical point of view.
We find that in some cases refitting PS results in larger flux attributed to the GC excess at energies below a few GeV, 
however for the \textit{pointlike} algorithm, refitting the PS leads to smaller GC excess flux for all energies.
The conclusion is that although all PS detection algorithms are based on maximizing a local likelihood, 
the details on how the PS are selected are important and may have a significant influence on the GC excess.

\section{Summary of the Spectral Results}
\lb{sec:spectrum}

\begin{figure}[htbp]
\begin{center}
\includegraphics[scale=\twopic]{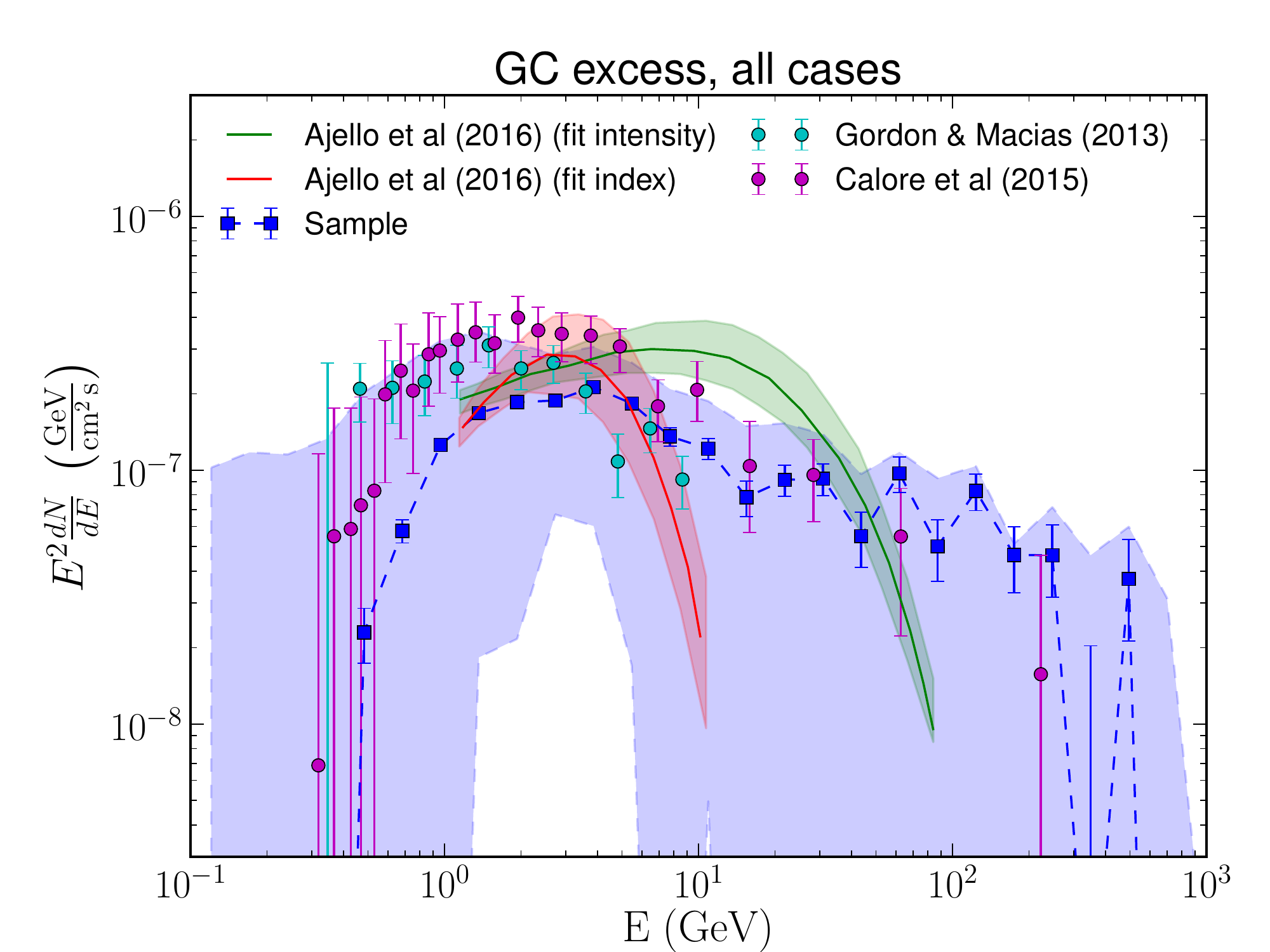}
\noindent
\caption{\small Spectrum of the GC excess. Points are derived using the \SM described in Section \ref{sec:baseline}.
The systematic uncertainty band is derived from taking the envelope of the GC excess fluxes for different analysis configurations, and different models of diffuse gamma-ray emission and sources in Sections from \ref{sec:analysis_setup} to~\ref{sec:PS}. 
Our results are compared to previous determinations of the GC excess spectrum from the literature.
Note, that the area of integration varies in different cases. In this analysis we mask some bright PS, which effectively masks the GC within about $2^\circ$ radius.
\cite{Gordon:2013vta} have a $7^\circ \times 7^\circ$ square around the GC.
The flux from \cite{Calore:2014xka} is obtained by taking the intensity in Figure 14 and multiplying by the area of the ROI
($2^\circ < |b| < 20^\circ$ and $|\ell | < 20^\circ$) in their analysis.
The ROI in \cite{2016ApJ...819...44A} is a $15^\circ \times 15^\circ$ square around the GC.
The two cases that we consider here correspond to the model with the CR sources traced by the distribution of pulsars \citep{2004A&A...422..545Y} where either only overall intensity (``fit intens") or both intensity and index (``fit index") for the diffuse components spectra are fit to the data \citep[cf. Figure 13 of][]{2016ApJ...819...44A}.
}
\label{fig:GCE_fluxband}
\end{center}
\end{figure}

Table~\ref{tab:variations} summarizes the impact on the measured spectrum of the excess of the different sources of uncertainties considered in the previous sections. In Figure~\ref{fig:GCE_fluxband} we show the spectrum of the GC excess with the uncertainty band encompassing all the variations in analysis setup and modeling of other gamma-ray emission components considered in our study. 
The GC excess peaks at $\sim$ 3~GeV as reported in the literature. 
Large uncertainties of various nature affect the determination of the GC excess spectrum. 
The upper edge of the uncertainty band for energies below 2 GeV is due to uncertainties in PS modeling,
while above 2 GeV the upper band is driven by the model with CR sources traced by OB stars.
The lower edge of the uncertainty band below 1 GeV and above $\sim$ 6 GeV is consistent with zero, 
which is due to additional CR sources near the GC, the \Fermi bubbles, and modeling of PS.
The excess remains significant in all cases in the energy range from 1~GeV to a few~GeV, although its flux is found to vary by a factor of 
$\gtrsim 3$ owing to uncertainties in the modeling of IC emission, additional CR sources near the GC, 
and a contribution of the low-latitude emission from the \Fermi~bubbles.

Figure~\ref{fig:GCE_fluxband} also shows that our determination of the GC excess spectrum is generally consistent with previous determinations in the literature, but our assessment of systematic uncertainties is generally larger than that reported in other studies.
We note that the ROIs used to determine the flux and the flux profiles assumed are different for different analyses, thus the curves cannot be compared quantitatively. 
The main purpose of the figure is to show that there is a qualitative agreement.

\begin{table}
\footnotesize
\centering
\begin{tabular}[t]{cccc}
\hline
Variation & Parameters & Effect on GC excess & Energy range\\
\hline
\hline
Choice of the data sample & Clean, UltraCleanVeto,  & Minor & all \\
& UltraCleanVeto PSF 2 and 3 & Slightly larger &  below 1 GeV\\
\hline
Choice of the ROI & $|b|,\; |\ell| < 10^\circ$  & Significantly larger & below 1 GeV \\
 & $|b|,\; |\ell| < 20^\circ$  & Slightly smaller & all \\
 & $|b|,\; |\ell| < 30^\circ$  & Smaller & below a few GeV \\
\hline
Tracers of CR sources & OB stars & Larger & all, especially below 1 GeV\\
 & Pulsars and SNRs & Minor & all \\
Propagation halo size  & $z = 4$ kpc & Smaller & below a few GeV \\
  & $R = 30$ kpc & Minor & all \\
Spin temperature & Optically thin & Minor & all \\
\hline
IC models & Split in 5 rings & Smaller & all\\
 & Combine all rings and && \\
 & ISRF components & Smaller & all, especially below a few GeV \\
\hline
Gas distribution & \Planck, GASS surveys & Slightly smaller & below 1 GeV\\
 & SL extinction & Larger & below 1 GeV\\
\hline
GC CR sources & Bulge electron source & Smaller & between 1 and 10 GeV\\
 & CMZ, z = 2, 4 kpc & Minor & all\\
 & CMZ, z = 8 kpc & Smaller & below a few GeV\\
\hline
\Fermi bubbles &  & Excess vanishes & below 1 GeV, above 10 GeV\\
 &  & Smaller & between 1 GeV and 10 GeV\\
\hline
PS templates within $10^\circ$ & 3FGL, Pointlike & Slightly larger & below 1 GeV\\
 & Fermipy & Larger & below 1 GeV\\
Refit PS within $10^\circ$ & 3FGL, Fermipy, 1FIG & Larger & below a few GeV\\
& Pointlike & Smaller & all, especially below a few GeV\\
& 3FGL, Pointlike, Fermipy, 1FIG & Smaller & above 10 GeV\\
\hline
\end{tabular}
\caption{\small
\label{tab:variations}
Summary of the effect on the GC excess spectrum of variations of data selections and inputs in the Sample Model.
The effect is relative to the GC excess spectrum in the Sample model.
More details can be found in Figures \ref{fig:GC_escess_vars}, \ref{fig:bubbles_spectra}, \ref{fig:SCA_cutoff_spectra}, and 
\ref{fig:PS_syst}.
Sample Model GALPROP parameters: CR production traced by pulsars \citep{Lorimer:2006qs};  propagation halo $z = 10$ kpc, $R = 20$ kpc; spin temperature 150 K (see Section \ref{sec:baseline} for details).
}
\end{table}

\section{Morphology of the Galactic Center Excess}
\label{sec:morphology}

Characterizing the morphology of the GC excess is important to understand its nature.
In particular, spherical symmetry is expected for DM annihilation as well as, to a good approximation, for a population of MSPs in the bulge of the Milky Way
\citep[e.g.,][]{2015ApJ...812...15B} or young pulsars produced as a result of star formation near the GC \citep{2015arXiv150402477O},
while a continuation of the \Fermi bubbles to the Galactic plane may have a bi-lobed shape \citep[e.g.,][]{2016ApJS..223...26A}.
There are claims of both spherical \citep[e.g.,][]{Daylan:2014rsa, Calore:2014xka} and 
bi-lobed \citep[e.g.,][]{2016A&A...589A.117Y, 2016arXiv161106644M} morphology of the excess.

In Section~\ref{sec:FBtmpl}, we derived an all-sky template for the \Fermi bubbles, assuming that the bubbles spectrum
at low latitudes is the same as the spectrum at high latitudes in the energy range between 1 GeV and 10 GeV.
We have also shown in Section~\ref{sec:FBtmpl} that there is excess emission near the GC remaining after accounting in that way for a low-latitude component of the \Fermi bubbles.
Thus, it is plausible that a separate emission component is present near the GC with a spectrum that differs from the 
\Fermi bubbles at high latitude.
In the rest of this section we will discuss the morphological properties of the GC excess, in the light of the uncertainties in the models of foreground emission and with special focus on the differences when accounting or not for low-latitude emission from the \Fermi bubbles.

A detailed study of the morphology is complicated by the modeling uncertainties in the Galactic plane 
(Appendix \ref{app:unc_maps}), so we focus on integrated quantities, namely: the GC excess spectrum in quadrants; longitude, latitude, and radial profiles; the centroid position and radial index of the gNFW GC excess template.

\subsection{Galactic Center Excess Spectrum in Quadrants}

\begin{figure}[htbp]
\begin{center}
\includegraphics[scale=\twopic]{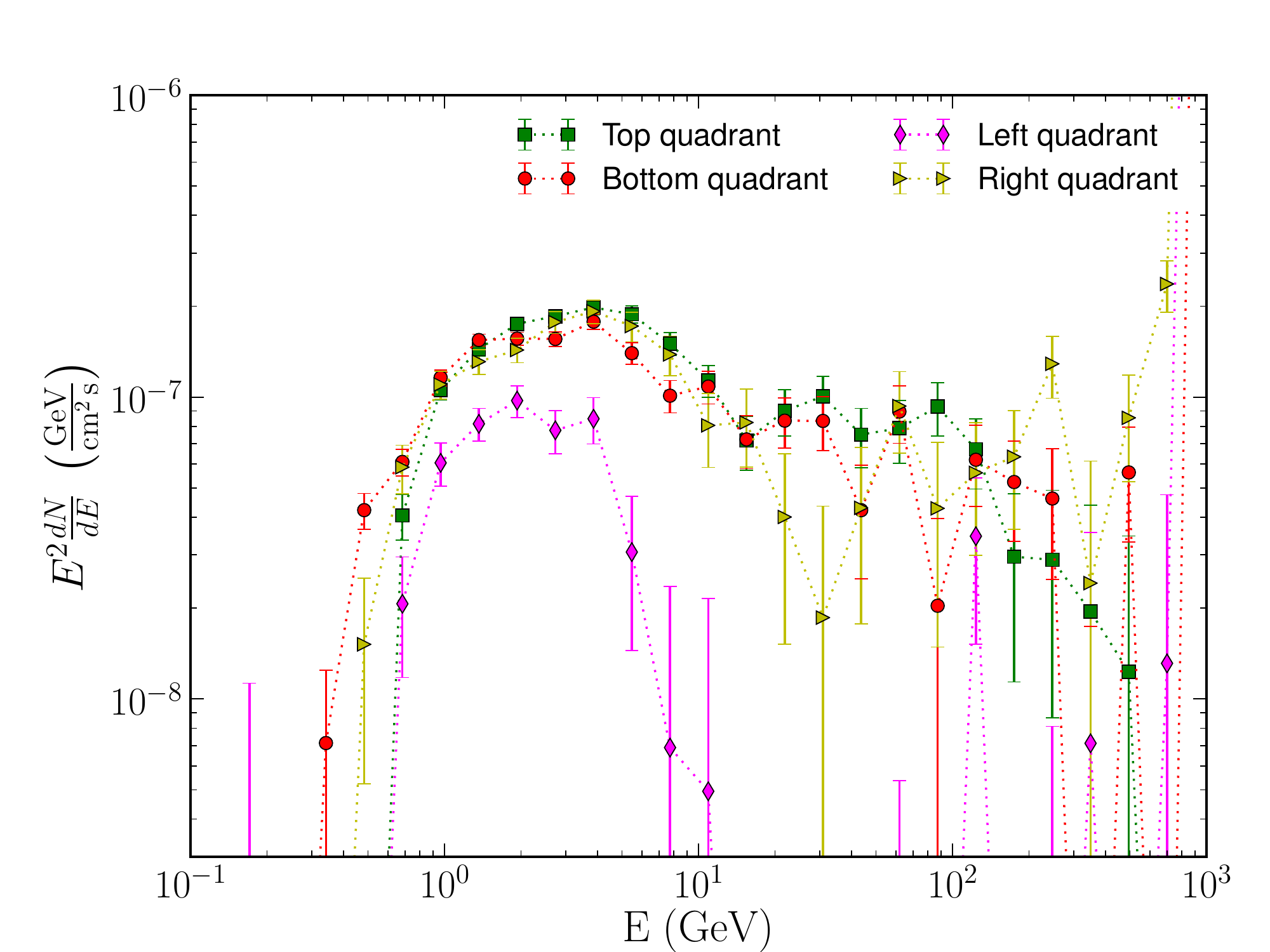}
\includegraphics[scale=\twopic]{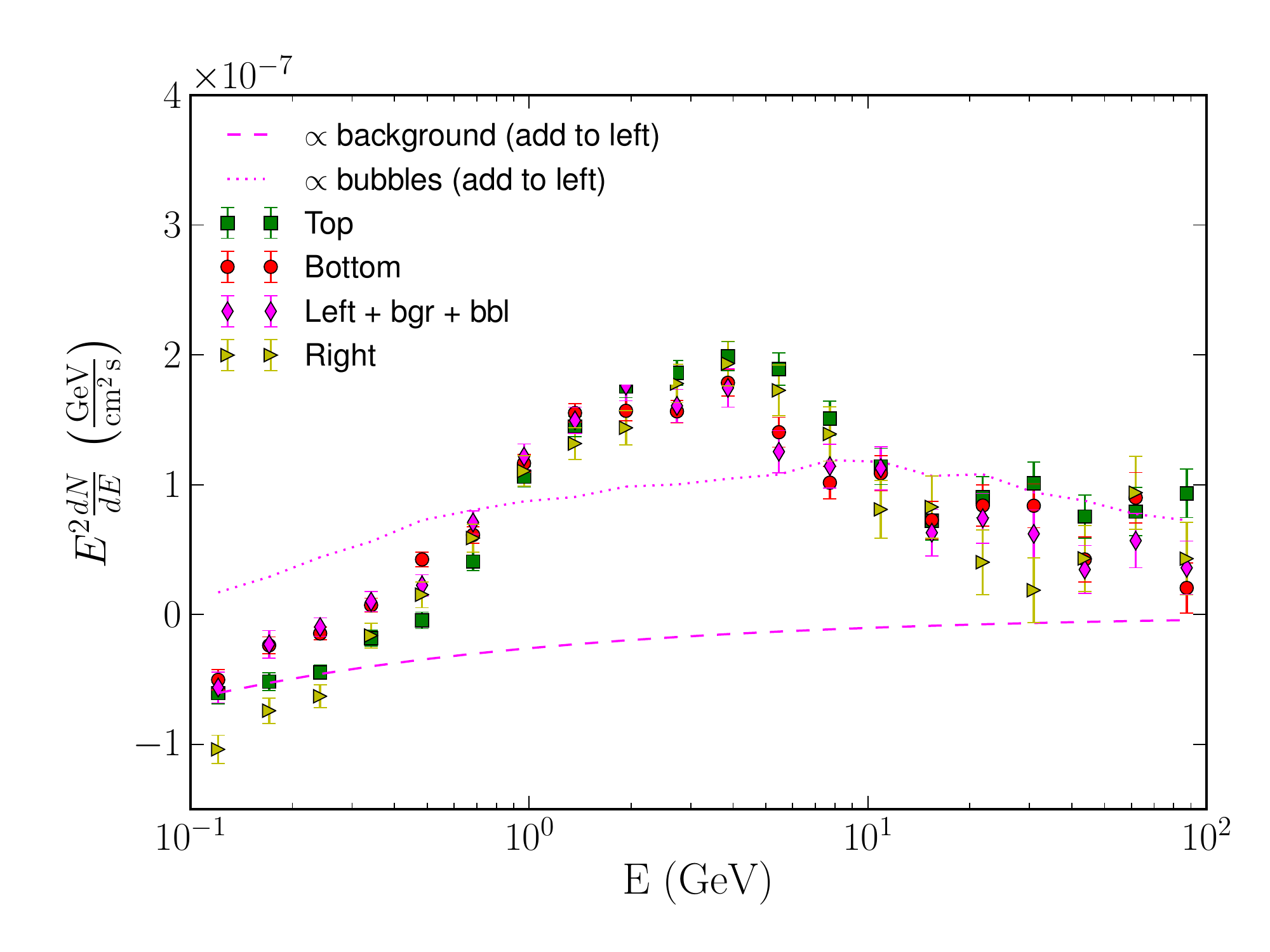}
\noindent
\caption{\small 
Spectra of the GC excess template split into quadrants when all the other components are modeled as in the Sample Model \ref{sec:baseline}. 
Left: fluxes of components integrated over the whole sky.
Right: fluxes of the quadrant templates where we add bubble-like (dotted line) and background-like (dashed line) spectra to the left
quadrant spectrum to minimize the difference with the bottom quadrant spectrum (see text for details).
Note that the y-axis on the right plot has linear scale.
}
\label{fig:quads_sample}
\end{center}
\end{figure}

In this section, we derive the spectrum independently in sectors along and perpendicular to the Galactic plane  
by dividing the gNFW template into four templates centered at the GC with an opening angle of $90\degr$: top, bottom, left and right. 
The fit is all-sky and the only difference from the Sample Model is that now we have four independent templates for the excess in the four quadrants instead of a single one.
The spectra of the quadrant templates are shown in Figure \ref{fig:quads_sample} on the left.
The spectra of the top, bottom, and right quadrants are similar to each other, while the spectrum in the left quadrant is different from the other three.

Let us recall that in the Sample Model, the \Fermi bubbles template is defined
only for $|b| > 10^\circ$.
The difference in the spectra of the quadrant templates may be due to an asymmetry in emission from the \Fermi bubbles near the GC, cf. Figure \ref{fig:SCA_bubbles}.
To qualitatively investigate this hypothesis we ``correct'' the spectrum of the left quadrant by adding a bubble-like spectrum (as derived in the Sample Model Figure \ref{fig:baseline_spectra}) and a background-like spectrum ($\propto E^{-2.4}$, the same as the soft spectral component in the derivation of the all-sky bubbles template) with free normalizations adjusted so that differences with respect to the bottom quadrant are minimized. We take the bottom quadrant as reference since it is the least contaminated by emission from the Galactic plane and local gas to the North of the GC.
The ``corrected'' left quadrant spectrum is compared to the others in Figure \ref{fig:quads_sample} on the right. This shows that it is plausible to have a spherical excess on top of emission from the \Fermi~bubbles and other foregrounds. Another implication is that the bubbles are contributing mostly to the right, top, and bottom quadrants, i.e., the contribution is asymmetric with respect to the GC, which is consistent with the description of the bubbles by \cite{2016ApJS..223...26A}.

Negative values of the GC excess spectrum below 1 GeV in, e.g., Figure \ref{fig:quads_sample} on the right are troubling. 
Throughout the paper, we restrict the values of the diffuse emission components to be non-negative,
but since the presence of the GC excess is not known a priori, we treat it as a ``residual'' component and allow both positive and negative
values.
If the morphology of the foreground components is not modeled perfectly, the fit can try to compensate for imperfections by subtracting the GC excess template such that the correction can be larger than the otherwise positive flux from the excess component.
Indeed, we notice that the contribution from the background-like component in Figures \ref{fig:quads_sample} on the right is negative,
i.e., one needs to subtract the background-like spectrum from the left quadrant to minimize differences with the bottom quadrant,
and the GC excess spectrum in the four quadrants is mostly positive with respect to this ``corrected'' zero level.

\begin{figure}[htbp]
\begin{center}
\includegraphics[scale=\twopic]{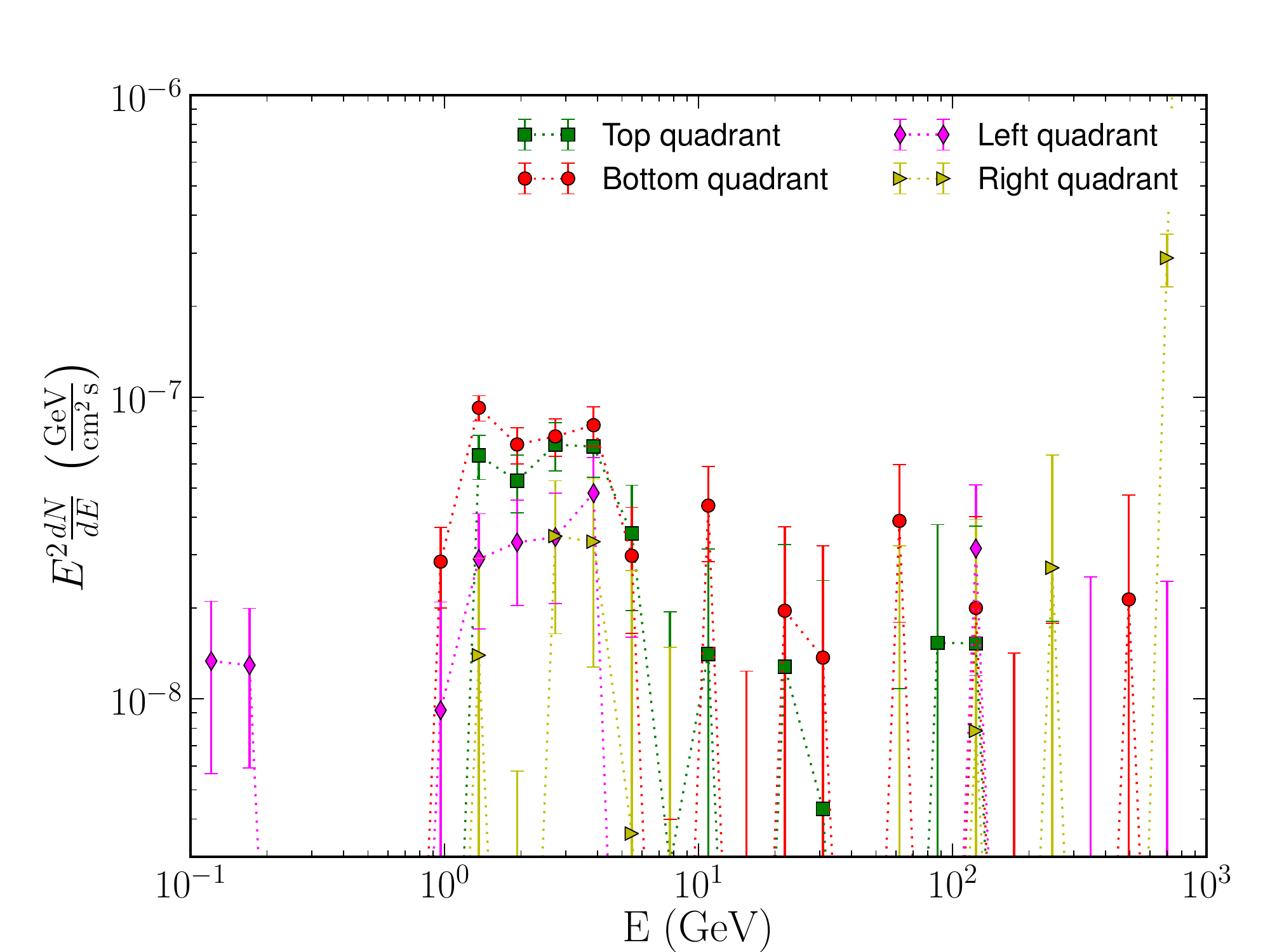}
\noindent
\caption{\small 
Same as Figure \ref{fig:quads_sample} left but with the diffuse model including the all-sky bubbles template,
(Section~\ref{sec:Fermibubbletemplate}).
}
\label{fig:quads_sca}
\end{center}
\end{figure}

Figure \ref{fig:quads_sca} shows the spectra of the four quadrants once the all-sky bubbles template derived in Section~\ref{sec:SCA} is included in the fit. 
Similarly to Figure \ref{fig:SCA_cutoff_spectra}, the spectrum of the quadrant templates changes dramatically, and emission remains significant only below 10 GeV. The spectra in the four quadrants are closer to each other, but yet not consistent within the statistical uncertainties. This inconsistency, however, may be due to an imperfect derivation of the \Fermi bubbles template.

In summary, we find that establishing whether the GC excess has spherical morphology is challenging due to uncertainties in the contribution from low-latitude emission from the \Fermi bubbles. However, at present we cannot exclude that a component with spherical morphology is present in addition to a continuation of the \Fermi bubbles.

\subsection{Longitude, Latitude, and Radial Profiles}

\begin{figure}[htbp]
\begin{center}
\includegraphics[scale=\twopic]{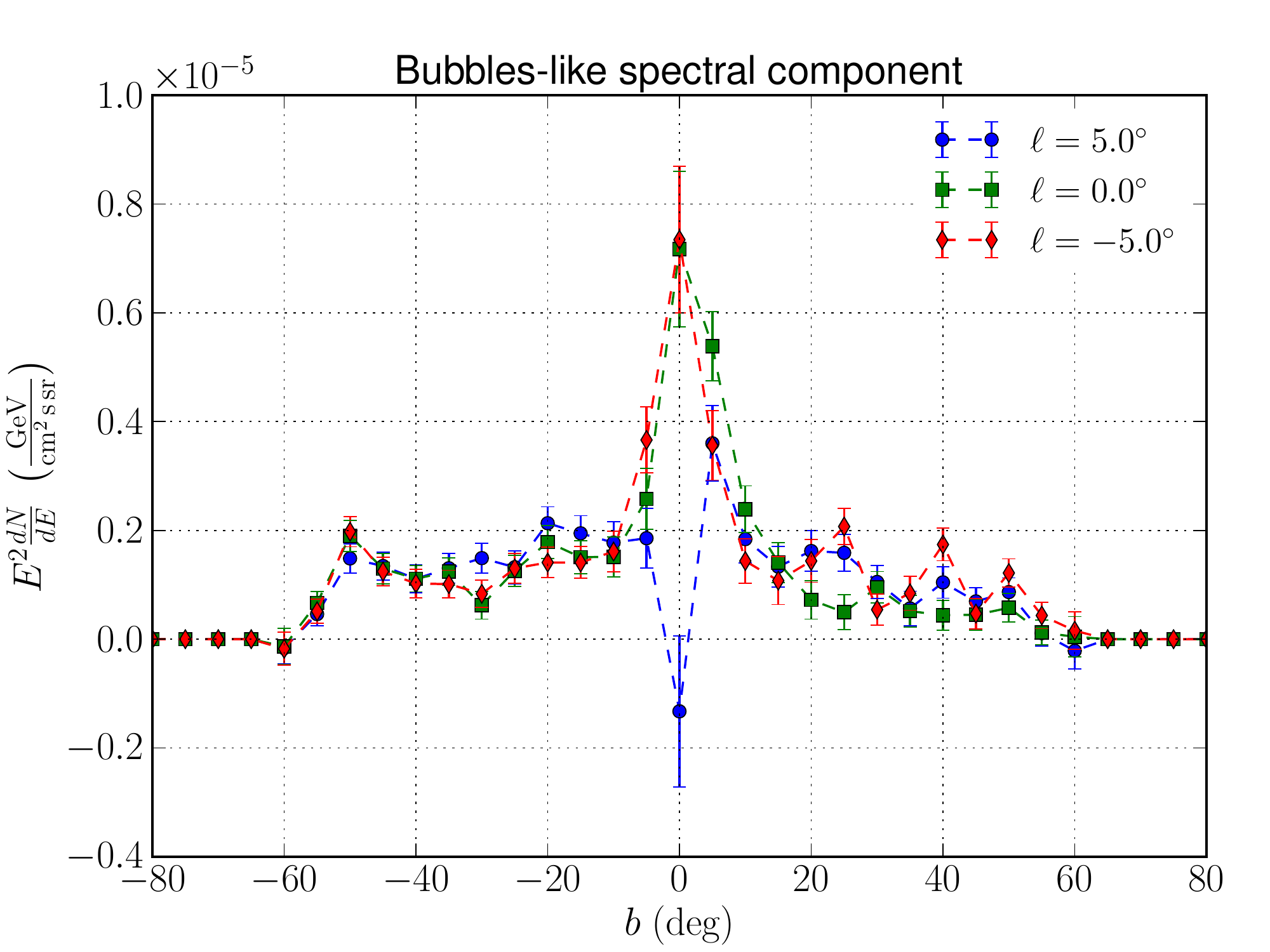}
\includegraphics[scale=\twopic]{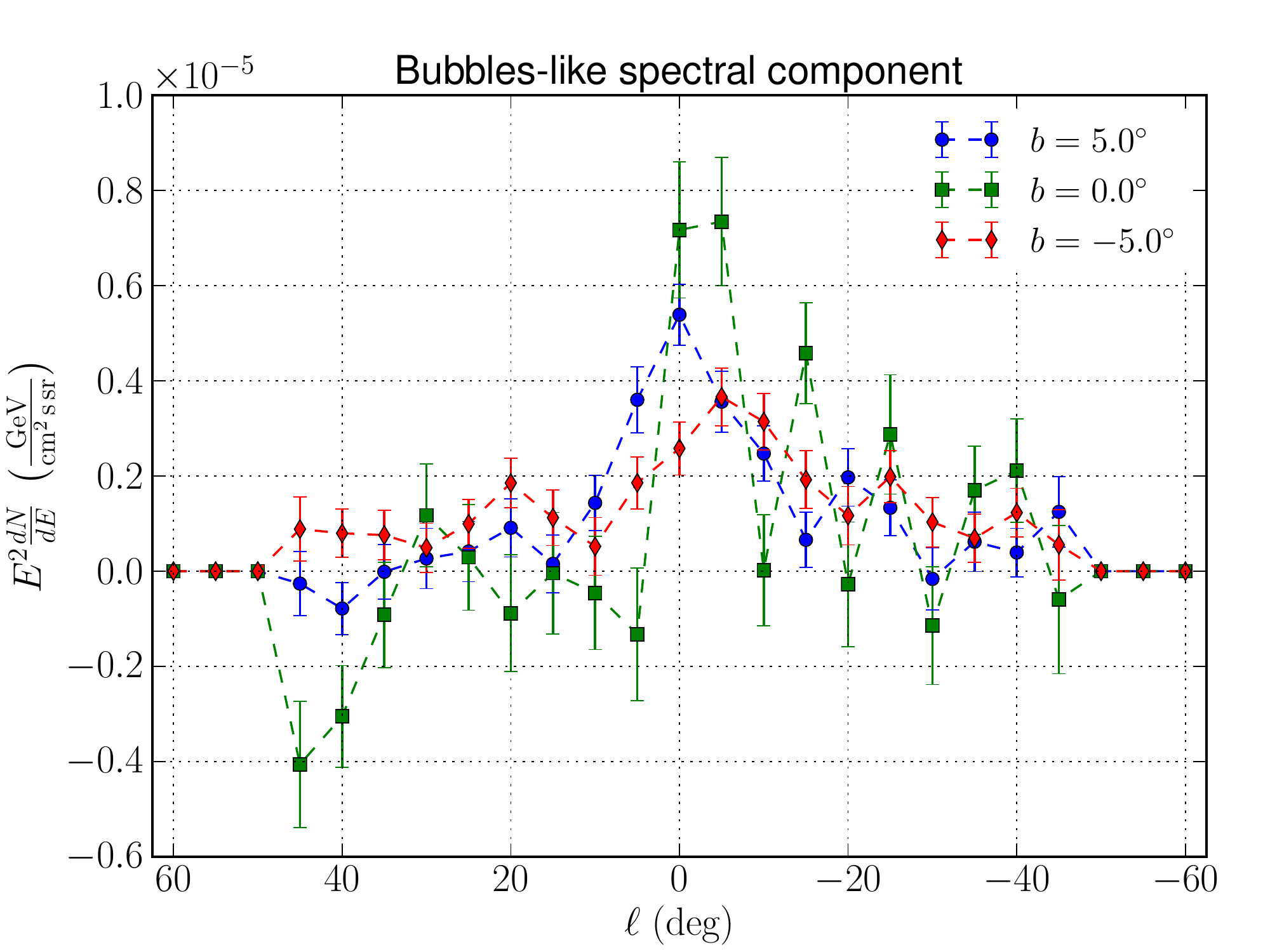}
\noindent
\caption{\small 
Latitude and longitude profiles of the bubble-like component (the medium component in Figure \ref{fig:SCA_cutoff_components}).
The normalization corresponds to the intensity of this component at 2 GeV.
}
\label{fig:bbllike_profiles}
\end{center}
\end{figure}

In Section~\ref{sec:3comp_sca}, we derived templates for the emission near the GC correlated with the \Fermi bubbles spectrum at high latitudes 
and with an average MSP spectrum.
Longitude and latitude profiles for the component with a bubble-like spectrum (Figure \ref{fig:SCA_cutoff_components} middle)
are presented in Figure \ref{fig:bbllike_profiles}.
The profiles are shown at a reference energy of 2 GeV.
The latitude profiles are relatively flat for $10^\circ \lesssim |b| \lesssim 50^\circ$,
but the intensity increases by a factor of $\sim$ 5 near the Galactic plane.
One can also see that the emission associated with the \Fermi bubbles in this model is shifted to the right (negative longitudes) 
relative to the GC.

\begin{figure}[htbp]
\begin{center}
\includegraphics[scale=\twopic]{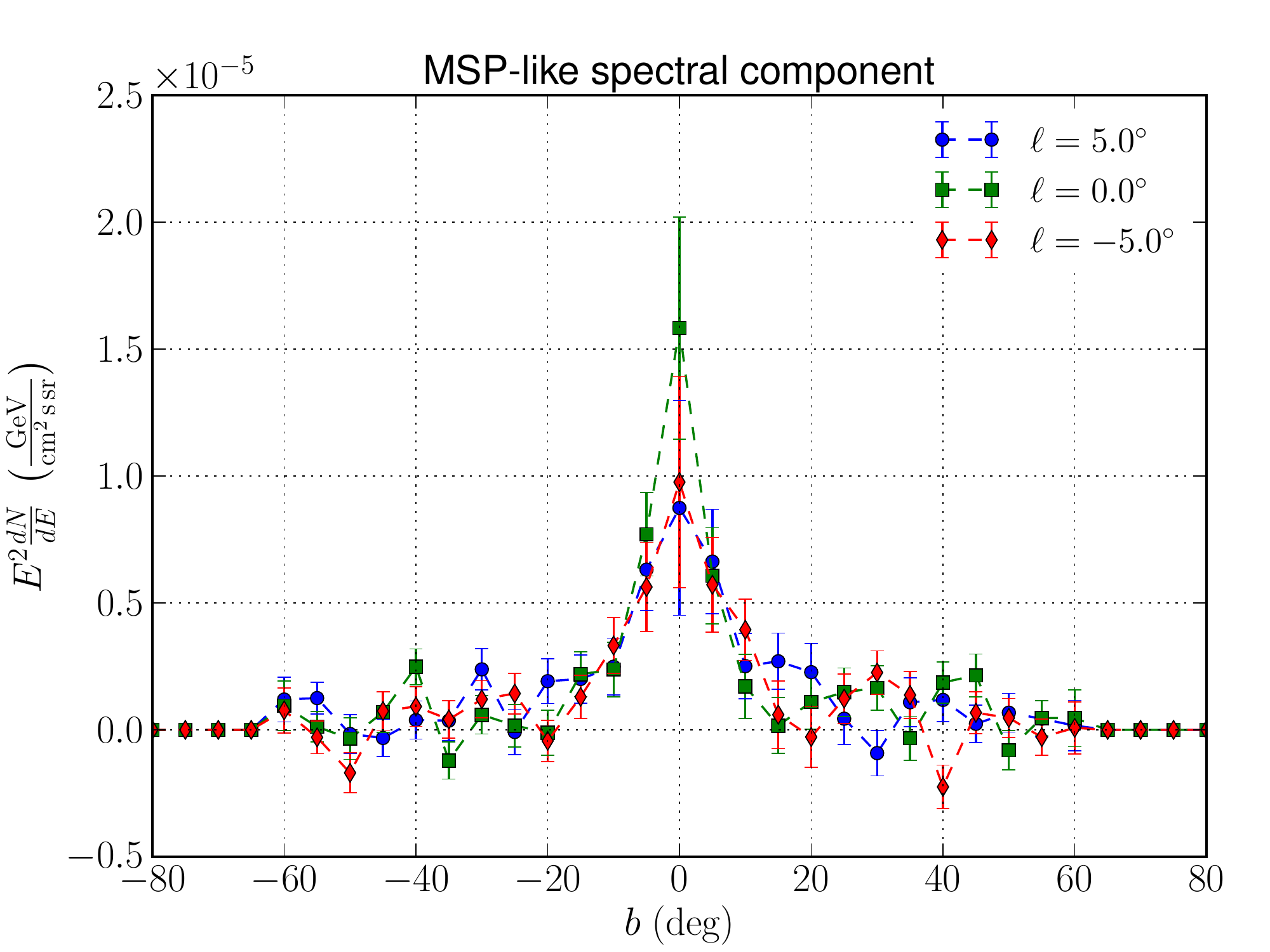}
\includegraphics[scale=\twopic]{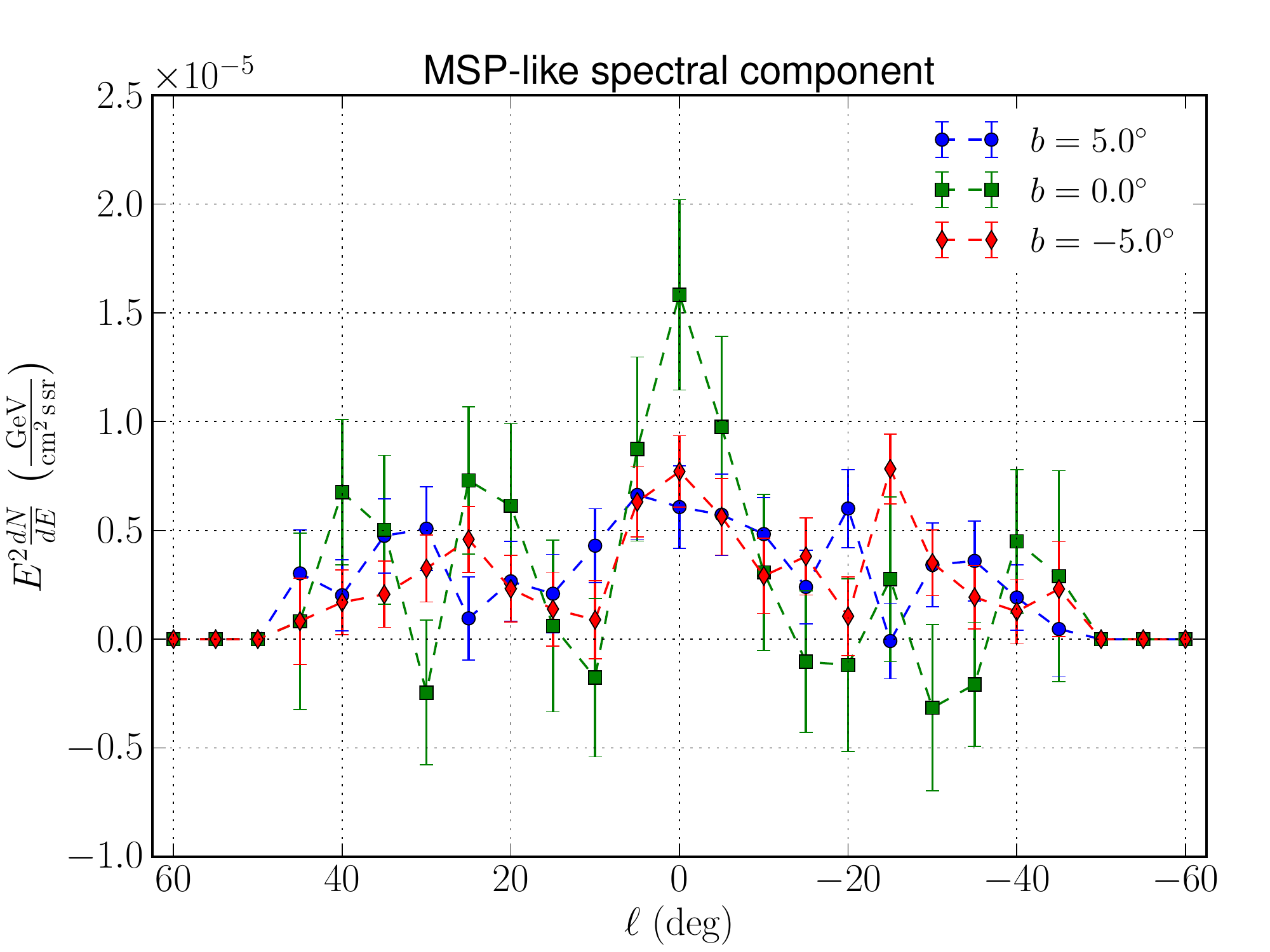}
\noindent
\caption{\small 
Latitude and longitude profiles of the MSP-like component (the hard component in Figure \ref{fig:SCA_cutoff_components}).
The normalization corresponds to the intensity of this component at 2 GeV.
}
\label{fig:MSPlike_profiles}
\end{center}
\end{figure}

Similarly, longitude and latitude profiles of the MSP-like component (Figure \ref{fig:SCA_cutoff_components} on the right)
are shown in Figure \ref{fig:MSPlike_profiles}.
The latitude and longitude profiles of this component are symmetric with respect to the GC, with a possible enhancement along the Galactic plane, which can be expected as a contribution from millisecond and regular pulsars in the Galactic disk
\citep[e.g.,][]{2010JCAP...01..005F, 2013A&A...554A..62G}. 

\begin{figure}[htbp]
\begin{center}
\includegraphics[scale=\twopic]{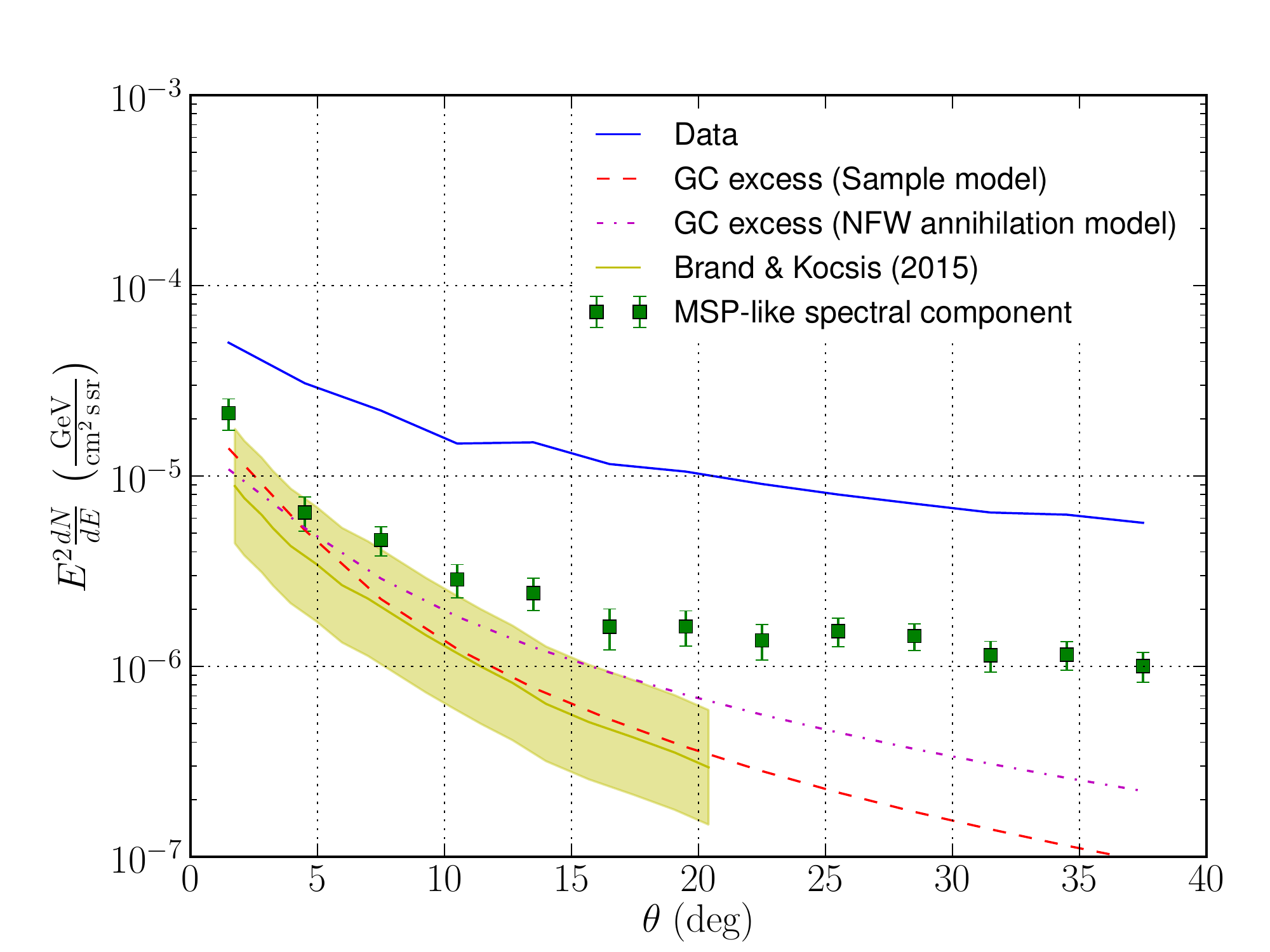}
\noindent
\caption{\small 
Solid blue line: radial profile as a function of distance from the GC for the total gamma-ray data at 2 GeV with bright PS masked. 
Squares: radial profile of the MSP-like spectral component (the hard component in Figure \ref{fig:SCA_cutoff_components}).
Dashed red line: GC excess in the Sample Model modeled by the gNFW profile ($\gamma = 1.25$).
Dash-dotted magenta line: GC excess profile in the Sample Model with the NFW profile ($\gamma = 1$) replacing the gNFW profile.
Yellow band: expectation for a population of MSPs in the Galactic bulge from disrupted globular clusters \citep{2015ApJ...812...15B}.
All values correspond to intensity at 2 GeV.
}
\label{fig:MSPlike_radial}
\end{center}
\end{figure}

Finally, in Figure \ref{fig:MSPlike_radial} we compare the profile as a function of radial distance from the GC at 2 GeV for the MSP-like spectral component with the total gamma-ray data 
and the gNFW profiles in the Sample Model, as well as for a standard NFW annihilation profile.
The MSP-like profile is similar to the DM annihilation profiles (gNFW with $\gamma = 1.25$ in the Sample Model
and the NFW profile)
within $\sim 5^\circ$ of the GC but it flattens at a higher intensity than
the gNFW profile, which is likely related to the positive values of the MSP-like component along the disk, cf. the longitude profile 
in Figure \ref{fig:MSPlike_profiles} on the right.

We also checked that using alternative PS templates within $10^\circ$ from the GC derived for UltraCleanVeto data with \textit{pointlike} and \textit{Fermipy} tools (Section \ref{sec:PS_refitting}) does not significantly affect any of the profiles for the MSP-like component.

In summary, the profiles in latitude, longitude, and radial distance from the GC corroborate the hypothesis that the excess is not obviously consistent with expectations from DM annihilation with gNFW/NFW density profiles, but such a component may exist in addition to emission from the \Fermi bubbles and from sources in the Galactic disk/bulge such as MSPs.

\subsection{Position and Index of the Generalized NFW Profile}
\label{sec:NFWindex}

\begin{figure}[htbp]
\begin{center}
\includegraphics[scale=\twopic]{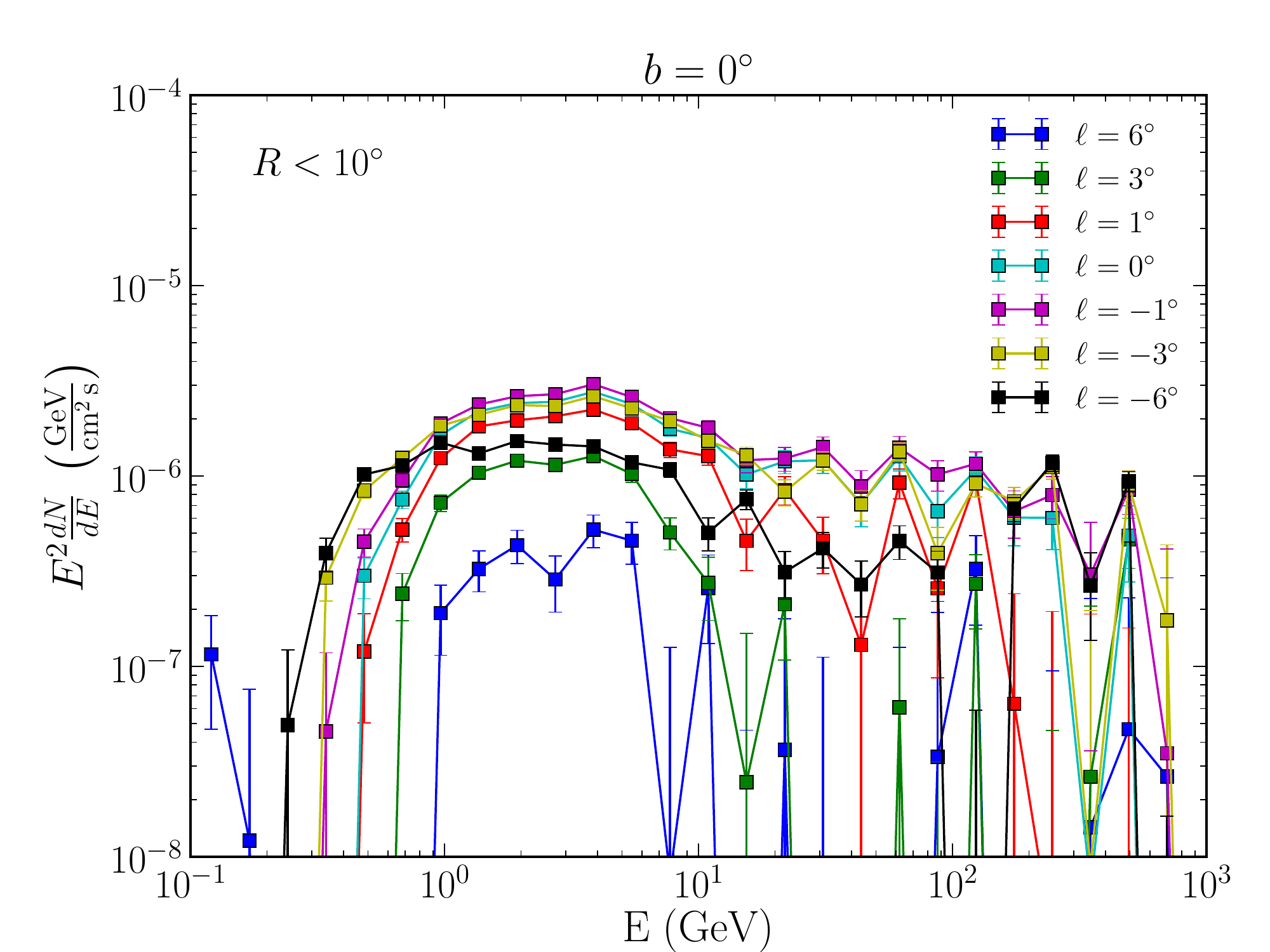}
\includegraphics[scale=\twopic]{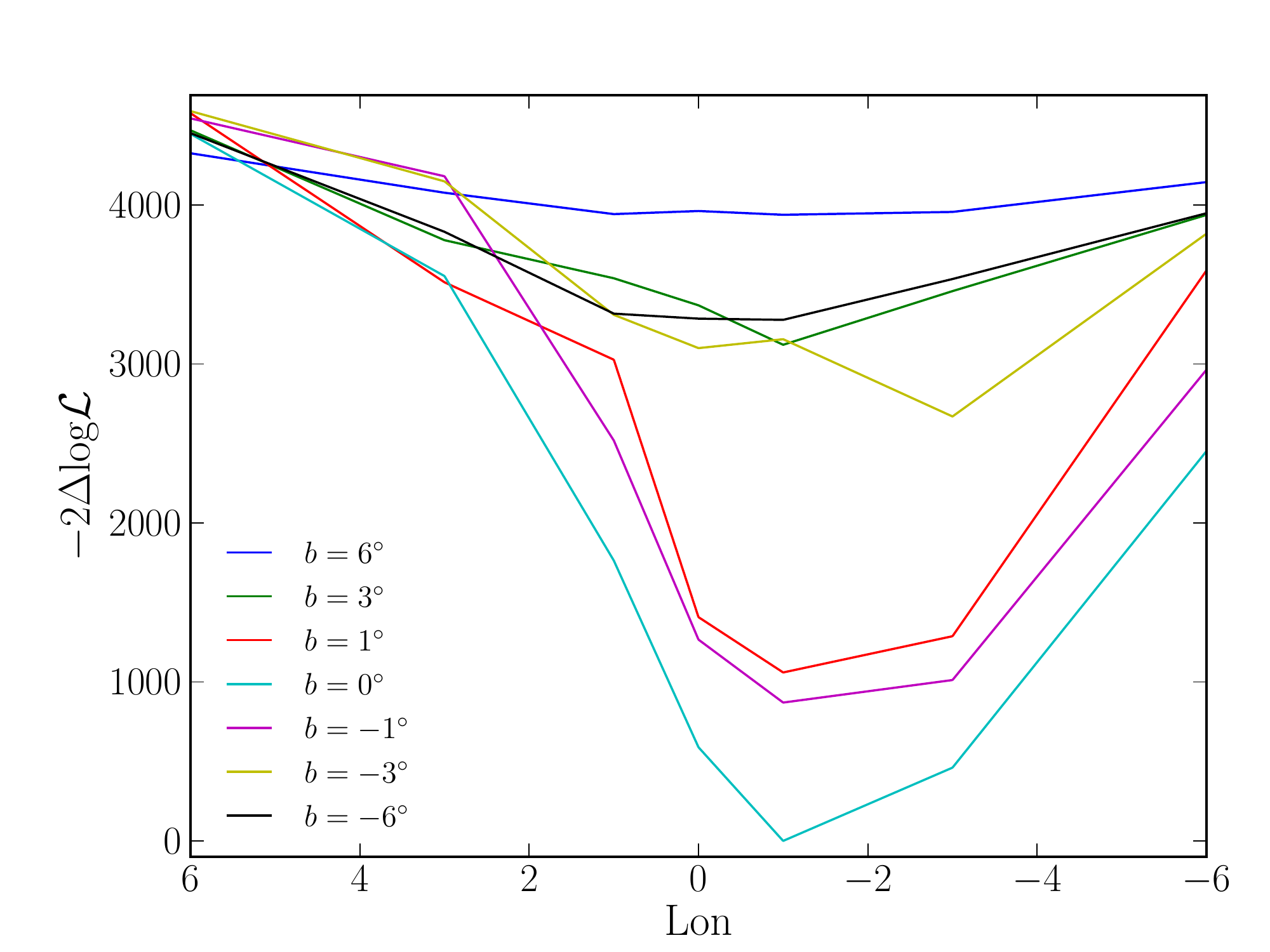}
\noindent
\caption{\small 
Scan of the position of the gNFW template center near the GC.
Left: spectra of the GC excess for the gNFW template center at $b = 0^\circ$.
Right: best-fit $-2 \Delta {\rm \log \La}$ for different positions of the gNFW center around the GC.
The minimum is at $b = 0^\circ$, $\ell = -1^\circ$, 
which corresponds to the largest flux associated with the gNFW template on the left-hand plot
(this is where the GC excess contributions is the most significant).
We also note that the improvement in log likelihood is relatively symmetric as a function of latitude.
}
\label{fig:BLscans}
\end{center}
\end{figure}

In this section, we assess the relative likelihoods of models in which we vary the centroid position of the gNFW annihilation template around the GC as well as its radial index $\g$.
Results from the scan in the position of the center of the gNFW template are shown in Figure \ref{fig:BLscans}.
The spectra of the excess for cases with the component centered at $b = 0^\circ$ and with various longitudes are presented in Figure \ref{fig:BLscans} on the left, while the $-2 \Delta {\rm \log \La}$ values for different locations around the GC are shown in Figure \ref{fig:BLscans} on the right.
The best-fit position is at $l \approx -1^\circ$. 
The spectrum of the excess depends on the location of the centroid. 
The spectra for the center at positive longitudes look similar to the GC excess in the left quadrant in Figure \ref{fig:quads_sample} left,
while the spectra for negative longitudes resemble more the spectrum of the \Fermi bubbles
with a less-pronounced bump at a few GeV and the spectrum extending to lower energies.
These findings are consistent with the possibility that the GC excess to the right (negative longitudes) from the GC
is mixed with a contribution from low-latitude emission of the \Fermi bubbles above 10 GeV.
This is also consistent with the observation by \cite{Calore:2014xka} that the best-fit longitude of the gNFW profile
is at $l \approx -1^\circ$ below 10 GeV and it shifts to $l \lesssim -2^\circ$ above 10 GeV.

\begin{figure}[htbp]
\begin{center}
\includegraphics[scale=\twopic]{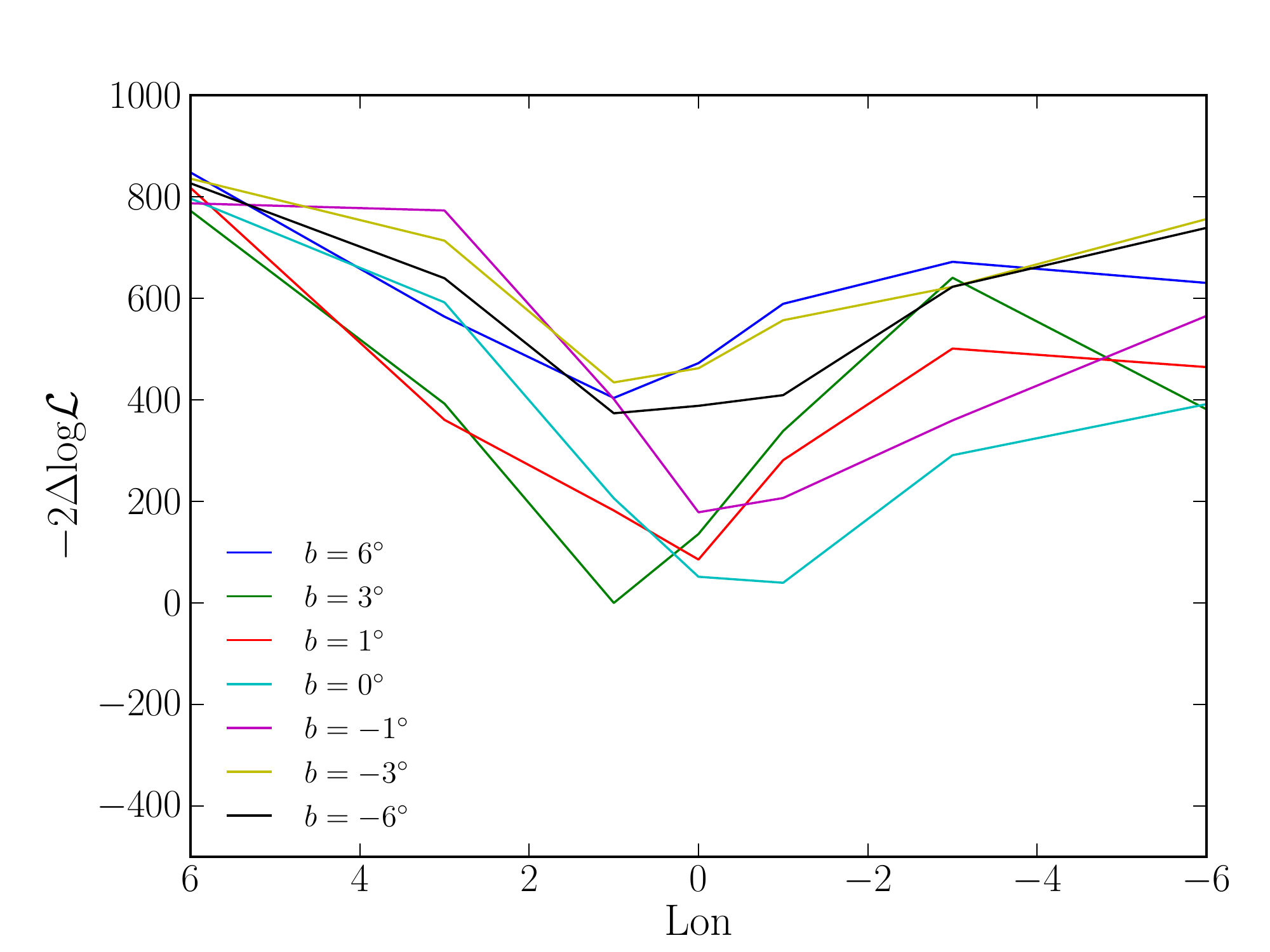}
\includegraphics[scale=\twopic]{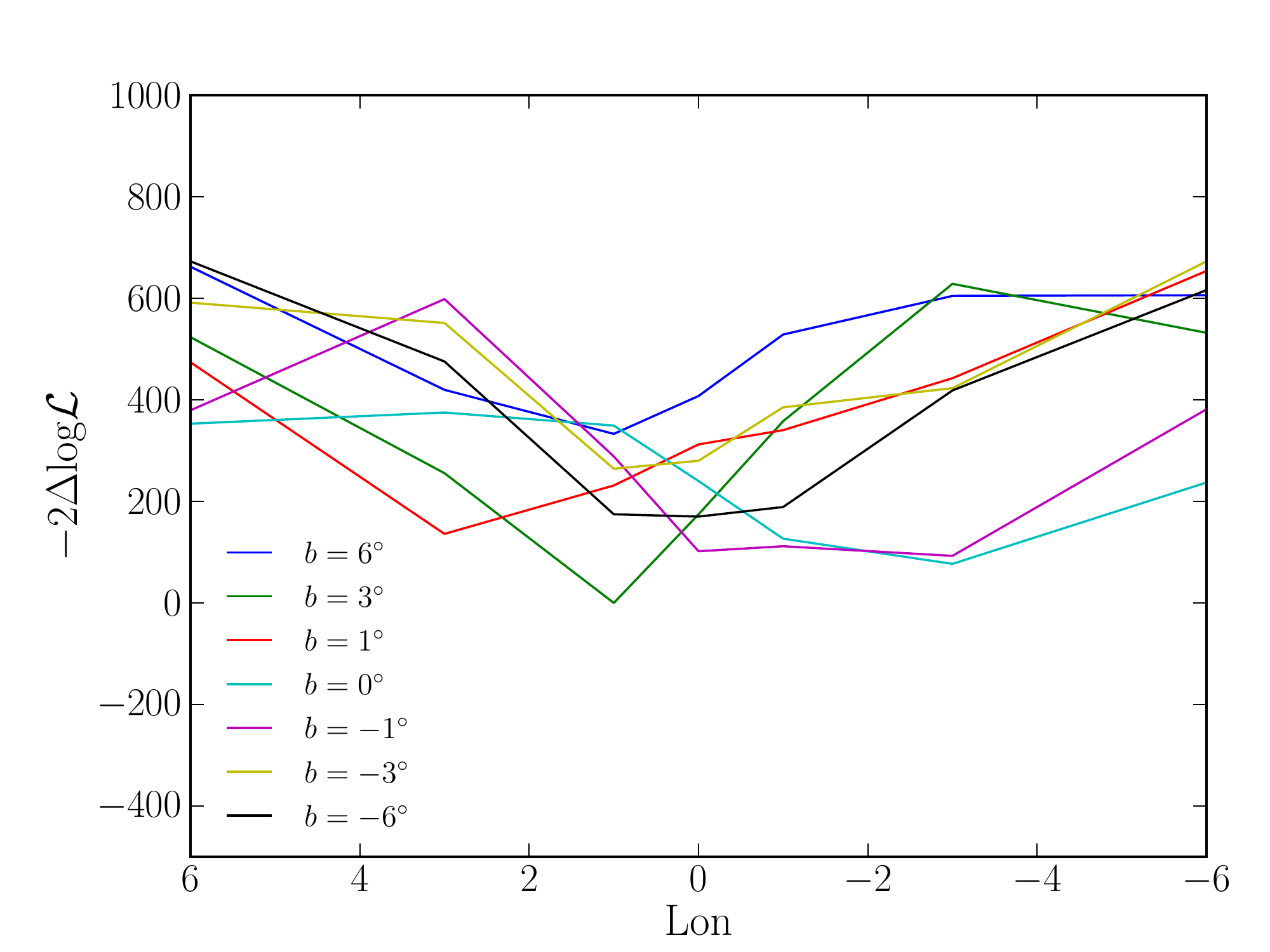}
\noindent
\caption{\small 
Scan of the position of the center of the gNFW template near the GC including the all-sky bubbles template.
Left: all-sky gNFW template.
Right: gNFW template truncated at $10^\circ$ from the center.
}
\label{fig:BLscans_bbl}
\end{center}
\end{figure}

The $-2 \Delta {\rm \log \La}$ values for the variations of the gNFW center when the model includes the all-sky bubbles template (i.e., including the component at low latitudes) are shown in Figure \ref{fig:BLscans_bbl}.
On the left we show results for all-sky gNFW templates, while on the right we truncate the gNFW template at $10^\circ$ from its center to test whether the difference in the best-fit location of the center
is due to residuals away from the GC.
Both the truncated and the all-sky gNFW templates have the best-fit position at $b = 3^\circ$, $\ell = 1^\circ$.
The range of $-2 \Delta {\rm \log \La}$ values is smaller when the all-sky bubbles component is included in the model.

\begin{figure}[htbp]
\begin{center}
\includegraphics[scale=\twopic]{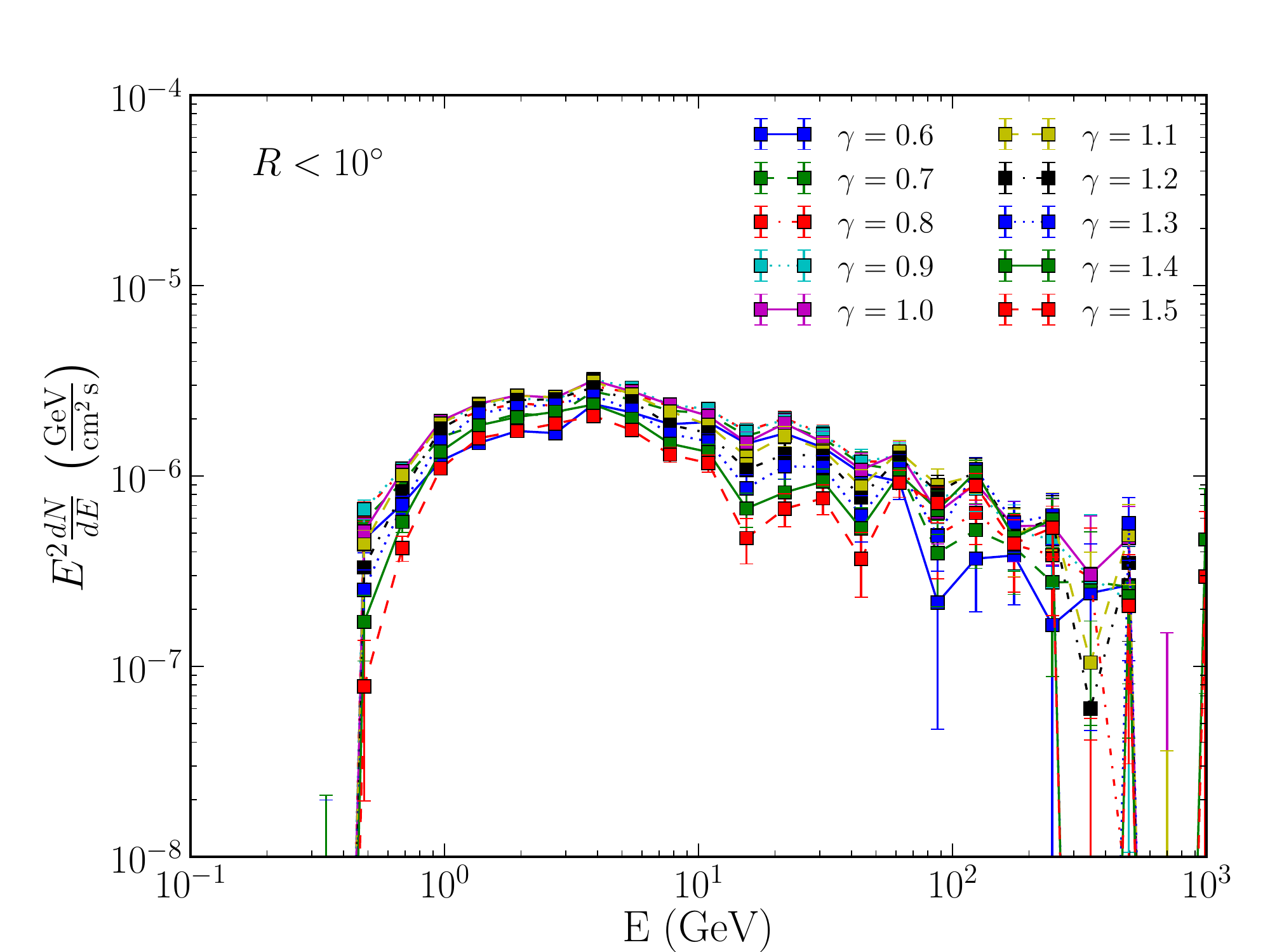}
\includegraphics[scale=\twopic]{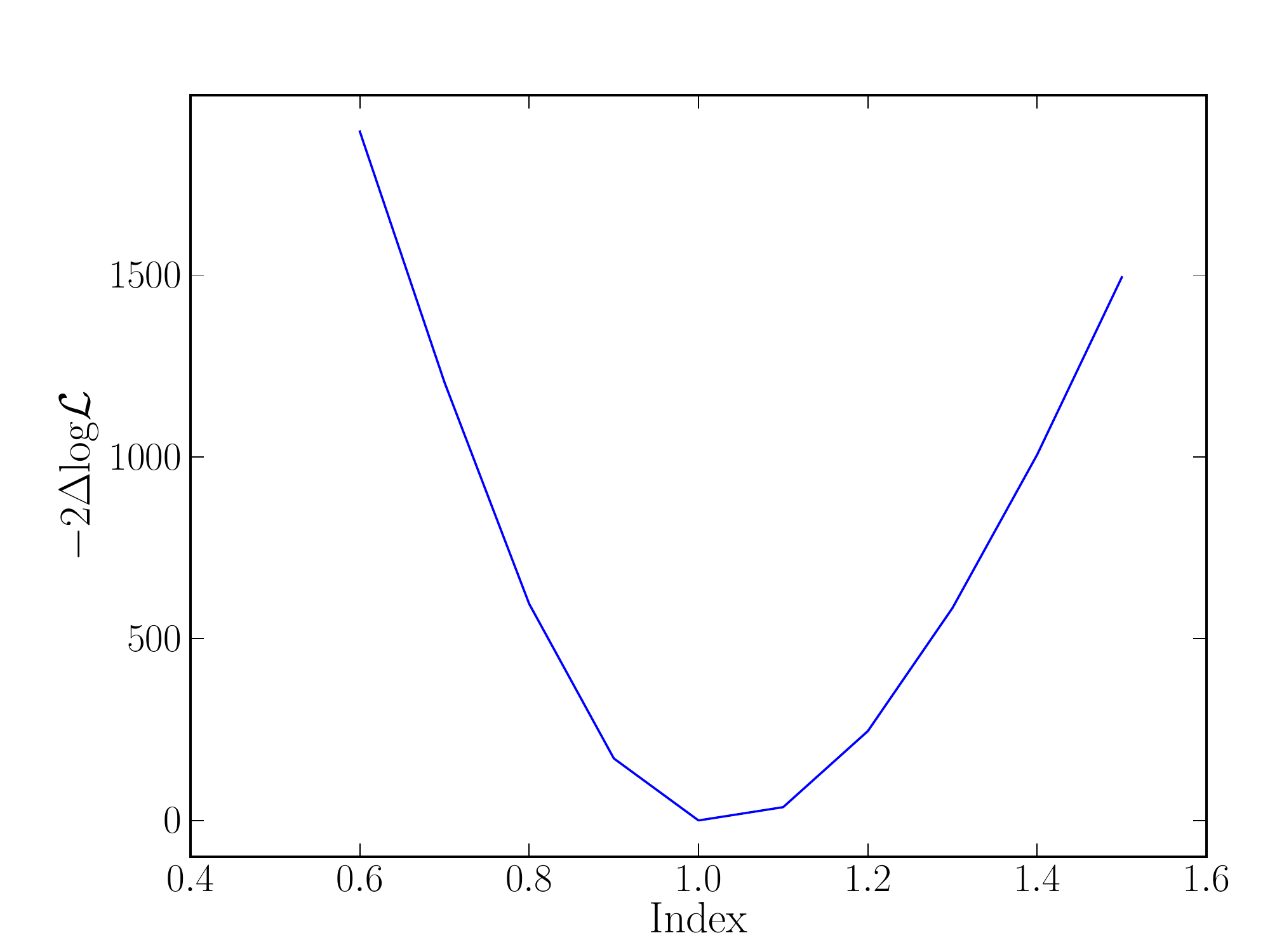}
\noindent
\caption{\small 
Spectra and $-2 \Delta {\rm \log \La}$ for different choices of the index of gNFW DM annihilation profile.
The best-fit index for the Sample Diffuse Model is $\g = 1$, which is the standard NFW profile. 
}
\label{fig:ind_scans}
\end{center}
\end{figure}

The spectra of the GC excess and the $-2 \Delta {\rm \log \La}$ for different indices of the gNFW profile when all other components are accounted for as in the Sample Model are shown in Figure \ref{fig:ind_scans}.
The step in the index scan was 0.1.
The best likelihood is obtained for the standard NFW profile with radial index $\gamma = 1$.
Scanning the profile with other models from Section \ref{sec:CRmodel} and from Section \ref{sec:PS} (including independent PS templates within $10^\circ$ from the GC), we find that 
for most of the models the best-fit indices are between 0.9 and 1.2 (the scan step is 0.1 in each case),
which overlaps with the range of indices found by \cite{Calore:2014xka}
and with the best-fit index of 1.2 found by \cite{2013PDU.....2..118H}.

\begin{figure}[htbp]
\begin{center}
\includegraphics[scale=\twopic]{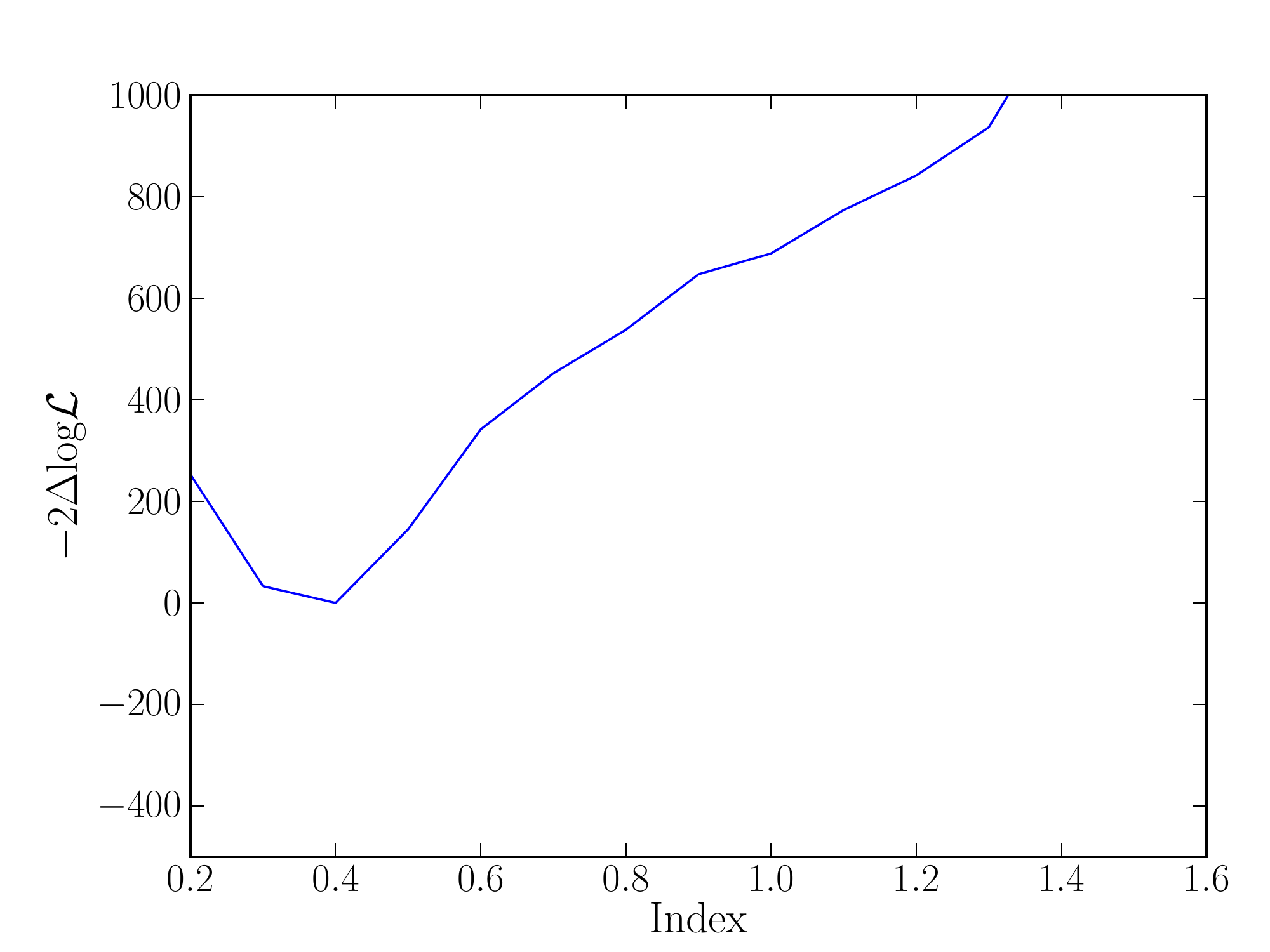}
\includegraphics[scale=\twopic]{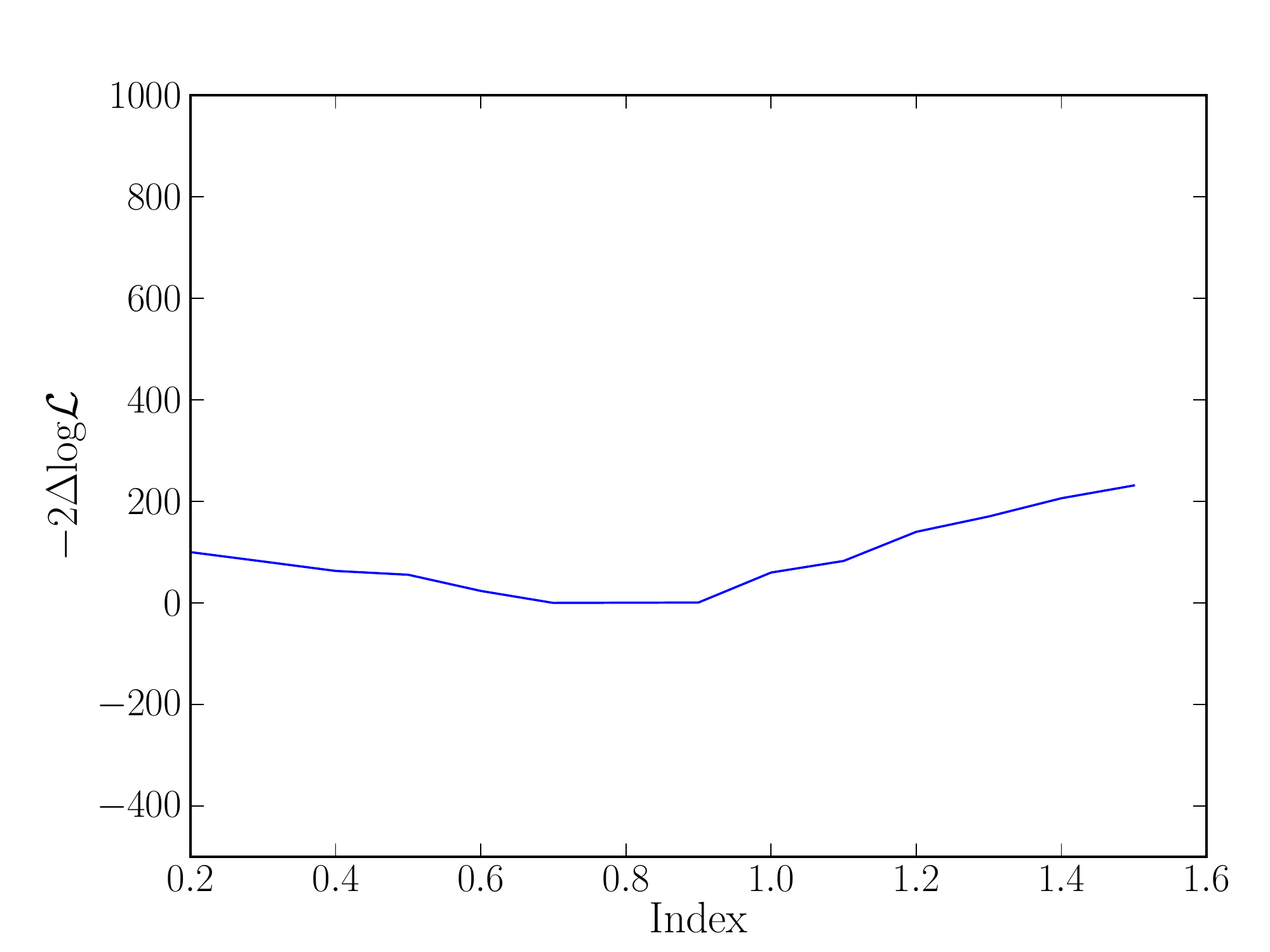}
\noindent
\caption{\small 
$-2 \Delta {\rm \log \La}$ from a scan of the gNFW index in a model with the all-sky bubbles template derived in Section \ref{sec:FBtmpl}.
Left: all-sky gNFW profile. Right: gNFW profile truncated at $10^\circ$ from the GC.
The difference in the best-fit index value is interpreted as being due to the influence of residuals away from the GC.
}
\label{fig:ind_scan_bbl}
\end{center}
\end{figure}

The results of a scan of the all-sky gNFW profile in the presence of the all-sky bubbles template derived in Section \ref{sec:FBtmpl}
are shown in Figure \ref{fig:ind_scan_bbl} on the left. The best-fit index is equal to 0.4,
which is significantly smaller than best-fit index $\gamma = 1$ obtained in the scan with the Sample Model. 
The results of the scan for the truncated gNFW template are shown in Figure \ref{fig:ind_scan_bbl} on the right.
The best-fit index for the truncated template is 0.9.

The rather large variations of the best-fit index and the gNFW centroid in the presence of the all-sky bubbles template
show that the inferred morphology
of the excess critically depends on the model of the \Fermi bubbles near the GC that can bias the derivation of the morphology and, as we have shown in Section \ref{sec:SCA},
the GC excess spectrum. Because of these large uncertainties, at present it is not possible to firmly associate the centroid of the excess with the GC itself or precisely determine its density profile.

\section{Investigation of Dark Matter Interpretation of the GeV Galactic Center Excess}
\lb{sec:DMlimits}

The predicted \g-ray signal from DM annihilation is strongest in the GC
owing to its proximity and the enhanced density of DM.
However, searches for \g-ray emission from DM annihilating in the GC are complicated by foreground
and background emission along the line of sight, and also from other processes
that can produce \g rays near the GC.
In the previous sections we explored several issues related to the uncertainties in foreground/background modeling.
We also introduced several non-DM templates to account for \g-ray emission in the inner Galaxy.
In all cases, we continued to find significant \g-ray emission correlated with the gNFW annihilation template.
However, this type of investigation necessarily remains incomplete. In this section to explore the robustness of a DM interpretation of the GC excess, we contrast the region of the GC
with control regions along the Galatic plane (GP) where no DM signal is expected.

\subsection{The Galactic Center and Control Regions along the Galactic Plane}
\lb{sec:beff_scan}

Many groups have shown that the spectral energy distribution of the GC excess peaks at energies around a few
GeV and can be fit with a model of either $\sim40$~GeV DM annihilating to \bbbar
or $\sim10$~GeV DM annihilating to \tautau~
\citep{Goodenough:2009gk, Hooper:2010mq,Hooper:2011ti, Gordon:2013vta,2013PDU.....2..118H, Daylan:2014rsa, Abazajian:2014fta, 
Calore:2014xka}.

We used DM annihilation spectra for a variety of DM masses and two representative DM annihilation channels, \bbbar and \tautau, to model the GC excess spectrum that we found using our Sample Model (see Appendix~\ref{app:fitDM} for details).
To estimate the uncertainty level of the DM-like signal, 
we repeat the analysis by placing the gNFW template at different locations along the GP instead of the GC.
Since we compare fits from many regions across the GP with varying levels of \g-ray intensity, 
we quantify the best-fit DM component as a fraction of the effective
background:
\begin{equation}
\lb{eq:fracSig}
f = \frac{N_{\rm sig}}{\beff},
\end{equation}
where $N_{\rm sig}$ is the number of signal counts integrated over the energy bins and \beff is the ``number of counts" in the effective background. 
If the signal were localized in a small region on the sky with expected intensity much smaller than the background intensity,
then the statistical variance of the signal measurement would be proportional to the number of background counts in that region.
In general, for a signal that covers a large portion of the sky (or possibly the entire sky) with a varying intensity,
one can determine a weighted sum of the background counts, the effective background, so that the statistical variance of the signal is still
proportional to this weighted sum
\citep[see][and Appendix~\ref{app:beff} for details about the evaluation of the effective background]{Ackermann:2015lka, Buckley:2015doa, Caputo:2016ryl}:
\begin{equation}
\label{eq:beff}
b_{\rm eff} = \frac{N}{ \sum_{i,\al}\frac{(P^{(\rm sig)}_{i\al})^2}{P^{(\rm bkg)}_{i\al}} - 1},
\end{equation}
where the sum is over energy indices $\al$ and pixel indices $i$, $N$ is the total number of counts,
$P^{(\rm sig)}_{i\al}$ ($P^{(\rm bkg)}_{i\al}$) are the signal (background) intensity distributions normalized to $1$:
$\sum_{i,\al} P^{(\rm sig)}_{i\al} = \sum_{i,\al} P^{(\rm bkg)}_{i\al} = 1$.
In a particular case, if there is one energy bin and the background is uniformly distributed over the whole sky 
$P^{(\rm bkg)}_{i} = 1 / N_{\rm pix}$,
while the signal is uniformly distributed over a small number of pixels $k_{\rm pix}$ so that 
$P^{(\rm sig)}_{i} = 1 / k_{\rm pix}$ and $N_{\rm pix} / k_{\rm pix} \gg 1$,
then Equation (\ref{eq:beff}) gives
$\beff \approx N \cdot k_{\rm pix} / N_{\rm pix}$, which is the number of background counts in the signal region.

\begin{figure*}[t]
\includegraphics[width=0.5\linewidth]{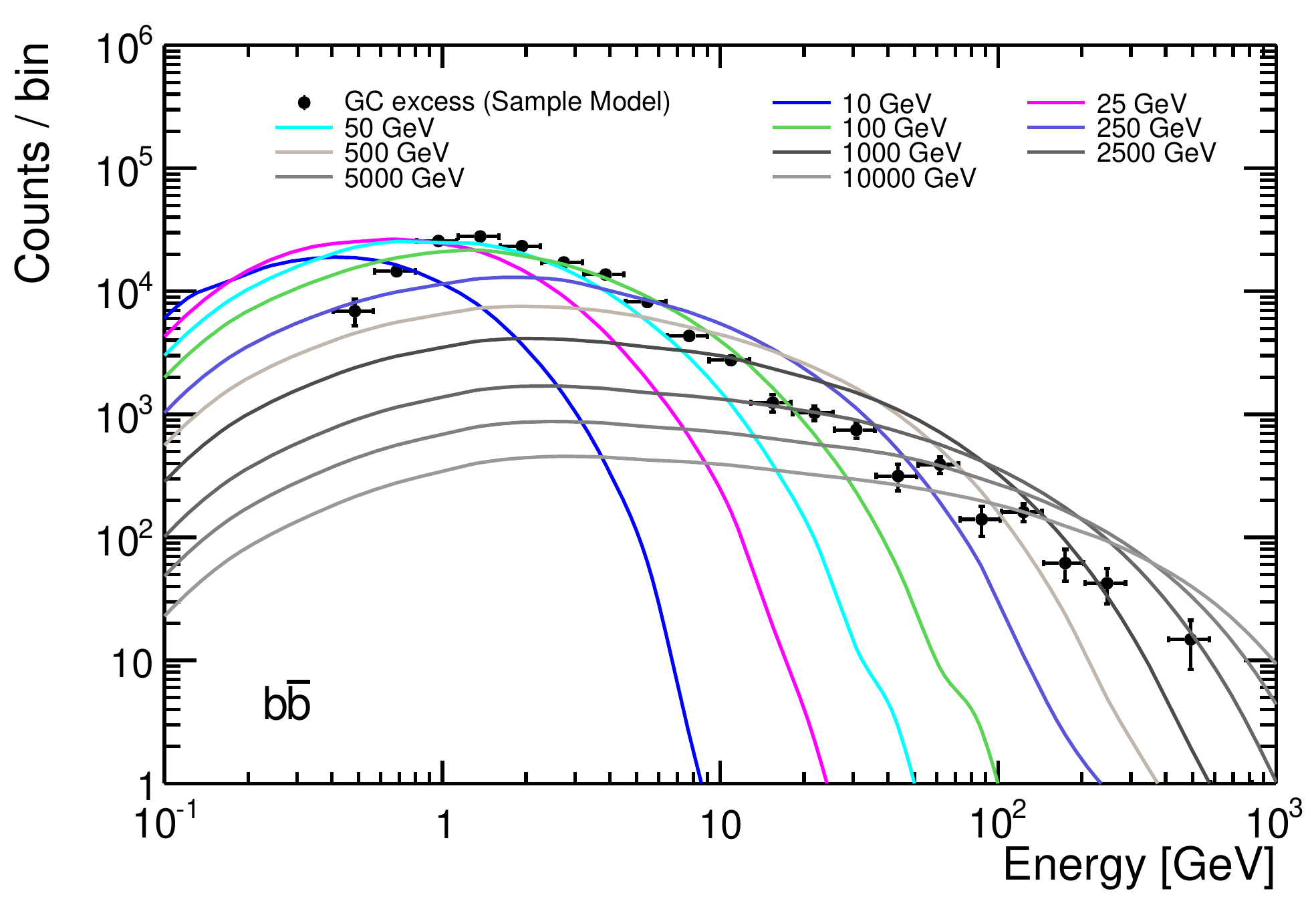}
\includegraphics[width=0.5\linewidth]{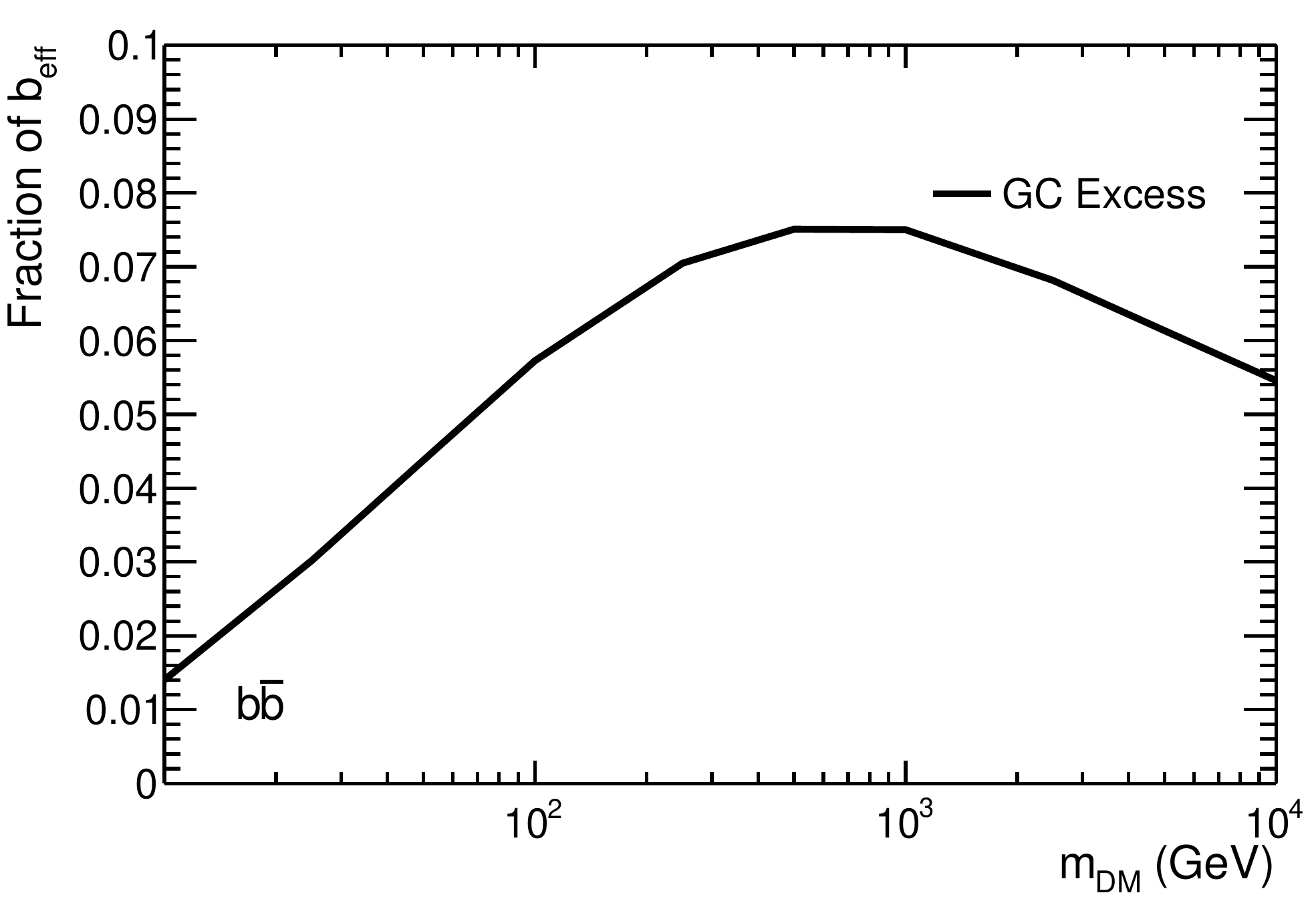}
\caption{\label{GCE_specDM} Left: Best-fit DM model for the GC excess energy spectrum in the Sample Model (Section \ref{sec:baseline})
transformed to counts. Different curves correspond to different masses of DM particles.
Right: size of each best-fit DM model to the GC excess spectrum in the Sample Model as a fraction of \beff.}
\end{figure*}

Figure \ref{GCE_specDM} (left) shows the GC excess spectrum in the Sample Model (Section \ref{sec:baseline}) in counts space. In this fit, the normalization of the GC excess template is fit to the data together with the other templates in each energy bin. 
The first four energy bins had negative best-fit normalizations of $(-3.4\pm0.3)\times10^4, (-2.3\pm0.3)\times10^4, (-1.5\pm0.3)\times10^4, (-1.3\pm1.8)\times10^3$ respectively. 
The errors given are the statistical errors only.
Figure \ref{GCE_specDM} also shows the best-fit DM annihilation spectrum for the \bbbar channel for various values of the mass of the DM particle,
$m_{\rm DM}$.
The DM annihilation counts spectra were calculated from their corresponding flux spectra using the DMFitFunction within the standard \Fermi Science Tools and the detector exposure for this data set.
The signal counts $N_{\rm sig}$ for each DM model are simply the integral of the best-fit counts spectrum. Therefore, we
can evaluate the strength of the best-fitting DM model relative to the effective background \beff
(see Figure \ref{GCE_specDM} right).

We note that none of the DM fits to the GC excess spectrum are very good (the reduced $\chi^2$ is $>30$ for all fits). In particular, the sample spectrum has a high-energy ($>50$ GeV) tail that cannot be explained by, e.g., $< 50$~GeV DM annihilating to \bbbar.  However, as we have shown, the high-energy tail may be consistent with a low-latitude extension of the \Fermi bubbles (Section \ref{sec:Fermibubbletemplate}). 
Nevertheless, we consider the results of these fits as we are interested in quantifying the `DM-like' component of the spectrum for various DM models.

\begin{figure*}[t]
\includegraphics[width=0.5\linewidth]{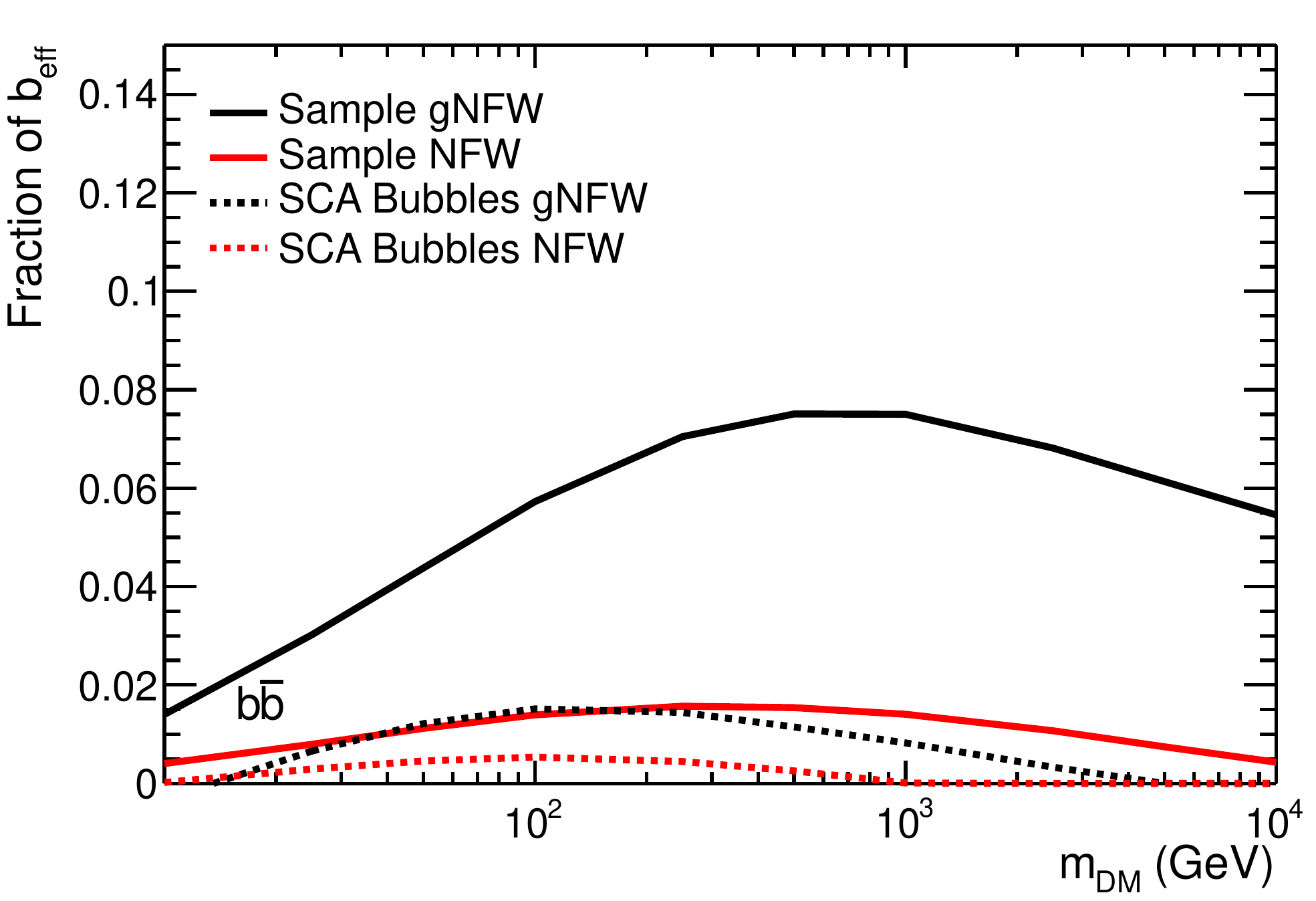}
\includegraphics[width=0.5\linewidth]{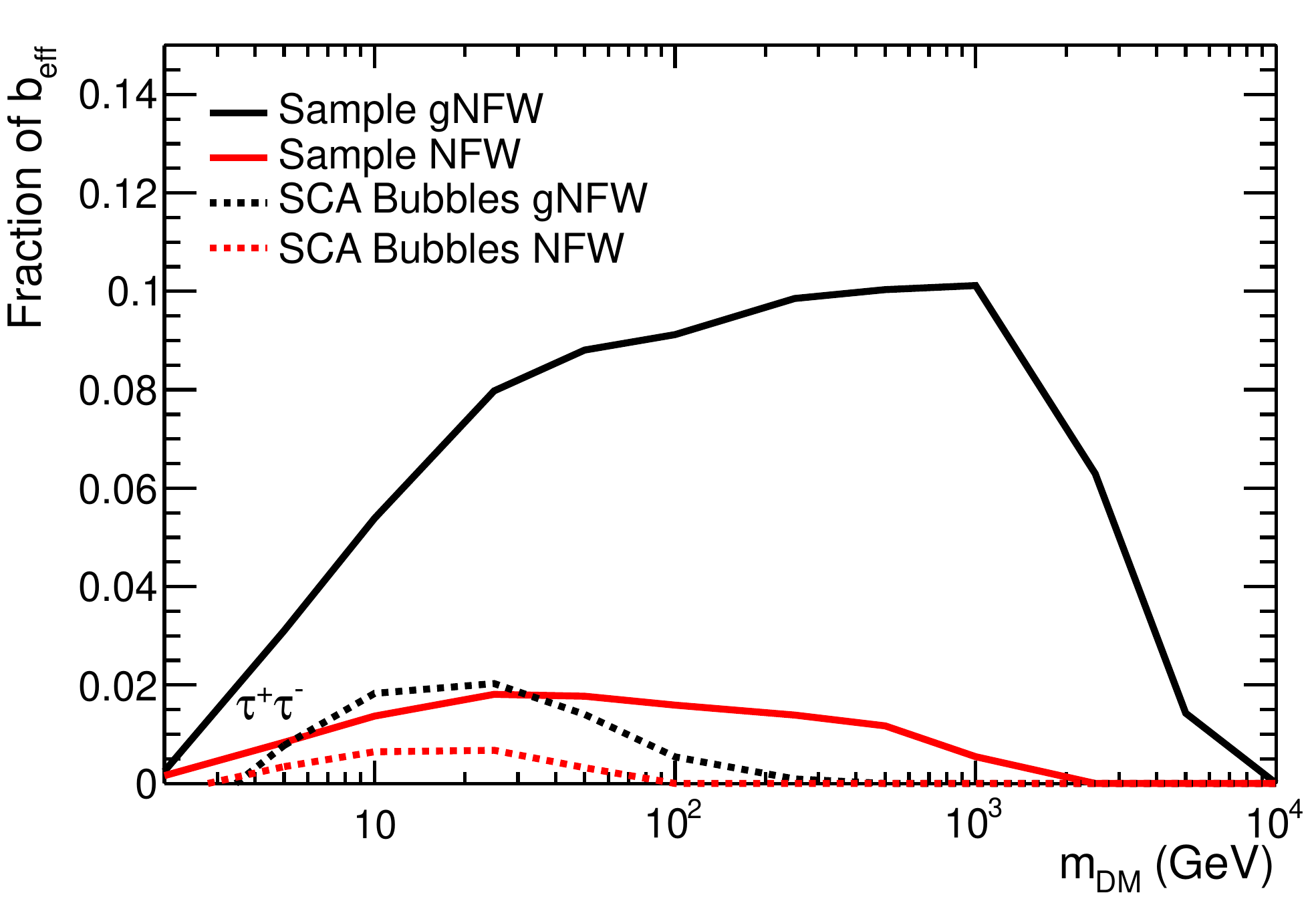}
\caption{\label{GCE_specDM_comp} 
Fraction of best-fit DM model counts relative to \beff for various GC excess spectra, as a function of dark matter mass. 
The curves show DM models fits to the GC excess spectral points (see an example of the fits in Figure \ref{GCE_specDM})
using the Sample (solid lines) and SCA bubbles (dashed lines) background models, and gNFW (black lines) and standard NFW (red lines) spatial templates used for the GC excess. Left: fits for DM annihilation to \bbbar. 
Right: fits for DM annihilation to \tautau.}
\end{figure*}

In addition to the gNFW excess template, we also fit the standard NFW ($\gamma=1.0$) DM annihilation template since it is the best-fit template for the excess in our analysis (Section~\ref{sec:NFWindex}). 
Figure \ref{GCE_specDM_comp} shows the best-fit $N_{\rm sig}$ of various DM models to the GC excess spectra both in the Sample Model and in the model including the SCA bubbles (Section~\ref{sec:Fermibubbletemplate}) as a fraction of \beff. 
The bubbles template derived using the SCA method can account for a large amount of the GC excess, especially at high energies. Therefore, the reduction of the amplitudes of the best-fit DM models (especially at high masses) is expected.

Because the GC is very bright in \g rays, many of the DM models we test have very small statistical errors in inferred $N_{\rm sig}$ ($\delta N_{\rm sig} < 0.01$).
However, we are not able to model the \g-ray sky to a similar level of precision (recall that the fractional residuals from our fits are typically in the range $-0.2$ to $0.2$; see Figure~\ref{fig:baseline_resid}). Therefore, systematic uncertainties that may mimic or mask a DM signal need to be accounted for.

To assess these systematic uncertainties beyond what was already done with model variations, we estimate $\delta N_{\rm sig,syst}$ by fitting for DM-like signals in control regions along the GP, based on two assumptions:  that the expected DM signal is approximately zero for $30^{\circ} \leq l \leq 330^{\circ}$, and the systematic uncertainty scales with \beff for effects that can induce or mask a DM-like signal.
An excess may be a fraction of the background if it is caused by a single (or a few) 
errors in the modeling of the gamma-ray intensity,
which are proportional to the ``average" emission,
or when the uncertainty is dominated by errors in a single component that also dominates the overall emission.
Fluctuations due to several small effects, such as uncertainties in emission components where each component contributes a small fraction of the total emission, would be best estimated as a square root of the $\beff$; in this case the characteristic values would be 
$N_{\rm sig} / \sqrt{\beff}$.
Fluctuations in emission which are caused by one or a few components which are not directly correlated with the overall gamma-ray emission,
such as a local SNR, or an AGN-like activity, would be best characterized by their absolute values.
Since the gamma-ray emission towards the GC is the largest, taking the fractional excess as a figure of merit to estimate its significance
is the most conservative assumption, which we will adopt for our analysis.

Control regions along the Galactic plane to estimate the modeling uncertainty were used before by \citet{Calore:2014xka}.
They fit a DM-like spatial profile along the Galactic plane and represented the results as a covariance matrix in energy bins, 
which is used to determine the expected level of modeling uncertainty at the GC.
Our approach is to fit DM-like excess along the Galactic plane including both the spatial profile and the energy spectrum of a DM annihilation channel.
We then express the uncertainty as a ratio of the signal to the local effective background.
Both of these differences are likely to increase the estimate of the modeling uncertainty, since we get the maximal possible DM-like signal in each location, and then we divide by the local background, which is smaller along the plane than at the GC.

\begin{figure*}[t]
\includegraphics[width=0.5\linewidth]{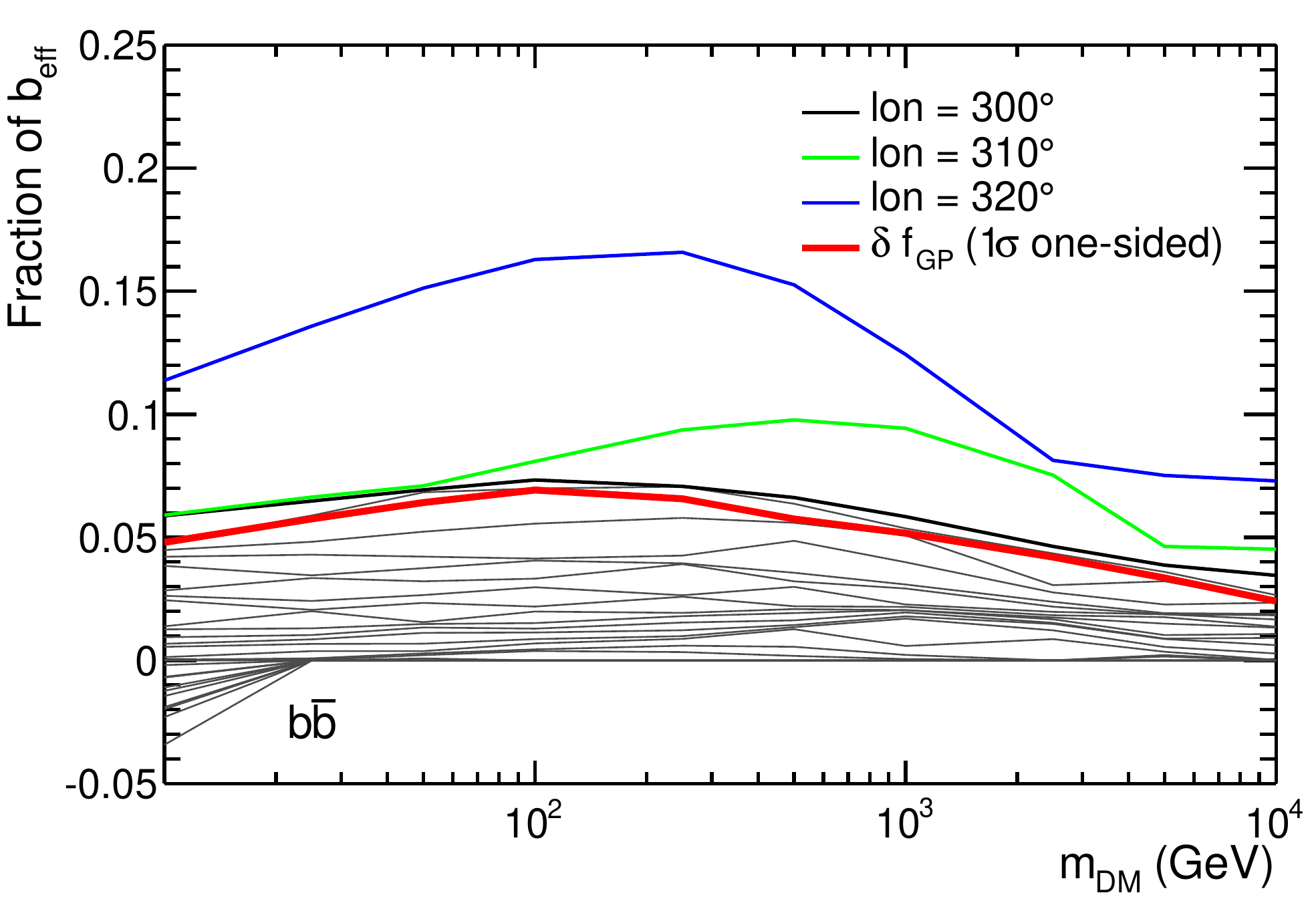}
\includegraphics[width=0.5\linewidth]{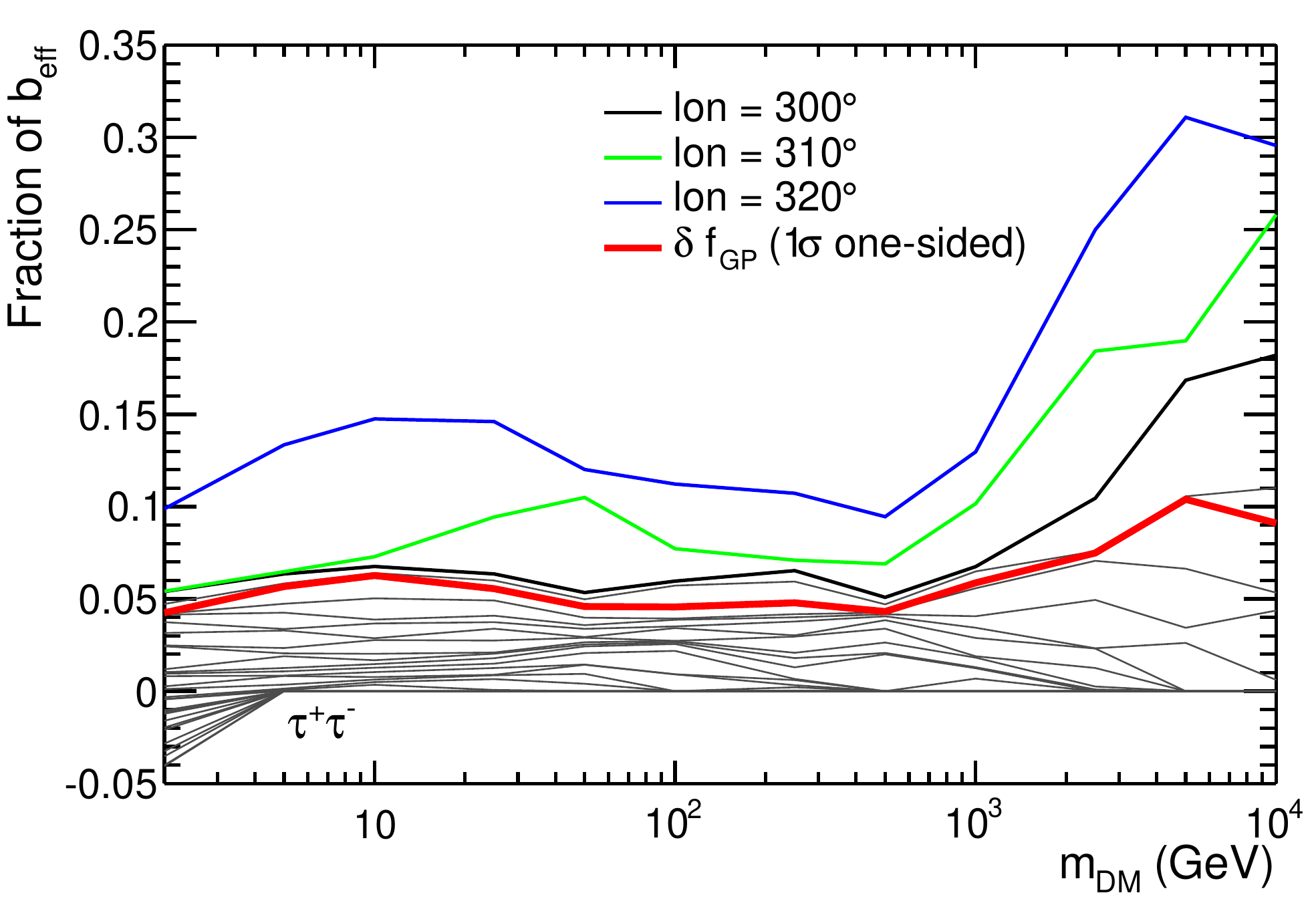}
\caption{
\label{fig:fracSig_GP} 
Size of best-fit DM models as a fraction of \beff (see text) evaluated for the gNFW template shifted in steps of $10^{\circ}$ for $30^{\circ} \leq l \leq 330^{\circ}$ at $b=0^{\circ}$. The red curve is the value chosen as an estimate of our systematic uncertainty (see text).  Only positive signals are shown. Small negative amplitudes are found only below 200 MeV in few control regions. The four largest excesses are represented by colored lines and the corresponding longitude is given in the legend.}
\end{figure*}

We perform all-sky fits using the same diffuse emission components as in the Sample Model, but shift the gNFW excess template 
in steps of $10^\circ$ in longitude at $b=0$ for $30^{\circ} \leq l \leq 330^{\circ}$. 
Figure~\ref{fig:fracSig_GP} shows the amplitudes of the best-fit DM model spectra (as a fraction of \beff) measured in the control regions.

Though the fits were allowed to be negative, most longitudes preferred a positive DM normalization.
Also, about half of the total fits had $|f| < 0.01$.
We define $\delta f_{\rm GP}$ as the value of $f$ for which we obtain larger values in only 16\% of the fits along the GP, which corresponds to a one-sided $1\sigma$ exclusion in the case of a Gaussian distribution.
We take $\delta f_{\rm GP}$ as representative of the amplitude 
of the components of the diffuse \g-ray residuals consistent with DM signals, and therefore a measure of the systematic uncertainty for our DM search in the GC:
\bea
\delta f_{\rm syst} &=& \delta f_{\rm GP}; \\ 
\lb{eq:dNsig}
\delta N_{\rm sig, syst} &=& \beff \times \delta f_{\rm syst}.
\eea
See Appendix~\ref{app:beff} for more on the motivation of this choice and the application of $\delta f_{\rm GP}$ in our fitting.

\begin{figure*}[t]
\includegraphics[width=0.5\linewidth]{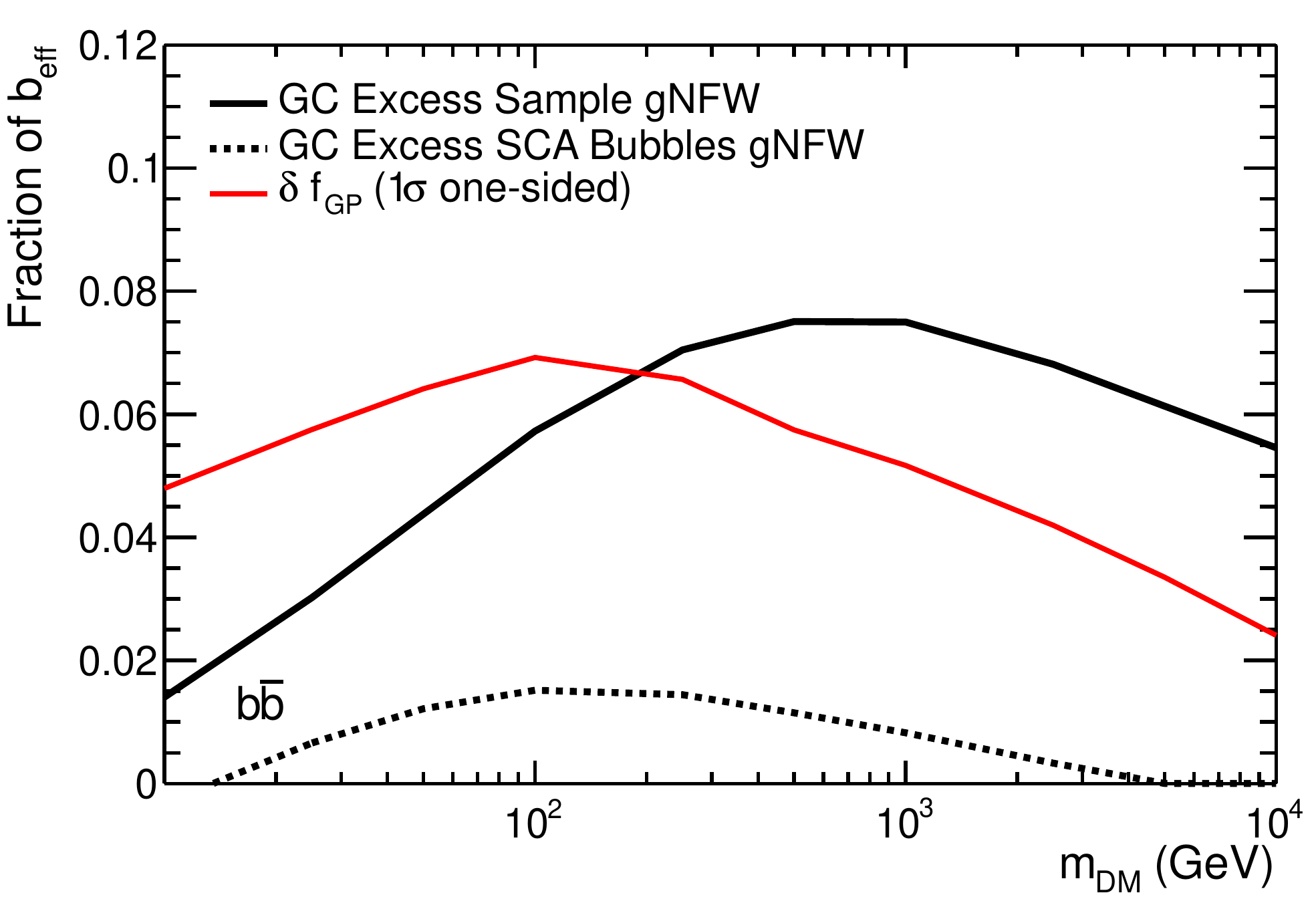}
\includegraphics[width=0.5\linewidth]{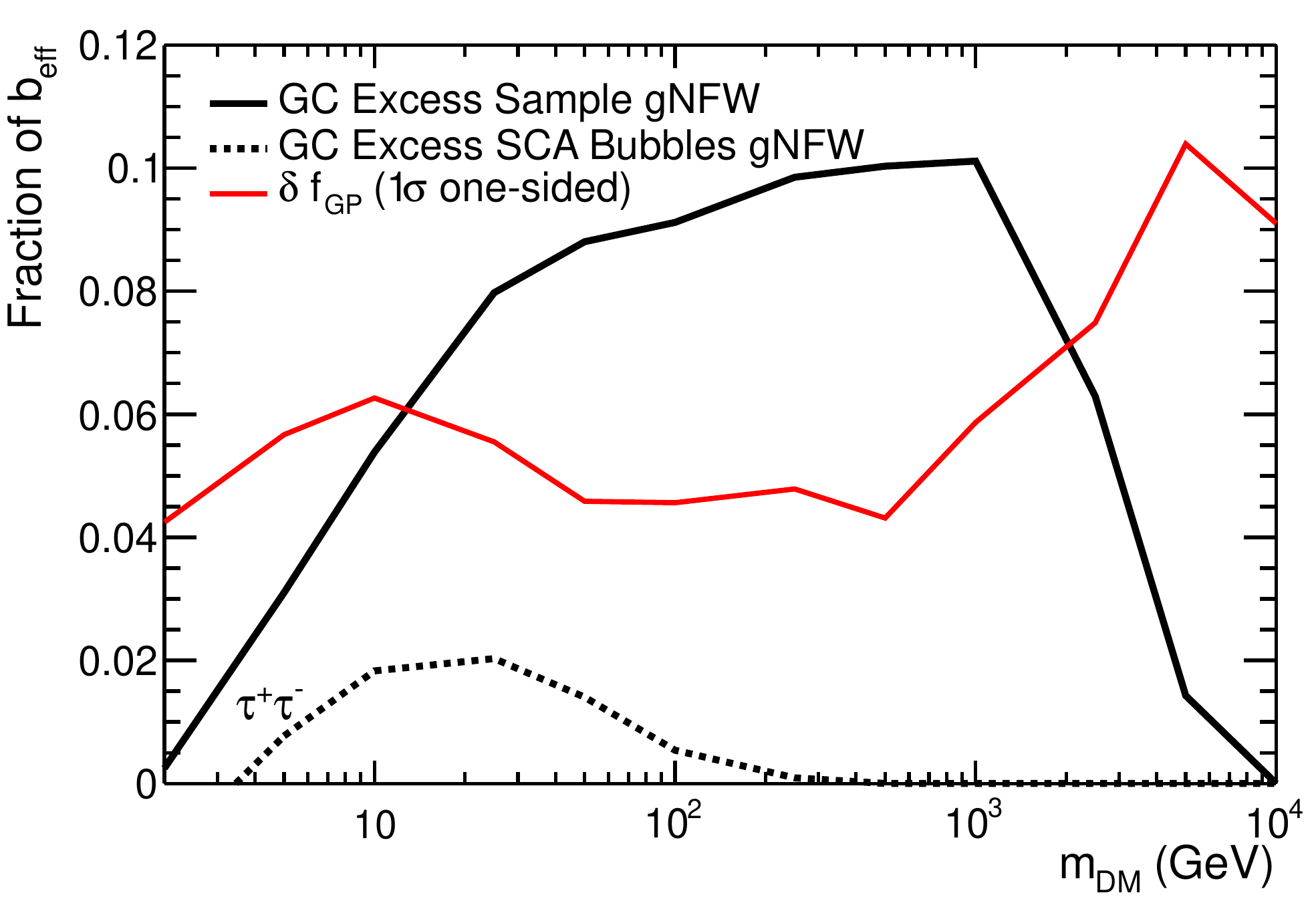}
\caption{ 
\label{fig:fracSig_GPvGC}
Comparison of the size of best-fit DM models as a fraction of \beff (see text) evaluated for the gNFW template in the GC compared to the systematic uncertainty determined from the Galactic plane scan (see text).}
\end{figure*}

A comparison of the fits in the GC to the characteristic $\delta N_{\rm sig, syst}$ in Equation (\ref{eq:dNsig})
is shown in Figure~\ref{fig:fracSig_GPvGC}. 
The largest DM signal as a fraction of \beff in the GC in the Sample Model is similar to the characteristic uncertainty level from the Galactic plane scans. 
Consequently, we cannot claim that the DM interpretation of the signal in the GC is robust when we compare it with the DM-like signals in the control regions along the GP.
We note that the same conclusion holds when an NFW profile is adopted for the spatial distribution of the DM, and for the model that includes all-sky SCA bubbles. 
In the latter case, the GC signal is much smaller than those seen in the GP scan. Furthermore, the same conclusions are reached if we adopt higher energy thresholds of 300~MeV and 1~GeV for the analysis.

As a corollary from these results, we conclude that the model variations discussed earlier in the paper may not capture the complete range of systematic uncertainties that can mimic a DM signal. 
Either the variations considered do not cover the full range of the associated model uncertainties or there are other sources of gamma rays that play a significant role. A noteworthy candidate for the latter is a population of unresolved sources, such as MSPs, that other analyses \citep{2016PhRvL.116e1102B, 2016PhRvL.116e1103L, 2015ApJ...812...15B} are indicating as a likely cause of the GC excess.

\subsection{Limits on Dark Matter Annihilation}

In the previous subsection we have found that, although the GC excess is statistically significant and it is present in all models that we have considered, 
a similar level of fractional excesses are found at other locations along the Galactic plane, where no DM annihilation is expected.
As a consequence, a DM interpretation of the GC excess is not robust.
Since we cannot claim a detection of DM annihilation at the GC, we derive upper limits on the DM annihilation cross section.
In previous studies, e.g., \citet{Ackermann:2015lka} and \citet{Buckley:2015doa},
the systematic uncertainty inferred from control regions was used also to set upper limits on the DM cross section. 
In Appendix~\ref{app:fitDM} we use the modeling uncertainty derived from the scan along the Galactic plane in a derivation of upper limits on DM annihilation.
This derivation assumes that the probability of having negative fluctuations (i.e., masking a real DM signal) is as large as the positive fluctuations seen in GP scan. However, this is likely to be too conservative since in almost all fits in control regions along the Galactic Plane we found positive DM-like excesses (Figure \ref{fig:fracSig_GP}).

\begin{figure*}[t]
\includegraphics[width=0.5\linewidth]{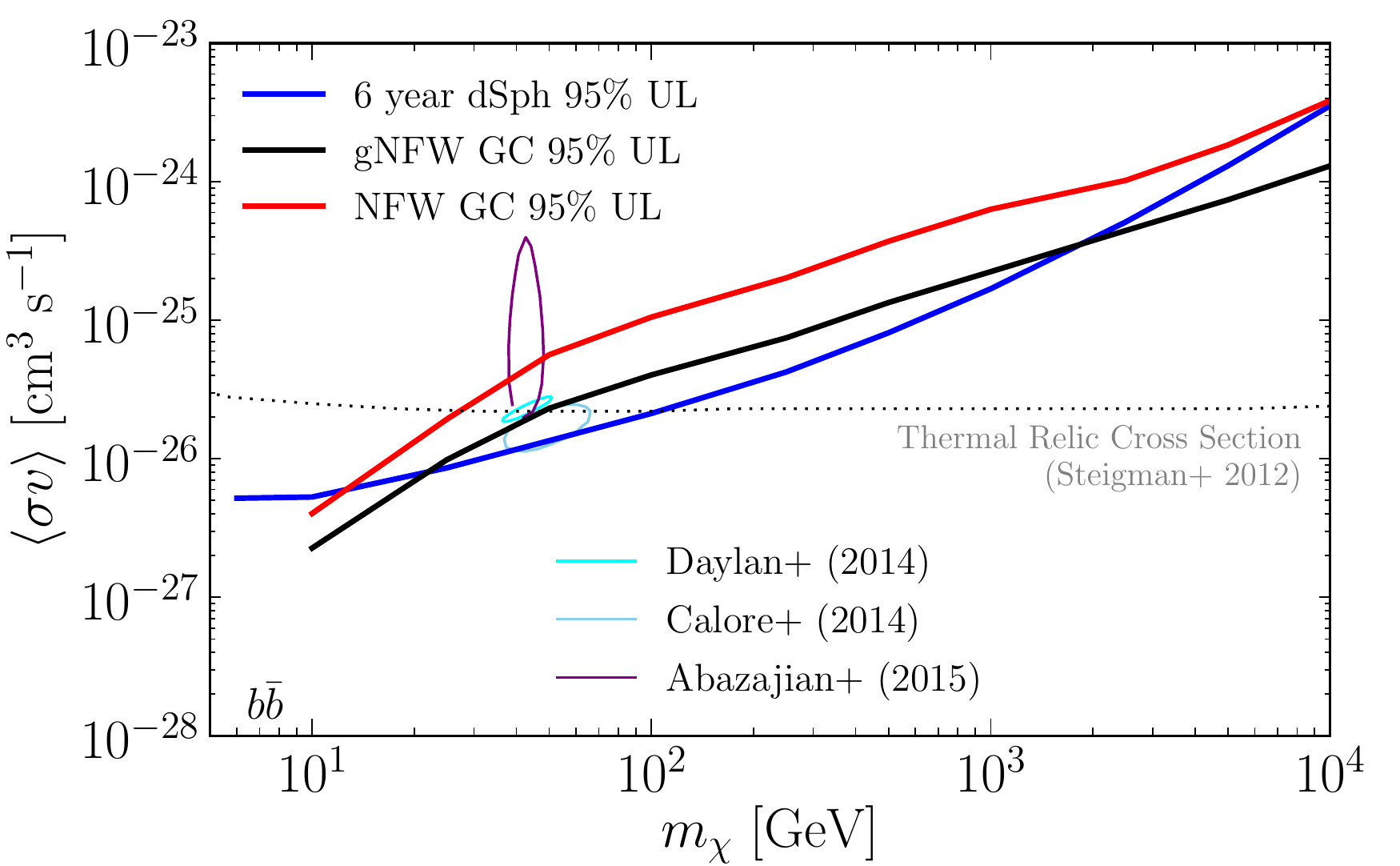}
\includegraphics[width=0.5\linewidth]{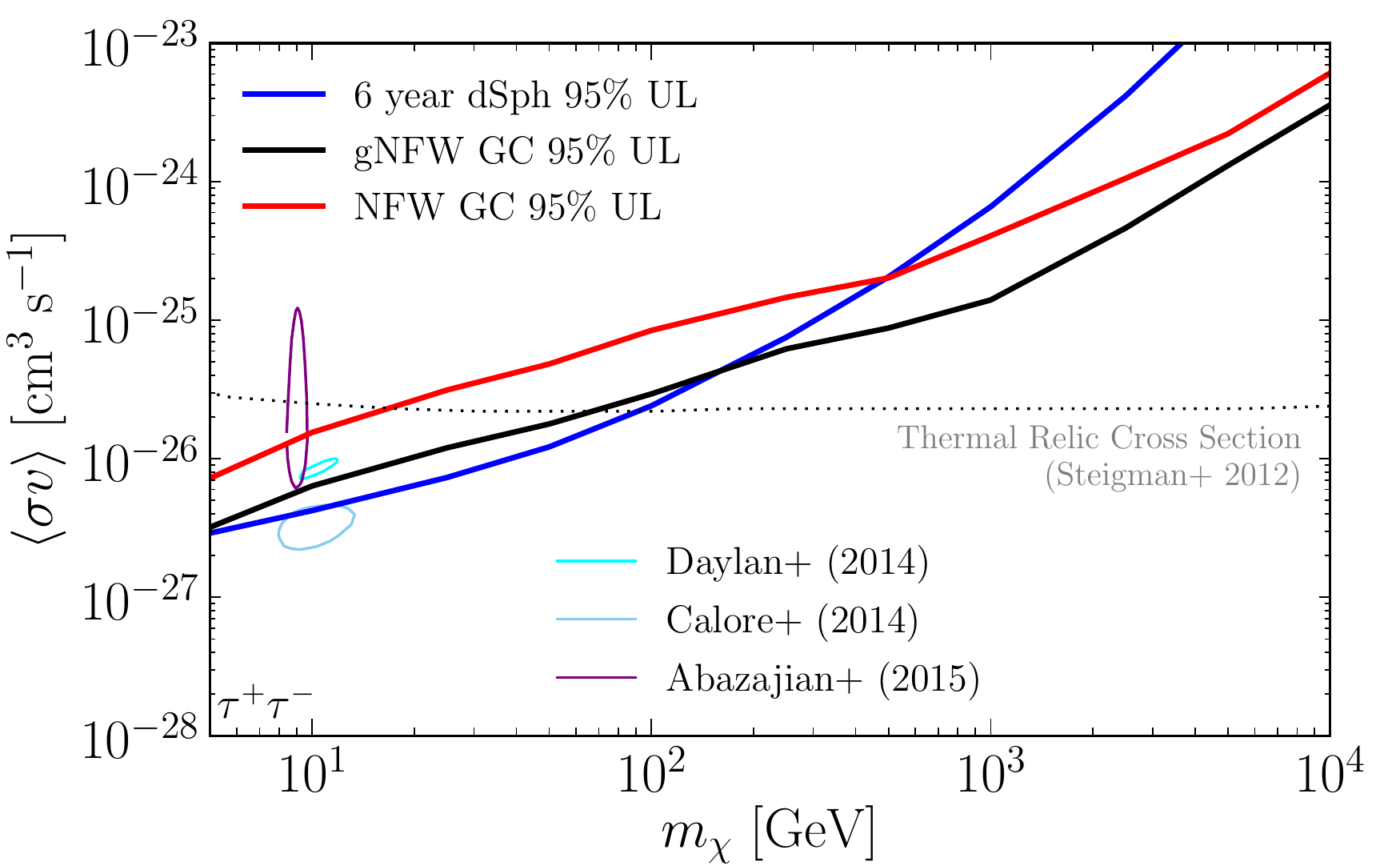}
\caption{
\label{fig:DM_limits}
DM upper limits obtained by requiring that a DM signal not exceed the largest GC excess found in each energy bin in all background model scenarios considered in this work at the 95\% confidence level assuming the gNFW (black) and NFW (red) DM profiles. Shown in blue are the upper limits from the recent analysis of 15 dwarf spheroidal galaxies using 6 years of \Fermi LAT data \citep{Ackermann:2015zua}. The contours show the signal regions from recent analyses of the GC region \citep{Daylan:2014rsa,Calore:2014xka,Abazajian:2014fta}.
The dotted line represents the thermal relic cross section \citep{2012PhRvD..86b3506S}.
}
\end{figure*}

Therefore, here we proceed to set limits on \sigmav by requiring a tentative DM signal to not exceed the largest GC excess flux value in each energy bin for all the background models considered in this work (i.e., the upper edge of the blue band in Figure \ref{fig:GCE_fluxband}) at the 95\% statistical confidence level. Results are shown in Figure~\ref{fig:DM_limits} and compared with other relevant DM limits from the literature.

Our limits are similar to the DM parameters found in works
where the GC excess has been interpreted as DM.  This is not surprising since all the independent GC excess analyses use similar data sets and interstellar emission models. We also see that the gNFW limits are stronger than the NFW limits since the gNFW profile has a steeper inner slope of the DM density towards the GC. 
Both gNFW profiles provide stronger limits than the dSph analysis at masses above a few TeV (few hundred GeV) for the \bbbar (\tautau) chanel.
Although for the Sample Model the GC analysis also provides stronger than dSph DM limits below few tens of GeV for the \bbbar chanel,
these limits are subject to larger modeling uncertainties.
This is the regime where the analyses become statistics limited. Since the expected signal in our GC analysis is much larger than that of the dSph analysis, the statistical uncertainties are smaller, resulting in more constraining limits.

\section{Conclusions}
\lb{sec:concl}
We have characterized the so-called ``\Fermi GC GeV excess'' using 6.5 years of Pass 8 \Fermi LAT data. We investigated the uncertainties in the spectrum and morphology of the excess due to the analysis procedure and the modeling of other components of emission near the GC, including interstellar emission and resolved point sources. Specifically, we have:
\begin{itemize}
\item examined different choices for the event selection and analysis region (Section~\ref{sec:data_selection});
\item varied assumptions on CR production and propagation in the Galaxy (Section~\ref{sec:galprop_vars}) and allowed more degrees of freedom in the fit of IC emission (Section~\ref{sec:IC});
\item considered alternative distributions of interstellar gas along the LOS to the GC (Section~\ref{sec:gasunc});
\item included sources of CR near the GC (Section~\ref{sec:GC_CR_sources});
\item derived a model for the \Fermi bubbles extending to low latitudes (Section~\ref{sec:SCA});
\item tested different lists of PS near the GC based on different background models and analyses (Section~\ref{sec:PS}).
\end{itemize}
Some of these tests had already been discussed in the literature, 
and we repeat them here with different parameter choices and a different dataset  for completeness. 
Several tests are presented in this work for the first time. In particular, we exploit a decomposition of the gas along the line of sight based on SL extinction,
we self-consistently determine PS lists from our analysis with Pass 8 data in the energy range from 100 MeV to 500 GeV, 
and we determine a template for the \textit{Fermi} bubbles at low latitudes and for the excess itself using the SCA method.
In addition, to test the robustness of a DM interpretation of the GC excess, we perform a systematic search  for excesses with (generalized) NFW annihilation profile and with DM annihilation spectra from different channels in control regions along the Galactic plane, where we do not expect a DM signal.

The main conclusions are:
\begin{itemize}
\item an excess at the GC around a few GeV is statistically significant in all the cases considered;
\item the spectrum of the excess varies significantly depending on the analysis method/assumptions:  the flux changes by a factor of $\sim$3 at few GeV, and even more dramatically at energies $<1$~GeV and $>10$~GeV (Figure \ref{fig:GCE_fluxband});
\item emission from the \Fermi~bubbles is one of the major sources of uncertainty: in the presence of low-latitude emission from the bubbles the excess can vanish above 10~GeV, and below 10~GeV the flux is reduced by a factor $\gtrsim 2$ (Figure~\ref{fig:SCA_cutoff_spectra});
\item characterization of resolved point sources may significantly affect the spectrum of the GC excess, especially below 1~GeV (Figure~\ref{fig:PS_syst});
\item the excess has a complex morphology, and the simplest interpretation is that it is composed of two contributions: bi-lobed emission from the \Fermi~bubbles that is displaced from the GC to negative longitudes, and residual emission, azimuthally symmetric around the GC with a spectrum that peaks around 3 GeV (Section~\ref{sec:morphology});
\item excesses with fractional amplitude similar to the one in the GC are found to be fairly common in control regions along the Galactic plane (Figure~\ref{fig:fracSig_GP}).
\end{itemize}

The range of explored uncertainties, albeit larger than in any other study to date, is yet not a full representation of the uncertainties in the modeling, because residuals persist in all cases considered. The spectrum and morphology of the excess are not obviously consistent with the expectations for DM annihilation, or at least suggest an underlying astrophysical component on top of a potential DM component. This is also consistent with the presence of similar fractional excesses along the Galactic plane where no DM signal is expected.

Therefore, we derive stringent limits on the annihilation cross section of DM particle candidates by requiring that the DM annihilation signal does not exceed the upper bound
on the GC excess spectrum from the variations of conventional emission components models plus 95\% statistical confidence.
We find that the limit on the annihilation cross section is sensitive to the profile of the DM distribution in the Galaxy.
For the $b\bar{b}$ ($\tau^+\tau^-$) channel the thermal cross section can be excluded for $M < 50$ GeV ($M < 100$ GeV)
in the case of gNFW profile with $\gamma = 1.25$ while for the standard NFW annihilation profile the thermal cross section
is excluded for $M < 25$ GeV ($M < 20$ GeV). Due to larger expected signal 
and better statistics near the GC, the limits are more constraining than those from dwarf galaxies for $M > 1$~TeV ($M > 100$~GeV) in the case of gNFW profile with $\gamma = 1.25$, and for $M > 8$~TeV ($M > 600$~GeV) for the standard NFW annihilation profile.

In summary, we find that the GC excess is present in all models that we have tested, but its origin remains elusive: part of it may be attributed to the \Fermi bubbles, but there may also be a contribution from interactions of CRs from sources in the proximity of the GC, from a population of yet unresolved point sources such as MSPs, or from annihilation of dark matter particles.
The fractional size of DM-like excesses relative to background in other locations along the GP is comparable to that in the GC. Therefore, a DM interpretation of the GC excess cannot be robustly claimed. We consequently derive limits on the DM particle properties.
Future gamma-ray studies and multi-wavelength observations will be essential in searches for new point sources and
for the characterization of the CR distribution and the structure of the \Fermi bubbles near the GC.

\acknowledgments

The \textit{Fermi} LAT Collaboration acknowledges generous ongoing support
from a number of agencies and institutes that have supported both the
development and the operation of the LAT as well as scientific data analysis.
These include the National Aeronautics and Space Administration and the
Department of Energy in the United States, the Commissariat \`a l'Energie Atomique
and the Centre National de la Recherche Scientifique / Institut National de Physique
Nucl\'eaire et de Physique des Particules in France, the Agenzia Spaziale Italiana
and the Istituto Nazionale di Fisica Nucleare in Italy, the Ministry of Education,
Culture, Sports, Science and Technology (MEXT), High Energy Accelerator Research
Organization (KEK) and Japan Aerospace Exploration Agency (JAXA) in Japan, and
the K.~A.~Wallenberg Foundation, the Swedish Research Council and the
Swedish National Space Board in Sweden.
 
Additional support for science analysis during the operations phase is gratefully acknowledged from the Istituto Nazionale di Astrofisica in Italy and the Centre National d'\'Etudes Spatiales in France.

We made use of data products based on observations obtained with \textit{Planck} (\url{http://www.esa.int/Planck}), an ESA science mission with instruments and contributions directly funded by ESA Member States, NASA, and Canada.

This work was partially funded by NASA grants NNX14AQ37G and NNH13ZDA001N.

\facility{\textit{Fermi}, \textit{Planck}}.

\software{Astropy \citep[\url{http://www.astropy.org},][]{2013A&A...558A..33A}, 
Fermipy (\url{http://fermipy.readthedocs.io/}),
GALPROP \citep[\url{http://galprop.stanford.edu/},][]{Vladimirov:2010aq}, HEALPix \citep[\url{http://healpix.jpl.nasa.gov/},][]{2005ApJ...622..759G}, ROOT \citep[\url{http://root.cern.ch/},][]{1997NIMPA.389...81B}, matplotlib \citep{Hunter:2007}}.

\appendix
\section{Derivation of Gas Distribution with Starlight Extinction Data}
\lb{app:SLext}

One of the most important uncertainties in the derivation of the GC excess flux is the distribution
of the gas along the line of sight to the GC.
Since the rotation of the Galactic disc is perpendicular to this line of sight,
the usual derivation of the gas distribution based on the red and blue shifts of the atomic and molecular lines is
not applicable.
In this Appendix we use the starlight extinction data to derive a distribution of dust along the line of sight to the GC,
which can be used as a tracer of the total gas density distribution.
We modified the original maps from \citet{FermiLAT:2012aa} using the extinction maps from \citet{schultheis2014} in the region at $|l|<10\arcdeg$ and $|b|<5\arcdeg$ through the following procedure:
\bi
\item For each line of sight on the grid of \citet{schultheis2014} we determined the fraction of the total extinction belonging to each heliocentric distance bin, then averaged over the angular bins subtended by each angular bin in the gas maps. The heliocentric bin fractions were converted into Galatocentric annuli fractions using the annuli definition of the gas maps.
\item The fractional extinctions in Galactocentric annuli are partitioned into extinction associated with atomic and molecular gas based on the atomic/molecular fractions from the gas maps in \citet{FermiLAT:2012aa} through interpolation from two adjacent regions outside of $|l|<10\arcdeg$ with $\Delta l = 3\arcdeg$ at the same latitude. For this purpose the CO intensities were converted into \hd column densities using the values of the \xco ratio in \citet{FermiLAT:2012aa} corresponding to our \SM. For the innermost ring (enclosed in the region $|l|<10\arcdeg$) the fractions are assumed to be the same as in the closest ring. 
\item The maps in \citet{schultheis2014} cover only heliocentric distances $\lesssim 10$~kpc. Therefore, we corrected the fractional atomic and molecular extinctions to take into account gas missing in the extinction maps at distances $>10$~kpc from the Earth. For Galactocentric radii $<8.5$~kpc we upscaled the content of each \hi or CO annulus to account for material on the other side of the Galactic center assuming that the distribution of matter in the Galaxy is axisymmetric. We took into account the different physical distances from the plane on the two sides of the Galaxy by assuming that the density of \hi as a function of distance from the plane is described by a Gaussian with full width at half maximum derived from \citet{kalberla2009} and plateauing at 210 pc for Galactocentric radii $<5$~kpc, and the density of \hd as a function of distance from the plane is described as a Gaussian with full width at half maximum of 146 pc \citep{ferriere2001}. For Galactocentric radii $>8.5$~kpc the fraction of extinction associated to either atomic or molecular gas was extrapolated from the closest neighbor annulus based on the fractions in the two adjacent bands outside of $|l|<10\arcdeg$ with $\Delta l = 3\arcdeg$ at the same latitude. The extrapolation was performed iteratively to determine the content in each annulus at radius $>8.5$~kpc from the closest to the farthest from the Sun.
\item The resulting fractional extinctions associated with atomic and molecular gas in the annuli are multiplied by the total along each line of sight derived from the original maps in \citet{FermiLAT:2012aa}, producing the modified maps. 
\ei 

\begin{figure}[htbp]
\begin{center}
\includegraphics[scale=\onepic]{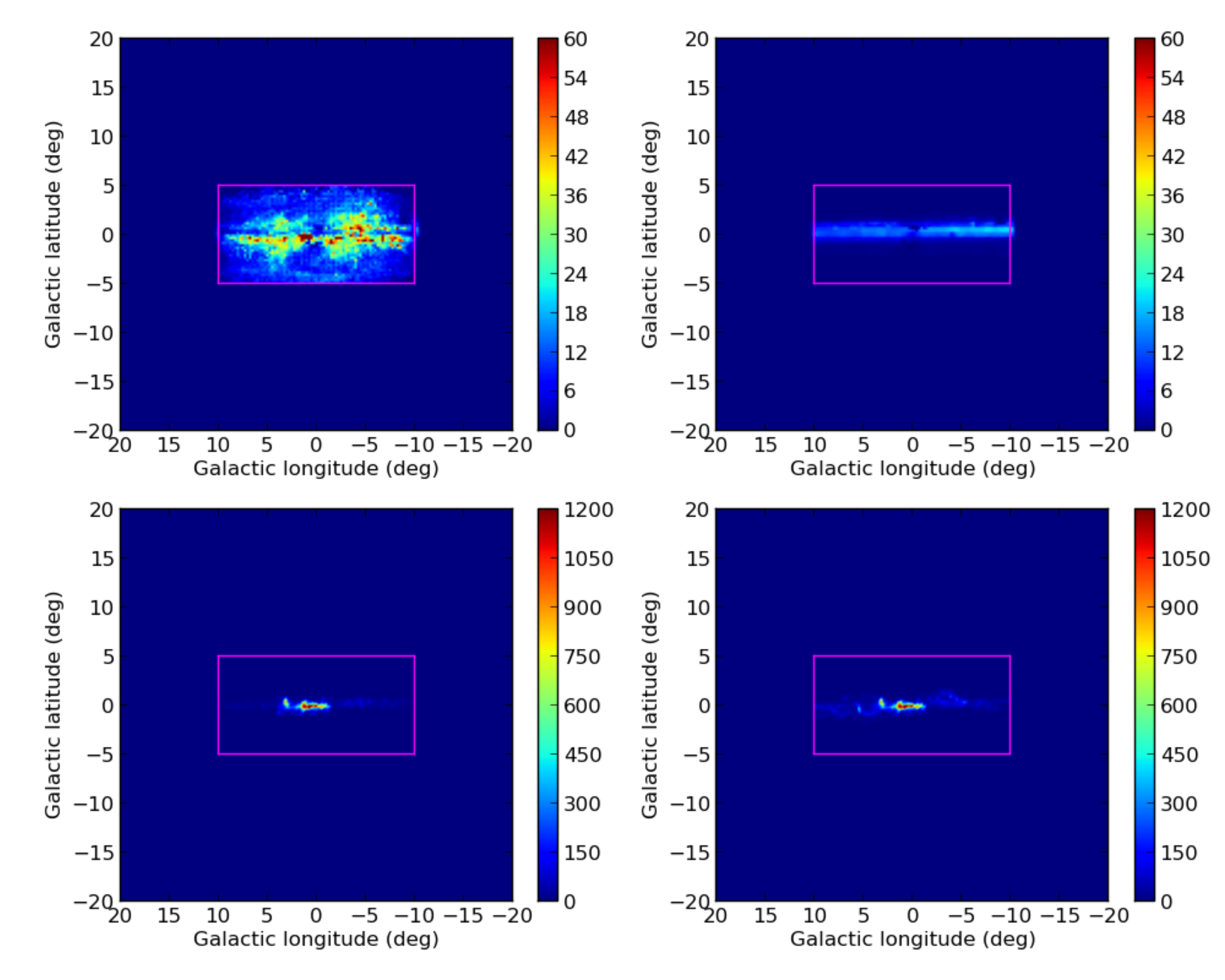}
\noindent
\caption{\small 
Maps of the innermost Galactocentric annulus (0~kpc to 1.5~kpc) obtained from partitioning the gas column densities along the line of sight within the area spanned by the magenta box using the starlight extinction method (left), and the interpolation method (right).
Top: \hi column density in $\rm 10^{20}\; cm^{-2}$, bottom: Integrated CO line inensity $W_{\rm CO}$ in $\rm K\, km\, s^{-1}$.}
\label{fig:alt_gas_examples}
\end{center}
\end{figure}

\begin{figure}[htbp]
\begin{center}
\includegraphics[scale=\onepic]{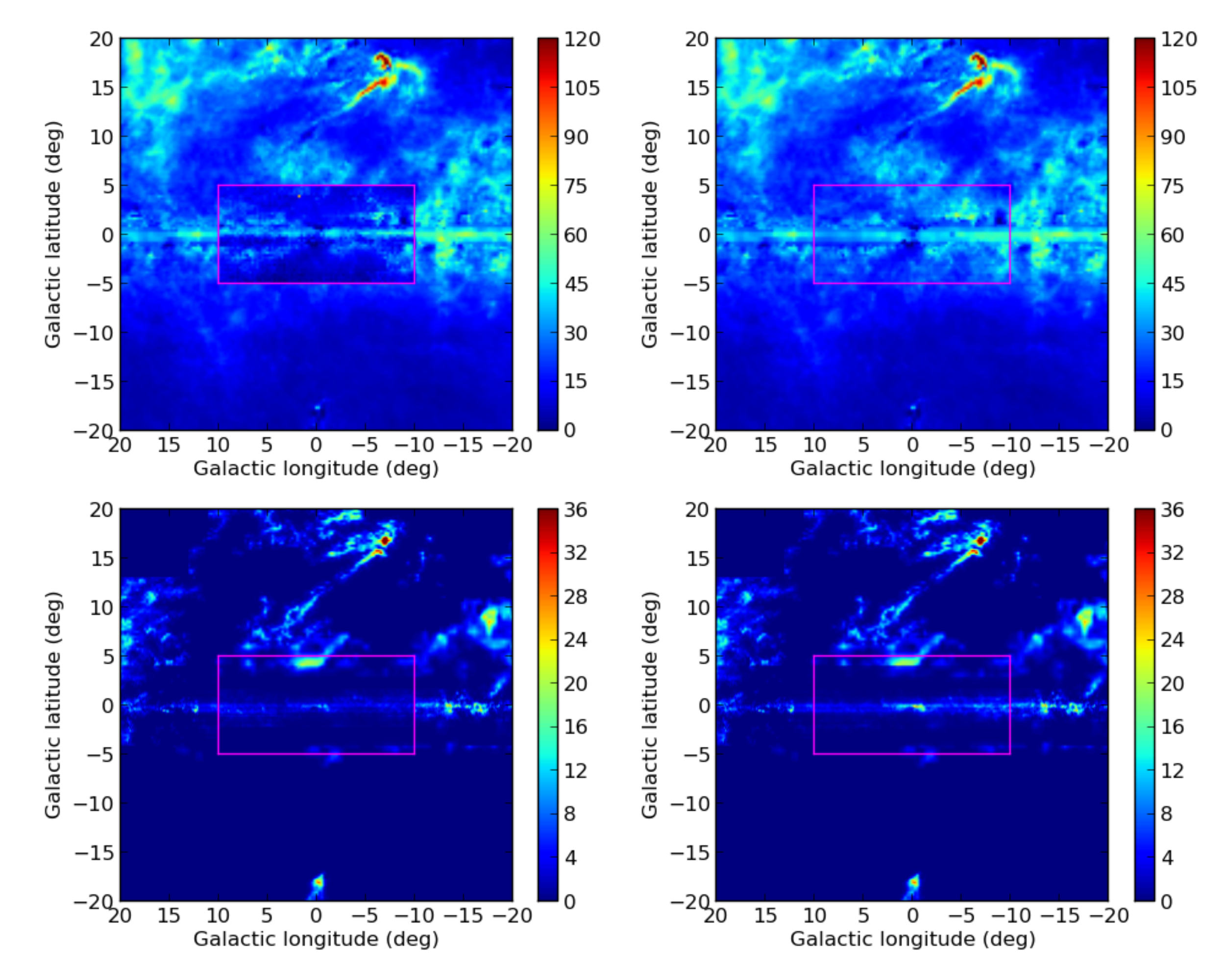}
\noindent
\caption{\small 
Maps of the local annulus (8~kpc to 10~kpc) obtained from partitioning the gas column densities along the line of sight within the area spanned by the magenta box using the starlight extinction method (left), and the interpolation method (right).
Top: \hi column density in $\rm 10^{20}\; cm^{-2}$, bottom: Integrated CO line inensity $W_{\rm CO}$ in $\rm K\, km\, s^{-1}$.}
\label{fig:alt_gas_examples2}
\end{center}
\end{figure}

This alternative scheme to partition the material along the line of sight in the region where the Doppler shift of gas lines is not available is not necessarily providing a more precise estimate of the real distribution of the gas in the galaxy, due to the many uncertain assumptions. However, it provides an alternative to explore whether this aspect has a significant impact on the results of our analysis.
Some examples of the alternative gas distribution are shown in Figures~\ref{fig:alt_gas_examples} and~\ref{fig:alt_gas_examples2}.

\section{Amplitude of the Galactic Center GeV Excess Relative to Statistical and Systematic Uncertainties}
\lb{app:unc_maps}

\begin{figure}[htbp]
\begin{center}
\includegraphics[scale=\twopic]{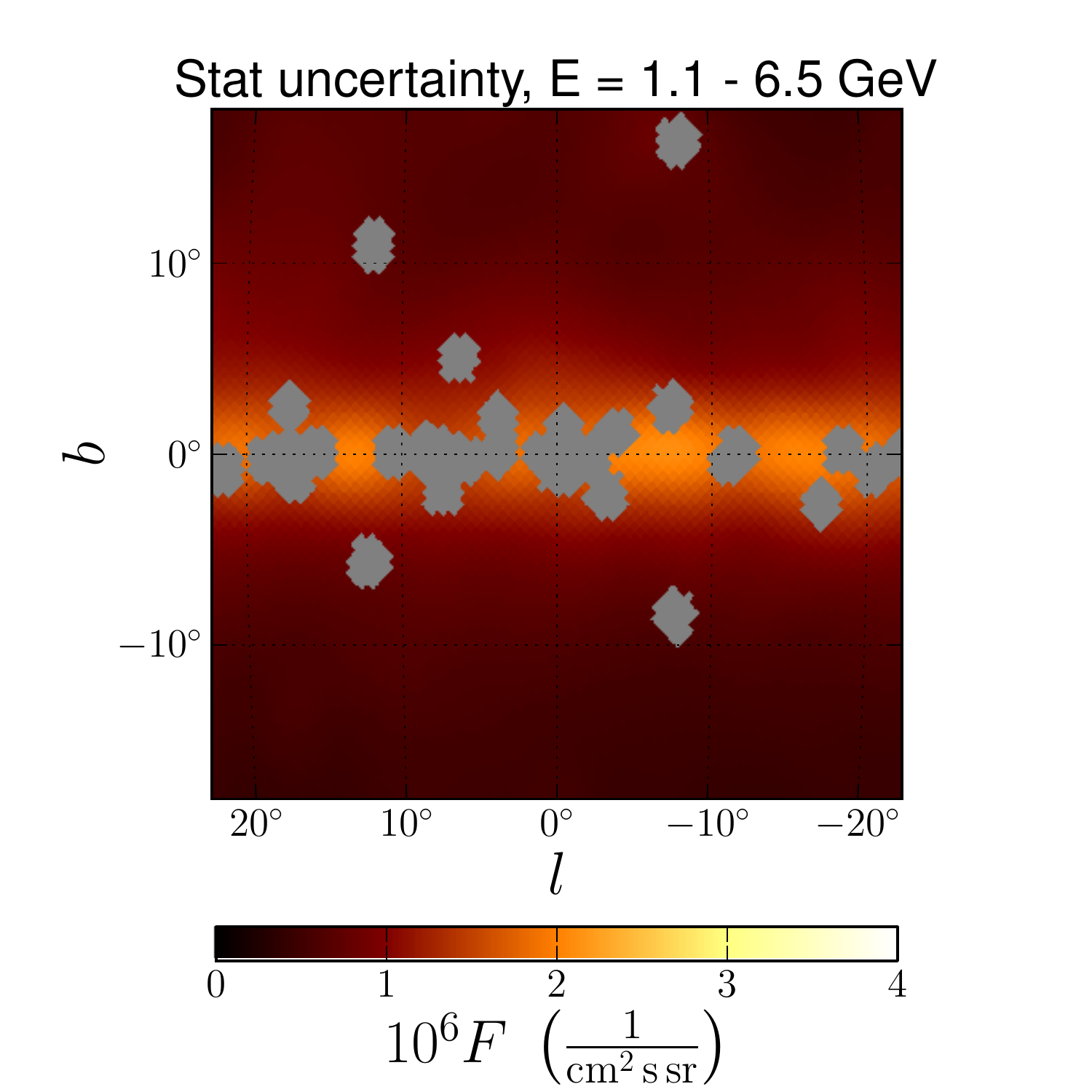}
\includegraphics[scale=\twopic]{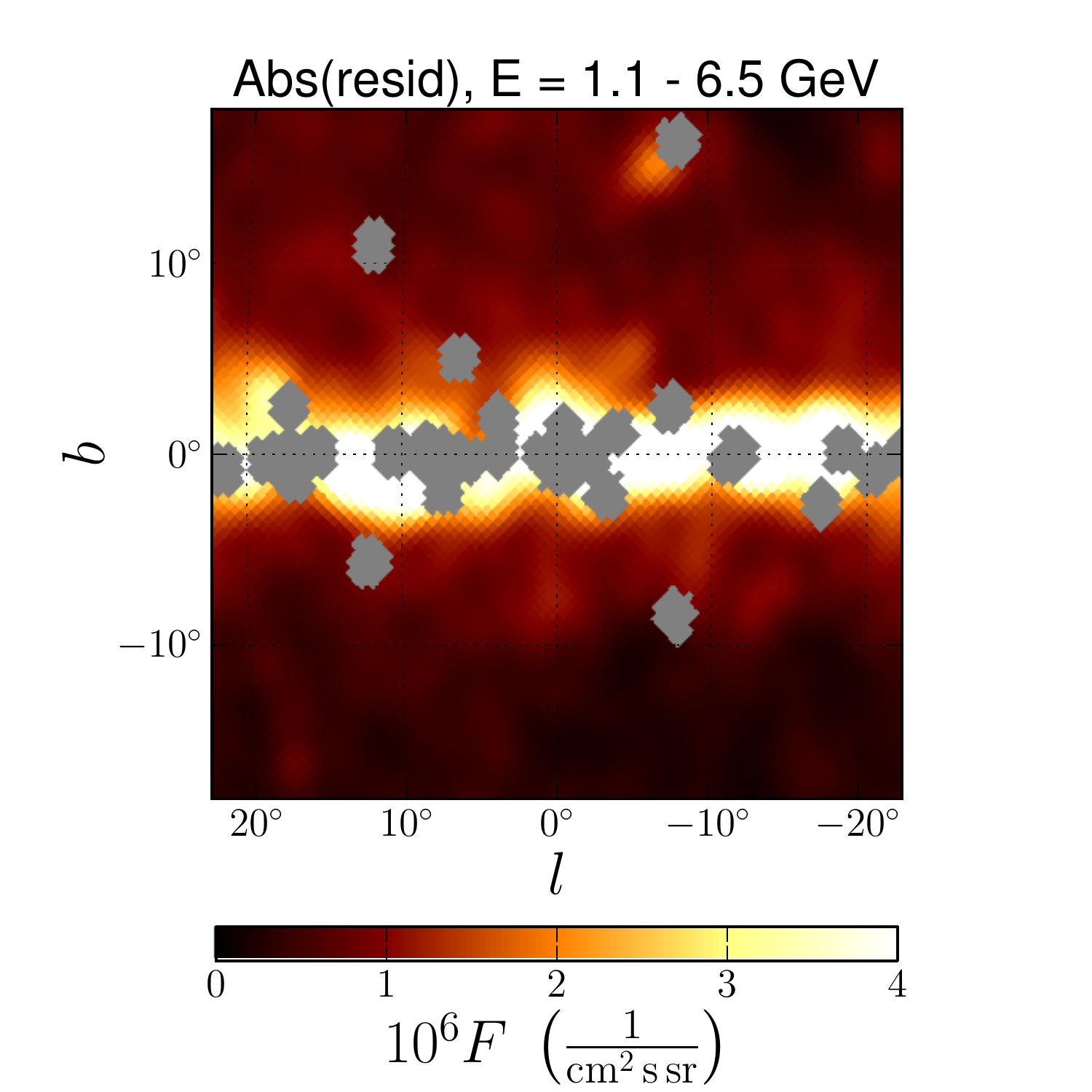}
\includegraphics[scale=\twopic]{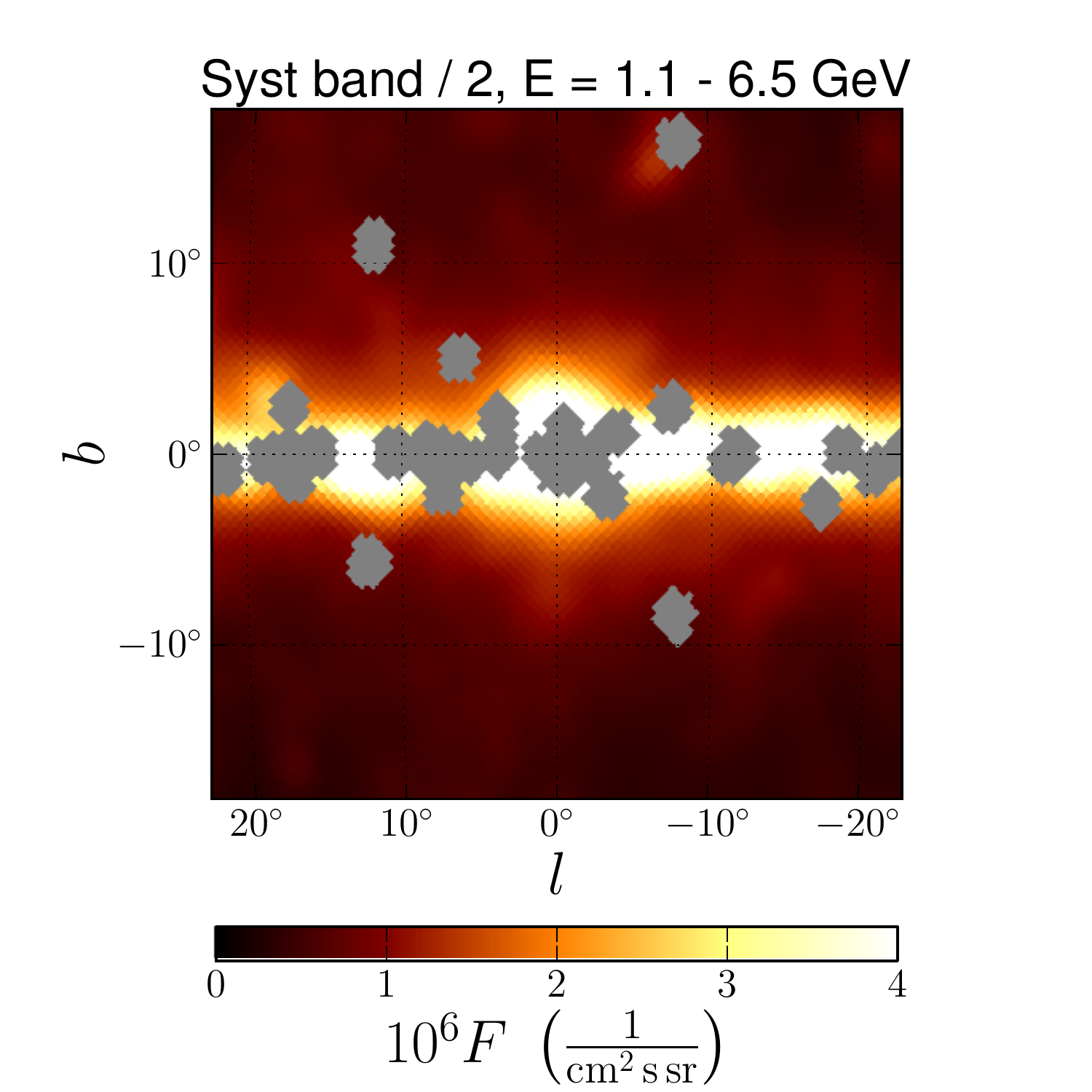} \\
\includegraphics[scale=\twopic]{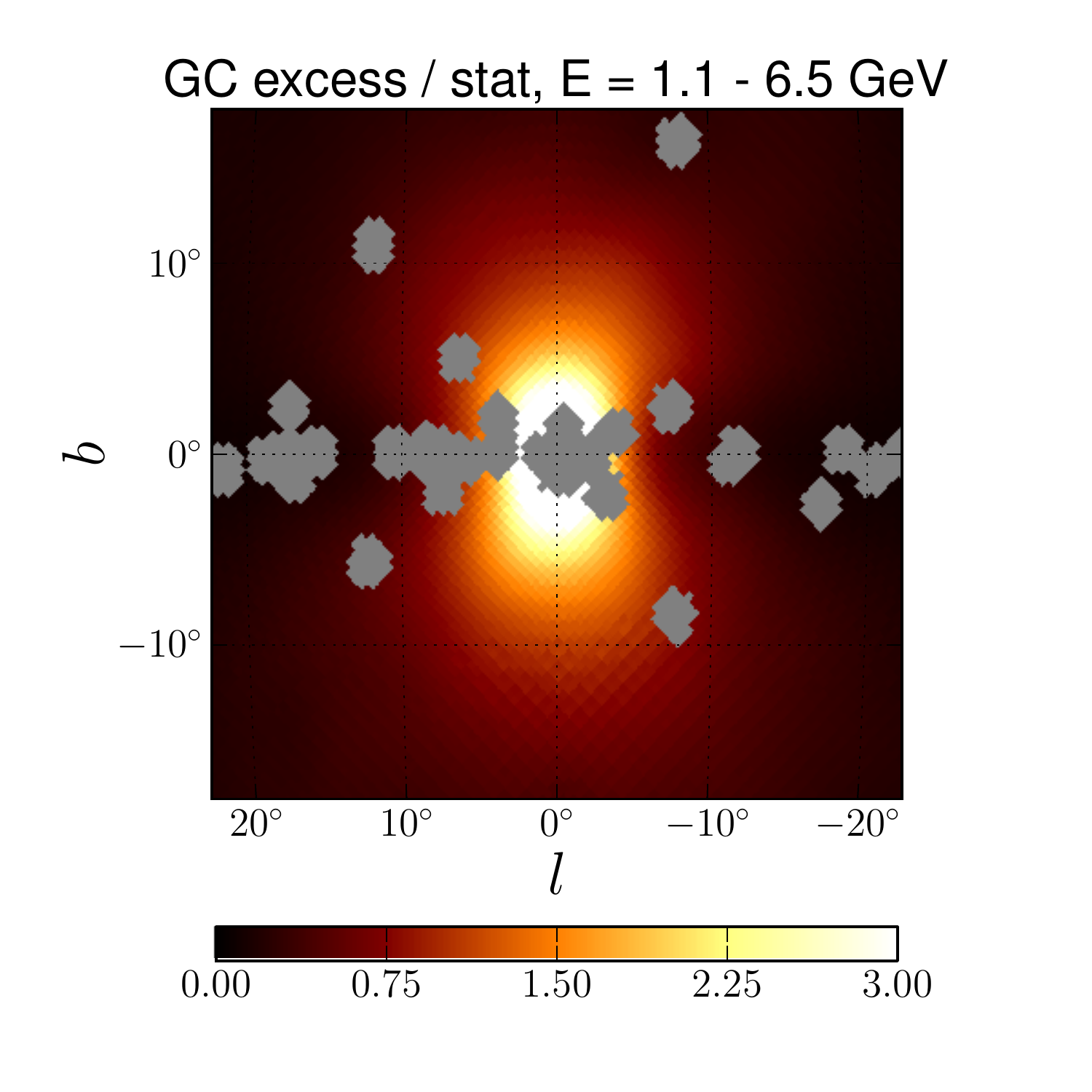}
\includegraphics[scale=\twopic]{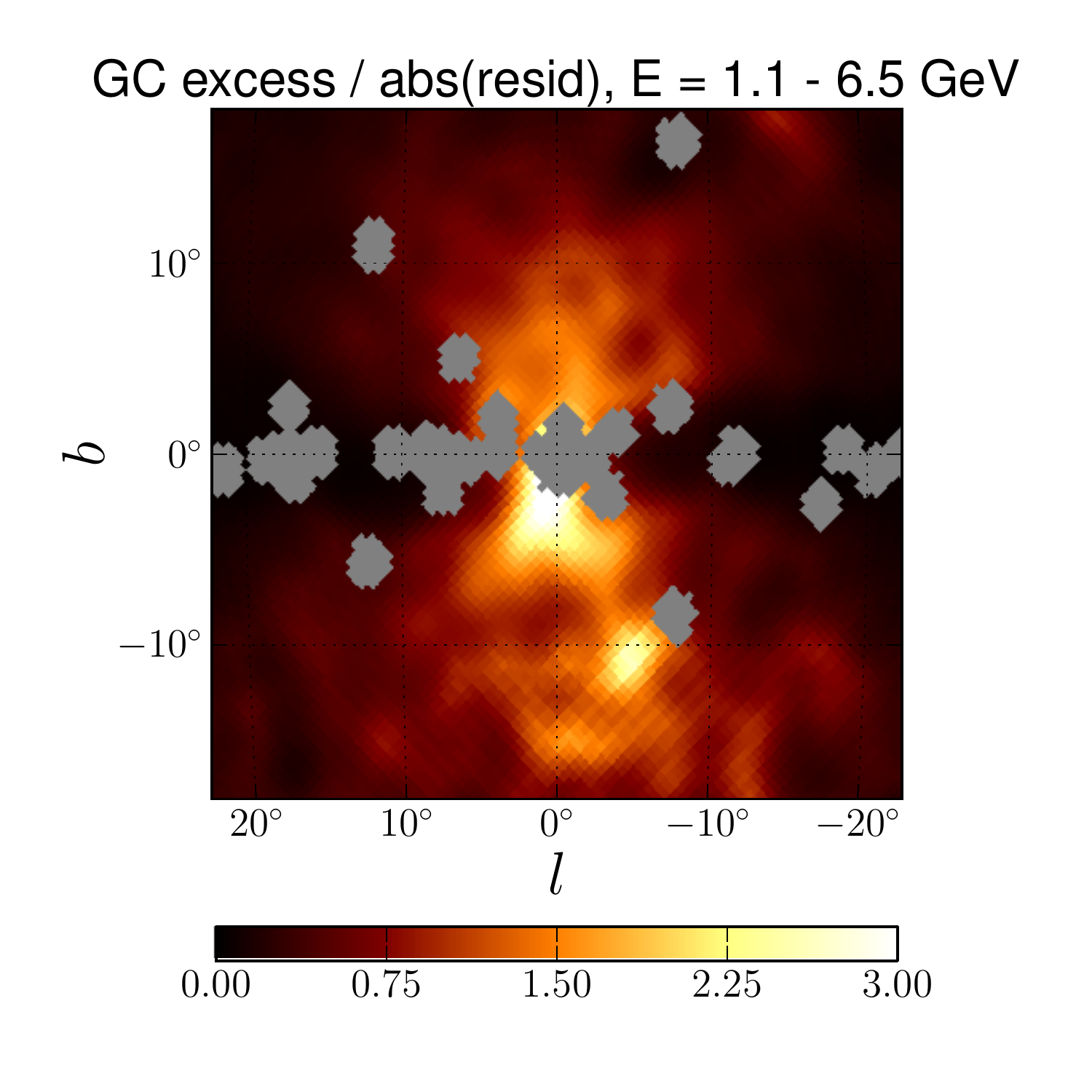}
\includegraphics[scale=\twopic]{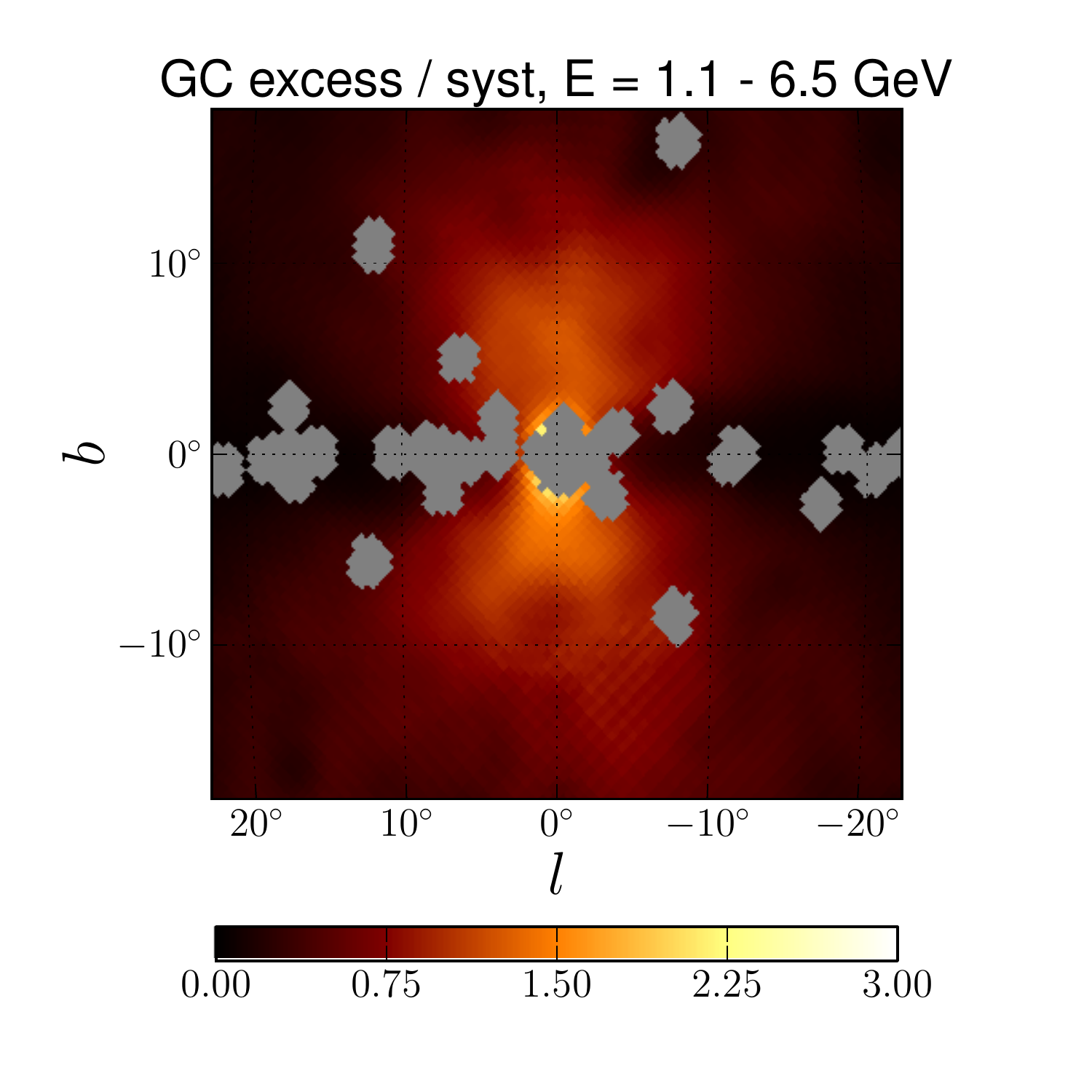}
\noindent
\caption{\small 
Top: statistical and modeling uncertainties, bottom: ratio of the GC excess in the Sample Model
to the uncertainty maps.
The units for the top row are intensity integrated over the energy range between 1.1 GeV and 6.5 GeV.
All uncertainty maps are smoothed with $1\degr$ Gaussian kernel for better presentation of large-scale features.
Left: statistical uncertainty (square root of the data), middle: absolute value of the residual,
right: half of the envelope of the diffuse models.
The grey circles indicate the masked locations of the bright point sources (200 sources from the 3FGL catalog over the whole sky).
}
\label{fig:signal2syst}
\end{center}
\end{figure}

In this appendix, we compare the GC excess signal to the statistical uncertainties as well as uncertainties in the models of foreground emission.
The level of statistical uncertainty and the ratio of the signal in the Sample Model to the statistical uncertainty 
are shown in Figure \ref{fig:signal2syst} left.
We combine 5 energy bins between 1.1 and 6.5 GeV, where the excess is most significant.
The statistical uncertainty is calculated as the square root of the photon counts in pixels.
The map is smoothed with $1\degr$ Gaussian kernel for presentation purposes to highlight the large-scale features.

To estimate the modeling uncertainties we consider two approaches: 1) comparing the residuals 
left after subtracting the Sample Model from the data as a measure of how incomplete/inadequate the model is; 2) evaluating the envelope of the diffuse models considered in the previous sections as an estimate of modeling uncertainties. 
The absolute value of the residuals (smoothed with a $1\degr$ Gaussian kernel) and the ratio of the signal to the smoothed residuals are shown 
in Figure \ref{fig:signal2syst} middle. 
The envelope of the diffuse models divided by two and the ratio of the signal to the envelope half width are shown 
in Figure \ref{fig:signal2syst} right.

The statistical uncertainties are smaller than the modeling uncertainties in the Galactic plane
(notice the scale on the color bars).
The GC GeV excess is also smaller than the modeling uncertainties in the Galactic plane,
while above and below the GC the ratio of the signal to the modeling uncertainties is larger than one.

\section{Effective Background}
\lb{app:beff}

In this appendix we give an abbreviated summary of the effective background method used to quantify systematic uncertainties in setting limits on DM annihilation 
(see~\citealt{2015PhRvD..91l2002A}, \citealt{Caputo:2016ryl}, and especially Section VB of \citealt{Buckley:2015doa} for more details).

Suppose that the data are represented as a combination of two components: background $P^{(\rm bkg)}$ and signal $P^{(\rm sig)}$. Then the overall normalizations $N_{\rm bkg}$ and $N_{\rm sig}$ for the components can be found by minimizing the $\chi^2$
\be
\chi^2 = \sum_{i,\al} \frac{(d_{i\al} - \sum_{m=\rm bkg,sig} N_m P^{(m)}_{i\al})^2}{\sm_{i\al}^2},
\ee
where the summation is over pixel indices $i$ and energy bin indices $\al$, $d$ is the data, and $\sigma$ is the statistical uncertainty.
It is convenient to introduce the scalar product with the metric $1/\sm_{i\al}^2$: 
$\bra a, b \ket_\sm = \sum_{i\al} a_{i\al} b_{i\al} / \sm^2_{i\al}$.
Then the Hessian (the matrix of the second derivatives of $\chi^2$ with respect to $N_m$) is $H_{nm} = \bra P^{(n)}, P^{(m)} \ket_\sm$.
The best-fit solutions and the covariance matrix of coefficients $N_m$ are
\bea
N_m &=& \sum_n {H^{-1}}_{mn} \bra P^{(n)}, d \ket_\sm; \\
{\rm Cov}(N_m, N_n) &=& {H^{-1}}_{mn}.
\eea
If the signal is expected to be small, then we can approximate $d_{i\al} = N_{\rm bkg} P^{(\rm bkg)}_{i\al} = \sm_{i\al}^2$.
We choose the normalization of templates $P^{(m)}_{i\al}$ such that 
$\sum_{i,\al} P^{(m)}_{i\al} = 1$, then $N_{\rm bkg} \approx N$, where $N$ is the total number of counts.
In this case, the Hessian takes the form
\be
H = \frac{1}{N}\left(
\ba{cc}
1 & 1 \\
1 & \sum_{i,\al} \frac{(P^{(\rm sig)}_{i\al})^2}{P^{(\rm bkg)}_{i\al}}
\ea
\right)
\ee
and the statistical uncertainty of $N_{\rm sig}$ is
\be
\lb{eq:sstat}
(\dl N_{\rm sig})^2 \equiv {\rm Cov}(N_{\rm sig}, N_{\rm sig})  = \frac{N}{\sum_{i,\al} \frac{(P^{(\rm sig)}_{i\al})^2}{P^{(\rm bkg)}_{i\al}} - 1}.
\ee
Using an analogy with the Poisson distribution, we define the ``background under the signal region" or the ``effective background"
as $b_{\rm eff} = (\dl N_{\rm sig})^2$, where the statistical uncertainty is determined in Equation (\ref{eq:sstat}). Thus
\begin{equation}
\label{eq:beff_app}
b_{\rm eff} = \frac{N}{ \sum_{i,\al}\frac{(P^{(\rm sig)}_{i\al})^2}{P^{(\rm bkg)}_{i\al}} - 1}.
\end{equation}
It is interesting to note that the square roots of the terms in the sum, ${P^{(\rm sig)}_{i\al}} / {\sqrt{P^{(\rm bkg)}_{i\al}}}$, are approximately equal to the statistical significance of the signal plotted in Figure \ref{fig:signal2syst}, left.
We use the ratio of the signal counts to \beff as a figure of merit in estimating the modeling uncertainties.
Note that the actual statistical uncertainty of the GC excess flux is different from $\sqrt{b_{\rm eff}}$ because we have more than one background component.
The derivation presented here is a motivation for $b_{\rm eff}$ in Equation (\ref{eq:beff_app}), which we use as an estimate of the 
number of background counts under the signal.

Note that the greater the correlation between the signal and the background model, the closer to 1 the summation term in Equation (\ref{eq:beff_app}) is.  
For perfect correlation \beff would in fact diverge. If the two models are not very correlated, then the summation term becomes much larger than 1, and the resulting \beff is less than $N$. When the background and the signal models are essentially uncorrelated, i.e., more easily distinguished, \beff becomes smaller, resulting in a smaller statistical error on $N_{\rm sig}$.
Also, for larger values of $N$, the relative statistical error on $N_{\rm sig}$ decreases.

\begin{figure*}[t]
\includegraphics[width=0.5\linewidth]{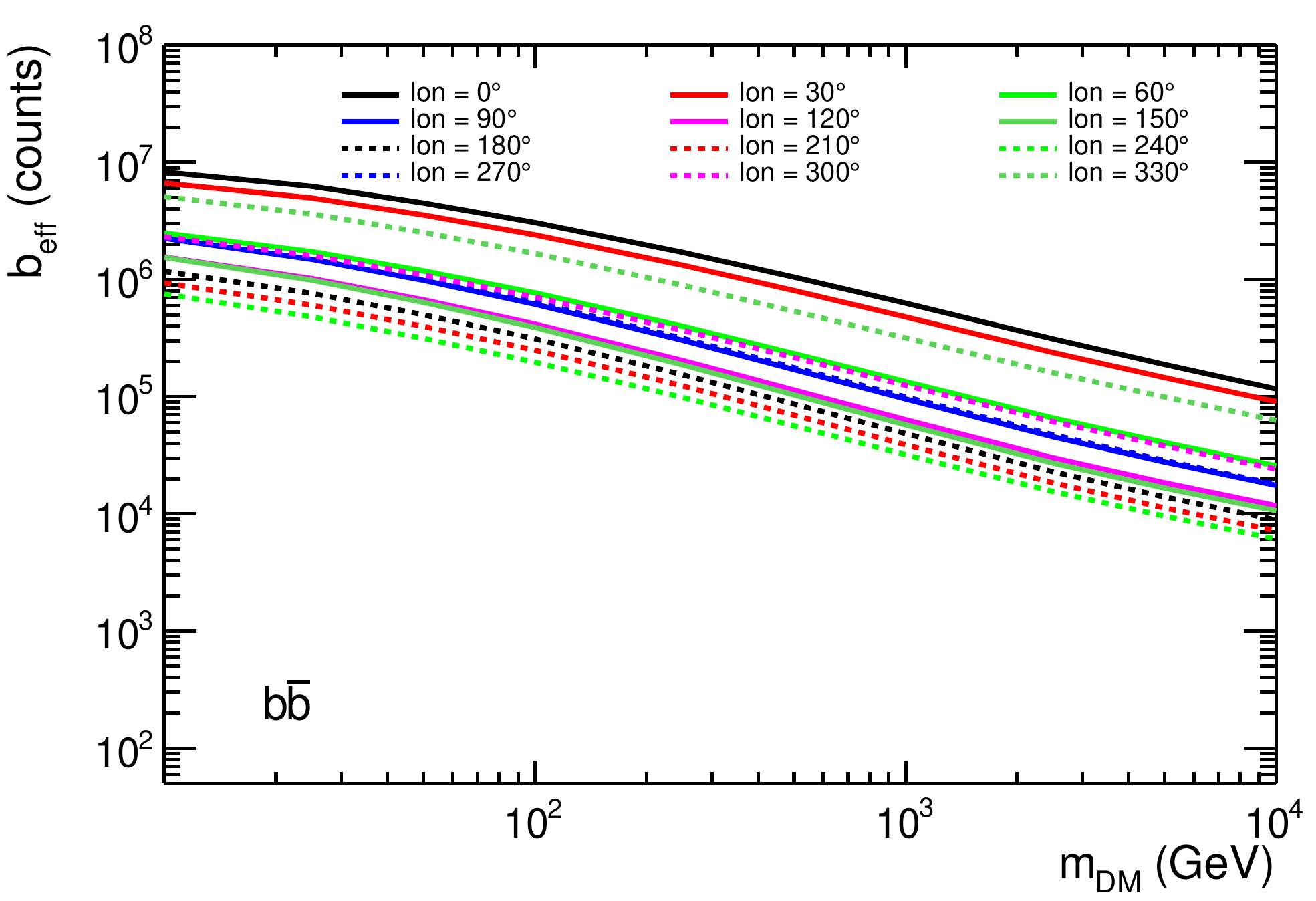}
\includegraphics[width=0.5\linewidth]{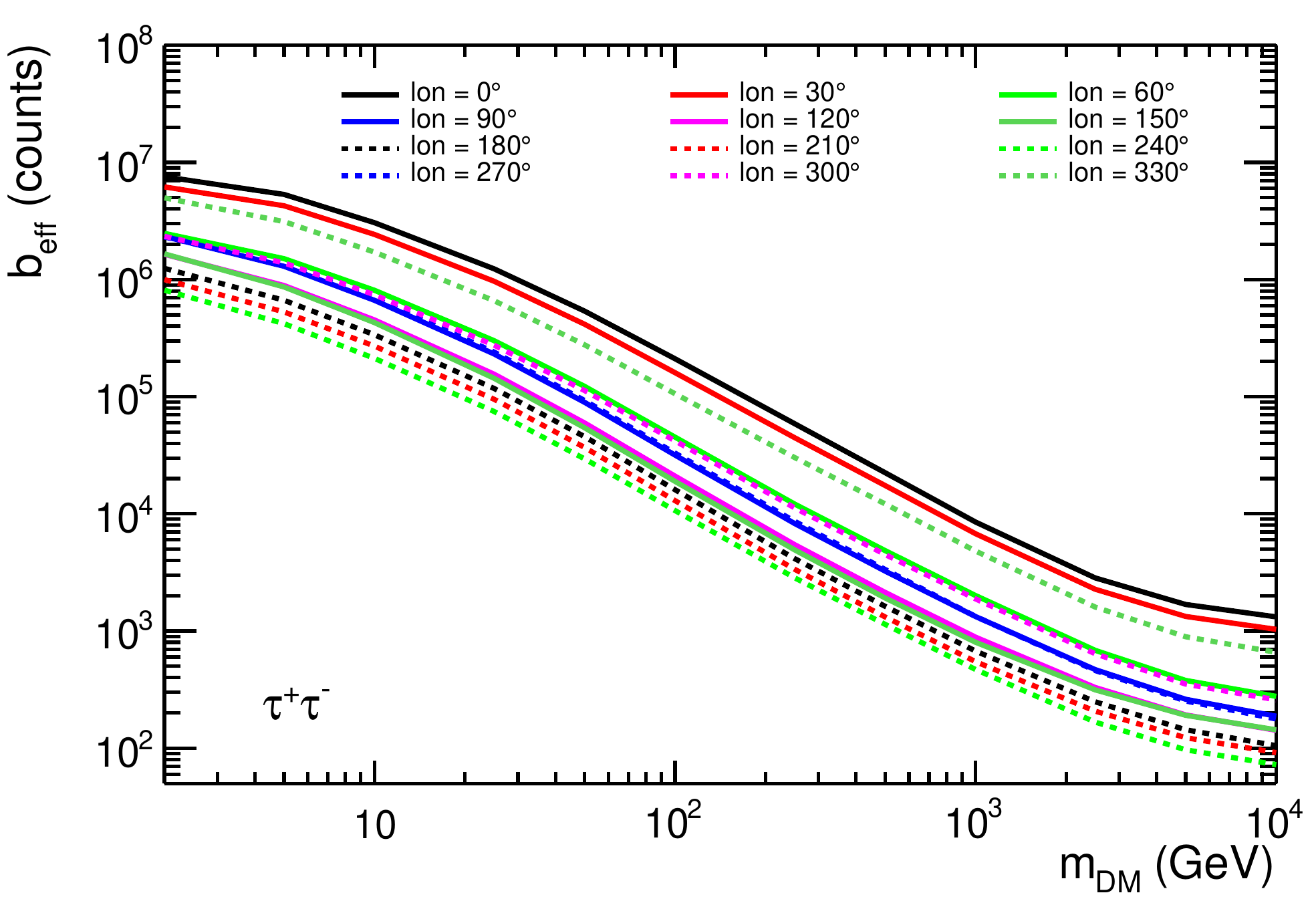}
\caption{
\label{fig:beff}
Effective background \beff vs. DM mass \mDM for DM annihilating to \bbbar (left) and \tautau (right) using the Sample Model 
(without the gNFW template) as the background model.
The spatial DM template is centered at $b=0$ and at longitudes
along the Galactic plane.
}
\end{figure*}

In Figure \ref{fig:beff} we show \beff calculated using the Sample Model as $P_{\rm bkg}$ and our standard gNFW template centered at $b=0$ and various longitudes as $P_{\rm sig}$. The value of \beff is largest when it is computed for the gNFW template centered at the GC. 
Also, \beff decreases as $m_{\rm DM}$ is increased in the signal model,
because the resulting model \g-ray spectra are harder than the typical non-DM astrophysical emission, making $P_{\rm bkg}$ less correlated with $P_{\rm sig}$.

We quantify systematic uncertainties that mask or induce DM signals as a fraction of \beff (Section \ref{sec:beff_scan}). 
Therefore we can now also define a systematic uncertainty on $N_{\rm sig}$ as

\begin{equation}
\lb{eq:Nsyst}
\delta N_{\rm sig, syst} = \delta f_{\rm syst} \times \beff.
\end{equation}

The total uncertainty on $N_{\rm sig}$ is $\delta N_{\rm sig} = \sqrt{\delta N_{\rm sig,stat}^2 + \delta N_{\rm sig,syst}^2}$.

\subsection{Fitting Procedure and Upper Limits on Dark Matter Annihilation Models}
\lb{app:fitDM}

In Section \ref{sec:DMlimits}, we fit a variety of DM models to the GC excess spectrum that we found using our sample background model. The differential \g-ray flux expected from DM annihilation in a given region of interest ($\Delta\Omega$) is
\begin{equation}
\lb{eq:DM}
\frac{d\phi}{dE_\gamma} = \left( \frac{ \sigmav}{8\pi}\frac{dN_\gamma}{dE_\gamma}\frac{1}{m_\chi^2}\right) \left( \int_{\Delta\Omega} d\Omega \int_\text{l.o.s.} d\ell ~\rho_\chi^2(\vec{\ell}\,)\right),
\end{equation}
where $dN_\gamma/dE_\gamma$ is the differential counts spectrum of gamma rays from annihilation of a pair of DM particles, $m_\chi$ is the mass of each DM particle, \sigmav is the velocity-averaged annihilation cross section, and $\rho^2_{\chi}$ is the density distribution of the DM. $dN_\gamma/dE_\gamma$ depends on the relative strength of each annihilation channel for a specific DM model. 
In this work, we calculate limits on DM annihilation for two representative annihilation channels: \bbbar and \tautau. The value in the first set of parentheses in Equation (\ref{eq:DM})
depends on the particle nature of the DM; specifically on the particle mass, annihilation channel, and annihilation cross section. The value in the second set of parentheses is the so-called `J factor', which depends on the distribution of DM. We consider both a standard NFW profile ($\gamma=1$) and a slightly contracted generalized NFW profile ($\gamma=1.25$). The J-factors integrated over the whole sky (including the point source masking) assuming a local DM density of $0.4 \rm{GeV}/\rm{cm}^3$ are $2.15\times10^{22} \rm{GeV}^2\rm{cm}^{-5}$ and $1.53\times10^{22} \rm{GeV}^2\rm{cm}^{-5}$ for the gNFW and NFW profiles respectively.

We fit using the $\chi^2$ method to the GC excess spectrum obtained by fitting the excess spatial template (gNFW) in each energy bin independently (Section \ref{sec:data_fitting}). The differential number of counts assuming a set of DM parameters is simply 
Equation (\ref{eq:DM}) convolved with the PSF and multiplied by the instrument exposure. If we assume a specific mass, annihilation channel, and J-factor then the normalization of the DM component will be proportional to \sigmav, which is left free in the fit.  The $\chi^2$ of each DM fit is
\begin{equation}
\lb{eq:chi2_DM}
\chi^2 = \sum_j \frac{(s_j - \lambda_j(\sigmav,\vec{p}))^2}{\epsilon^2_j} + \left(\frac{\sigmav}{\delta \sigmav}\right)^2
\end{equation}
where $s_j$ is the GC excess counts in energy bin j, $\lambda_j$ is the expected photon counts for a given DM model, $\vec{p}$ are the assumed DM model parameters $[m_\chi,dN_\gamma/dE_\gamma,\rm{J-factor}]$ and $\epsilon_j$ is the statistical uncertainty on $s_j$ from the spatial fit. 
The second term is the log likelihood of the prior probability on annihilation cross section.
We conservatively assume that the prior $\sigmav = 0$. The uncertainty $\delta \sigmav$ is derived by scanning the cusp profile along the Galactic plane, see Equations  (\ref{eq:Nsyst}) and (\ref{eq:DM}).  If the value of the cross-section necessary to minimize the first term were large compared to the uncertainty in the second term denominator, then the $\chi^2$ in Equation (\ref{eq:chi2_DM}) would be large, which could have been interpreted as an exclusion of $\sigmav = 0$ prior. However, due to relatively large uncertainty on $\sigmav$, the first term in Equation (\ref{eq:chi2_DM}) can be minimized without a large increase in the second term, i.e., the value of $\sigmav$ that minimizes the first term is consistent with the uncertainty.
Based on this $\chi^2$ accounting for systematic uncertainties from the GP in Section \ref{sec:DMlimits} 
we concluded that a DM interpretation of the GC excess at present is not robust.

We note that $\delta \sigmav$ in Equation~(\ref{eq:chi2_DM}) equally accounts for systematic uncertainties that mask or induce DM-like signals. 
However, this was not what was seen in the control region fits, which were more like a one-sided Gaussian distribution with about half the normalizations being zero. 
Therefore we chose $\delta \sigmav$ 
by finding the 84-th percentile of the fit profiles, i.e., the fractional \beff for which only 16\% of the fit results are larger.  
In the case of a Gaussian distribution this corresponds to a one-sided one sigma exclusion.
The use of a nuisance parameter of this form is valid when calculating the TS since 
systematic uncertainties that would induce a DM-like signal, and so would reduce the TS of a DM annihilation component, are properly represented.

\begin{figure*}[t]
\includegraphics[width=0.5\linewidth]{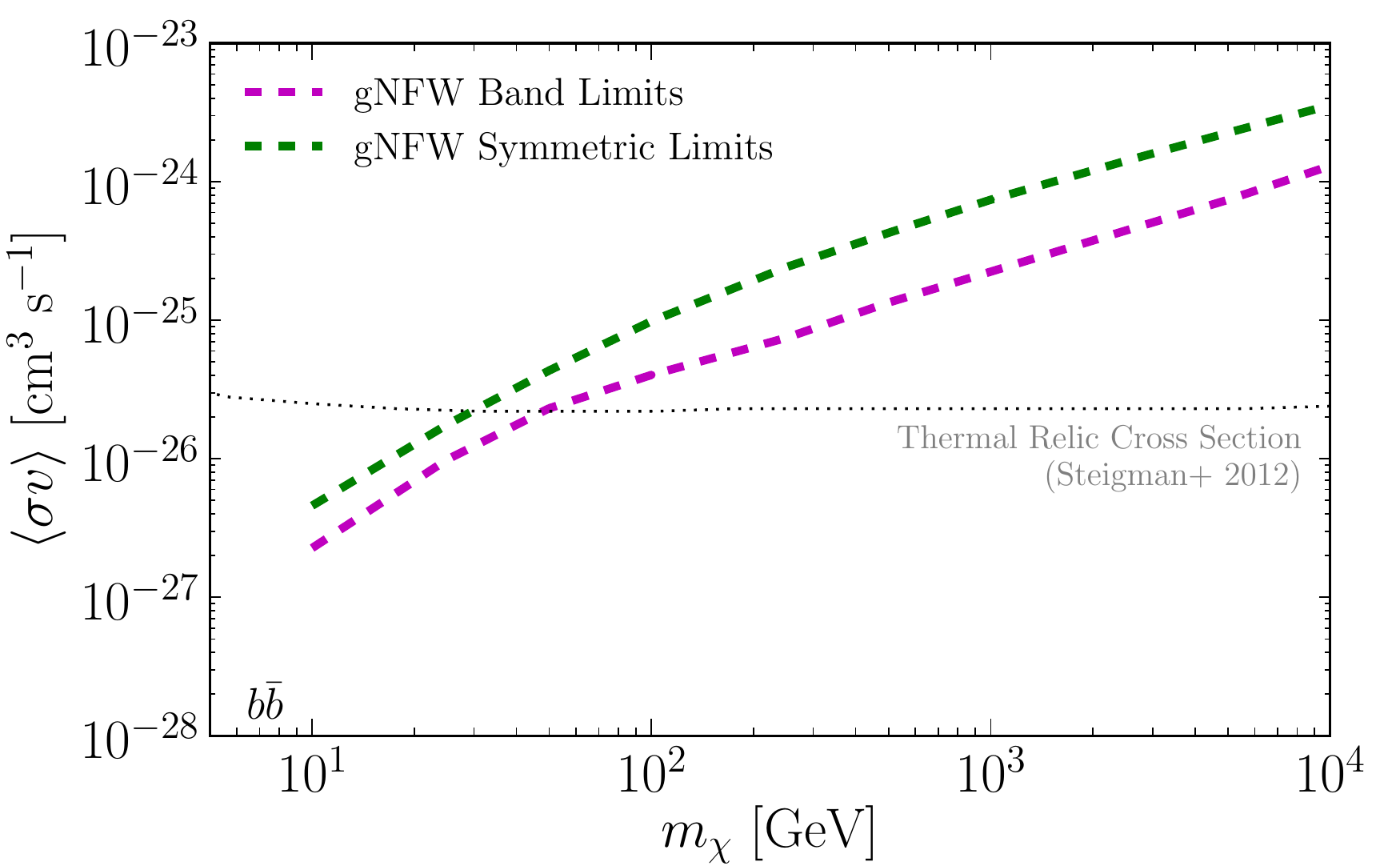}
\includegraphics[width=0.5\linewidth]{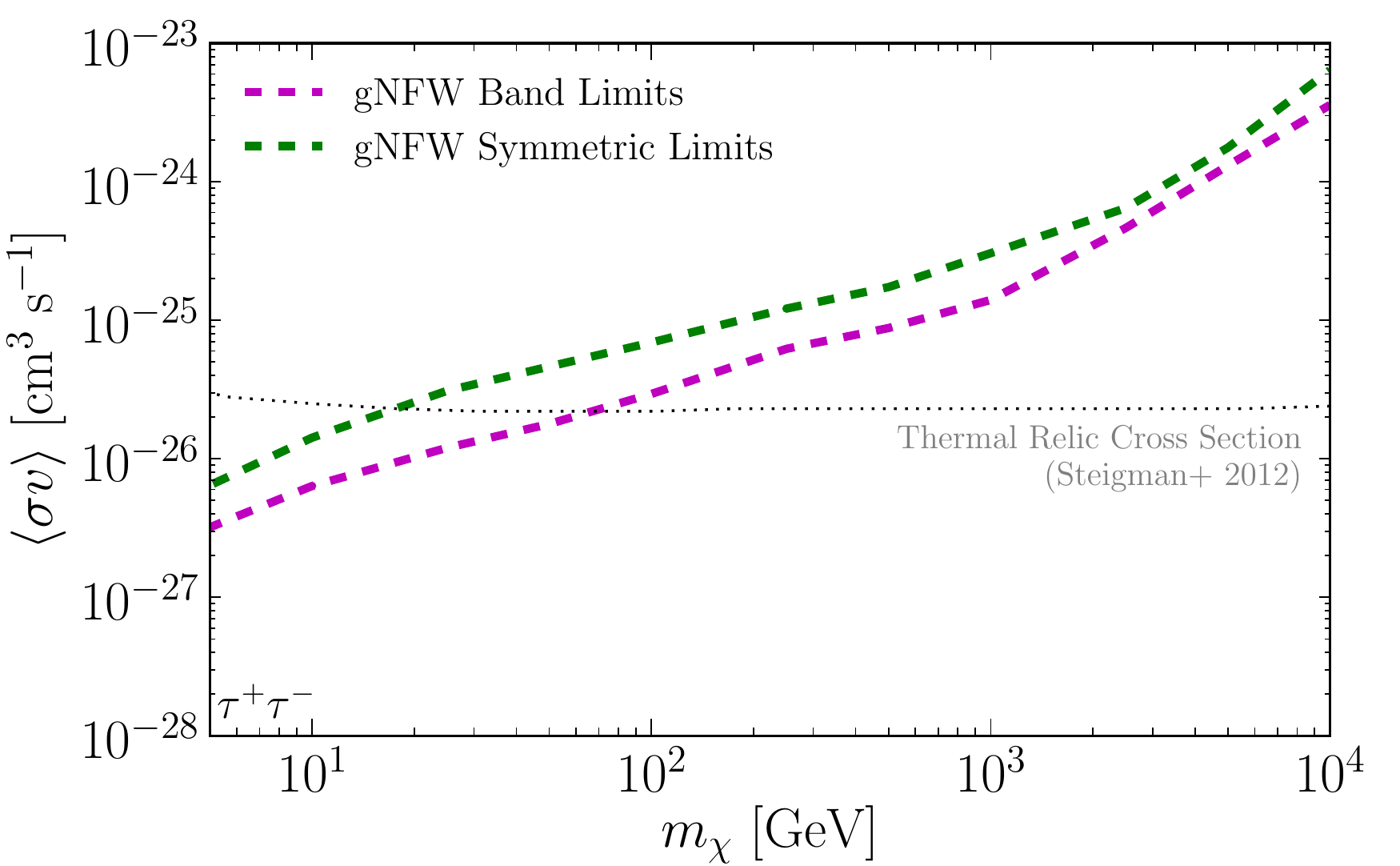}
\caption{
\label{fig:DM_asymComp}
Comparison of limits derived incorporating different systematic uncertainty estimates using the gNFW for the \bbbar (left) and \tautau (right) channels. The symmetric limits are based on the assumption that systematic uncertainties that could mask DM signal are as large as those that could mimic a DM signal derived from the GP scan. The band limits are based on requiring that a DM signal does not exceed the upper bound on the GC flux from the model variation approach (see text for details). The dotted line represents the thermal relic cross section \citep{2012PhRvD..86b3506S}.
}
\end{figure*}

Then, in Section \ref{sec:DMlimits} we derive upper limits by requiring the DM signal to not exceed the largest value of the excess spectrum found in each energy bin under all the modeling scenarios considered.
Alternatively, we can use Equation \ref{eq:chi2_DM} to derive 95\% confidence level limits based on the results of the GP scan by calculating when $\Delta \chi^2 = 3.84$.
We assume that  $\delta f_{\rm syst}$ equally represents the amplitude of systematic uncertainties that mask or induce DM-like signals.  Both versions of limits are shown in Figure \ref{fig:DM_asymComp}. The limits based on the symmetric nuisance parameter from the GP scan are less constraining.
However, the symmetric assumption results in limits that are likely too conservative, since a positive excess was found in the GC in every modeling scenario considered, and the vast majority of the fits in control regions along the Galactic Plane found positive DM-like excesses. 

\newpage
\bibliography{gce_papers}

\end{document}